\documentclass[aps,prd,amsmath,amssymb,amsfonts,showpacs,superscriptaddress,floatfix,twocolumn]{revtex4-1}
\usepackage{amsmath}
\usepackage{amssymb}
\usepackage{graphicx}
\usepackage{caption}
\usepackage{subcaption}
\usepackage{epsfig}
\usepackage{braket}
\usepackage{slashed}
\usepackage{bm}
\usepackage[usenames,dvipsnames]{color}
\usepackage{color}
\usepackage{float}
\usepackage{hyperref}

\def\be{\begin{equation}}
\def\ee{\end{equation}}

\def\bea{\begin{eqnarray}}
\def\eea{\end{eqnarray}}
\numberwithin{equation}{section}
\renewcommand\theequation{\arabic{section}.\arabic{equation}}
\begin{document}

\title{Hamiltonian simulation of the Schwinger model at finite temperature} 

\author{Boye Buyens}
\affiliation{Department of Physics and Astronomy,
Ghent
 University,
  Krijgslaan 281, S9, 9000 Gent, Belgium}

\author{Frank Verstraete}
  \affiliation{Department of Physics and Astronomy,
Ghent
 University,
  Krijgslaan 281, S9, 9000 Gent, Belgium}
  \affiliation{Vienna Center for Quantum Science and Technology, Faculty of Physics, University of Vienna, Boltzmanngasse 5, 1090 Vienna, Austria}

 \author{Karel Van Acoleyen}
\affiliation{Department of Physics and Astronomy,
Ghent
 University,
  Krijgslaan 281, S9, 9000 Gent, Belgium}

\begin{abstract}
\noindent Using Matrix Product Operators (MPO) the Schwinger model is simulated in thermal equilibrium. The variational manifold of gauge-invariant MPO is constructed to represent Gibbs states. As a first application, the chiral condensate in thermal equilibrium is computed and agreement with earlier studies is found. Furthermore, as a new application the Schwinger model is probed with a fractional charged static quark-antiquark pair separated infinitely far from each other. A critical temperature beyond which the string tension is exponentially suppressed is found, and is in qualitative agreement with analytical studies in the strong coupling limit. Finally, the CT symmetry breaking is investigated and our results strongly suggest that the symmetry is restored at any nonzero temperature.    
\end{abstract}

\maketitle
\section{Introduction}
\noindent Completing the phase diagram of quantum chromodynamics (QCD) is one of the major challenges of theoretical physics since the Seventies of the previous century. Huge efforts combining different techniques already led to great insights in QCD. Perturbative computations \cite{Sterman1995} using Feynman diagrams, which were very successful for quantum electrodynamics (QED), enabled experimentalists to test QCD in the lab, for instance by predicting the R ratio of electron-positron annihilation. Unfortunately, due to the asymptotic freedom of QCD, perturbative computations are limited to high-energy phenomenons excluding the study of confinement, dynamical mass generation and chiral symmetry breaking.

At the numerical level Lattice QCD \cite{Bali1999}, based on Monte Carlo sampling, led to a big breakthrough in completing the phase diagram of QCD \cite{Philipsen2007,Fukushima2011,Sharma2014,Philipsen2013}. For vanishing baryon density a good understanding of the phase diagram has been achieved. Unfortunately when including a finite number of baryons, which is relevant for the physics of heavy-ion collisions, neutron stars and supernovae, simulations are troubled by the notorious sign problem. Recent efforts using the Taylor extrapolation method, reweighting or analytical continuation of the chemical potential lead to results for small baryon densities and/or high temperatures \cite{Fukushima2011,Philipsen2012}. However, for large chemical potential an accurate description is still lacking. 

In the last decade the Tensor Networks States (TNS) \cite{Orus2014,Cirac2009} formalism has become a very popular method to tackle strongly correlated many body systems.  Their construction is motivated by the area law for ground states of local Hamiltonians \cite{Hastings2007,Eisert2008,Masanes2009} for which the von Neumann entropy scales with the boundary instead of being extensive. As a Hamiltonian method the TNS approach is free of any sign problem, see for instance \cite{Kraus2010,Corboz2010} for the implementation of fermions in two spatial dimensions, and enables the difficult simulation of out-of-equilibrium physics, e.g. \cite{Vidal2003, Banuls2009, Haegeman2014a,Murg2007}. The most famous example of TNS are the Matrix Product States (MPS) \cite{Fannes1992,Schollwoeck2011} in one dimension that underlie Steve White's Density Matrix Renormalization Group (DMRG) \cite{White1992}. Different applications of MPS on (1+1)-dimensional gauge theories \cite{Buyens2013, Montangero2015, Kuhn2015,Byrnes2002,Banuls2013a,Banuls2013b,Sugihara,Rico,Kuehn2014,Silvi,Buyens2014,Buyens2015, Milsted2015} demonstrated the potential of MPS to tackle gauge theories. In higher dimensions, the TNS formalism is at present less developed, although some first promising results have been obtained for (2 + 1)-dimensional gauge theories \cite{Luca,Luca2,Milsted2014,gaugingStates,2015arXiv150708837Z}.

When coupling a system to a heat bath it is described by mixed density operators. In one dimension the straightforward generalization of MPS to operators are the Matrix Product Operators (MPO) \cite{Verstraete2004,Zwolak2004}. Just like MPS are an efficient and faithful representation for ground states of local gapped Hamiltonians \cite{Hastings2007}, MPO are an efficient approximation for Gibbs states \cite{Hastings2006,Molnar2015} which describe the system in thermal equilibrium. Therefore we expect TNS to be useful for investigating canonical and grand canonical ensembles for gauge field theories. This has recently been confirmed by a successful study of the chiral condensate of the Schwinger model at finite temperature \cite{Saito2014,Banuls2015,Saito2015, Banuls2016}. 

In this paper we also study the Schwinger model in thermal equilibrium, but now also focus on asymptotic confinement and CT symmetry breaking, thereby continuing our recent work at zero temperature \cite{Buyens2015}. Historically, Schwinger considered this model \cite{Schwinger1962} as an example of a gauge vector field that can have a nonzero mass \cite{Schwinger1961}. Not so coincidentally, a few months later Anderson published his proposal for the Higgs mechanism where gauge fields acquire mass without breaking gauge invariance \cite{Anderson1963}. Other interesting physical features like, for instance, confinement and chiral symmetry breaking made this model very attractive to test new methods. In the last decade this model also gained interest from the experimentalists in the context of quantum simulators \cite{Cirac2010,Zohar2011,Zohar2012,Banerjee2012,Banerjee2013,Hauke2013,Stannigel2013,Tagliacozzo2013a,Wiese2013,Zohar2013,Zohar2013a,Zohar2013b,Zohar2013c,Kosior2014,Kuehn2014,Marcos2014,Wiese2014,Mezzacapo2015,Notarnicola2015,Zohar2015a,Martinez2016}. 

In the next section we discuss the setup of our simulations using MPO. To test our method, we compute in section \ref{sec:CC} the chiral condensate and compare our results with earlier studies \cite{Sachs1991,Banuls2016}. In section \ref{sec:asCon} we turn our attention to the asymptotic aspects of confinement for a static quark-antiquark pair with fractional charge. At high temperatures we find that the string tension becomes exponentially small. Furthermore, we also study the CT symmetry and find strong indications that the spontaneous symmetry breaking of the ground state at zero temperature vanishes, as soon as a nonzero temperature is turned on.  

\section{Setup}\label{sec:Setup}
\subsection{Hamiltonian and gauge invariance}
\noindent The Schwinger model is (1+1)-dimensional QED with one fermion flavor. We start from the Lagrangian density in the continuum:
\be \mathcal{L} = \bar{\psi}\left(\gamma^\mu(i\partial_\mu+g A_\mu) - m\right) \psi - \frac{1}{4}F_{\mu\nu}F^{\mu\nu}\,.\label{Lagrangian} \ee
One then performs a Hamiltonian quantization in the time-like axial gauge ($A_0=0$), which can be turned into a lattice system by the Kogut-Susskind spatial discretization \cite{Kogut}. The two-component fermions are sited on a staggered lattice. These fermionic degrees of freedom can be converted to spin-1/2 degrees of freedom by a Jordan-Wigner transformation with the eigenvectors $\{\ket{s_n}_n: s_n \in \{-1,1\} \}$ of $\sigma_z(n)$ as basis of the local Hilbert space at site $n$. The compact gauge fields $\theta(n)=a g A_1(n)$, live on the links between the sites. Their conjugate momenta $E(n)$, with $[\theta(n),E(n')]=ig\delta_{n,n'}$ correspond to the electric field. The commutation relation determines the spectrum of $E(n)$ up to a constant: $E(n)/g = \alpha(n)+p$ , with $\alpha(n)\in \mathbb{R}$ corresponding to the background electric field at link $n$ and $p\in\mathbb{Z}$. 

In this formulation the gauged spin Hamiltonian derived from the Lagrangian density (\ref{Lagrangian}) reads (see \cite{Banks,Kogut} for more details):
\bea\label{equationH0} H&=& \frac{g}{2\sqrt{x}}\Biggl(\sum_{n \in \mathbb{Z}} \frac{1}{g^2} E(n)^2 + \frac{\sqrt{x}}{g} m \sum_{n \in \mathbb{Z}}(-1)^n\sigma_z(n) \nonumber
\\ &&+ x \sum_{n \in \mathbb{Z}}(\sigma^+ (n)e^{i\theta(n)}\sigma^-(n + 1) + h.c.)\biggl) \eea
where $\sigma^{\pm} = (1/2)(\sigma_x \pm i \sigma_y)$ are the ladder operators. Here we have introduced the parameter $x$ as the inverse lattice spacing in units of $g$: $x \equiv 1/(g^2a^2)$. The continuum limit will then correspond to $x\rightarrow \infty$.  
Notice the different second (mass) term in the Hamiltonian for even and odd sites which originates from the staggered formulation of the fermions. In this formulation the odd sites are reserved for the charge $-g$ `quarks', where spin up, $s = +1$, corresponds to an unoccupied site  and spin down, $s = -1$, to an occupied site. The even sites are reserved for the charge $+g$ `antiquarks' where now conversely spin up corresponds to an occupied site and spin down to an occupied site. 

In the timelike axial gauge the Hamiltonian is still invariant under the residual time-independent local gauge transformations generated by:
\begin{align} g\mathcal{G}(n) = & E(n)-E(n-1)-\frac{g}{2}( \sigma_z(n) + (-1)^n )\,.\label{gauss} \end{align}
As a consequence, if we restrict ourselves to physical gauge-invariant operators $O$, with $[O,\mathcal{G}(n)]=0$, the Hilbert space decomposes into dynamically disconnected superselection sectors, corresponding to the different eigenvalues of $\mathcal{G}(n)$. In the absence of any background charge the physical sector then corresponds to the $\mathcal{G}(n)=0$ sector. Imposing this condition (for every $n$) on the physical states is also referred to as the Gauss law constraint, as this is indeed the discretized version of $\partial_z E - j^0=0$, where $j^0$ is the charge density of the dynamical fermions. 

The other superselection sectors correspond to states with background charges. Specifically, if we want to consider two probe charges, one with charge $-gQ$ at site $0$ and one with opposite charge $+gQ$ at site $k$, we have to restrict ourselves to the sector:
\begin{align} g\mathcal{G}(n) =  gQ(\delta_{n,0} - \delta_{n,k})\label{spingauss0}\,.\end{align} Notice that we will consider both integer and noninteger (fractional) charges $Q$.  

As in the continuum case \cite{ColemanCS}, we can absorb the probe charges into a background electric field string that connects the two sites. This amounts to the substitution $E(n) = g[L(n) + \alpha(n)]$ where $\alpha(n)$ is only nonzero in between the sites: $\alpha(n) = -Q\Theta(0 \leq n < k)$; and $L(n)$ has an integer spectrum: $L(n)=p\in \mathbb{Z}$. In terms of $L(n)$ the Gauss constraint now reads:
 \be\label{spingauss} G(n) = L(n) - L(n-1) -\frac{ \sigma_z(n) + (-1)^n }{2} = 0\,,  \ee and we finally find the Hamiltonian:
 \bea\label{equationH} H&=& \frac{g}{2\sqrt{x}}\Biggl(\sum_{n \in \mathbb{Z}} [L(n) + \alpha(n)]^2 + \frac{\sqrt{x}}{g} m \sum_{n \in \mathbb{Z}}(-1)^n\sigma_z(n) \nonumber
\\ &&+ x \sum_{n \in \mathbb{Z}}(\sigma^+ (n)e^{i\theta(n)}\sigma^-(n + 1) + h.c.)\biggl),\eea
in accordance with the continuum result of \cite{Coleman1976}. 

In this paper we will consider the system coupled to a heat reservoir with fixed temperature $T$. If the system only exchanges energy with this reservoir and it reaches thermal equilibrium it is represented by the canonical ensemble. The density operator that describes this canonical ensemble is the Gibbs state $e^{-\beta H}$, where $\beta = 1/T$ is the inverse temperature. The probability of finding the system in a particular eigenstate $\ket{E}$ of $H$ is $e^{-\beta E}/Z(\beta)$ where $Z(\beta) = \mbox{tr}(e^{-\beta H})$ is the partition function.  Because the physical sector corresponds to states with $G(n) = 0$, we need to exclude the (micro)states that are not gauge invariant. In particular, the probability to find the system in an eigenstate $\ket{E}$ of $H$ which is not gauge invariant, $G(n) \ket{E} \neq 0$, should be zero: $\braket{E\vert \rho(\beta) \vert E} = 0$. Therefore we need to project $H$ onto the $(G(n) = 0)-$subspace. If $P$ is the projector onto the $(G(n) = 0$)-subspace, the canonical ensemble is thus described by the density operator $\rho(\beta) = Pe^{-\beta H} (= Pe^{-\beta H}P = e^{-\beta H} P)$. The ensemble average of a given gauge-invariant observable $Q$ is computed as
$$ \langle Q \rangle_{\beta} =  \frac{\mbox{tr}(PQPe^{-\beta H})}{Z(\beta)} \mbox{ with }Z(\beta) = \mbox{tr}\left(Pe^{-\beta H}\right) $$
the partition function. Note that this expectation value indeed corresponds to the expectation value obtained from the Wilsonian path integral \cite{Baaquie1986}. 

\subsection{Gauge-invariant MPO}\label{subsec:MPO}
\noindent Here we will construct a Matrix Product Operator (MPO) to approximate the Gibbs state $Pe^{-\beta H}$. Therefore we will use the method discussed in \cite{Verstraete2004}. The main idea is that we purify the MPO ansatz by a MPS in a higher-dimensional Hilbert space. Starting from the identity on the $(G(n) = 0)$-subspace for $\beta = 0$, we obtain the state for finite $\beta$ by evolving this purification in imaginary time. In addition to \cite{Verstraete2004}, we will take gauge invariance into account when constructing the MPS purification by imposing a block structure similar to \cite{Buyens2013} on the tensors describing the MPS.
\\
\\
Consider now the lattice spin-gauge system (\ref{equationH}) of $2N$ sites. On site $n$ the matter fields are represented by the spin operators with basis $\{ \ket{s_n}_n: s_n \in \{-1,1\}\}$. The gauge fields live on the links and on link $n$ their Hilbert space is spanned by the eigenkets $\{\ket{p_n}_{[n]}: p_n \in \mathbb{Z}[p_{min},p_{max}] \}$ of the angular operator $L(n)$. Note that we only retained a finite range for the eigenvalues $p_n$ of $L(n)$ for our numerical scheme. We will address the issue of which values to take for $p_{min}$ and $p_{max}$ in subsection \ref{subsec:iTEBD}. It will be convenient to block site $n$ and link $n$ into one effective site with local Hilbert space $\mathcal{H}_n$ spanned by $\{\ket{s_n,p_n}_n \}$. 
Writing $ \kappa_n = (s_n,p_n)$ we introduce the multi-index 
$$\bm{\kappa} = \bigl((s_1,p_1),(s_2,p_2),\ldots,(s_{2N},p_{2N})\bigl) = (\kappa_1,\ldots,\kappa_{2N}).$$ 
With these notations we have that the local Hilbert space $\mathcal{H}_n$ on the effective site $n$ is spanned by $\{\ket{\kappa_n}_n \}$. Therefore the Hilbert space of the full system of $2N$ sites and $2N$ links, $\mathcal{H} = \bigotimes_{n = 1}^{2N} \mathcal{H}_n,$ has basis $\{\ket{\bm{\kappa}}= \ket{\kappa_1}_1\ldots \ket{\kappa_{2N}}_{2N} \}$. A general operator thus takes the form:
$$\sum_{\bm{\kappa}} C^{(\kappa_1,\kappa_1'),\ldots,(\kappa_{2N},\kappa_{2N}')}\ket{\bm{\kappa}}\bra{\bm{\kappa}'}$$
 with $C^{(\kappa_1,\kappa_1'),\ldots,(\kappa_{2N},\kappa_{2N}')} \in \mathbb{C}$. 

In this basis, the projector $P$ on the $(G(n) = 0)$-subspace reads
\be \label{eq:rho0}P = \sum_{\bm{\kappa}}\left(\prod_{n=1}^{2N} \delta_{p_n - p_{n-1}, \frac{s_n + (-1)^n}{2}}\right) \ket{\bm{\kappa}}\bra{\bm{\kappa}}\ee
where we take periodic boundary conditions ($p_{0} = p_{2N}$). For $\beta = 0$ we have that $\rho(0) = P$. We will now write this state as a MPO \cite{Verstraete2004,Zwolak2004}:
\be\label{eq:MPO} \rho(0) = \sum_{\bm{\kappa}, \bm{\kappa'}} \mbox{tr}\left(W_1^{\kappa_1\kappa_1'}\ldots W_{2N}^{\kappa_{2N}\kappa_{2N}'}\ \right)\ket{\bm{\kappa}}\bra{\bm{\kappa'}} \ee
where $W_n^{\kappa_n\kappa_n'} \in \mathbb{C}^{D \times D}$ are complex matrices. Thereto we put
$$ W_n^{\kappa_1,\kappa_2} = \sum_{\kappa^a} A_n^{\kappa_1,\kappa^a}\otimes \bar{A}_n^{\kappa_2,\kappa^a} $$
with
$ \kappa^a = (s^a,p^a)$ and $A_n^{(\kappa),(\kappa^a)} = A_n^{(s,p),(s^a,p^a)} \in \mathbb{C}^{D \times D}$ complex matrices.  In order that $\rho(0) = P$ we give, similar as in \cite{Buyens2013}, the virtual indices $(\alpha,\beta)$ of $[A_n^{(s,p),(s^a,p^a)}]_{\alpha,\beta}$ a multiple index structure:  $\alpha \rightarrow (q,\alpha_q)$, $\beta \rightarrow (r,\beta_r)$ where $q,r \in \mathbb{Z}$ label the eigenvalues of $L(n)$. One can now check that if we put
\begin{multline} \label{eq:GIS} [A_n^{(s,p),(s^a,p^a)}]_{(q,\alpha);(r,\beta)} = [a_n]_{\alpha_q,\beta_r}
\\ \delta_{r,q + [s+(-1)^n]/2}\delta_{p,r} \delta_{s,s^a} \delta_{p^a,q + [s^a+(-1)^n]/2},\end{multline}
where  $a_n \in \mathbb{C}^{D_q \times D_r}$ can be any nonzero matrix, that $\rho(0)$ equals (\ref{eq:rho0}) up to a normalization factor.

To obtain a purification of the state $\rho(0)$ we need to consider the Hilbert space
$$\mathcal{H}_{full} = \bigotimes_{n = 1}^{2N}\mathcal{H}_n\otimes\mathcal{H}_n^a $$
where $\mathcal{H}_n^a = \mbox{span}\{\ket{\kappa_n^a}_n = \ket{s_n^a}_n\ket{p_n^a}\}$ is an auxiliary Hilbert space with the same dimension as $\mathcal{H}_n$. Then we introduce the MPS \cite{Verstraete2004} 
$$\ket{\Psi[A]} = \sum_{\bm{\kappa},\bm{\kappa}^a}\mbox{tr} \left(A_1^{\kappa_1,\kappa_1^a}\ldots A_{2N}^{\kappa_{2N},\kappa_{2N}^a} \right)\ket{\bm{\kappa},\bm{\kappa}^a} \in \mathcal{H}_{full},$$
$$\ket{\bm{\kappa},\bm{\kappa}^a} = \ket{\kappa_1}_1 \ket{\kappa_1^a}_1 \ldots, \ket{\kappa_{2N}}_{2N}\ket{\kappa_{2N}^a}_{2N}, $$
where $A_n^{\kappa_n,\kappa_n^a}$ is defined in (\ref{eq:GIS}). By contracting the $\kappa_n^a$ we obtain up to a normalization factor $P$:
$$ \mbox{tr}_{\mathcal{H}^a}\left(\ket{\Psi[A]}\bra{\Psi[\bar{A}]}\right)  \varpropto P.$$

Because $\rho(0)$ is the projector $P$ on the ($G(n) = 0$)-subspace, $[H,P] = 0$ implies that $\rho(\beta)  = Pe^{-\beta H}= e^{-\beta H/2}Pe^{-\beta H/2}$. As a consequence, if we evolve the purification $\ket{\Psi[A(\beta)]}$ according to 
\be\label{eq:thermEvo} \ket{\Psi[A(\beta)]} = e^{-(\beta/2)H}\ket{\Psi[A(0)]}. \ee
we have for all values of $\beta$ that
$$\rho(\beta) =  \mbox{tr}_{\mathcal{H}^a}\left(\ket{\Psi[A(\beta)]}\bra{\Psi[\bar{A}(\beta)]}\right).$$
Note that the Hamiltonian $H$ here only acts on $\mathcal{H}_n$ but not on the auxiliary Hilbert spaces $\mathcal{H}_n^a$. 

Because $A_n(\beta  = 0)$ takes the form (\ref{eq:GIS}), gauge invariance of $H$ implies that during the evolution (\ref{eq:thermEvo}) $A_n(\beta)$ will have a similar form:
\begin{multline} \label{eq:GISn} [A_n^{(s,p),(s^a,p^a)}(\beta)]_{(q,\alpha);(r,\beta)} = [a_n^{s,p,s^a}(\beta) ]_{\alpha_q,\beta_r}
\\ \delta_{r,q + [s+(-1)^n]/2}\delta_{p,r} \delta_{p^a,q + [s^a+(-1)^n]/2} \end{multline}
where $a_n^{s,p,s^a} \in \mathbb{C}^{D_q \times D_r}$ represents the variational freedom of the MPS $\ket{\Psi[A(\beta)]}$. Note that contrary to (\ref{eq:GIS}) $a_n^{s,p,s_a}$ now also depends on $s,p$ and $s_a$. The total bond dimension of this MPS is $D = \sum_q D_q$. Finally we note that by restricting ourselves to finite eigenvalues of $L(n)$ we cannot represent the initial state $\rho(0)$ exactly. Fortunately, as we will see later this does not spoil our results for nonzero $\beta$, see subsection \ref{subsec:iTEBD} and in particular figs. \ref{fig:truncationa} and \ref{fig:truncationc}.

The MPS framework enables us to perform our simulations directly in the thermodynamic limit $(N \rightarrow + \infty)$. In this case the Hamiltonian (\ref{equationH}) is translation invariant over an even number of sites. By starting from a state which has this symmetry, i.e. by taking in (\ref{eq:GISn}) all $a_n$ equal for $\beta = 0$, we have for all values of $\beta$ that $a_n^{s,p,s^a}(\beta)$ depends only on the parity of $n$: $a_{2n-1}^{s,p,s^a}(\beta) = a_1^{s,p,s^a}(\beta)$ and $a_{2n}^{s,p,s^a}(\beta) = a_2^{s,p,s^a}(\beta), \forall n$.

\subsection{iTEBD for thermal evolution}\label{subsec:iTEBD}
\noindent In the previous subsection we purified the Gibbs state $\rho(\beta) = Pe^{-\beta H}$ by the MPS $\ket{\Psi[A(\beta)]}$. Using gauge invariance and translation invariance over two sites we identified the variational degrees of freedom $a_1^{s,p,s^a}, a_2^{s,p,s^a} \in \mathbb{C}^{D_q \times D_r}$ of $\ket{\Psi[A(\beta)]}$, see eq. (\ref{eq:GISn}). There now only remains to solve equation (\ref{eq:thermEvo}) within the MPS manifold which is performed by using the infinite time-evolving block decimation (iTEBD) algorithm \cite{Vidal2003,Buyens2013}. At the core of this method lies the Trotter decomposition \cite{Hatano2005} which decomposes $e^{- d\beta H/2}$ into a product of local operators, the so-called Trotter gates. We performed a fourth-order Trotter decomposition of $e^{-(d\beta/2)H}$ for small steps $d\beta$ which yields an error of order $(d\beta)^5$ for each step $d\beta$.  By applying these Trotter gates to the MPS $\ket{\Psi[A(\beta)]}$ another MPS is obtained with larger bond dimension. This MPS is projected to a MPS with smaller bond dimensions $D_q$ in order to avoid the bond dimensions to increase exponentially with $\beta$. 

This projection is performed as an effective truncation in the Schmidt spectrum of $\ket{\Psi[A(\beta)]}$ with respect to the bipartition $\{\mathcal{A}_1(n) = \mathbb{Z}[-\infty, n],\mathcal{A}_2(n) = \mathbb{Z}[n+1,+\infty]\}$. The Schmidt decomposition with respect to the bipartition $\{\mathcal{A}_1(n),\mathcal{A}_2(n)\}$ reads
\be\ket{\Psi[A(\beta)]} = \sum_{q = p_{min}}^{p_{max}}\sum_{\alpha_q = 1}^{D_{q}} \sqrt{\lambda_{q,\alpha_q}^n} \ket{\psi_{q,\alpha_q}^{\mathcal{A}_1(n)}}\ket{\psi_{q,\alpha_q}^{\mathcal{A}_2(n)}} \label{eq:Schmidtdec} \ee
where $\ket{\psi_{q,\alpha_q}^{\mathcal{A}_k(n)}}$ are orthonormal unit vectors in the Hilbert space $\mathcal{H}_{\mathcal{A}_k(n)} = \bigotimes_{j \in \mathcal{A}_k(n)}(\mathcal{H}_j \otimes \mathcal{H}_j^a)$ $(k = 1,2)$ and $\lambda_{q,\alpha_q}^n$, called the Schmidt values, are non-negative numbers that sum to one. Note that the Schmidt values are labeled by the eigenvalues of $L(n)$ which is a consequence of (\ref{eq:GISn}). Due to translation symmetry over two sites $\lambda_{q,\alpha_q}^n$ only depends on the parity of $n$: $\lambda_{q,\alpha_q}^{2n-1} = \lambda_{q,\alpha_q}^1$ and $\lambda_{q,\alpha_q}^{2n} = \lambda_{q,\alpha_q}^2, \forall n$. From eq. (\ref{eq:Schmidtdec}) one observes that the limit $D_q \rightarrow + \infty,$ $p_{min}\rightarrow - \infty$ and $p_{max} \rightarrow + \infty$ yields an exact representation of the state $\ket{\Psi[A(\beta)]}$ and thus of the Gibbs state. The success of the approach using MPO is explained by the fact that by using relatively small values of $D_q$ we can obtain very accurate approximations of the Gibbs state \cite{Hastings2006,Molnar2015}. After every Trotter step the iTEBD algorithm discards all Schmidt values $\lambda_{q,\alpha_q}^{n} < \epsilon^2$ with $\epsilon$ a preset tolerance.  In particular, when all Schmidt values $\lambda_{q,\alpha_q}^n$ corresponding to an eigenvalue $q$ are smaller than $\epsilon^2$ this eigenvalue sector is discarded and $p_{min}$ is increased or $p_{max}$ is decreased. In this way $p_{min}$, $p_{max}$ and $D_q$ are adapted dynamically. 

\begin{figure}
\begin{subfigure}[b]{.24\textwidth}
\includegraphics[width=\textwidth]{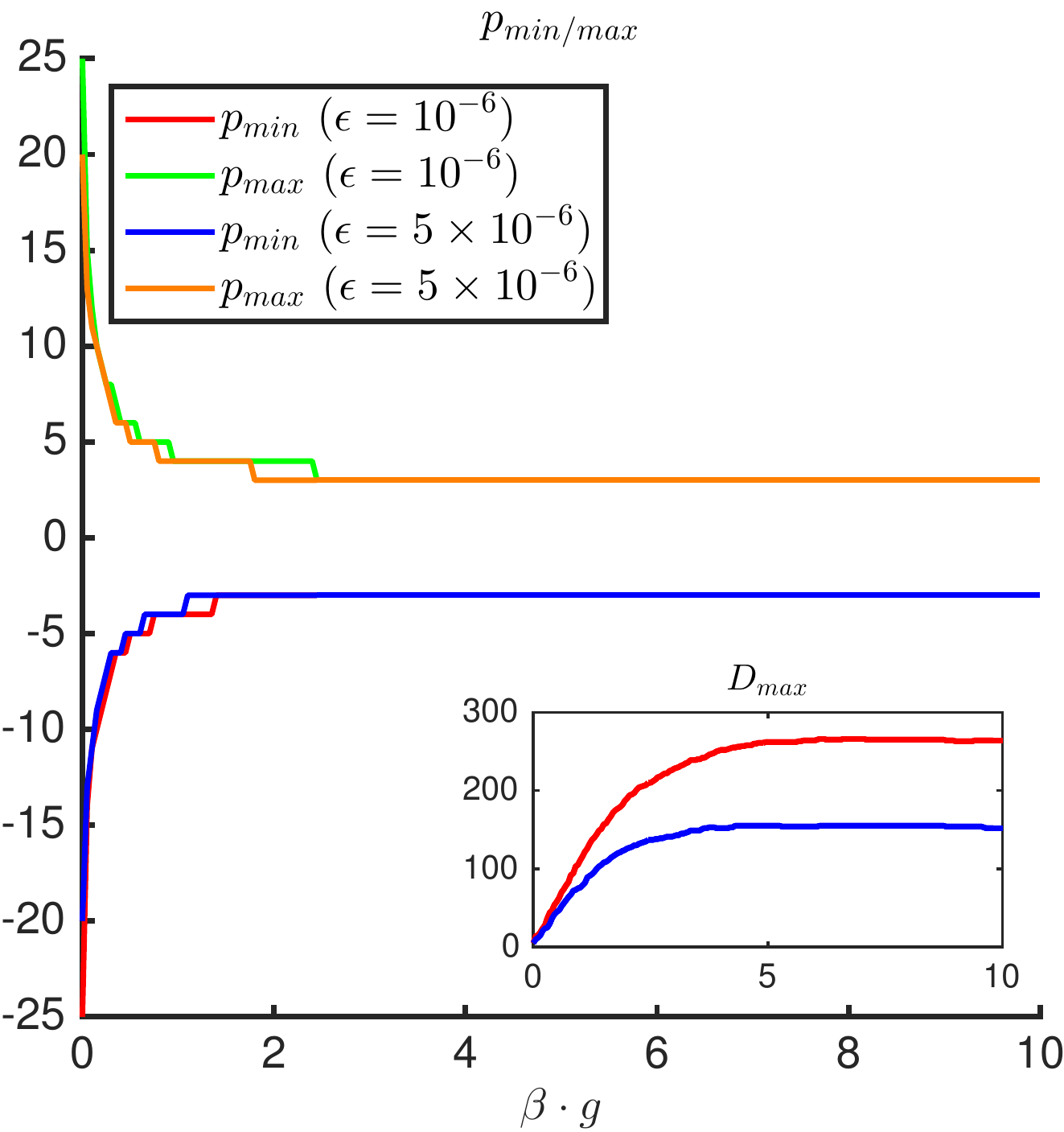}
\caption{\label{fig:truncationa}}
\end{subfigure}\hfill
\begin{subfigure}[b]{.24\textwidth}
\includegraphics[width=\textwidth]{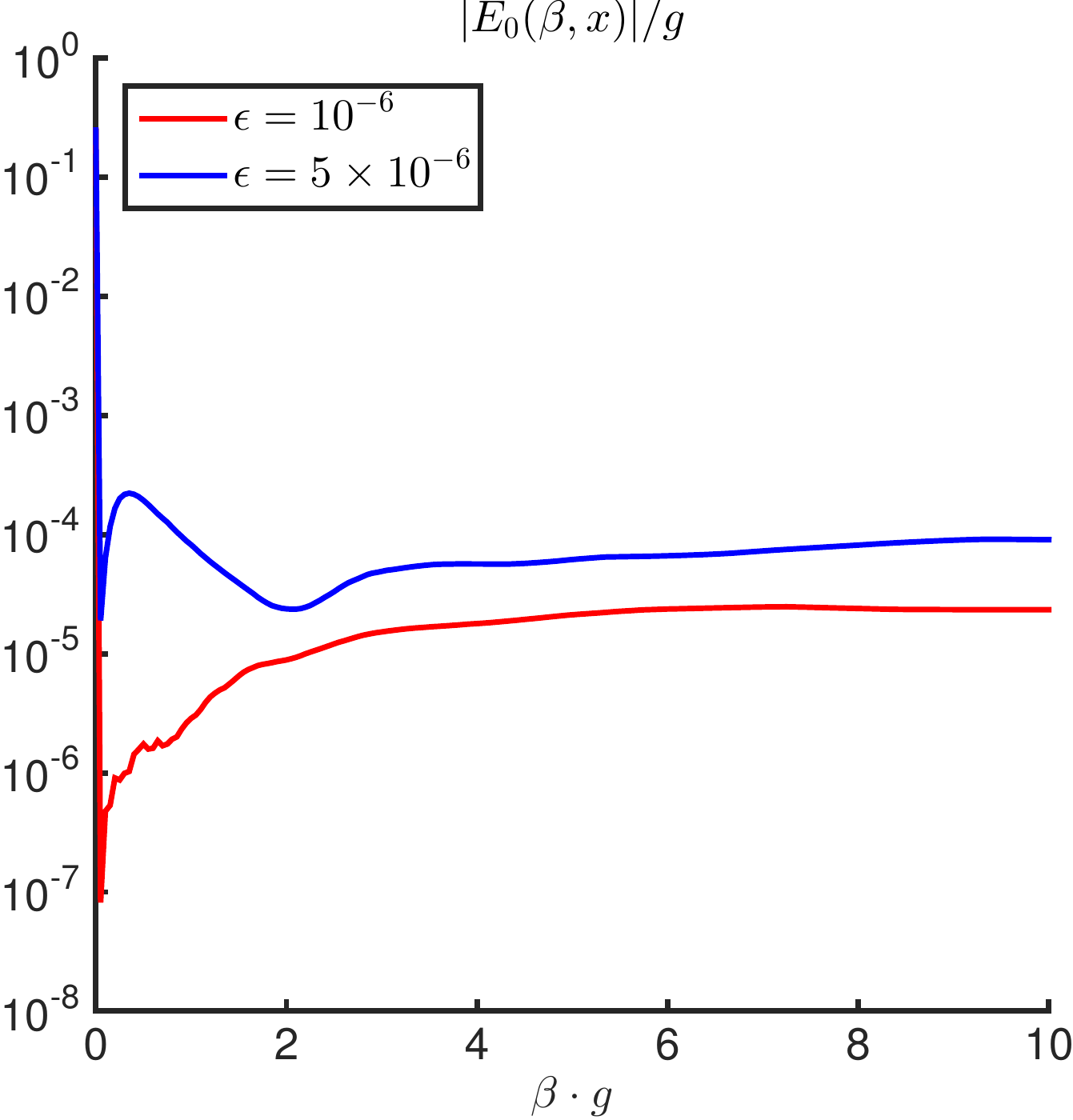}
\caption{\label{fig:truncationb}}
\end{subfigure}\hfill
\vskip\baselineskip
\captionsetup{justification=raggedright}
\caption{$m/g = 0.25, x = 200, \alpha = 0$. (a) $p_{min}$ and $p_{max}$ for $\epsilon = 10^{-6}$ and $\epsilon = 5\times 10^{-6}$. Inset: maximum bond dimension over all the charge sectors. (b): Electric field per site $E_0(\beta,x)$. Because $\alpha = 0$ we should have  $E_0(\beta,x) = 0$.  }
\end{figure}

In fig. \ref{fig:truncationa} we plot $p_{max}$ and $p_{min}$ for our simulations with $m/g = 0.25$, $x = 200$, $\alpha = 0$, once with preset tolerance $\epsilon = 10^{-6}$ and once with preset tolerance $\epsilon = 5\times 10^{-6}$. For $\epsilon = 10^{-6}$ we started with $p_{max} = -p_{min} = 25$ and for $\epsilon = 5 \times 10^{-6}$ we started with $p_{max} = -p_{min} = 20$. We observe that $p_{max}$ and $p_{min}$ decrease very quickly in magnitude as a function of $\beta$ to $p_{max} = 3$ and $p_{min} = -3$. The fact that we can accurately describe the system with a finite range of eigenvalues of the electric field should not come as a surprise. Physically, we do not expect it to be very likely to observe the system, which is in thermal equilibrium, in a state with extremely large electric field compared to the temperature. This follows from the first term in the Hamiltonian (\ref{equationH0}) that appears in the Gibbs state $\rho(\beta) = Pe^{-\beta H}$. Note that the $p_{min}$ and $p_{max}$ at $\beta g = 10$ corresponds to the values of $p_{max}$ and $p_{min}$ in our simulations at zero temperature in \cite{Buyens2013,Buyens2015}.  In the inset we show the evolution of the maximum bond dimension over all the charge sectors. The bond dimension is an almost linearly increasing function of $\beta g$ for $\beta g \lesssim 5$. When $\beta g \gtrsim 5$ the bond dimension remains almost constant, indicating that for these parameters $\beta g \gtrsim 5$ is already very close to zero temperature. If we want better accuracy we need smaller values of $\epsilon$. In the inset of \ref{fig:truncationa} one observes that this requires more variational freedom in the MPS representation of $\ket{\Psi[A(\beta)]}$ and thus longer computation time.

As a first check on our method we show in fig. \ref{fig:truncationb} the electric field $E_{\alpha}(\beta,x)$ at lattice spacing $a = 1/\sqrt{g x}$ where 
$$E_{\alpha}(\beta,x)= \frac{g}{2}\mbox{tr}\left(\frac{Pe^{-\beta H}P}{\mbox{tr}(Pe^{-\beta H})}(L(1) + L(2) + 2\alpha)\right).$$
For zero background field, $\alpha = 0$, this quantity should be zero which follows from CT symmetry of the Hamiltonian ($C$ is charge conjugation: $\sigma^z \rightarrow -\sigma^z, L \rightarrow - L$, and $T$ is translation over one site). For very small values of $\beta g$ the errors on $E_0(\beta,x)$ are relatively large. This is a consequence of taking finite values for $p_{min}$ and $p_{max}$; as we discussed in the previous section, for $\beta g = 0$ one should consider all possible electric field values ($p\in[-\infty,+\infty]$) to represent the Gibbs state $\rho(0)=P$. Fortunately, discarding these Schmidt values for small values of $\beta g$ does not spoil the results for larger values of $\beta g$. Indeed, for $\beta g > 0$ the errors on $E_0(\beta,x)/g$ are only of order $10^{-4}$. In this plot one also observes that taking a smaller value for $\epsilon$ leads to better accuracy. From this example it is clear that, unless one is interested in the $\beta\rightarrow 0$ limit, one can safely neglect eigenvalue sectors $q$ with $q + \alpha$ larger than $20$ in magnitude. 

\begin{figure}
\begin{subfigure}[b]{.24\textwidth}
\includegraphics[width=\textwidth]{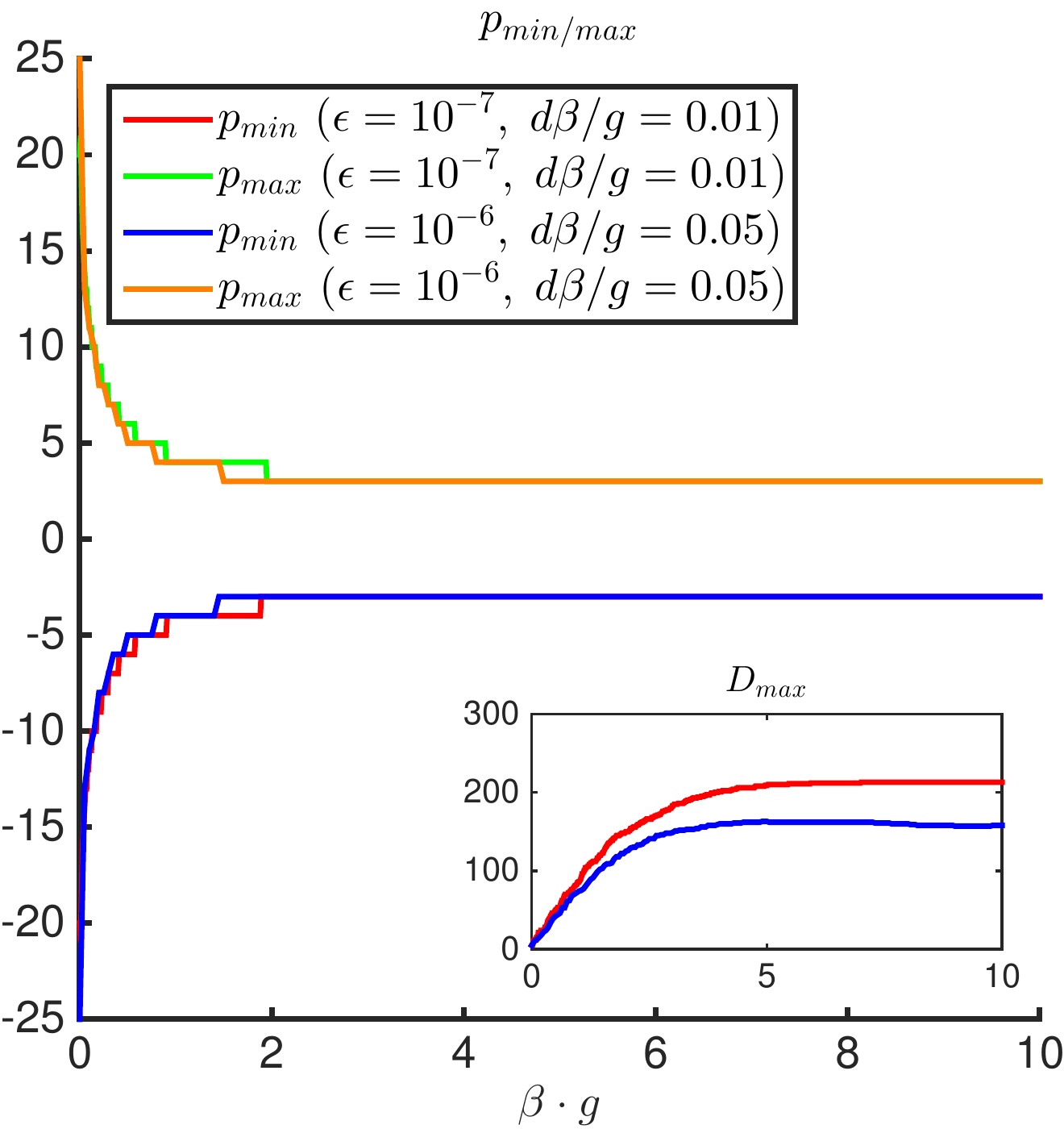}
\caption{\label{fig:truncationc}}
\end{subfigure}\hfill
\begin{subfigure}[b]{.24\textwidth}
\includegraphics[width=\textwidth]{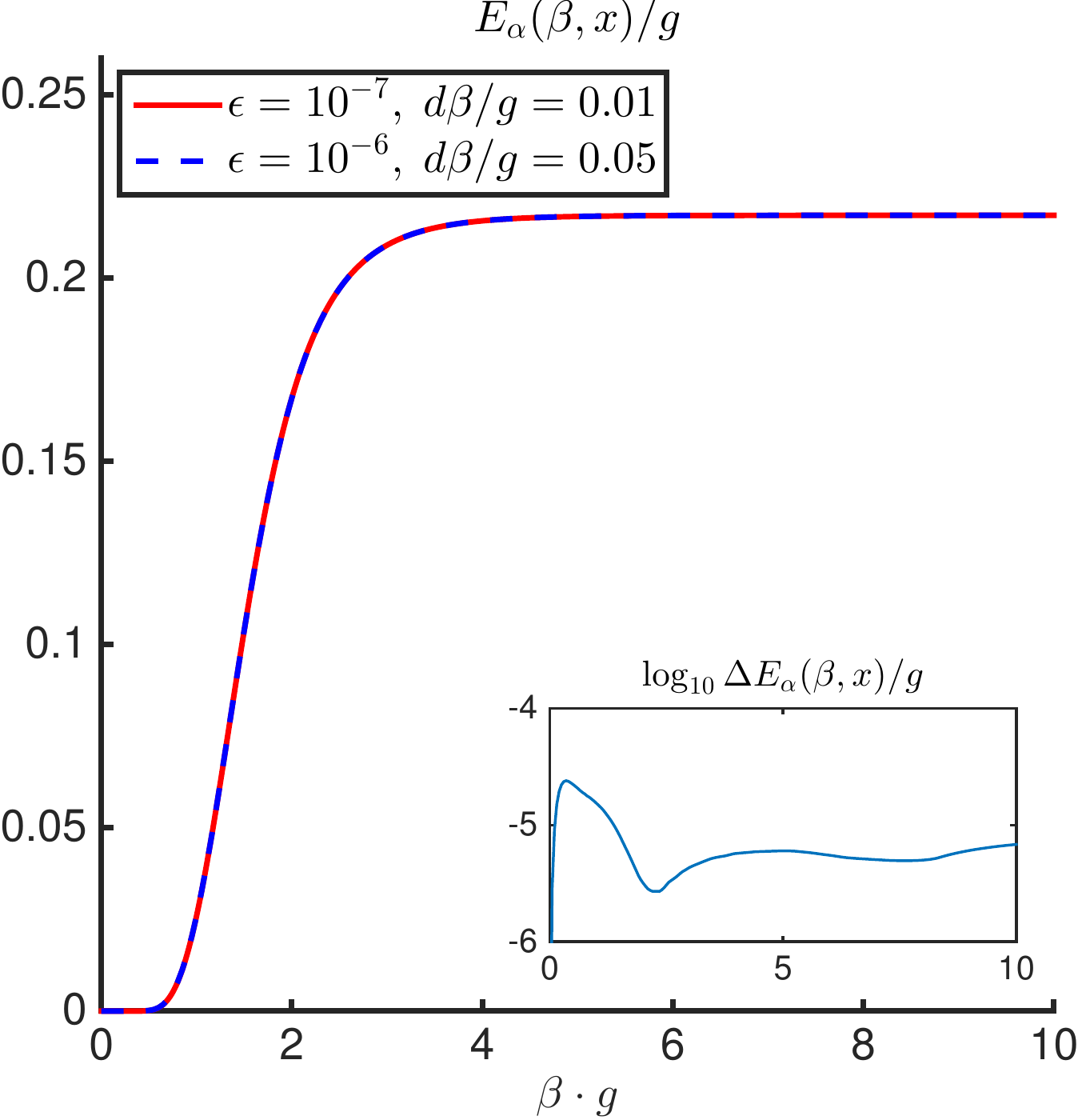}
\caption{\label{fig:truncationd}}
\end{subfigure}
\vskip\baselineskip
\captionsetup{justification=raggedright}
\caption{$m/g = 0.5, x = 100, \alpha = 0.25$. (a) $p_{min}$ and $p_{max}$ for $\epsilon = 10^{-7}, d\beta = 0.01$ and $\epsilon = 10^{-6}, d\beta = 0.05$. Inset: maximum bond dimension over all the charge sectors.  (b): Electric field per site $E_{\alpha}(\beta,x)/g$. Inset: 10-base logarithm of the difference of $E_{\alpha}(\beta,x)/g$ between simulations with $\epsilon = 10^{-7}, d\beta = 0.01$ and $\epsilon = 10^{-6}, d\beta = 0.05$. }
\end{figure}

In fig. \ref{fig:truncationc} and fig. \ref{fig:truncationd} we compare two simulations for different values of the step size $d\beta$ and $\epsilon$, now for nonzero background field $\alpha \neq 0$. The first one has $(d\beta, \epsilon) = (0.05, 10^{-6})$ and the second one has $(d\beta, \epsilon) = (0.01, 10^{-7})$.  In fig. \ref{fig:truncationd} we compare the electric field, which is nonzero for $\alpha \neq 0$, for both simulations and observe that the results  are the same up to order $10^{-5}$. For $\beta g \lesssim 2$ the difference is slightly larger, but still sufficiently small. This slightly larger error mainly originates from ignoring large eigenvalues of $L(n)$ at $\beta g = 0$. The fact that the results for different choices of $d\beta$ and $\epsilon$ are in agreement indicates that taking $(d\beta,\epsilon) = (0.05,10^{-6})$ is sufficient for most of our simulations. An extended discussion on how quantify the errors made of taking nonzero $(\epsilon,d\beta)$ is given in appendices \ref{subsec:errorsFinD} and \ref{subsec:errFinDalphaneq0}.

\section{Chiral condensate}\label{sec:CC}
\noindent  In QCD with massless up and down quarks, the nonzero chiral quark condensate signals spontaneous symmetry breaking of the chiral symmetry. This spontaneous symmetry breaking occurs for relatively low temperatures and explains the existence of pions \cite{Nambu2009,Kawamoto1981}. For physical quark masses this chiral symmetry is explicitly broken. One can still distinguish two phases separated by a pseudocritical temperature $T_c \approx 150 - 190$ MeV. For temperatures $T \ll T_c$ thermal expectation values are dominated by the pions which are a `remnant' of the chiral symmetry, while at high temperatures the thermodynamics are well described by the quarks and the gluons. Hence, the pions can be interpreted as an example of confined quark bound states that dominate the physics only below $T_c$. Therefore, not so surprisingly, it is also suggested that around this pseudocritical temperature QCD changes from the confined phase to the deconfined phase, although this is still a subject of debate \cite{Glozman2012,Bowman2011}. In the confined phase the gluons confine the quarks to baryons and mesons while in the deconfined phase QCD should resemble a quark-gluon plasma.

Here we will consider the chiral condensate of the Schwinger model to benchmark our method. In the one-flavor massless Schwinger model the nonzero chiral condensate is a consequence of the chiral symmetry being anomalous.  The nonzero chiral condensate also determines the confining behavior of external charges in mass perturbation theory \cite{Rodriguez1996}. For $m/g = 0$, the chiral condensate is computed analytically by Sachs and Wipf \cite{Sachs1991}. Besides the studies in the exactly solvable case ($m/g = 0$) \cite{Sachs1991,Steele1995}, there are results available in mass perturbation theory ($m/g \ll 1$) \cite{Rodriguez1996,Hosotani1998}. Furthermore, in \cite{Hosotani1998} an approach using a generalized Hartree-Fock method beyond mass perturbation theory was studied. Recently, MPO simulations succeeded in recovering the analytical result of Sachs and Wipf for $m/g = 0$ \cite{Saito2014,Banuls2015} and also obtained the chiral condensate in the nonperturbative regime \cite{Saito2015, Banuls2016}. 

On the lattice with spacing $ga = 1/\sqrt{x}$ and $2N$ sites the chiral condensate $\Sigma(\beta) = \langle \bar{\Psi}(z)\Psi(z)\rangle_{\beta}$ equals
\be
\Sigma(\beta,x) = \frac{g\sqrt{x}}{2N}\sum_{n = 1}^{2N}(-1)^n \mbox{tr}\left( \frac{Pe^{-\beta H}}{\mbox{tr}(Pe^{-\beta H})}\frac{\sigma_z(n) + 1}{2}\right)\ee
and can be computed directly in the thermodynamic limit ($N \rightarrow + \infty$) using the ansatz (\ref{eq:thermEvo}), see \cite{Verstraete2004,Haegeman2011,Haegeman2013} for details. For $m/g = 0$ we compare our simulations with the analytical result in \ref{subsec:mdivg0}. In \ref{subsec:CCmdivgNonZero} we compute a subtracted chiral condensate in the non-perturbative regime: $m/g \sim \mathcal{O}(1)$. The results are compared with the recent simulations of Ba\~{n}uls \emph{et al.} \cite{Banuls2016}. 

\subsection{The chiral limit $m/g = 0$}\label{subsec:mdivg0}
\begin{figure}
\begin{subfigure}[b]{.24\textwidth}
\includegraphics[width=\textwidth]{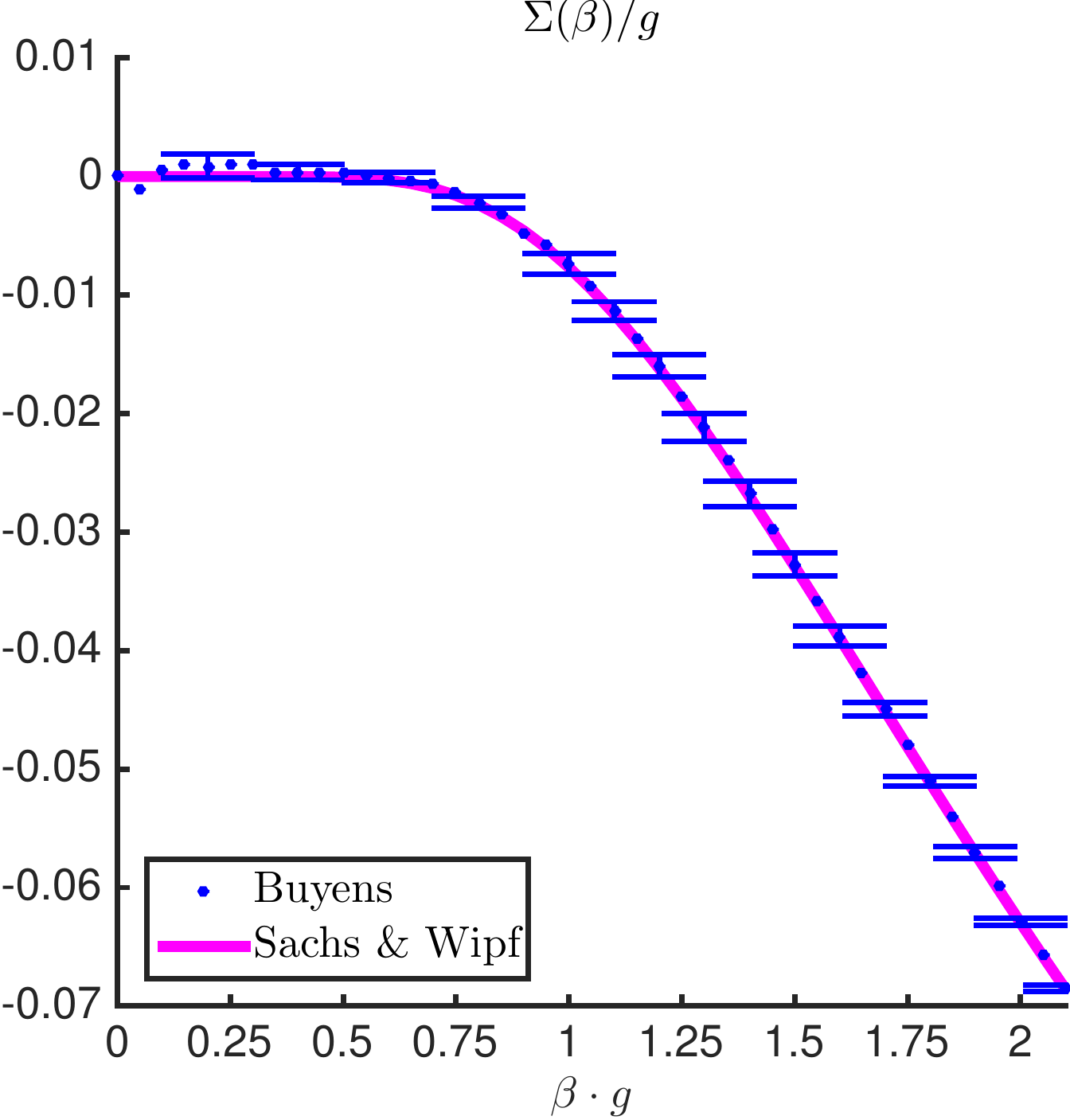}
\caption{\label{fig:CCmdivgzeroa}}
\end{subfigure}\hfill
\begin{subfigure}[b]{.24\textwidth}
\includegraphics[width=\textwidth]{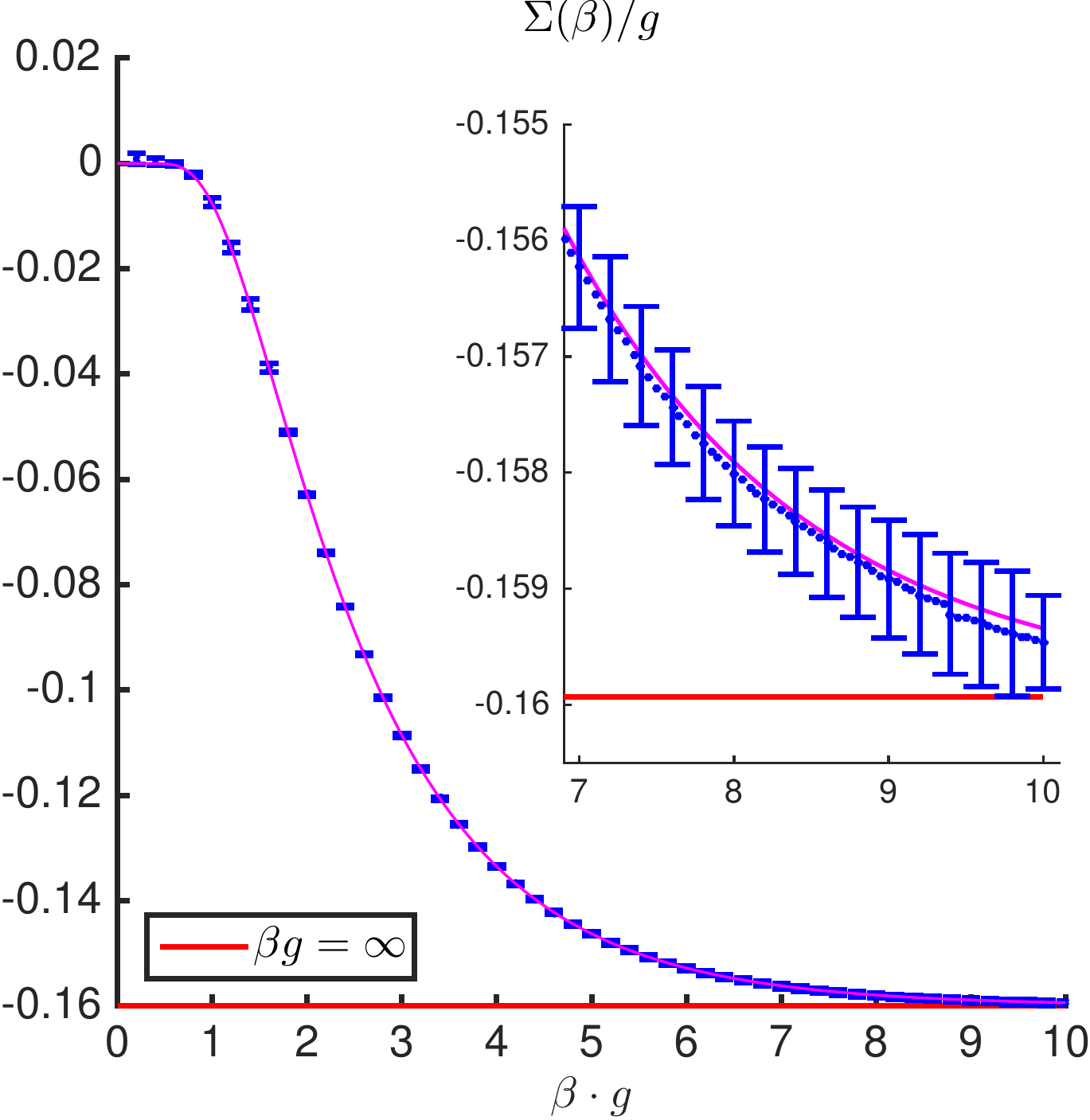}
\caption{\label{fig:CCmdivgzerob}}
\end{subfigure}\hfill
\vskip\baselineskip
\captionsetup{justification=raggedright}
\caption{\label{fig:CCmdivgzero}$m/g = 0$. Our continuum result $\Sigma(\beta)$ (blue) compared with the analytical result $\Sigma_{SW}(\beta)$ (magenta). (a) Zooming in on the interval $\beta g \in [0,2]$. (b) Results for $\beta g \in [0,10]$ and convergence toward the result at $\beta g = +\infty$ (red). Inset: zooming in on the interval $\beta g \in [7,10]$.}
\end{figure}

\noindent As for $m/g = 0$ the chiral condensate is known analytically, we can check our method by comparing the exact result with our simulations. Using path integral methods Sachs and Wipf \cite{Sachs1991} found that for $m/g = 0$:
\be\label{eq:SachsWipf}\Sigma_{SW}(\beta) = \frac{g}{2\pi^{3/2}}e^{\gamma}e^{2I(\beta g/\sqrt{\pi})}, I(u) = \int_0^{+\infty}\frac{dt}{1 - e^{u \cosh(t)}}, \ee 
with $\gamma \approx 0.57721\ldots$ the Euler-Mascheroni constant. In this subsection we will explain in detail how we compute $\Sigma(\beta)$ from our simulations and assign an appropriate uncertainty on this result by taking into account different sources of errors. As we will see, our estimate for the error is larger than the actual difference between our continuum estimate and the exact result. Therefore we can be confident that our extrapolation method is reliable in the sense that the exact results lies within the error bars. We refer to appendix \ref{subsec:appendixCCcontinuuma} for even more details.
\\
\\ On the lattice we computed the chiral condensate for $16 \leq x \leq 600$ and $\beta g \in [0,10]$ with steps $d\beta = 0.05$. Our tolerance for discarding the Schmidt values was set to $\epsilon = 10^{-6}$, i.e. after applying a Trotter gate we discarded the Schmidt values smaller than $\epsilon^2$. As discussed in subsection \ref{subsec:iTEBD}, taking $(\epsilon,d\beta) \neq 0$ introduces an error which includes the error originating from taking a finite bond dimension in the MPS representation of $\ket{\Psi[A(\beta)]}$. As explained in appendix  \ref{subsec:errorsFinD}, an estimate $\Delta^{(\epsilon,d\beta)}\Sigma(\beta,x)$ of this error is obtained by considering the difference of the chiral condensate computed for $(\epsilon,d\beta) = (10^{-6},0.05)$ with the chiral condensate computed for $(\epsilon,d\beta) = (5 \times 10^{-6},0.05)$ and $(\epsilon,d\beta) = (10^{-6},0.01)$, see eq. (\ref{eq:defDeltaepsdbSigma}). For our simulations we will only retain the $x-$values for which $\Delta^{(\epsilon,d\beta)}\Sigma(\beta ,x)/g \leq 4 \times 10^{-4}$. In practice it turns out that with this criterium for $\beta g \geq 2$ we only include the $x-$values with $16 \leq x \leq 300-400$. We refer to appendix \ref{subsec:errorsFinD} for the details. 

Once we computed for a fixed value of $\beta g$ the chiral condensate $\Sigma(\beta,x)$ and the corresponding errors $\Delta^{(\epsilon,d\beta)}\Sigma(\beta,x)$ for a range of $x-$values, we can use them to obtain a continuum estimate $\Sigma(\beta)$. We explain the procedure here briefly and refer to appendix \ref{subsec:appendixCCcontinuuma} for the details. Similar to \cite{Saito2014,Banuls2015,Saito2015,Banuls2016} we fit $\Sigma(\beta,x)$ to 
\begin{subequations}\label{eq:fitfunction}
\be\label{eq:fitfunctiona} f_1(x) = A_1 + B_1 \frac{\log(x)}{\sqrt{x}} + C_1\frac{1}{\sqrt{x}} \ee
\be\label{eq:fitfunctionb}f_2(x) = A_2 + B_2 \frac{\log(x)}{\sqrt{x}} + C_2\frac{1}{\sqrt{x}} + D_2\frac{1}{x},\ee
and to
\be\label{eq:fitfunctionc}f_3(x) = A_3 + B_3\frac{\log(x)}{\sqrt{x}} + C_3\frac{1}{\sqrt{x}} + D_3\frac{1}{x} + E_3\frac{1}{x^{3/2}} .\ee
\end{subequations}
For all the fitting functions $f_n$ ($n = 1,2,3$) we compute all possible fits against at least $n+5$ data points $\Sigma(\beta,x)$ of consecutive $x-$values with $\Delta^{(\epsilon,d\beta)}\Sigma(\beta ,x)/g \leq 4 \times 10^{-4}$. For every such fit the value of $A_n$ is an estimate for $\Sigma^{(n)}(\beta)$. We will only consider the significant fits, i.e. the fits that have for all coefficients a $p-$value smaller than $0.05$. In practice, this means that the error on each of the coefficients should be smaller than approximately half of the value of the coefficient itself. Note that our approach is less conservative than the one used in \cite{Banuls2016} where they call a fit statistically significant if the error on each of the coefficients is smaller than the value of the coefficient. For each fit we also compute its $\chi^2$-value (see eq. (\ref{eq:chisq}) in appendix \ref{subsec:appendixCCcontinuuma}). If for a fitting function $f_n$ we have more than $10$ significant fits with $\chi^2/N_{dof} \leq 1$, we take as the continuum estimate $\Sigma^{(n)}(\beta,x)$ for the fitting function $f_n$ the median of the distribution of all these estimates weighed by $\exp(-\chi^2/N_{dof})$, similar as in \cite{Banuls2013a}. Here $N_{dof}$ is the degrees of freedom,
$$ N_{dof} = \# \mbox{data points} - \underbrace{\# \mbox{coefficients to fit}}_{= n+2}.$$
The systematic error $\Delta^{(n)}\Sigma(\beta)$ for the choice of fitting interval comes from the $68,3 \%$ confidence interval. When we have less than $10$ significant fits with $\chi^2/N_{dof} \leq 1$
we take for $\Sigma^{(n)}(\beta,x)$ the estimate from the data points with $\chi^2/N_{dof} \leq 1$ which has the least error in $\Delta^{(\epsilon,d\beta)}\Sigma(\beta,x)$. Here the systematic error $\Delta^{(n)}\Sigma(\beta)$ is estimated by comparing this with the most outlying estimate coming from a statistically significant fit with $\chi^2/N_{dof} \leq 1$ from the same fitting ansatz $f_n$. 

So we now have for each of the fitting functions eq. (\ref{eq:fitfunction}) an estimate $\Sigma^{(n)}(\beta)$ and an error $\Delta^{(n)}\Sigma(\beta)$ which is the systematic error originating from the choice of $x$-interval. Of these three estimates we take as our final estimate the estimate corresponding to the fit which has the most significant fits with $\chi^2/N_{dof} \leq 1$. The error for the choice of fitting functions is obtained by comparing this value with the other $\Sigma^{(n)}(\beta)$.
\\
\\
In fig. \ref{fig:CCmdivgzero} we show our result for $\Sigma(\beta)/g$ and compare it with eq. (\ref{eq:SachsWipf}). The error bars are obtained as the maximum of the errors discussed above, i.e. the errors originating from taking a nonzero value for $(\epsilon,d\beta)$, the choice of fitting $x$-interval and the choice of fitting function $f_n$, see appendix \ref{subsec:appendixCCcontinuuma} for the details. For convenience we show in fig. \ref{fig:CCmdivgzero} the error bars after every step $d\beta = 0.2$ only. We indeed find very good agreement between our simulations and the exact result. In particular the exact result is always within the shown error bars. 

As is explained in appendix \ref{subsec:appendixCCcontinuuma} (see in particular fig. \ref{fig:extrapolationCCm0a0}), at higher temperatures there are more significant cutoff effects in $x$, which is reflected by relatively larger error bars for $\beta g \lesssim 1$ in fig. \ref{fig:CCmdivgzeroa}. We found indeed that at smaller values of $\beta g$ the reliable fits had higher order corrections in $1/\sqrt{x}$ or were through data points with large $x-$values. At lower temperatures, i.e. larger values of $\beta g$, we found that our continuum results were more robust against the choice of fitting function and fitting interval. In fig. \ref{fig:CCmdivgzerob} we observe that the chiral condensate converges to its result at $\beta g = \infty$, $\Sigma_{SW}(+\infty) = e^{\gamma} /(2\pi^{3/2} ) \approx  -0.1599288$ although there are still thermal correction at $\beta g \lesssim 10$ of order $10^{-3}$ (see inset). For large values of $\beta g$ the difference between our result and the exact result is of order $10^{-4}$ which is good enough but nevertheless two orders of magnitude larger than the difference we found at zero temperature for the chiral condensate $\Sigma = \Sigma(+\infty)$ in \cite{Buyens2014}. In that paper we reproduced the exact result up to $10^{-6}$. The accumulation of errors in $\epsilon$ and $d\beta$ during the imaginary time evolution has thus lead to less accuracy at smaller temperatures which indeed reflects the fact that direct optimization methods are better at determining the ground state than thermal evolution.   

As our continuum results are very close to the exact result and our error bars are large enough such that they always contain the exact result, we can be confident that the simulations and the extrapolation procedure to obtain the continuum limit are reliable. In particular we observe from fig. \ref{fig:CCmdivgzero} that our estimated errors in fact overestimate the real error.

\subsection{$m/g \neq 0:$ renormalization of $\Sigma(\beta)$}\label{subsec:CCmdivgNonZero}
\noindent At zero temperature the chiral condensate diverges for $m/g \neq 0$ when we approach the continuum limit $x \rightarrow + \infty$. By subtracting the free chiral condensate ($g = 0$) a UV-finite quantity was obtained \cite{Adam1998,Banuls2013b, Buyens2014, Banuls2016}. In \cite{Banuls2016} it was pointed out that at finite temperature it is also sufficient to remove the free chiral condensate at zero temperature to obtain a UV-finite quantity. Hence we consider the subtracted chiral condensate 
$$\Sigma_{sub}(\beta,x) = \Sigma(\beta,x) - \Sigma_{free}(x) $$
with $\Sigma_{free}(x)$ the free chiral condensate at zero temperature \cite{Banuls2013b}:
$$ \Sigma_{free}(x) = \frac{m}{\pi g}\frac{1}{\sqrt{1 +  \frac{m^2}{g^2 x}}}K\Biggl(\frac{1}{1 +  \frac{m^2}{g^2 x}}\Biggl), $$
where $K$ is the complete elliptic integral of the first kind.

\begin{figure}
\begin{subfigure}[b]{.24\textwidth}
\includegraphics[width=\textwidth]{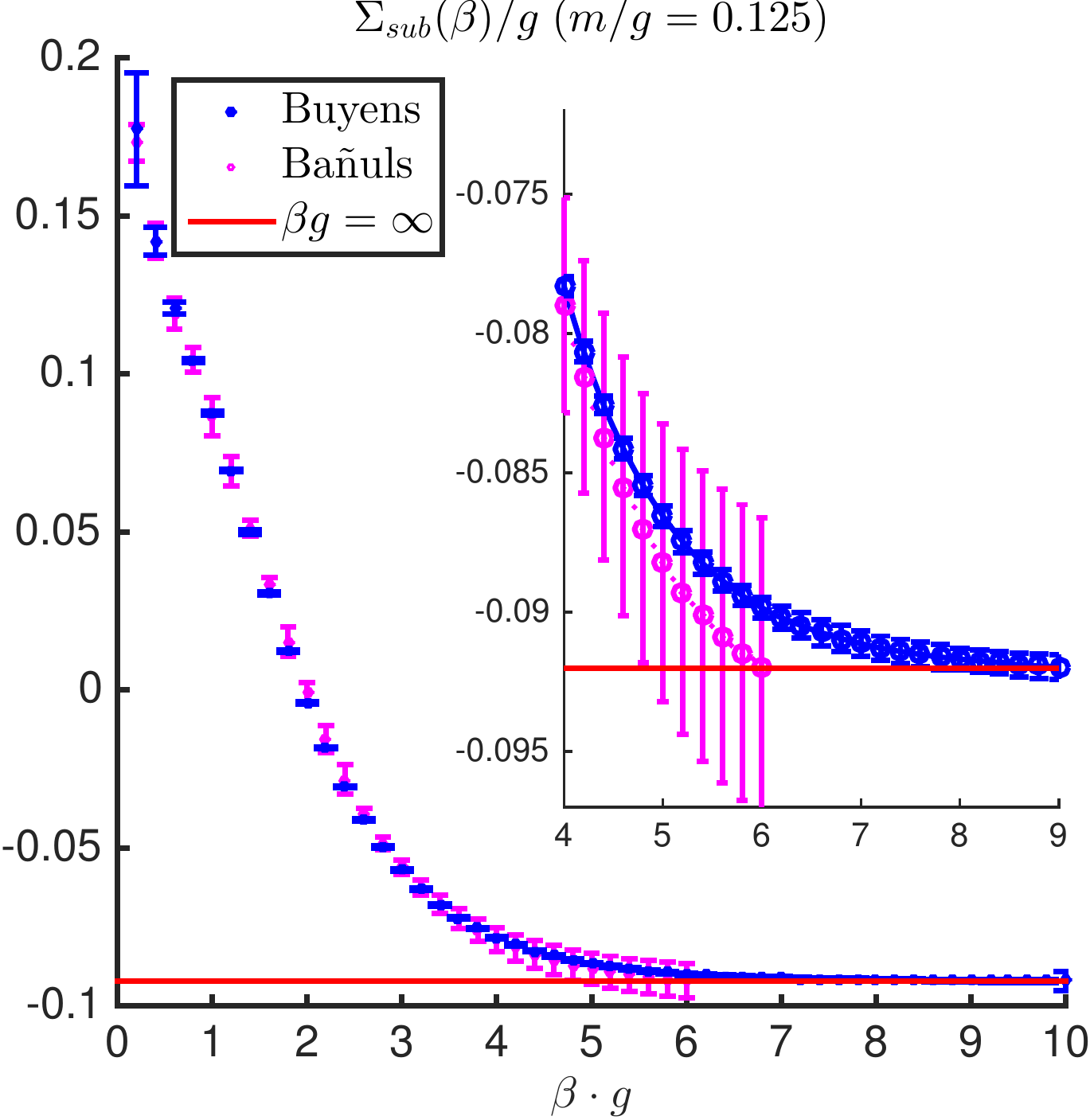}
\caption{\label{fig:CCmdivgNonzeroa}}
\end{subfigure}\hfill
\begin{subfigure}[b]{.24\textwidth}
\includegraphics[width=\textwidth]{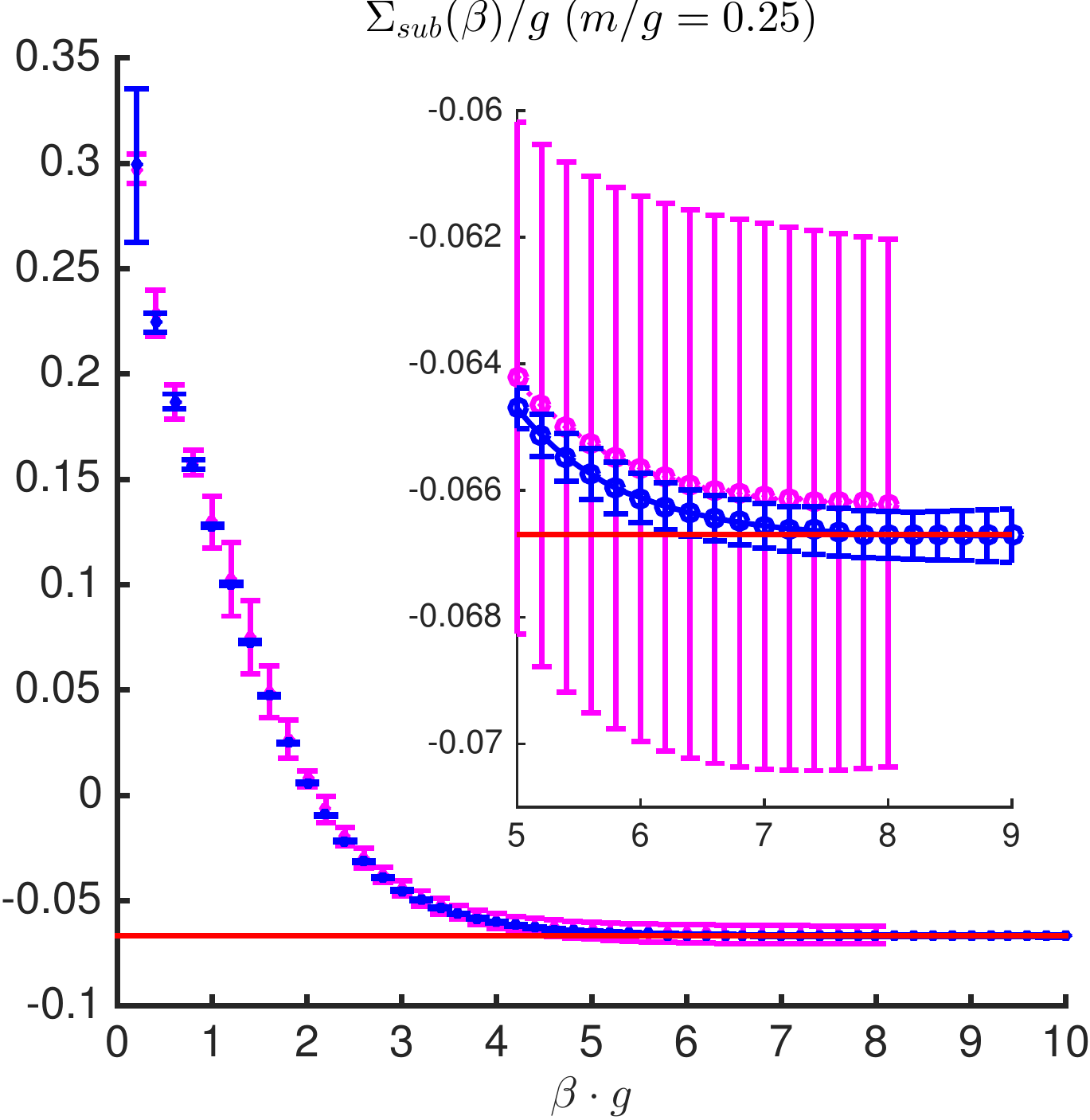}
\caption{\label{fig:CCmdivgNonzerob}}
\end{subfigure}\hfill
\vskip\baselineskip
\begin{subfigure}[b]{.24\textwidth}
\includegraphics[width=\textwidth]{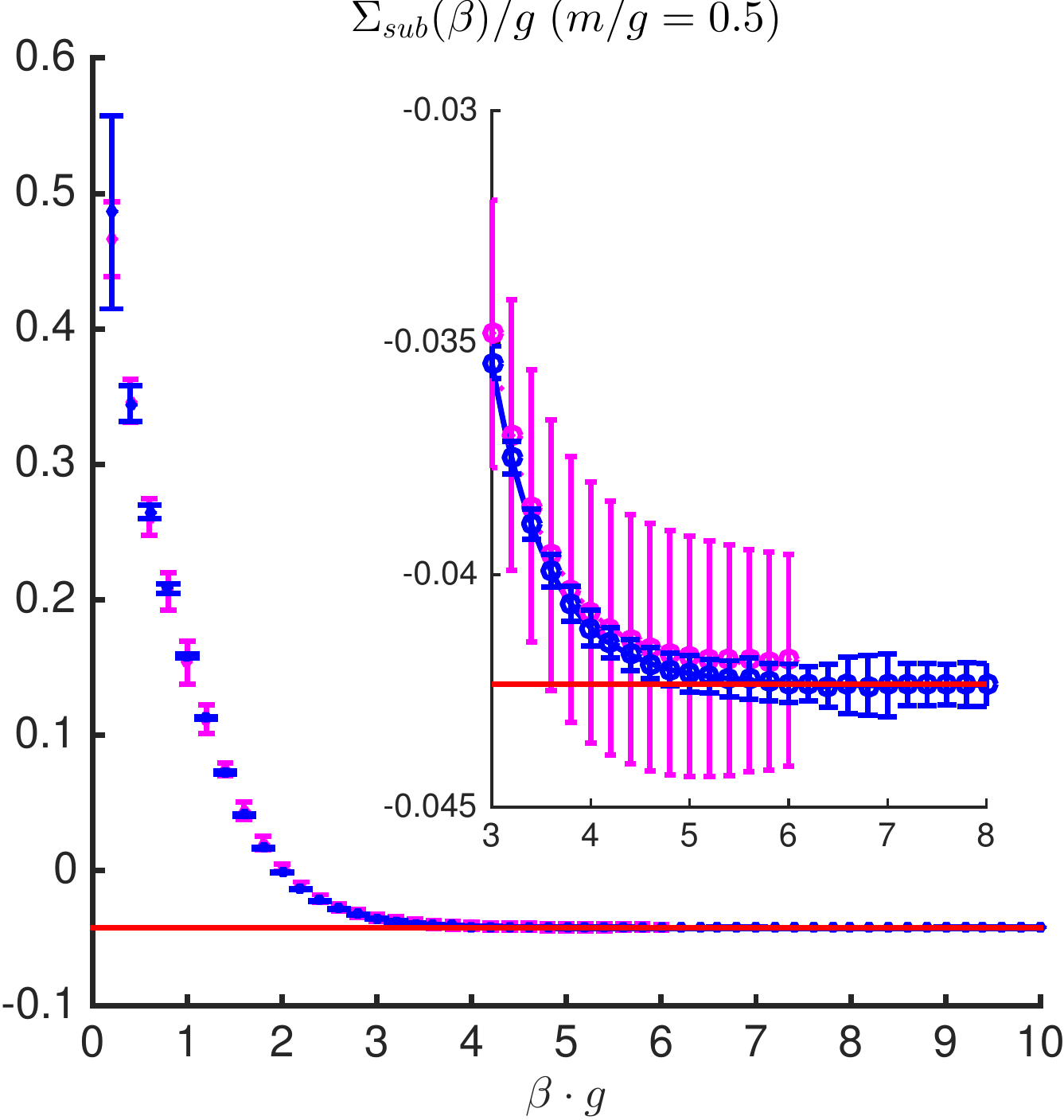}
\caption{\label{fig:CCmdivgNonzeroC}}
\end{subfigure}\hfill
\begin{subfigure}[b]{.24\textwidth}
\includegraphics[width=\textwidth]{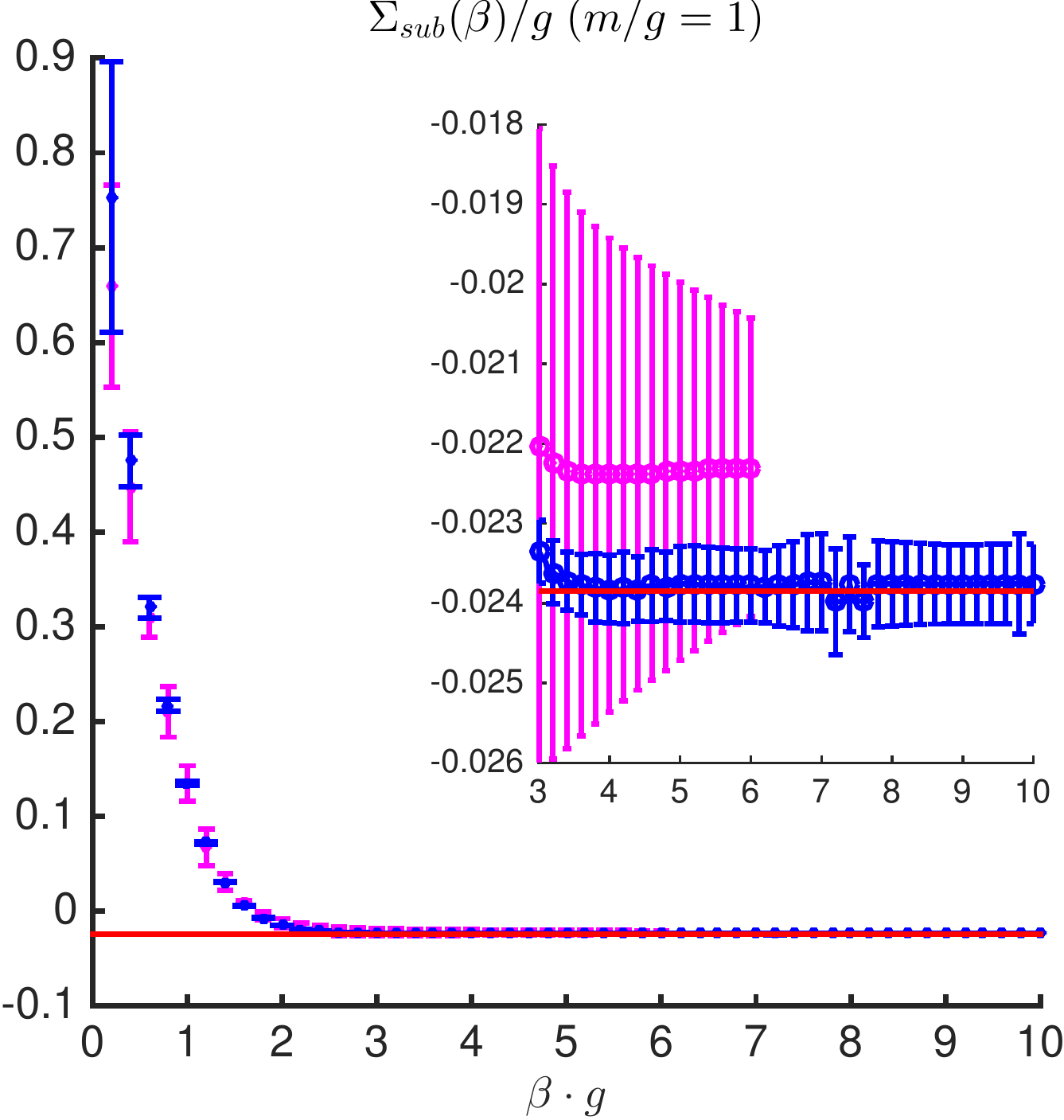}
\caption{\label{fig:CCmdivgNonzerod}}
\end{subfigure}\hfill
\vskip\baselineskip
\captionsetup{justification=raggedright}
\caption{\label{fig:CCmdivgNonzero} Subtracted chiral condensate $\Sigma_{sub}(m/g)$ for $m/g \neq 0$ for different values of $m/g$. Our results (blue) are compared with the simulations of Ba\~{n}uls \emph{et al.} \cite{Banuls2016} (magenta). The red line shows the results at $\beta g = +\infty$ obtained in \cite{Buyens2014}. In the inset we zoom in on the convergence toward this result for larger values of $\beta g$. (a): $m/g = 0.125$. (b) $m/g = 0.25$. (c) $m/g = 0.5$. (d) $m/g = 1$.}
\end{figure}

The procedure to take the continuum limit is exactly the same as for the massless case $m/g = 0$, but now we have to fit $\Sigma_{sub}(\beta,x)$ against $f_n(x)$ eq. (\ref{eq:fitfunction}). For a specific example we refer to fig. \ref{fig:extrapolationCCm25e2a0} in appendix \ref{subsec:appendixCCcontinuuma} where we discuss the continuum extrapolation of the subtracted chiral condensate for $m/g = 0.25$. Similar as for $m/g = 0$ cutoff effects in $x$ are more severe at higher temperatures. In contrast, at larger values of $\beta g$, the results are more robust against the choice of fitting interval and fitting function $f_n$. In fig. \ref{fig:CCmdivgNonzero} we show our results for $m/g = 0.125,0.25, 0.5,1$ (blue) and indeed find larger error bars for small values of $\beta g$. The error is estimated in the same way as for $m/g = 0$, it is the maximum of the error originating from taking nonzero values for $(\epsilon,d\beta)$, from the choice of fitting interval and from the choice of fitting function $f_n$ (see appendix \ref{subsec:appendixCCcontinuuma} for the details).  We computed $\Sigma_{sub}(\beta)$ with $d\beta = 0.05$ but show them here with steps $d\beta = 0.2$ for convenience. 
\\
\\ Our results are now compared with the simulations of Ba\~{n}uls \emph{et al.} \cite{Banuls2016} (magenta). We find that at all temperatures the error bars overlap, meaning that the results are in agreement with each other. This is a nontrivial check on MPO methods for gauge theories because in both approaches the optimization methods are different. In \cite{Banuls2016} they perform their simulations on a finite lattice and take the thermodynamic limit on the level of the expectation values. Also, instead of purifying the Gibbs state, they apply $Pe^{-\beta H}$ immediately to the MPO. After every step they project this MPO to an optimal MPO, in the sense of least squares, with smaller bond dimension $D$ which has been fixed before. In contrast, we did our simulations immediately in the thermodynamic limit and adapted the bond dimension by investigating the Schmidt spectrum of the purified state. One observes that the errors in  \cite{Banuls2016} are larger than ours which is partially explained by the fact that we use different extrapolation methods. The main difference is that we consider more possible fits through our data (i.e. more choices of the fitting interval) but are less conservative in the fits we call statistically significant (see also previous subsection). However our smaller error bars are also explained by the fact that we work immediately in the thermodynamic limit ($N\rightarrow + \infty$) which eliminates the uncertainty of the $N \rightarrow + \infty$ extrapolation.

Looking at the $\Sigma_{sub}(\beta)$, it seems to diverge for $\beta = 0$ which is a consequence of our renormalization scheme. Indeed, for $\beta = 0$ we have that $\Sigma(\beta,x) = 0$ while $\Sigma_{free}(x)$ diverges logarithmically in the limit $x \rightarrow + \infty$. For other values of $\beta g$ the chiral condensate is UV-finite and decreases to its ground-state expectation value (red line) as a function of $\beta g$, see insets in fig. \ref{fig:CCmdivgNonzero}. The chiral condensate tends faster to its ground-state expectation value for larger values of $m/g$. For instance, in fig. \ref{fig:CCmdivgNonzero} we observe that for $m/g = 1$ the chiral condensate is already very close to its ground-state expectation value for $\beta g \approx 2$ while for $m/g = 0$ even for $\beta g \approx 4$ there is still a significant difference with the ground-state expectation value. This is explained by the fact that the mass gap of $H$ grows with $m/g$, see \cite{Byrnes2002,Banuls2013a,Buyens2013}.

\section{Asymptotic confinement}\label{sec:asCon}

\noindent As mentioned in the previous section, QCD changes from the confined to the deconfined phase around a pseudo-critical temperature $T_c$. For infinitely heavy quark masses this phase transition is detected by the spontaneous breaking of the $SU(3)$-center symmetry or equivalently by examining the free energy of an infinitely separated probe quark-antiquark pair, which diverges in the confining phase and is finite in the deconfining phase. In the case of physical QCD, with finite quark masses, the notion of confinement versus deconfinement is less clear \cite{Greensite2003}: the infinitely separated probe pair will always be screened by charge production out of the vacuum, leading to a finite free energy, already at zero temperature.    

For the Schwinger model we have a similar situation: for integer probe charges the confining string will always be broken at large separation of the probe pair, due to screening by the dynamical fermions. However, this is not the case if we introduce fractional probe charges. In that case, at zero temperature, for $m/g\neq 0$ a confining string remains even at infinite separation \cite{ColemanCS}. So by probing the vacuum with fractional charge pairs at infinity we can examine the confining nature of the theory at finite temperatures.     

At zero temperature we already elaborated on this in \cite{Buyens2015} for finite and infinite distances $Lg$ between the quark and antiquark. Our simulations confirmed the known results that for $m/g = 0$ the quark-antiquark potential is never confining for large $Lg$ and that for $m/g \neq 0$ it is only confining if the charge of the heavy probe quarks is noninteger. A similar result has been shown when the system is in thermal equilibrium with a heath bath for $m/g = 0$ \cite{Hosotani1996} and $m/g \ll 1$ \cite{Fischler1979,Hetrick1988,Grignani1996}. But notice, that a critical temperature was found, above which the string tension is exponentially suppressed with the temperature. In the next subsection \ref{subsec:deconfinement} we will focus on this phenomenon in the nonperturbative regime $m/g \sim  \mathcal{O}(1)$. As we will discuss in subsection \ref{subsec:alphaclosetohalf}, our simulations in this mass regime also allow us to investigate the CT symmetry restoration in the $\alpha\rightarrow 1/2$ limit. But let us now first discuss the general setup of the simulations. 
\\
\\Assuming that the quark has charge $g\alpha$ and the antiquark has charge $-g\alpha$, this setup can be translated to a uniform background field $g\alpha$ in the Hamiltonian $H_{\alpha}$, see eq. (\ref{equationH}) \cite{Buyens2015,Coleman1976}. Note that we denoted the $\alpha$-dependence of $H$ in $H_{\alpha}$. The string tension at finite $x$, $\sigma_{\alpha}(\beta,x)$, is obtained from the partition function $Z_{\alpha}(\beta,x) = \mbox{tr}(e^{-\beta  H_{\alpha}} P)$ as 
\begin{align} \sigma_\alpha(\beta,x) &= -\frac{\sqrt{x}}{2N}\frac{1}{\beta}\log\left(\frac{Z_{\alpha}(\beta,x)}{Z_0(\beta,x)}\right) \nonumber \\
& = \frac{\sqrt{x}}{2N}\bigl( F_{\alpha}(\beta,x) - F_0(\beta,x)\bigl) \nonumber 
\end{align}
where $F_{\alpha}(\beta,x) = -(1/\beta)\log\bigl(Z_{\alpha}(\beta,x)\bigl)$ is the free energy for $L g = +\infty$. The MPO framework enables us to compute the partition function $Z_{\alpha}(\beta,x)$ and thus the free energy per unit of length $\mathcal{F}_{\alpha}(\beta,x) = \frac{\sqrt{x}}{2N}F_{\alpha}(\beta,x)$ directly, also in the thermodynamic limit ($N \rightarrow + \infty$). This is in contrast to Monte Carlo methods where the computation of the free energy is a difficult task \cite{Philipsen2013,Huscroft2000,Troyer2003,Sheng2015}. 

Other quantities that will be of interest here are the electric field and the Gibbs free entropy. The electric field, 
$$E_{\alpha}(\beta,x)= \lim_{N \rightarrow + \infty}\frac{g}{2N}\sum_{n = 1}^{2N}\mbox{tr}\left(\frac{e^{-\beta H_{\alpha}}P}{Z_{\alpha}(\beta,x)}(L(n) + \alpha)\right),$$ 
gives us more information about the $\alpha$-dependence of the string tension because it equals $E_{\alpha}(\beta,x) = \partial_\alpha \sigma_{\alpha}(\beta,x)$. The (Gibbs) entropy per unit of length, 
$$ \mathcal{S}_{\alpha}(\beta,x) = -\frac{\sqrt{x}}{2N}\mbox{tr}\left(\frac{e^{-\beta  H_{\alpha}} P}{Z_{\alpha}(\beta,x)}\log\left(\frac{e^{-\beta  H_{\alpha}} P}{Z_{\alpha}(\beta,x)}\right)\right),$$
is a measure for thermal fluctuations in the Gibbs state. When the canonical ensemble behaves as the ground state and corrections to ground-state expectation values are negligible, i.e. when the system is effectively at zero temperature, the entropy is very small and vice versa. 

\begin{figure}[t]
\begin{subfigure}[b]{.24\textwidth}
\includegraphics[width=\textwidth]{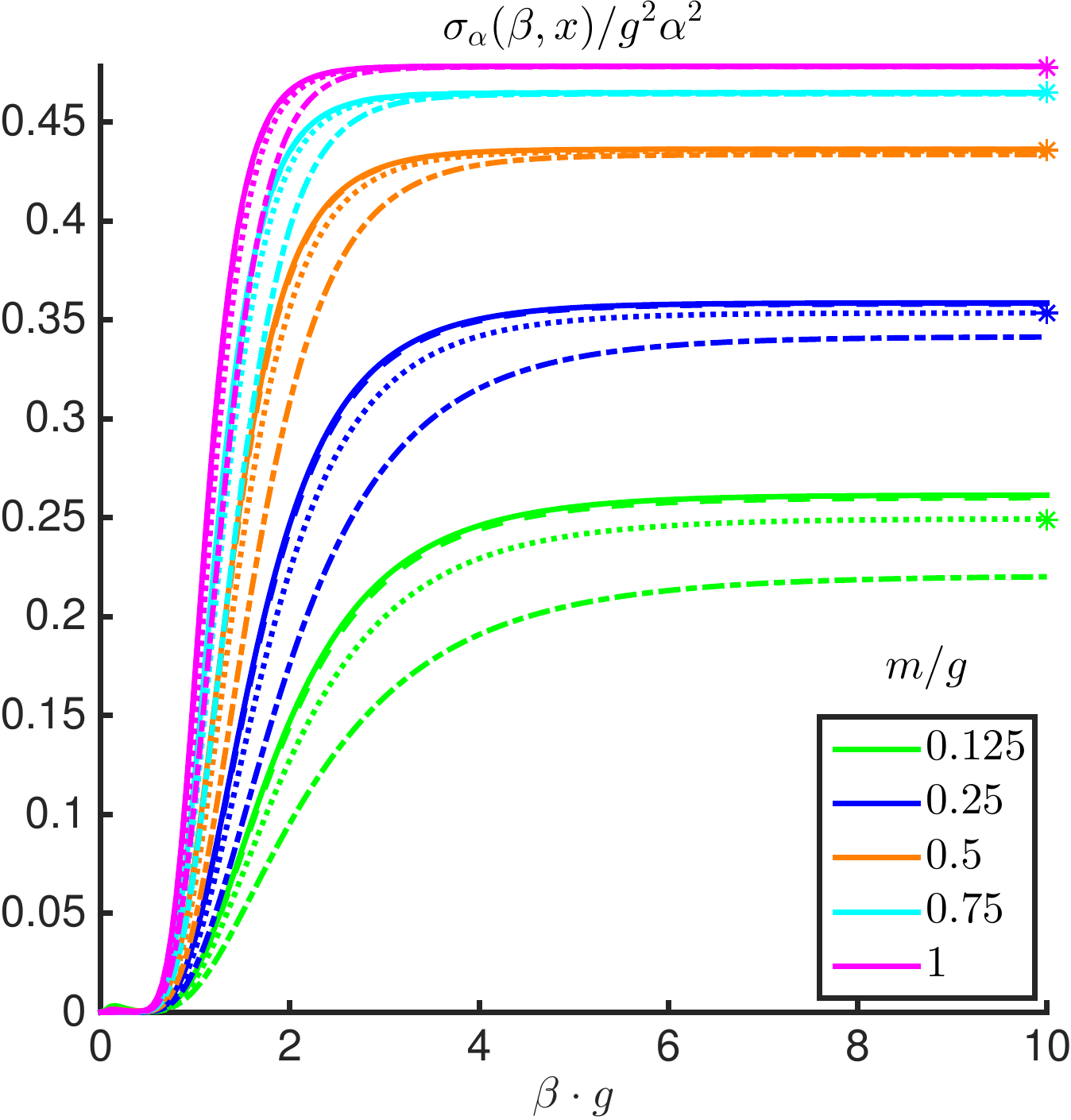}
\caption{\label{fig:FreeEnergyPert}}
\end{subfigure}\hfill
\begin{subfigure}[b]{.24\textwidth}
\includegraphics[width=\textwidth]{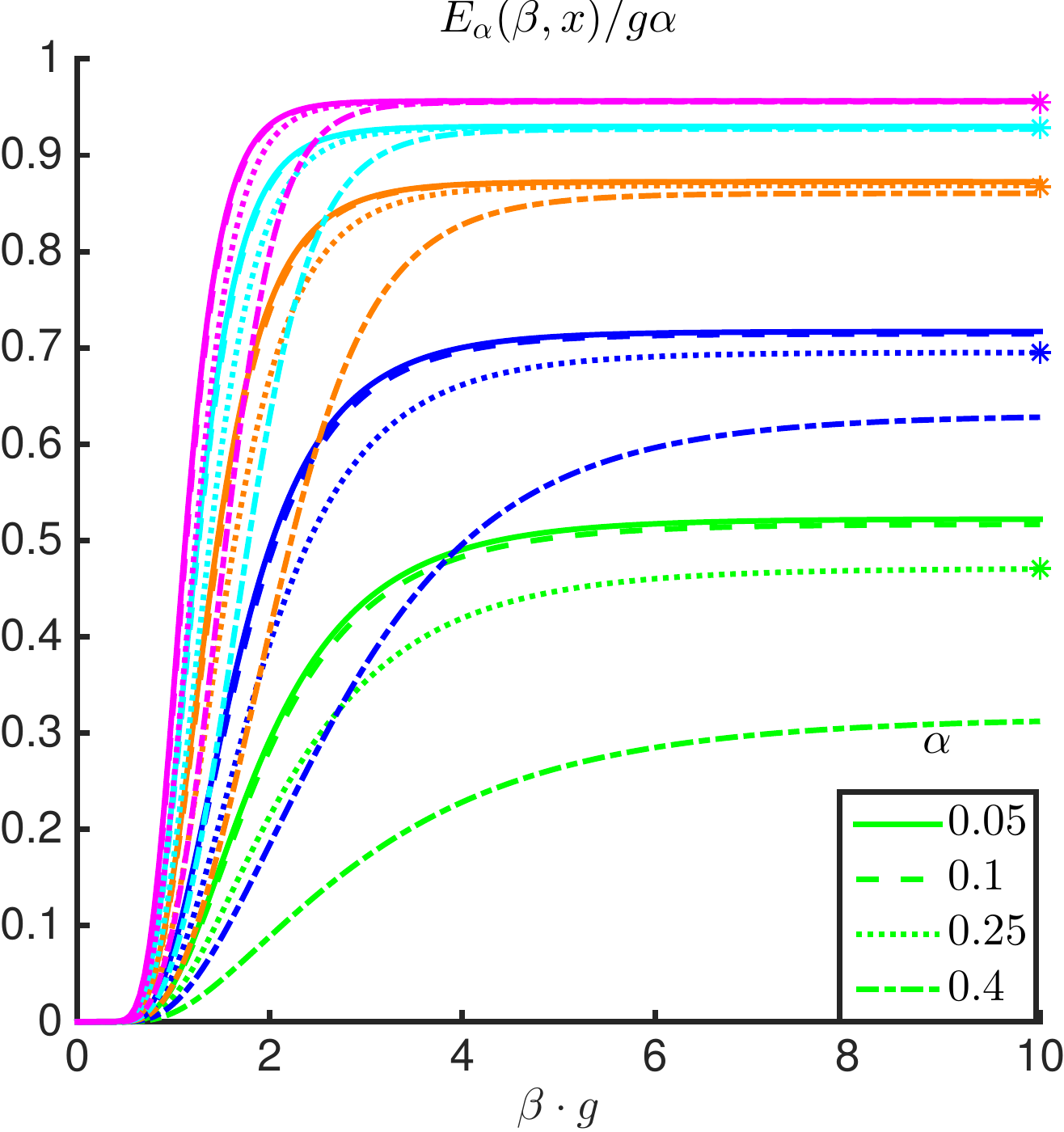}
\caption{\label{fig:EFPert}}
\end{subfigure}\hfill
\vskip\baselineskip
\begin{subfigure}[b]{.24\textwidth}
\includegraphics[width=\textwidth]{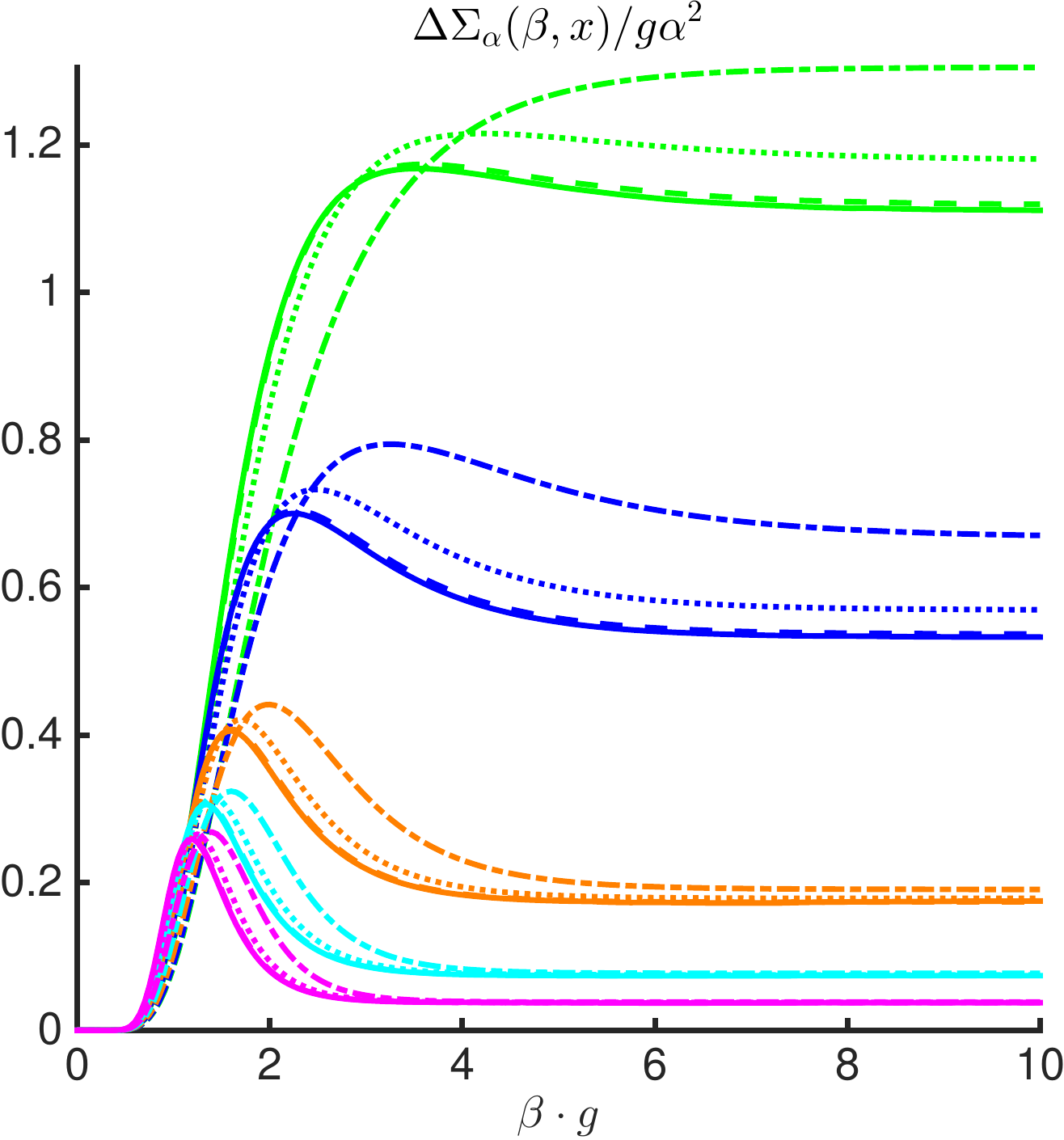}
\caption{\label{fig:CCPert}}
\end{subfigure}\hfill
\begin{subfigure}[b]{.24\textwidth}
\includegraphics[width=\textwidth]{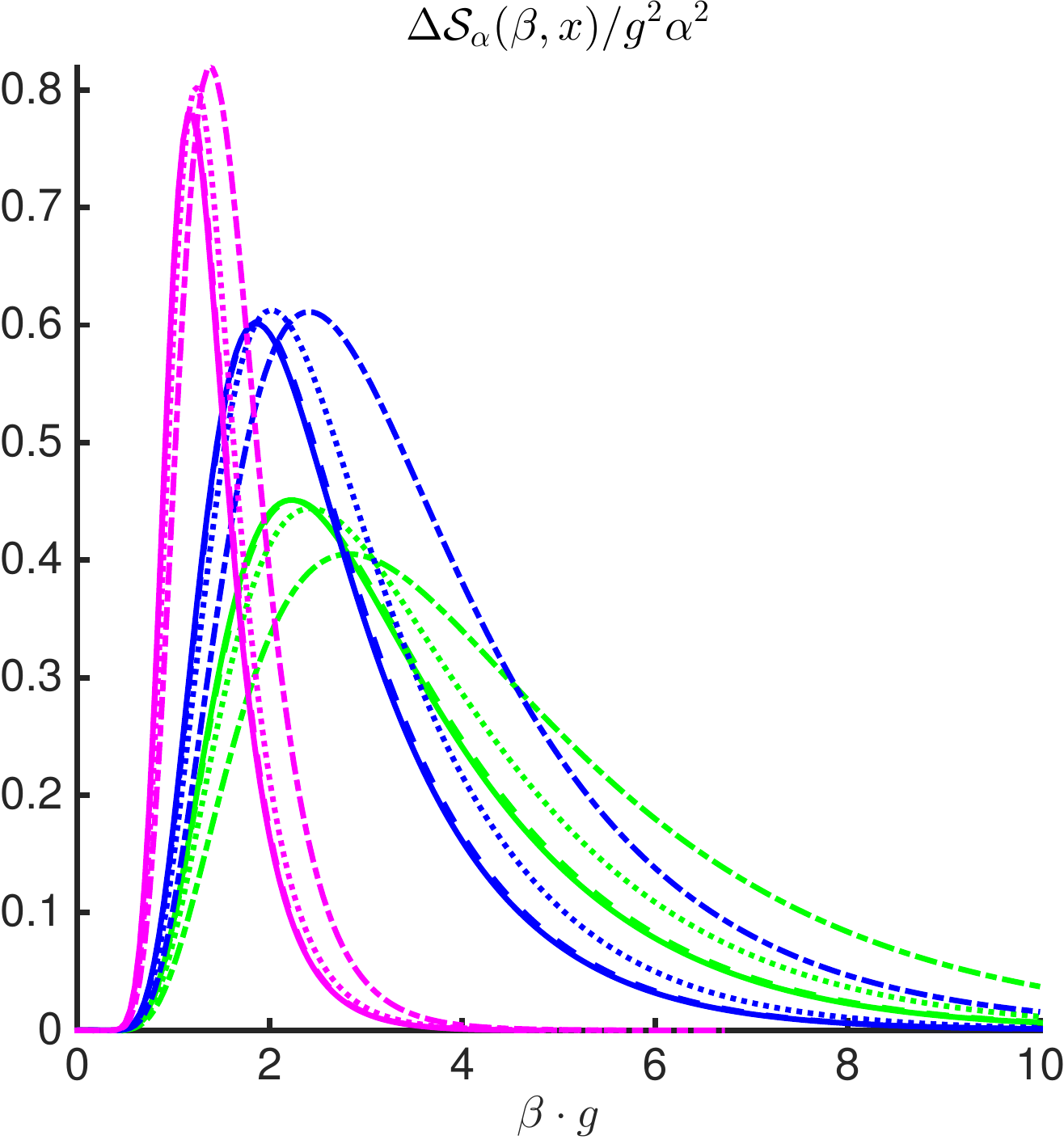}
\caption{\label{fig:EntrPert}}
\end{subfigure}\hfill
\vskip\baselineskip
\captionsetup{justification=raggedright}
\caption{\label{fig:quantPert} $x = 100$. $\alpha = 0.05$ (full line), $\alpha = 0.1$ (dashed line), $\alpha = 0.25$ (dotted line) and $\alpha = 0.4$ (dashed dotted line). (a) 
$\sigma_{\alpha}(\beta,x)/g^2\alpha^2$. (b) $E_{\alpha}(\beta,x)/g\alpha$. (c) $\Delta \Sigma_{\alpha}(\beta,x)/g\alpha^2$. (d) $\mathcal{S}_{\alpha}(\beta,x)/g^2\alpha^2$. The stars in (a) and (b) are the values at $\beta g = +\infty$ for $\alpha = 0.25$.}
\end{figure}

$\mathcal{S}_{\alpha}(\beta,x)$ is obtained from the average energy per unit of length $\mathcal{E}_{\alpha}(\beta,x) = \frac{\sqrt{x}}{2N}\frac{1}{Z_{\alpha}(\beta,x)}\mbox{tr}(H_{\alpha}e^{-\beta  H_{\alpha}} P)$ via the standard relation $\mathcal{S}_{\alpha}(\beta,x) = -\beta\bigl(\mathcal{F}_{\alpha}(\beta,x) - \mathcal{E}_{\alpha}(\beta,x)\bigl)$ for Gibbs states. For every value of $\beta g$ we subtract its $(\alpha = 0)$-value from it and we thus consider $\Delta \mathcal{S}_{\alpha}(\beta,x) = S_{\alpha}(\beta,x) - S_{0}(\beta,x)$. Because retaining only a finite range of eigenvalues of $L(n)$ leads to the same errors in $S_{\alpha}(\beta,x)$ and $S_{0}(\beta,x)$, the quantity $\Delta \mathcal{S}_{\alpha}(\beta,x)$ can be obtained accurately at all temperatures. We will see later that $\Delta \mathcal{S}_{\alpha}(\beta,x)$ is actually still a good measure for characterizing the transition from the effective zero temperature behavior at small temperatures toward a thermal behavior at larger temperatures, see subsection \ref{subsec:deconfinement} and in particular fig. \ref{fig:EntrPert}. 

One can also compute the chiral condensate $\Sigma_\alpha(\beta,x)$. Contrary to the previous section, we will now renormalize it by subtracting its ($\alpha = 0$)-value and thus consider $\Delta \Sigma_{\alpha}(\beta,x) = \Sigma_{\alpha}(\beta,x) - \Sigma_0(\beta,x)$. 
\\
\\As in the previous section we perform our simulations for $(\epsilon,d\beta) = (10^{-6},0.05)$ and an error is estimated by comparing this with results for $(\epsilon,d\beta) = (10^{-6},0.01)$ and $(\epsilon,d\beta) = (5 \times 10^{-6},0.05)$, see appendix \ref{subsec:errFinDalphaneq0} and in particular eq. (\ref{eq:defDeltaepsdbQ}) for the details. As is demonstrated there, for $\beta g \gtrsim 0.5$ and $\alpha \leq 0.45$ these errors are at most of order $10^{-4}$ which is sufficiently small for our purposes. When $\beta g \leq 0.5$, the values of the considered quantities are very small and we need to take smaller values for $\epsilon$. As is discussed in subsection \ref{subsec:deconfinement}, this can be traced back to the deconfinement transition at $T = +\infty$. In this regime we have set $(\epsilon, d\beta) = (10^{-9},5 \times 10^{-3})$. For $\alpha > 0.45$ and $m/g \gtrsim 0.33$ simulations are troubled by the spontaneous breaking of the CT symmetry, see section \ref{subsec:alphaclosetohalf}. Therefore we will also need a smaller tolerance in this regime. More specifically, we found that taking $\epsilon \approx 5 \times 10^{-7}$ produces results with errors in $(\epsilon,d\beta)$ smaller than $10^{-3}$. We refer to appendix \ref{subsec:sima1half} for more details.

In appendix \ref{app:alphaneq0continuum} we show explicitly for $m/g = (0.125, 0.25)$ and $\alpha = (0.1,0.25,0.45)$ that $\sigma_\alpha(\beta)$, $E_\alpha(\beta)$, $\Delta\mathcal{S}_\alpha(\beta)$ and $\Delta\Sigma_\alpha(\beta)$ are UV-finite quantities. This is done by extrapolating our results for $x = 100,125,\ldots,300$ to the continuum limit $(x \rightarrow + \infty)$ for all values of $\beta g$, thereby addressing all the systematic errors. In contrast to section \ref{sec:CC} the data now support a polynomial fit in $1/\sqrt{x}$ instead of the fits eq. (\ref{eq:fitfunction}) (see for instance figs. \ref{figapp:CCm25a25e2ExtrCC} and \ref{fig:continuumExtrapolationEFreea25e2} in appendix \ref{app:alphaneq0continuum}), i.e. we now plot our data against  
\begin{subequations}\label{eq:fitPolyfunctions}
\be\label{eq:fitfunctionPolya} f_1(x) = A_1 + B_1\frac{1}{\sqrt{x}} \ee
\be\label{eq:fitfunctionPolyb}f_2(x) = A_2 + B_2\frac{1}{\sqrt{x}} + C_2\frac{1}{x},\ee
and to
\be\label{eq:fitfunctionPolyc}f_3(x) = A_3 + B_3\frac{1}{\sqrt{x}} + C_3\frac{1}{x} + D_3\frac{1}{x^{3/2}} .\ee
\end{subequations}

The continuum limit and an estimate of the error (which includes the systematic errors originating from taking nonzero $(\epsilon,d\beta)$, from the choice of $x-$interval and the choice of fitting function) are found in the same way as for the chiral condensate (see subsection \ref{subsec:mdivg0}). In appendix \ref{app:alphaneq0continuum} one can find an extended discussion on that. The results for $m/g = (0.125, 0.25)$ and $\alpha = (0.1,0.25,0.45)$ are also plotted there (figs. \ref{fig:continuumResmdivg125e3} and \ref{fig:continuumResmdivg25e2}) and we found that all errors were under control (of order $10^{-4}$). As an extra check on our results we observed convergence to the results of \cite{Buyens2015} at zero temperature for the string tension and the electric field. At $x = 100$ our results were found to be already close to the continuum limit.  When $m/g \gtrsim 0.5$, the variation for different values of $x$ becomes even less, see fig. \ref{fig:varDiffx} in appendix \ref{app:alphaneq0continuum}. Hence, even though we will restrict our analysis here to $x = 100$, we can expect to be close to the continuum limit. 

Physics is periodic in $\alpha$ with period $1$ and due to CT symmetry, with $C$ charge conjugation
\begin{multline*} E(n) \rightarrow - E(n),\theta(n) \rightarrow - \theta(n),\\ \sigma_z(n) \rightarrow - \sigma_z(n),
 \sigma^{\pm}(n) \rightarrow \sigma^{\mp}(n) \end{multline*}
and $T$ translation over one site, we have that
\begin{multline*} \sigma_{\alpha}(\beta,x) = \sigma_{1-\alpha}(\beta,x), \Delta \Sigma_{\alpha}(\beta,x) = \Delta\Sigma_{1-\alpha}(\beta,x)
\\ E_{\alpha}(\beta,x) = - E_{1- \alpha}(\beta,x), \mathcal{S}_{\alpha}(\beta,x) = \mathcal{S}_{1-\alpha}(\beta,x). \end{multline*}
Therefore we can restrict ourselves to $\alpha \in [0,1/2]$. In subsection \ref{subsec:deconfinement} we will investigate the temperature dependence of the string tension. In the high temperature regime we focus on the deconfinement of the heavy quarks when the temperature $T$ becomes infinite: $T \rightarrow + \infty$. Then, in subsection \ref{subsec:alphaclosetohalf}, we treat the case when $\alpha$ tends to $1/2$. At zero temperature there is for this value of $\alpha$ and  $m/g = (m/g)_c \approx 0.33$ a phase transition \cite{Coleman1976,Byrnes2002} related to the spontaneous breaking of the $CT$ symmetry. Here we will investigate this spontaneous symmetry breaking at finite temperature. 

\subsection{Deconfinement transition at large $T$}\label{subsec:deconfinement}

\noindent When $\alpha$ is small one can expand the string tension into a series of powers of $\alpha$. Because $\sigma_{\alpha}$ is even in $\alpha$, this yields an expansion in $\alpha^2$:
\begin{subequations}\label{eq:PertExp}
\be \sigma_{\alpha}(\beta) \approx f_2(\beta,m) \alpha^2 + \mathcal{O}(\alpha^4).\ee 
For the electric field expectation value we then find:
\be E_{\alpha}(\beta,x)  = \frac{\partial }{\partial (g\alpha)}\left(\sigma_{\alpha}(\beta,x)\right)= 2\alpha \frac{f_2(\beta,m)}{g} + \mathcal{O}(\alpha^3)\, \ee
and similarly for the entropy and chiral condensate:
\begin{align} \Delta \mathcal{S}_{\alpha}(\beta,x) = \beta^2 \frac{\partial}{\partial \beta }\bigl(f_2(\beta,m)\bigl)\alpha^2+ \mathcal{O}(\alpha^4),  \end{align}
\begin{align} \Delta \Sigma_{\alpha}(\beta,x) 
= \frac{\partial f_2}{\partial m}(\beta,m)\alpha^2 + \mathcal{O}(\alpha^4).
\end{align}
\end{subequations}

\begin{figure}[t]
\begin{subfigure}[b]{.24\textwidth}
\includegraphics[width=\textwidth]{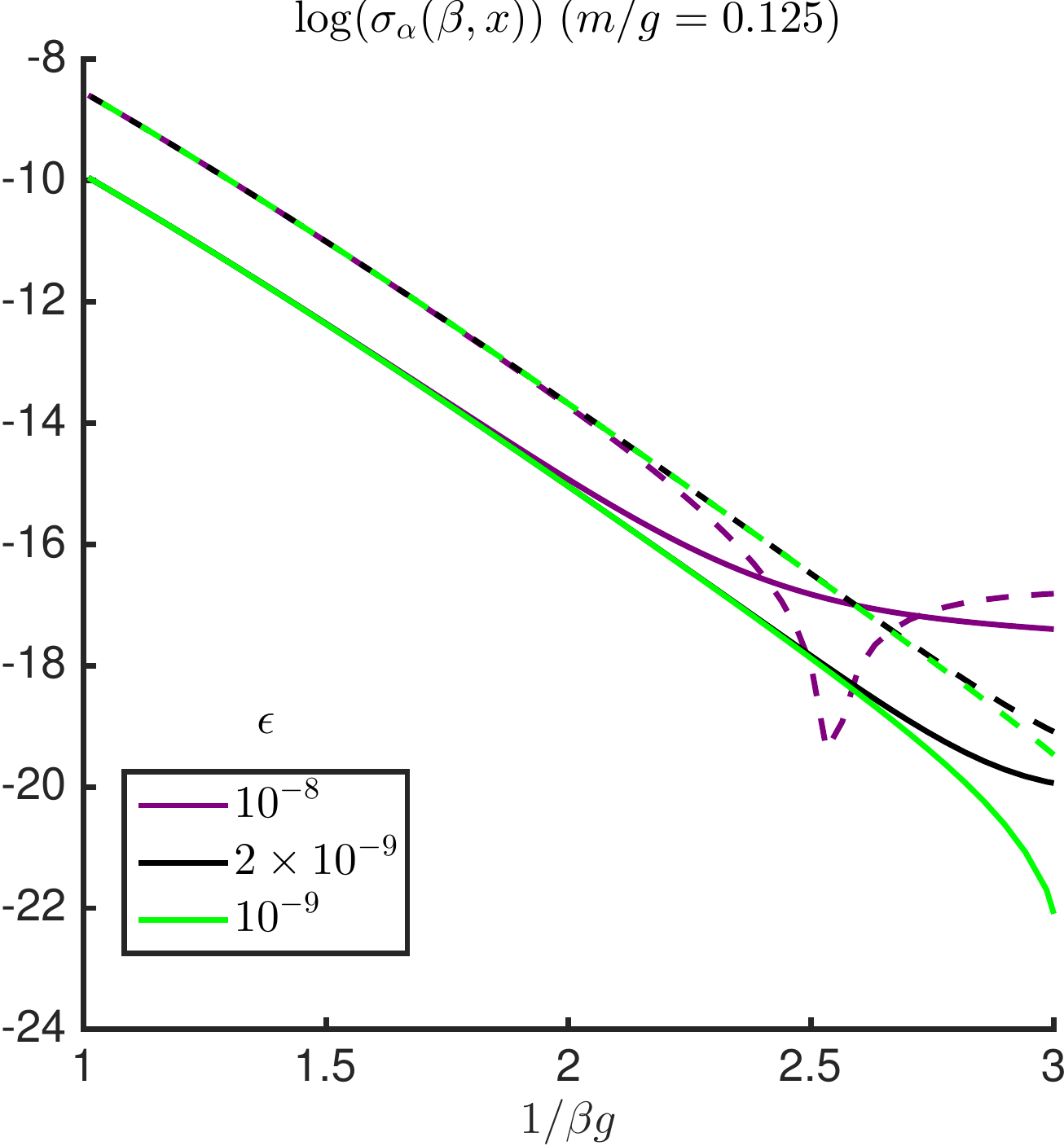}
\caption{\label{fig:STlargeTa}}
\end{subfigure}\hfill
\begin{subfigure}[b]{.24\textwidth}
\includegraphics[width=\textwidth]{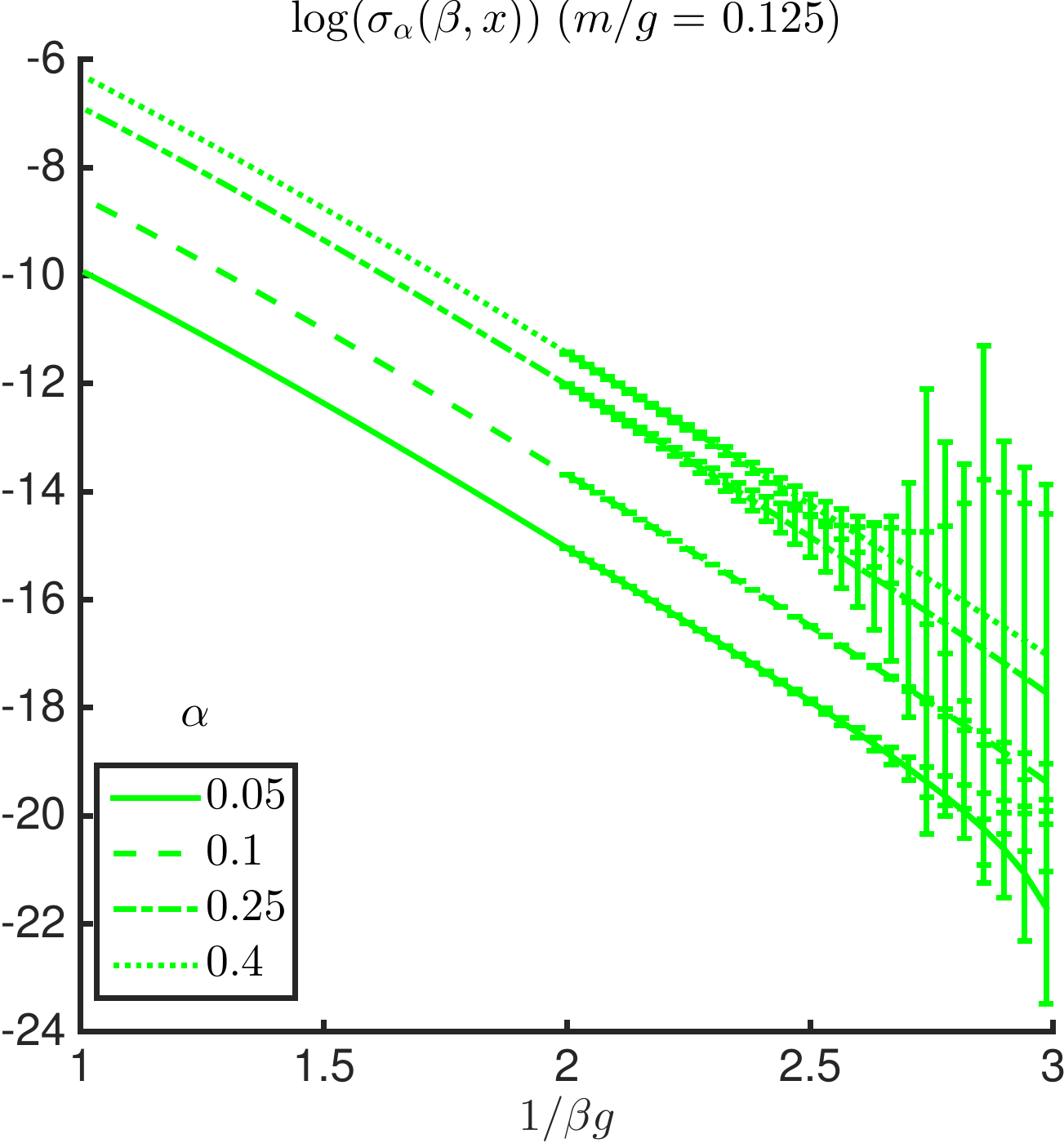}
\caption{\label{fig:STlargeTb}}
\end{subfigure}\hfill
\vskip\baselineskip
\begin{subfigure}[b]{.24\textwidth}
\includegraphics[width=\textwidth]{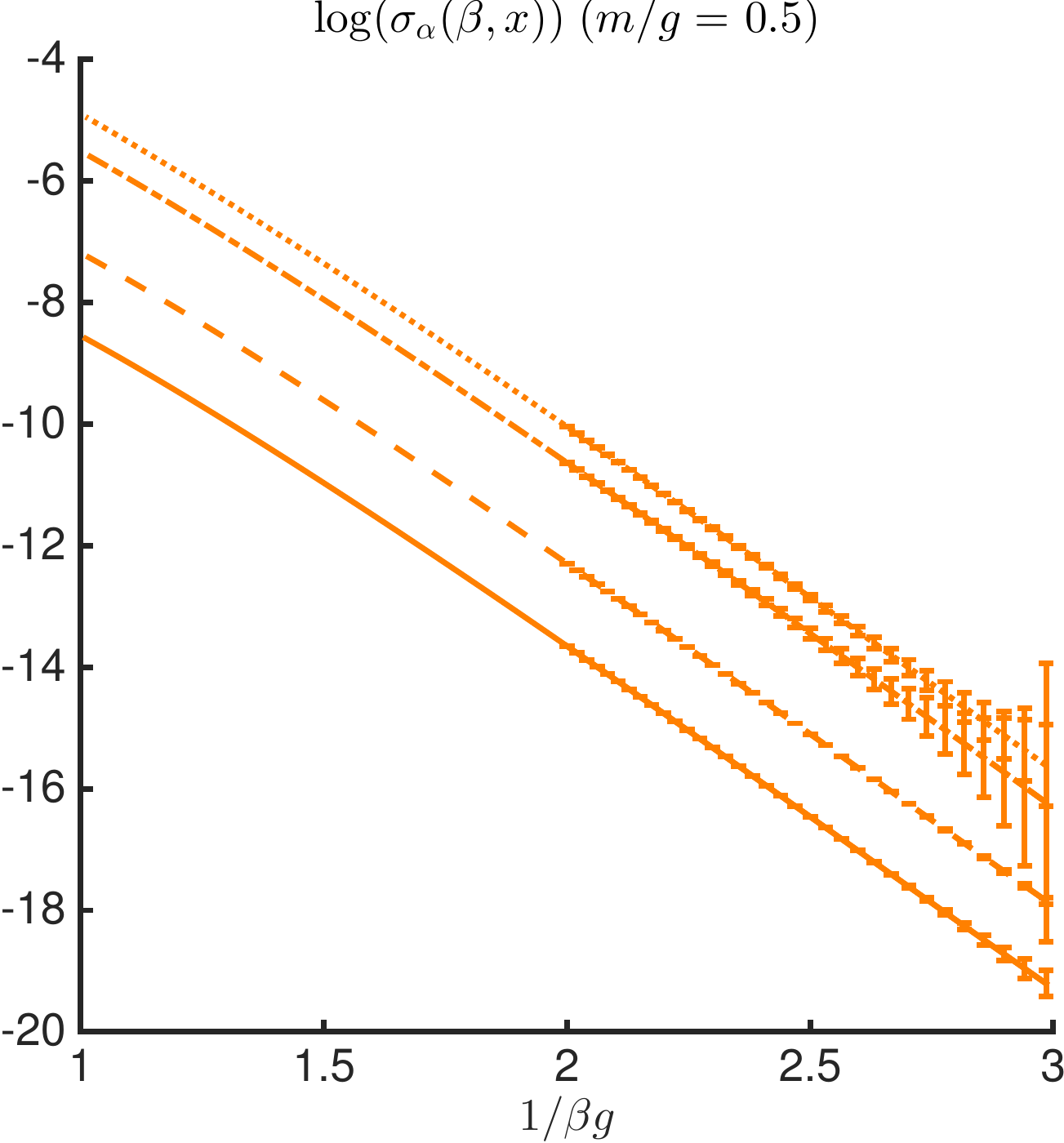}
\caption{\label{fig:STlargeTc}}
\end{subfigure}\hfill
\begin{subfigure}[b]{.24\textwidth}
\includegraphics[width=\textwidth]{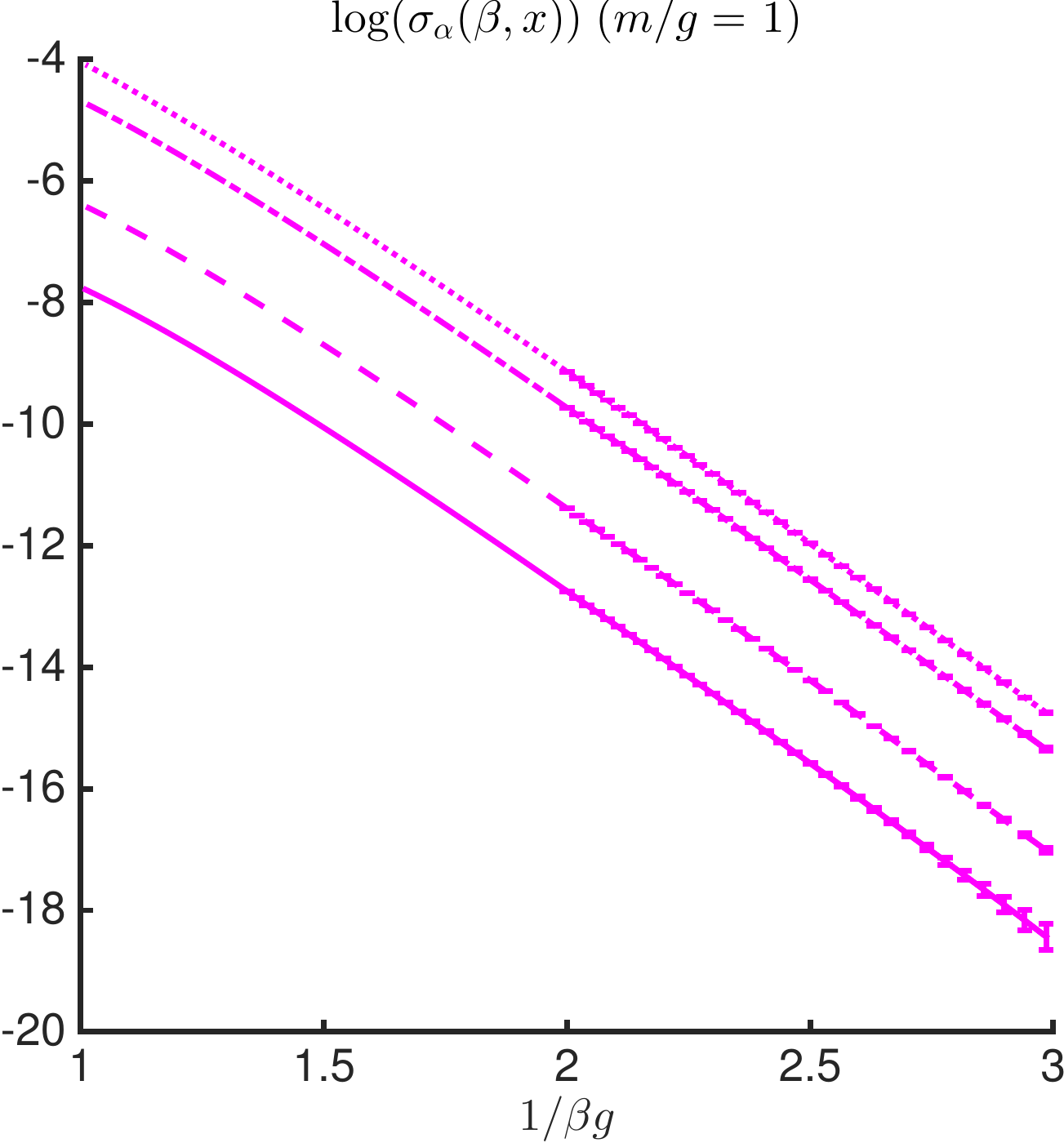}
\caption{\label{fig:STlargeTd}}
\end{subfigure}\hfill
\vskip\baselineskip
\captionsetup{justification=raggedright}
\caption{\label{fig:STlargeT} $x = 100$. Logarithm of the string tension at larger temperatures $T = 1/\beta g$ as a function of $T$. (a) $m/g = 0.125$, $\alpha = 0.05$ (full line), $\alpha = 0.1$ (dashed line). Convergence of our results when improving the precision $\epsilon$. (b) $m/g = 0.125$. Logarithm of string tension for $\alpha = 0.05$ (full line), $\alpha = 0.1$ (dashed line),$\alpha = 0.25$ (dotted line), $\alpha = 0.4$ (dashed dotted line). (c) Same as (b), now for $m/g = 0.25$. (d) Same as (b), now for $m/g = 1$. The error bars in (b), (c) and (d) for $\beta g \geq 2$ are obtained by comparing the simulation for $\epsilon = 10^{-9}$ with the simulation for $\epsilon = 2 \times 10^{-9}$. }
\end{figure}

We thus expect for $\alpha$ sufficiently small that the quantities $\sigma_{\alpha}(\beta,x)/g^2\alpha^2$, $E_{\alpha}(\beta,x)/g\alpha$, $\Delta \Sigma_{\alpha}(\beta,x)/g\alpha^2$ and $\Delta \mathcal{S}_{\alpha}(\beta,x)/g^2\alpha^2$ are independent of $\alpha$ up to order $\alpha^2$. This is precisely what we find in fig. \ref{fig:quantPert} for $\alpha = 0.05$ and $\alpha = 0.1$.  Note that as a consequence of (\ref{eq:PertExp}), the curves of the string tension and the electric field are similar, see figs. \ref{fig:FreeEnergyPert} and \ref{fig:EFPert}. For $\alpha \gtrsim 0.25$ the expansion (\ref{eq:PertExp}) is not in general valid anymore and higher-order corrections can become important. 

When $\beta g$ is large compared to the mass gap we find that all quantities have converged to their zero-temperature value, i.e. the system is effectively at zero temperature. The value of $(\beta g)_0$ above which all ensemble averages are close to their ground-state expectation agrees with the value of $\beta g$ above which $\Delta \mathcal{S}_{\alpha} \approx 0$.  This justifies taking the UV-finite renormalized entropy $\Delta \mathcal{S}_{\alpha}$ as a measure to quantify thermal fluctuations in the Gibbs state. We observe that $(\beta g)_0$  is larger for smaller values of $m/g$. This is explained by the fact that for the values of $\alpha$ we considered here, the mass gap of $H_\alpha$ increases with $m/g$ \cite{Buyens2015b,Coleman1976}. For $\alpha = 0.25$ we observe in figs. \ref{fig:FreeEnergyPert} and \ref{fig:EFPert} that the string tension and the electric field have converged to their ground-state expectation values (stars) for $\beta g= 10$.  In particular at low temperatures, large $\beta g$, the heavy probe charges are always confined when $\alpha$ is noninteger.   
 
For $\beta g \lesssim 0.5$, the string tension is very small which suggests a transition from the confined phase to a deconfined plasma phase. In fact, we expect the string tension to decay exponentially with temperature at large values of $T = 1/\beta$ and the transition to occur exactly at infinite temperature $T/g = 1/\beta g = + \infty$. This is corroborated by studies in the strong coupling limit \cite{Fischler1979,Rodriguez1996} where they found that 
\be \label{eq:SThightemp} \sigma_{\alpha}(\beta) \sim 2mT\sin(\pi \alpha)^2e^{- \pi^{3/2}T/g} + \mathcal{O}(m^2),\ee
at high temperatures ($T/g = 1/\beta g \gg 1$). In figs. \ref{fig:STlargeTb}, \ref{fig:STlargeTc}, \ref{fig:STlargeTd} we indeed find that the logarithm of the string tension is almost linear as a function of $T$ for $T/g \in[1,3]$ or equivalently $\beta g \in [0.33,1]$. Note that because for these values of $\beta g$ the string tension is very small, we needed very small values of the tolerance $\epsilon$ and the step $d\beta \leq 5\times 10^{-3}$ to investigate this regime, see fig. \ref{fig:STlargeTa}. To obtain an idea of the error in $\epsilon$ we can compare our simulations for $\epsilon = 10^{-9}$ with our simulations for $\epsilon = 2 \times 10^{-9}$. The errors are sufficiently small for $\beta g \in [1,2]$ (of order $0.1$ or smaller). For $\beta g \gtrsim 2$ errors are larger and represented by the error bars in figs. \ref{fig:STlargeTb}, \ref{fig:STlargeTc} and \ref{fig:STlargeTd}. 

\begin{table}
\begin{tabular}{| c|  c ||   c | c | c | }
        \hline
     $m/g$ &  $\alpha$ &  $\log(A_\alpha)$ &    $B_\alpha$  & $C_\alpha$\\
     \hline
     \hline
   0.125    &  0.05&    2.5 (2) & 1.8 (2)&  6.4 (3) \\
   		& 0.1 & 2.5 (1) & 1.9 (2) & 6.5 (2) \\
   		& 0.25 & 2.3 (1) & 1.9 (2) & 6.4 (2) \\
		& 0.4  & 1.9 (2) & 1.8 (3) & 6.6 (1) \\
		\hline 
   0.5 & 0.05 &  4.05 (3) & 2.15 (8) & 6.6 (1)\\
	 & 0.1 & 4.02 (2) & 2.13 (6) & 6.6 (1)\\
	& 0.25 &  3.81 (2) & 2.05 (3) & 6.55 (3)\\
	& 0.4 & 3.4 (1) & 1.9 (1) & 6.5 (1) \\
	\hline 
  1 & 0.05 & 5.1 (2) & 2.3 (4) & 6.7 (8) \\
     & 0.1&  4.9 (1) & 1.8 (3)&  6.4 (3) \\
     & 0.25  & 4.8 (1) & 2.1 (3) & 6.6 (5) \\
    & 0.4 & 4.57 (1) & 2.34 (1) &  6.74 (2) \\
    \hline
\end{tabular}
\captionsetup{justification=raggedright}
\caption{\label{table:STHighT} $x = 100$. Coefficients of the fit eq. (\ref{eq:fitSTHT}) to our data for $m/g = 0.125, 0.5$ and $\alpha = 0.05,0.1,0.25$.}
\end{table}

In table \ref{table:STHighT} we show the coefficients obtained by fitting our results for $\log(\sigma_{\alpha}(\beta,x)/g^2)$ against
\be \label{eq:fitSTHT} f(T/g) = \log(A_\alpha \alpha^2) +  B_\alpha \log(T/g) - C_\alpha T/g \ee
which is equivalent to
$$ \sigma_{\alpha}(\beta,x)/g^2 \approx \alpha^2 A_\alpha e^{-C_\alpha T/g} (T/g)^{B_\alpha} (T = 1/\beta).$$ 
More specifically we used our data of $\log(\sigma_{\alpha}(\beta,x)/g^2)$ for $\epsilon = 10^{-9}$ and $\beta g \in [1, 2.5]$ with step $d\beta = 0.005/g$ and considered all possible fits against at least $10$ consecutive $\beta$ values. Each fit $\theta$ gave us an estimate $\left(A_\alpha^{(\theta)}, B_\alpha^{(\theta)}, C_\alpha^{(\theta)}\right)$. The final estimates for these parameters are obtained similarly as for the chiral condensate (section \ref{sec:CC}): we take the median of all the estimates weighed by $\exp(-\chi_\theta^2/N_{dof}^\theta)$. The latter is now defined as 
$$\chi_\theta^2 = \sum_{\beta_j \in \mbox{fit } \theta} \left(\frac{f_\theta (1/\beta_j) - \log(\sigma_{\alpha}(\beta_j,x)/g^2)}{\Delta^{\epsilon}\log(\sigma_{\alpha}(\beta_j,x)/g^2)}\right)^2 $$
with 
$$ f_\theta(T/g) = \log(A_\alpha^{(\theta)} \alpha^2) +  B_\alpha^{(\theta)} \log(T/g) - C_\alpha^{(\theta)} T/g $$
and $\Delta^{\epsilon}\log(\sigma_{\alpha}(\beta_j,x)/g^2)$ the difference in magnitude of $\log(\sigma_{\alpha}(\beta_j,x)/g^2)$ computed for $\epsilon = 10^{-9}$ and $\epsilon = 2 \times 10^{-9}$. The error on our results shown in table \ref{table:STHighT} comes from the $68.3 \%$ confidence interval. 

For all values of $m/g$ we observe that $A_{0.05} \approx A_{0.01}$ which is a consequence of (\ref{eq:PertExp}). Furthermore, we also find that $B_{\alpha}$ is within $10 \%$ of $2$ for all values of $m/g$ considered here. The values of $C_\alpha$ obviously show that the string tension is exponentially suppressed for $T/g \gtrsim 2$. Note, however, that already for $m/g = 0.125$ the value of $C_\alpha$ deviates from the $C_\alpha^{0} =\pi^{3/2} \approx 5.56$ in the strong coupling limit (\ref{eq:SThightemp}).

 We conclude that we have confinement for all finite values of the temperature $T/g$, but for $T/g \gtrsim 2$, or equivalently $\beta g \lesssim 0.5$, the string tension is exponentially suppressed with $T/g$. At high temperatures, the string tension can thus only be observed if we would separate the heavy charges by a distance which scales exponentially in the temperature. In an experimental setting, this means that the heavy charges are actually deconfined for $\beta g \lesssim 0.5$.

\subsection{CT symmetry restoration at nonzero T}\label{subsec:alphaclosetohalf}  

\noindent At zero temperature there is a phase transition for $\alpha = 1/2$ and $m/g = (m/g)_c \approx 0.33$ \cite{Byrnes2002}. For $m/g \leq (m/g)_c$ the ground state is CT invariant whereas for $m/g \geq (m/g)_c$ the CT symmetry is spontaneously broken to $T^2$. The vacuum is still invariant under translation over two sites and is two-fold degenerate. A detailed study of this phase transition was performed by Byrnes \emph{et al.} \cite{Byrnes2002}. Their results for the critical indices, $\nu = 0.99(1)$ and $\beta/\nu = 0.125(5)$, gave strong evidence that the phase transition lies in the universality class of the transverse Ising model or equivalently of the 2D classical Ising model \cite{Onsager1944}. For the transverse Ising model the phase transition is determined by the $\mathbb{Z}_2$ symmetry. When this symmetry is spontaneously broken, the magnetization gains a nonzero expectation value.  Here the CT symmetry of the Schwinger model for $\alpha = 1/2$ plays the role of the $\mathbb{Z}_2$-symmetry and the electric field plays the role of the magnetization. 

Besides this phase transition the pattern of the eigenvalues of the Schwinger model at $\alpha = 1/2$ in the symmetry broken regime bears a remarkable resemblance to the transverse Ising model \cite{Byrnes2002}. Because of the similarities between both models, we might expect that also at finite temperature there are some analogies. In particular, because the spontaneous symmetry breaking in the transverse Ising model occurs only at zero temperature we might expect this also to be the case for the Schwinger model.  Furthermore, general theorems like for instance the Mermin-Wagner theorem \cite{Mermin1966} or Peierls argument \cite{Peierls1936} suggest that at finite temperature no spontaneous symmetry breaking occurs in one spatial dimension. However the former is not applicable as the CT transformation is discrete whereas the latter might not apply because the local dimension of the Hilbert space is infinite. Therefore it is a priori not sure whether the CT symmetry is restored at any finite temperature.  

\begin{figure}[t]
\begin{subfigure}[b]{.24\textwidth}
\includegraphics[width=\textwidth]{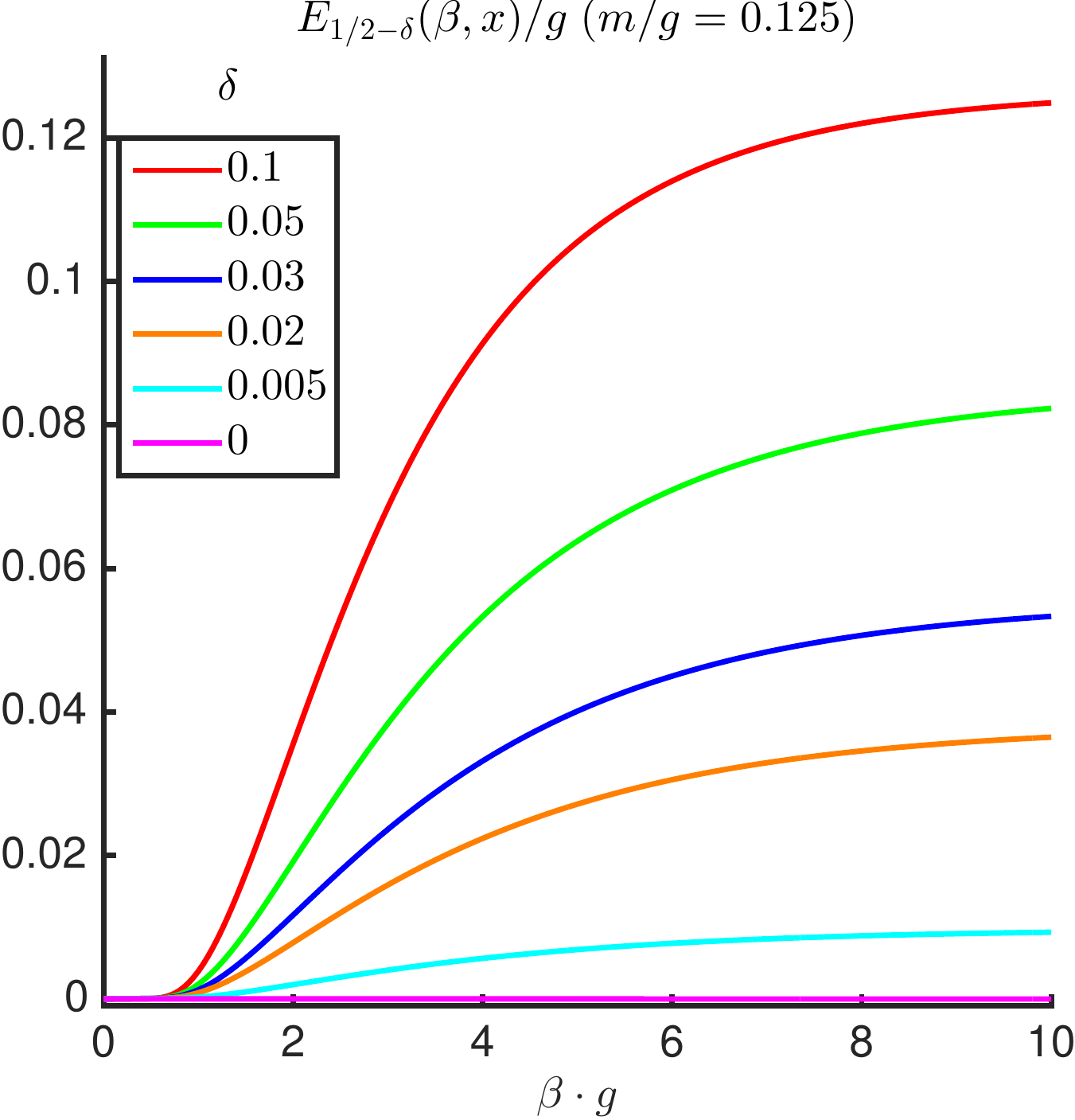}
\caption{\label{fig:ElectricFieldDiffQa}}
\end{subfigure}\hfill
\begin{subfigure}[b]{.24\textwidth}
\includegraphics[width=\textwidth]{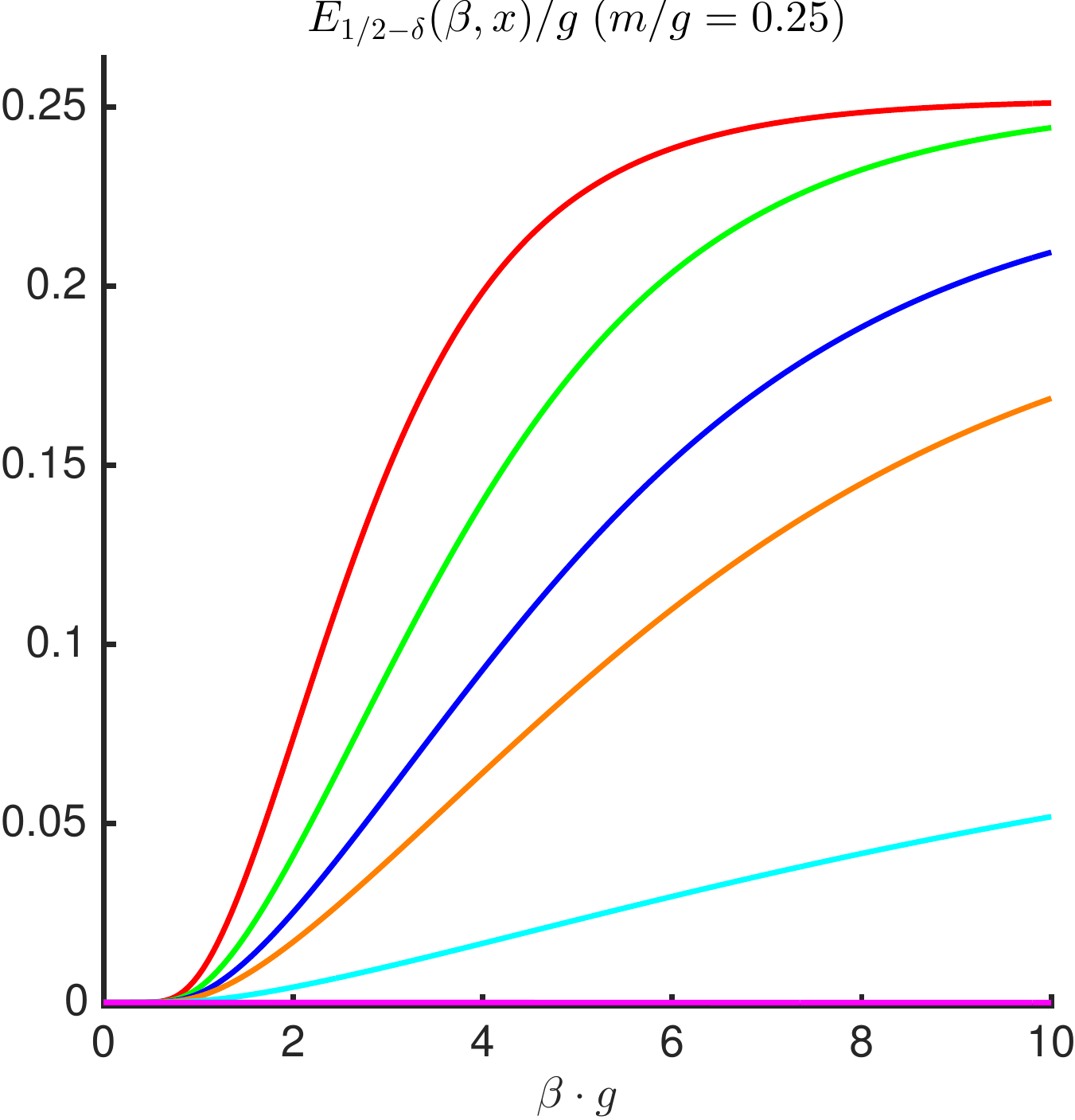}
\caption{\label{fig:ElectricFieldDiffQb}}
\end{subfigure}\hfill
\vskip\baselineskip
\begin{subfigure}[b]{.24\textwidth}
\includegraphics[width=\textwidth]{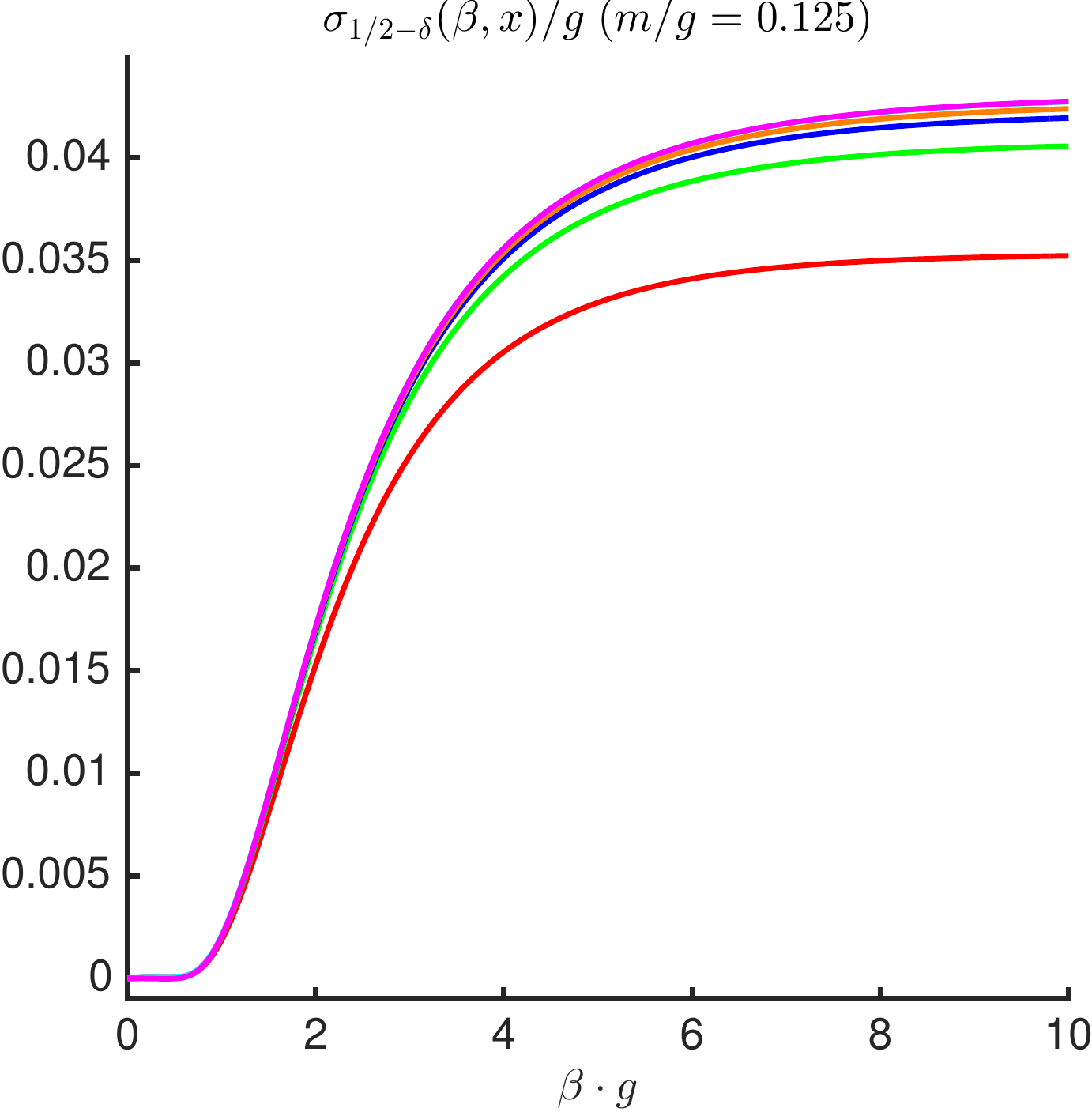}
\caption{\label{fig:FreeEnergyDiffQa}}
\end{subfigure}\hfill
\begin{subfigure}[b]{.24\textwidth}
\includegraphics[width=\textwidth]{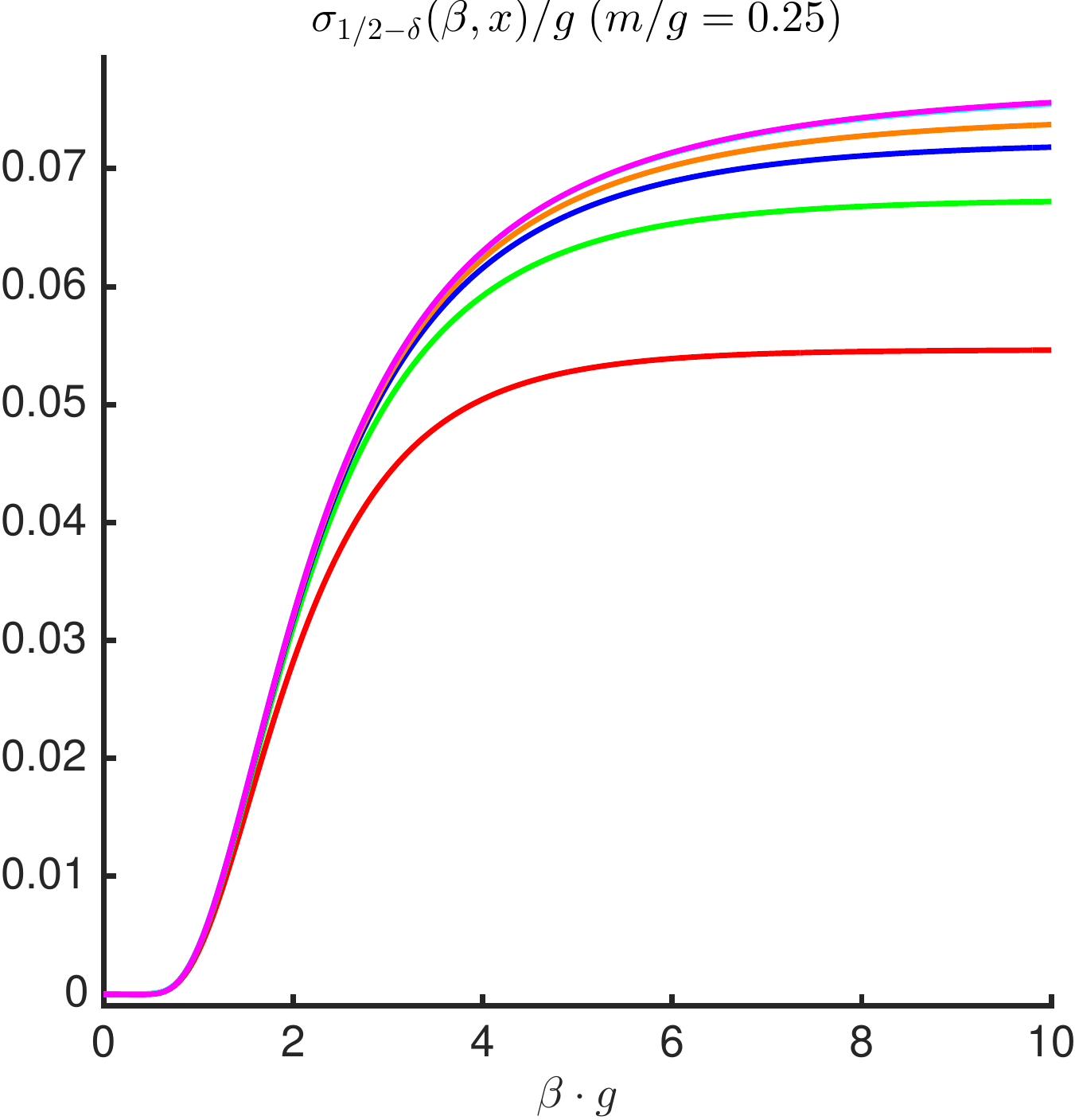}
\caption{\label{fig:FreeEnergyDiffQb}}
\end{subfigure}\hfill
\captionsetup{justification=raggedright}
\vskip\baselineskip
\caption{\label{fig:ElectricFieldDiffQ} $x = 100$. $\alpha = 1/2 - \delta$ for different values of $\delta$. Left: $m/g = 0.125$. (a) Electric field. (c) String tension.  Right: $m/g = 0.25$. (b) Electric field. (d) String tension. }
\end{figure}

The spontaneous symmetry breaking of the CT symmetry is detected by investigating the left/right limits ($\alpha=1/2\pm\delta$):
\be \label{eq:limitsCT}\lim_{\delta \rightarrow 0+} E_{1/2 + \delta}(\beta)  \mbox{ and } \lim_{\delta \rightarrow 0+} E_{1/2 - \delta}(\beta).   \ee
When these limits are different, the symmetry is spontaneously broken. Because of CT symmetry we thus need to check whether
$$ \lim_{\delta {\rightarrow} 0}E_{1/2 -   \delta}(\beta) = 0. $$
\\
\\
For $m/g = 0.125$ and $m/g = 0.25$ we find for all values of $\beta g$ that $E_{1/2 + \delta}(\beta)$ decreases to zero as $\vert \delta \vert \rightarrow 0$, see figs. \ref{fig:ElectricFieldDiffQa} and  \ref{fig:ElectricFieldDiffQb}. This leads us to the conclusion that $\lim_{\delta \rightarrow 0}E_{1/2 - \delta}(\beta,x) = 0$. We can actually perform direct simulations for $\alpha = 1/2$ and find numerically that  $\vert E_{1/2}(\beta,x) \vert \lesssim 5 \times 10^{-5}$. This can be improved by requiring better accuracy of our simulations, see appendix \ref{subsec:sima1half}. So, similarly as for zero temperature we observe for $m/g \lesssim (m/g)_c$ that $E_{1/2}(\beta,x) = 0$; implying that there is no CT symmetry breaking. Because $E_{\alpha}(\beta,x) = \partial \sigma_{\alpha}(\beta,x)/\partial{(g\alpha)}$, the string tension reaches its maximum for $\alpha = 1/2$. This is indeed what can be seen in figs. \ref{fig:FreeEnergyDiffQa} and \ref{fig:FreeEnergyDiffQb}: for all values of $\beta g$ the string tension increases monotonically as a function of $\alpha$ to its value at $\alpha = 1/2$ when $\delta$ tends to zero. At large temperatures, the string tension shows, similarly as in subsection \ref{subsec:deconfinement}, deconfinement for $T \rightarrow + \infty$. 

\begin{figure}[t]
\begin{subfigure}[b]{.24\textwidth}
\includegraphics[width=\textwidth]{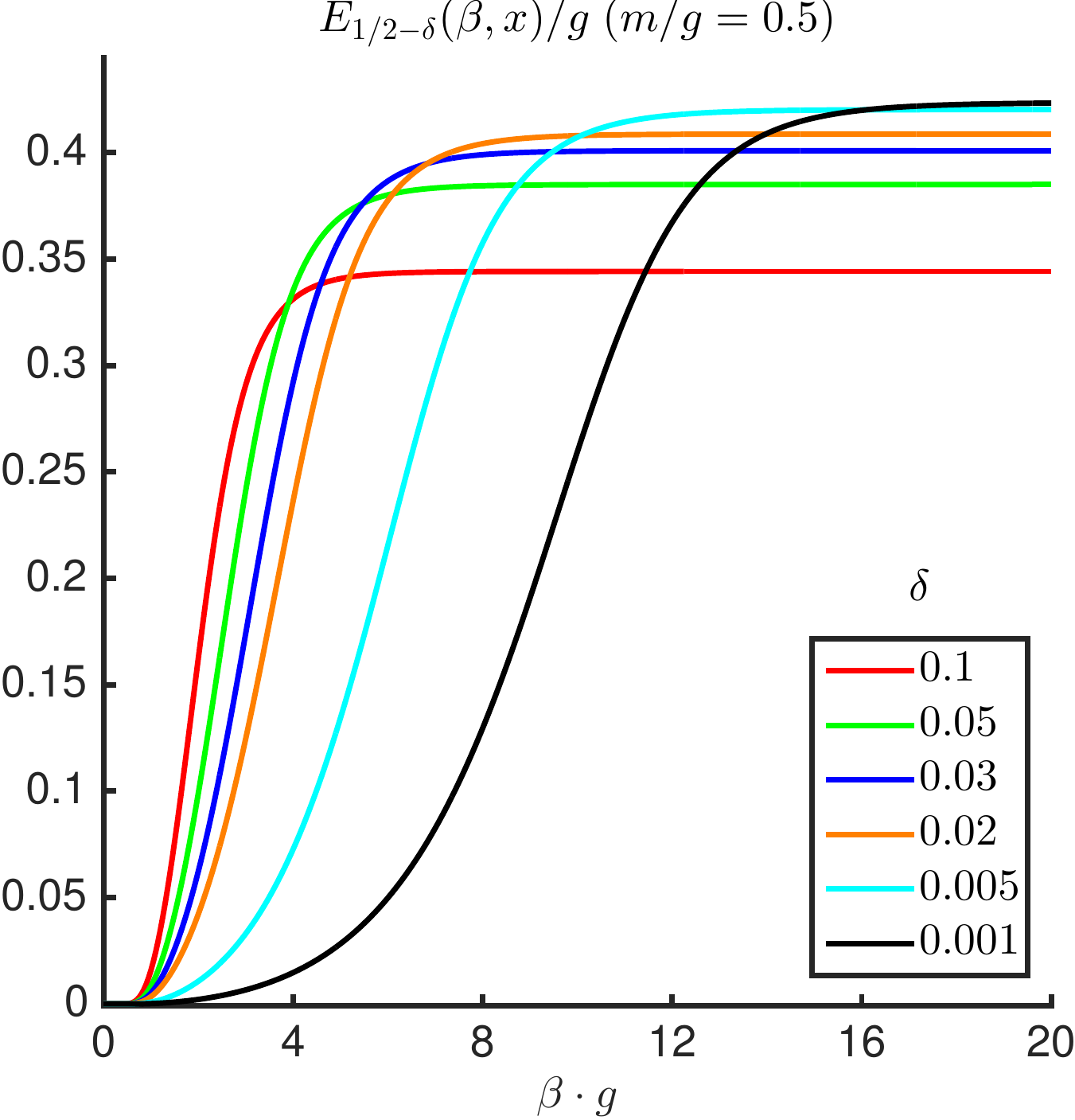}
\caption{\label{fig:ElectricFieldDiffQc}}
\end{subfigure}\hfill
\begin{subfigure}[b]{.24\textwidth}
\includegraphics[width=\textwidth]{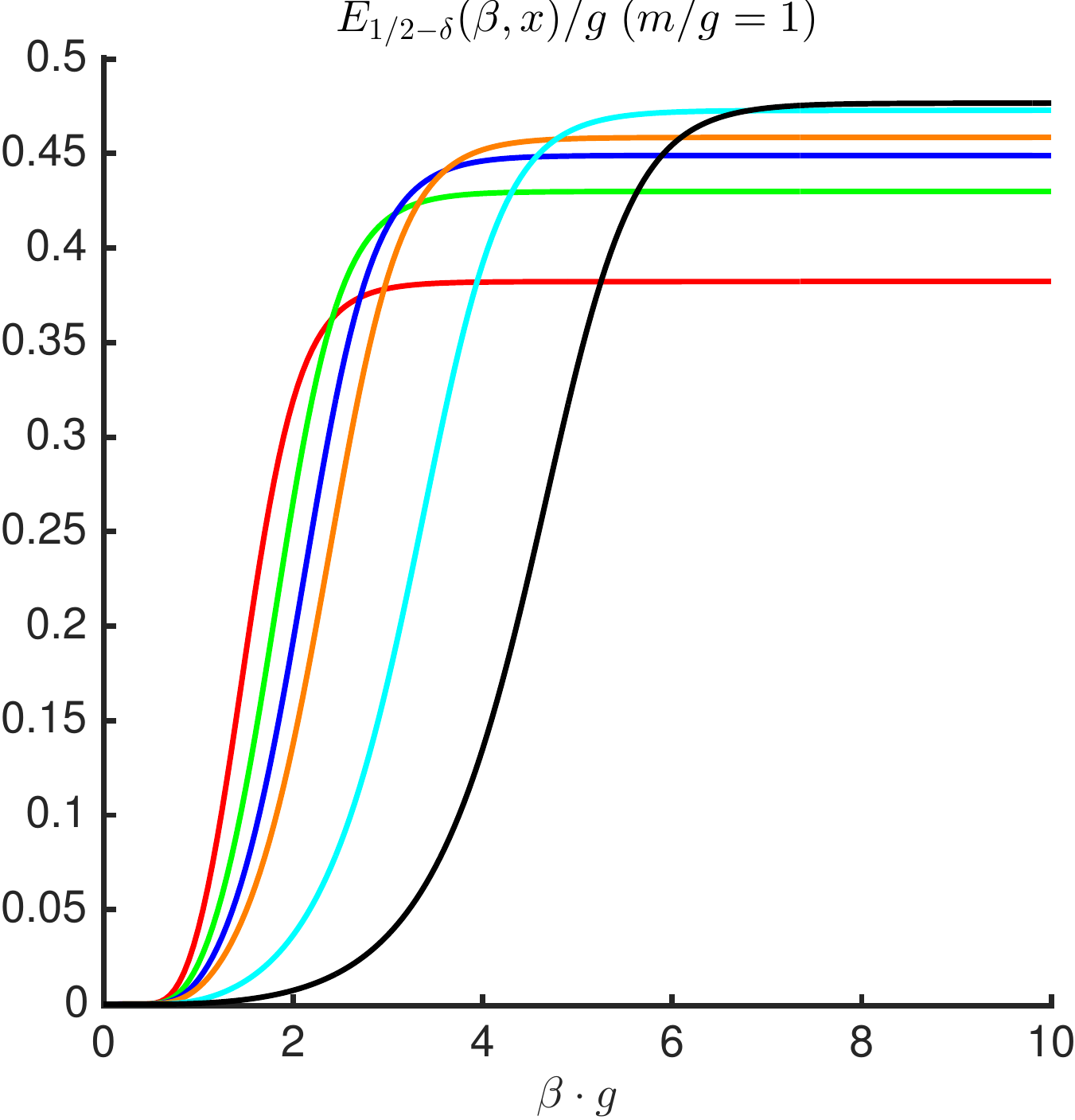}
\caption{\label{fig:ElectricFieldDiffQd}}
\end{subfigure}\hfill
\vskip\baselineskip
\begin{subfigure}[b]{.24\textwidth}
\includegraphics[width=\textwidth]{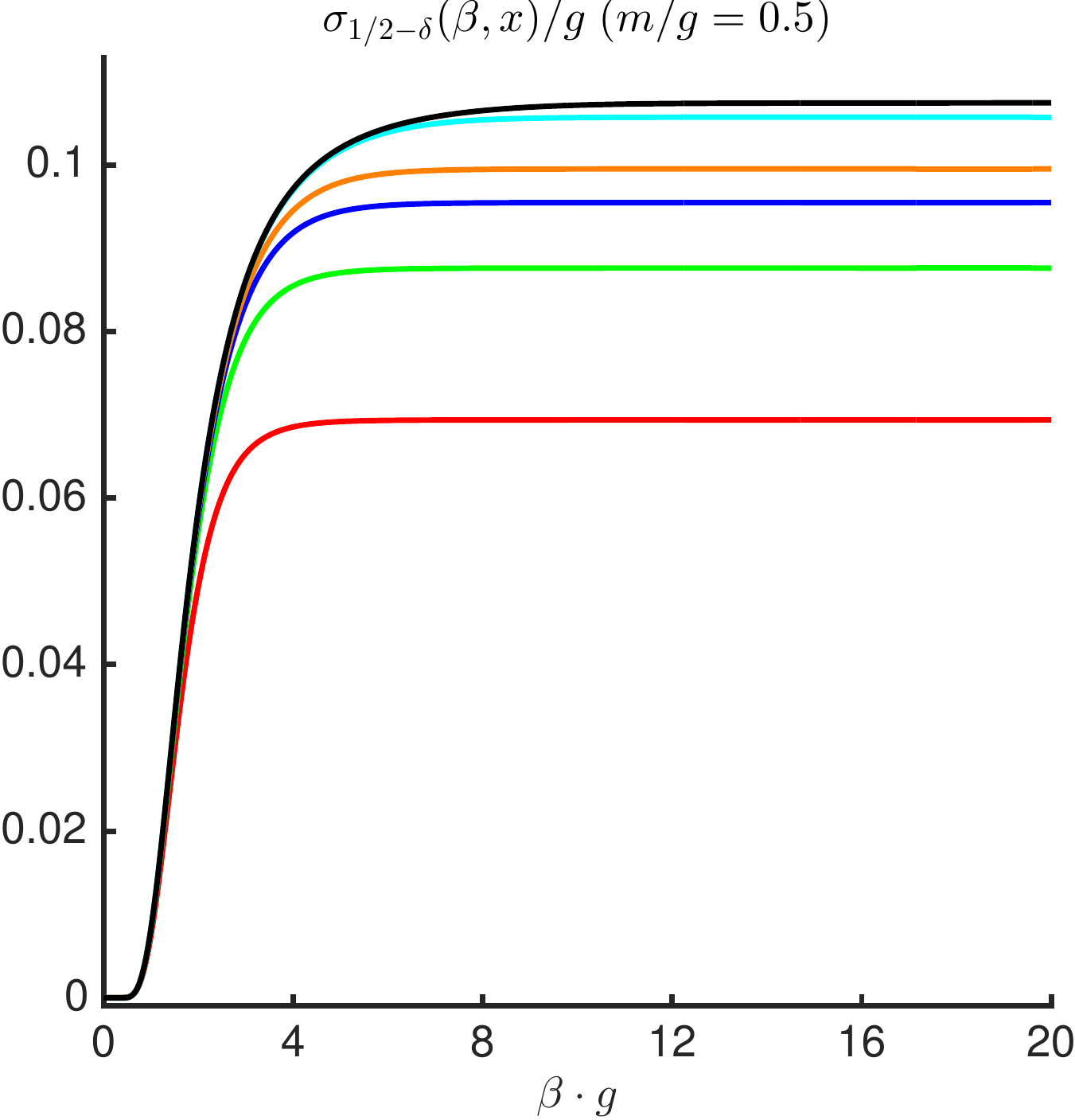}
\caption{\label{fig:FreeEnergyDiffQc}}
\end{subfigure}\hfill
\begin{subfigure}[b]{.24\textwidth}
\includegraphics[width=\textwidth]{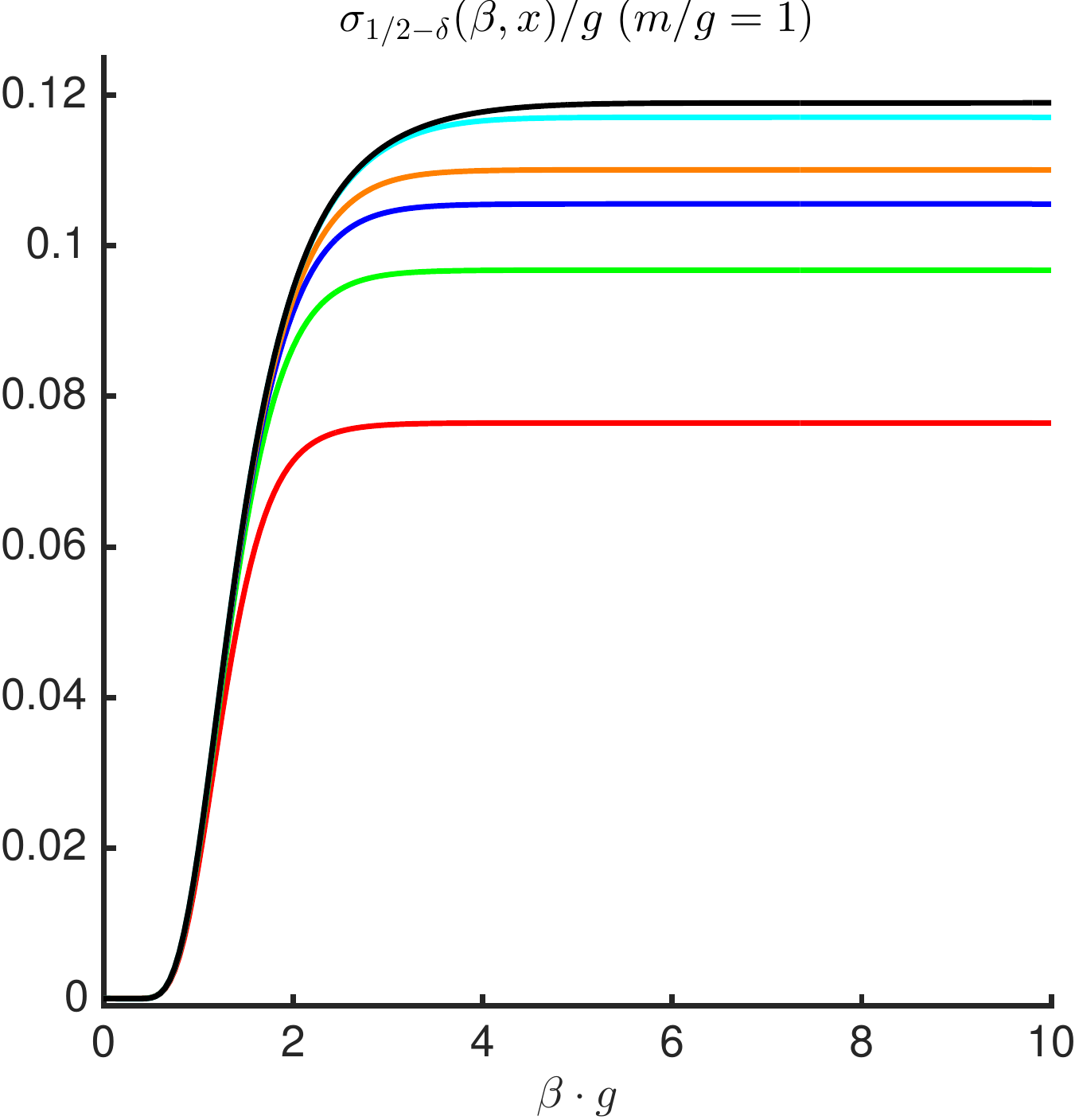}
\caption{\label{fig:FreeEnergyDiffQd}}
\end{subfigure}
\vskip\baselineskip
\captionsetup{justification=raggedright}
\caption{ $x = 100$. $\alpha = 1/2 - \delta$ for different values of $\delta$. Left: $m/g = 0.5$. (a) Electric field. (c) String tension.  Right: $m/g = 1$. (b) Electric field. (d) String tension. }
\end{figure}

Contrary to the case $\alpha = 0.1$ and $\alpha = 0.25$, we find that neither the free energy nor the electric field has entirely converged to its ground-state expectation value at $\beta g = 10$. This is what we would expect from our numerical simulations in \cite{Buyens2015b}. There we found that for $m/g = 0.125$ and $m/g = 0.25$ the mass gap decreases when $\alpha$ tends to $1/2$; consistent with the phase transition that occurs at $(m/g,  \alpha) = \bigl((m/g)_c,1/2)$. 
\\
For $m/g \geq (m/g)_c$, at the exact value $\alpha=1/2$, the simulations are not reliable anymore. Due to spontaneous symmetry breaking the ground state is now two-fold degenerate and for large values of $\beta g$ the iTEBD algorithm pushes the Gibbs state during the evolution either to the ground state $\ket{\Psi_{1/2-}}$ of $H_{1/2 - \delta}$ or to the ground state $\ket{\Psi_{1/2 +}}$ of $H_{1/2 + \delta}$ in the limit $\delta \rightarrow 0+$, see appendix \ref{subsec:sima1half}. 

To examine the CT symmetry breaking or restoration we need to consider nonzero $\delta > 0$. For small values of $\beta g$ we find that the electric field converges to zero when $\delta \rightarrow 0$, see figs. \ref{fig:ElectricFieldDiffQc} and \ref{fig:ElectricFieldDiffQd}. This indicates that there is no spontaneous symmetry breaking for small values of $\beta g$. 

For large values of $\beta g$ however, we find that even for $\delta = 0.001$ the electric field and string tension are still very close to the values in the spontaneous broken ground state $\ket{\Psi_{1/2 -}}$. But notice that the $\delta$-dependence of the observables in the intermediate temperature region suggests that the $\delta\rightarrow 0$ limit has not been reached yet.  This is corroborated by a study in the weak coupling limit, see appendix \ref{sec:thermalwc}, where we argue that thermal corrections to ground-state expectation values can only be relevant if
\be \label{eq:Km} \delta \lesssim K_m\frac{e^{-2\beta m}}{\beta}  \ee
with $K_m$ positive and independent of $\beta$. This implies that to observe thermal corrections to ground-state expectation values we should take $\delta$ exponentially small in $\beta m$.

\begin{figure}[t]
\begin{subfigure}[b]{.24\textwidth}
\includegraphics[width=\textwidth]{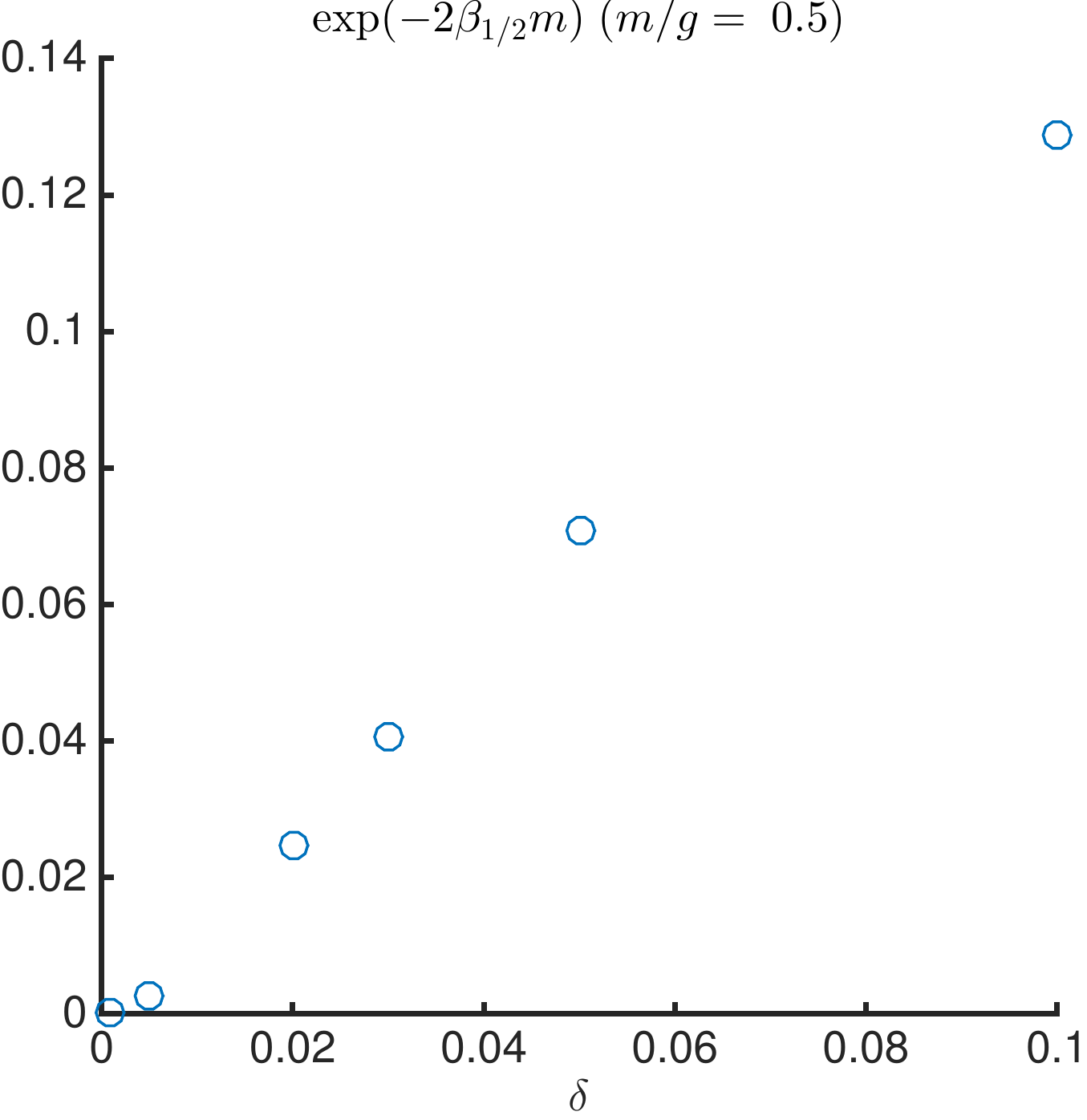}
\caption{\label{fig:betahalfa}}
\end{subfigure}\hfill
\begin{subfigure}[b]{.24\textwidth}
\includegraphics[width=\textwidth]{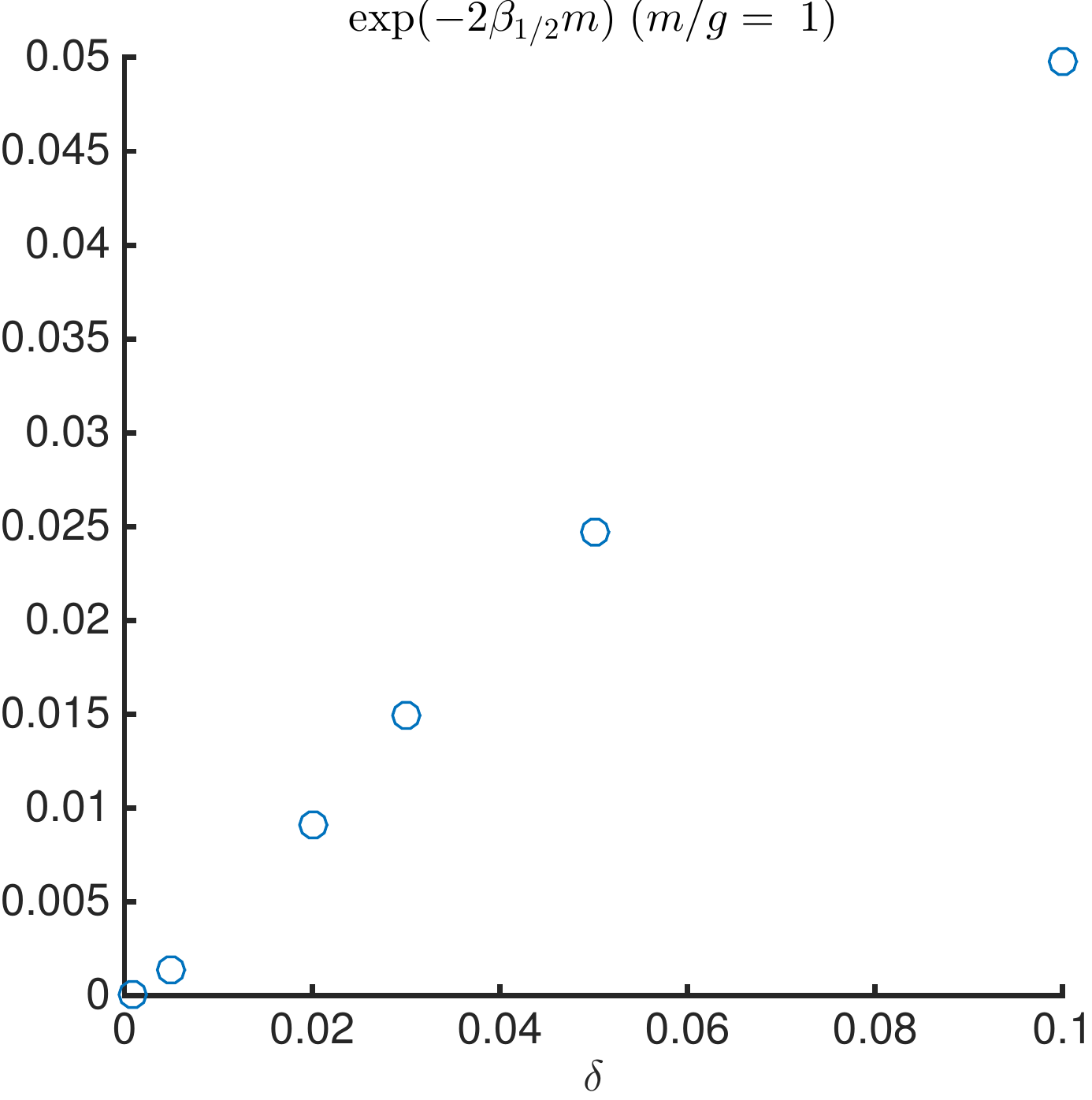}
\caption{\label{fig:betahalfb}}
\end{subfigure}\hfill
\vskip\baselineskip
\captionsetup{justification=raggedright}
\caption{\label{fig:betahalf} $x = 100$. $\exp(-2\beta_{1/2}m)$ as a function of $\delta$. (a) $m/g = 0.5$. (b) $m/g = 1$.}
\end{figure}

From the numerical point of view it is hard to simulate such small values of $\delta$. For instance, for $(m/g,\delta) = (1,0.001)$ we had to lower $\epsilon$ to $10^{-7}$ and during the evolution the bond dimension of our MPS representation already reached $267$. To examine the $\delta\rightarrow 0$ limit from our simulated $\delta$-values we investigate the scaling of $\beta_{1/2}$, the value of $\beta$ where the electric field $E_{1/2 - \delta}(\beta_{1/2},x)$ equals half of its ground-state expectation value $E_{1/2 - \delta}(\beta = +\infty,x)$. Motivated by (\ref{eq:Km}) we plot in fig. \ref{fig:betahalf} $\exp(-2\beta_{1/2}m)$ as a function of $\delta$. We seem to find there that $\exp(-2\beta_{1/2}m) \rightarrow 0$ as $\delta \rightarrow 0$, or equivalently that $\beta_{1/2} \rightarrow + \infty$ for $\delta \rightarrow 0$. This indicates that the curve of the electric field tends to a function which is zero for all finite values of $\beta$ as $\delta \rightarrow 0$.  Hence it seems, similar to the 1D quantum transverse Ising model, that the spontaneous symmetry breaking vanishes at all finite inverse temperatures and a phase transition would occur exactly at $T = 0$.

Similarly as for $m/g = 0.125,0.25$ the string tension, figs. \ref{fig:FreeEnergyDiffQc} and \ref{fig:FreeEnergyDiffQd}, converges nicely to its maximum for $\delta \rightarrow 0$ for all values of $\beta g$. At high temperatures we find again deconfinement for $T/g = 1/\beta g \rightarrow +\infty$.

To conclude, in this subsection we investigated the Schwinger model with an electric background field $\alpha$ close to $1/2$. We considered the values $m/g = 0.125,0.25,0.5$ and $1$. For all these values we find strong indications for the existence of CT symmetry at any nonzero temperature. In particular for $m/g>(m/g)_c\approx 0.33$, where the CT symmetry is broken at zero temperature \cite{Byrnes2002}, our results imply a restoration of the CT symmetry at any nonzero temperature. This is similar to what happens with the $\mathbb{Z}_2$ symmetry of the transverse Ising model and thus lends further support to the purported relationship between the Schwinger model at $\alpha = 1/2$ and the transverse Ising model, as suggested in \cite{Byrnes2002}.

\section{Conclusion}
\noindent In this paper we investigated the Schwinger model in thermal equilibrium within the framework of MPO. We computed the chiral condensate and found agreement with the analytical result for $m/g = 0$ and agreement with \cite{Banuls2016} in the nonperturbative regime. We also investigated the asymptotic aspects of confinement by considering a heavy quark-antiquark pair with fractional charge $g\alpha$, separated over an infinite distance. We find a nonzero string tension and therefore confinement for all values of $m/g$. However, at large temperatures $T\gtrsim 2 g$ we find that the string tension decays exponentially with the temperature. We also considered the case when $\alpha$ tends to $1/2$ and investigated the spontaneous breaking of the CT symmetry at finite temperature. Our results indicate that the spontaneous symmetry breaking vanishes at any nonzero temperature which implies that there is only a phase transition at zero temperature. We thus found two phase transitions that occur in limiting cases only: infinite temperature or zero temperature. 

Our simulations show that the MPO framework offers a reliable approach to study the non-perturbative regime of one dimensional gauge field theories. However, even within the Schwinger model there remain a lot of fascinating things to explore. For instance one can investigate string breaking at finite temperature between the probe charges when they are separated by a finite distance, similar to \cite{Buyens2015}. One can also explore confinement in a dynamical setting \cite{Prosen2009}. Another interesting path is to generalize this setup to non-Abelian gauge field theories.  From the theoretical point of view our approach can be generalized straightforwardly, but the implementation and simulations are a bit more tedious. 

More challenging is the step to higher dimensions. The analog of the MPO goes by the name of PEPS operators \cite{Orus2012}. Like for MPO, It has also been shown that they give a faithful and efficient representation of Gibbs states in two dimensions \cite{Molnar2015}. Progress in the last decade enabled the simulation of toy models \cite{Czarnik2012,Czarnik2015} at finite temperature. Given the potential of the PEPS operators it is certainly worthwhile exploring this direction further in the future to simulate gauge field theories at finite temperature.
\\
\\
\noindent \emph{Acknowledgements} We acknowledge interesting discussions with Jutho Haegeman, Hana Saito, Karl Jansen, Krzysztof Cichy and Mari Carmen Ba\~{n}uls. We also thank the authors of \cite{Banuls2016} for sharing their data with us. This work is supported by an Odysseus grant from the FWO, a PhD-grant from the FWO (B.B), the FWF grants FoQuS and Vicom, the ERC grant QUERG and the EU
grant SIQS.

\bibliography{paperFT}

\onecolumngrid
\newpage
\appendix
\numberwithin{equation}{section}
\renewcommand\theequation{\Alph{section}.\arabic{equation}}

\section{Details on the continuum extrapolation of the chiral condensate for $\alpha = 0$}
\noindent In this section we will explain the details on how to obtain a reliable continuum estimate for the (subtracted) chiral condensate from our results at finite $x$. In subsection \ref{subsec:errorsFinD} we discuss the errors originating from truncating the entanglement spectrum ($\epsilon > 0$) and taking finite steps for the imaginary time evolution ($d\beta > 0$) which includes the error of taking a finite bond dimension in the MPS representation. In subsection \ref{subsec:appendixCCcontinuuma} we go more into detail on the continuum extrapolations and the uncertainty originating from the choice of fitting interval and the choice of fitting ansatz. 

\subsection{Errors originating from taking finite values for $\epsilon$ and $d\beta$}\label{subsec:errorsFinD}
\begin{figure}[t]
\null\hfill
\begin{subfigure}[b]{.40\textwidth}
\includegraphics[width=\textwidth]{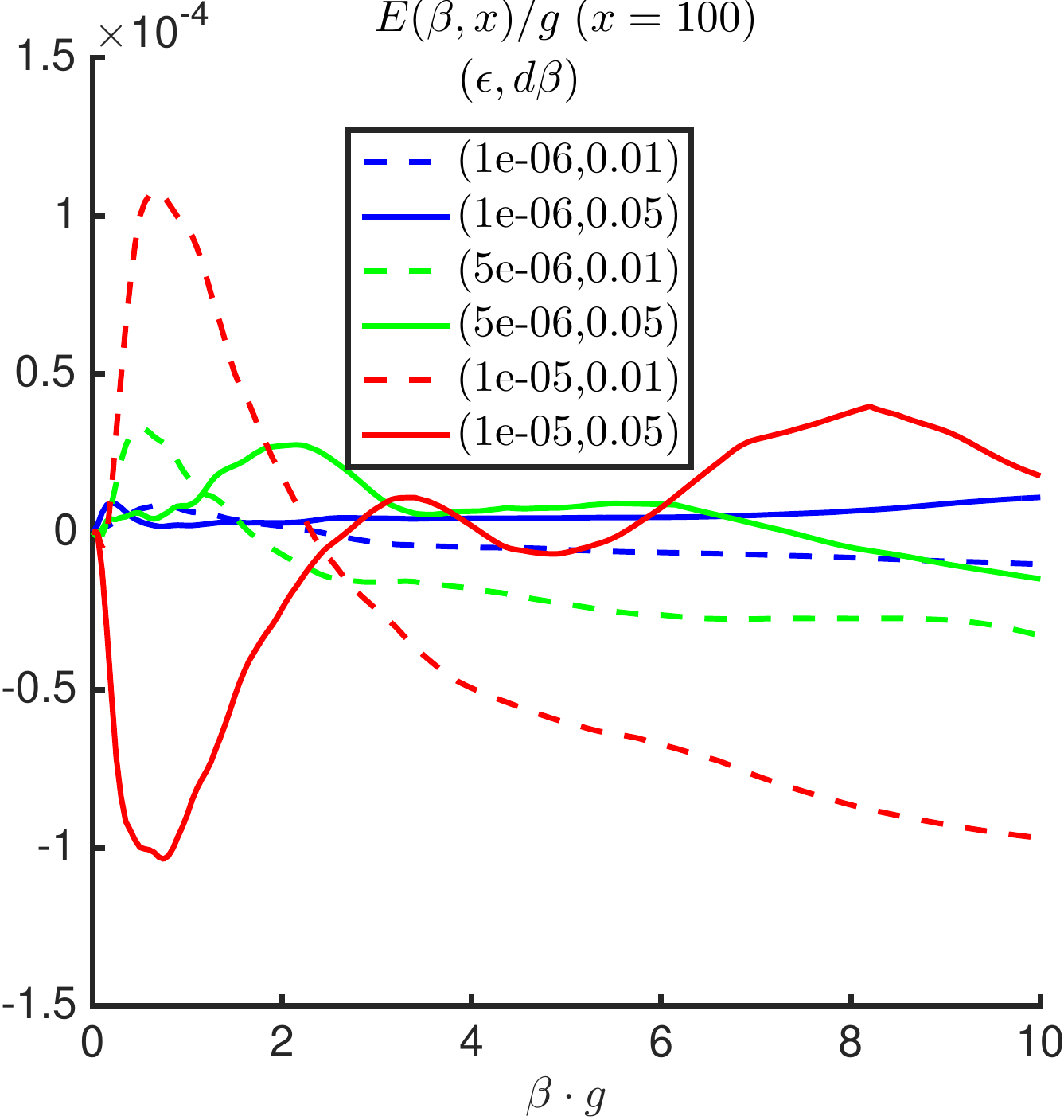}
\caption{\label{fig:diffbetadifftolEFa}}
\end{subfigure}\hfill
\begin{subfigure}[b]{.40\textwidth}
\includegraphics[width=\textwidth]{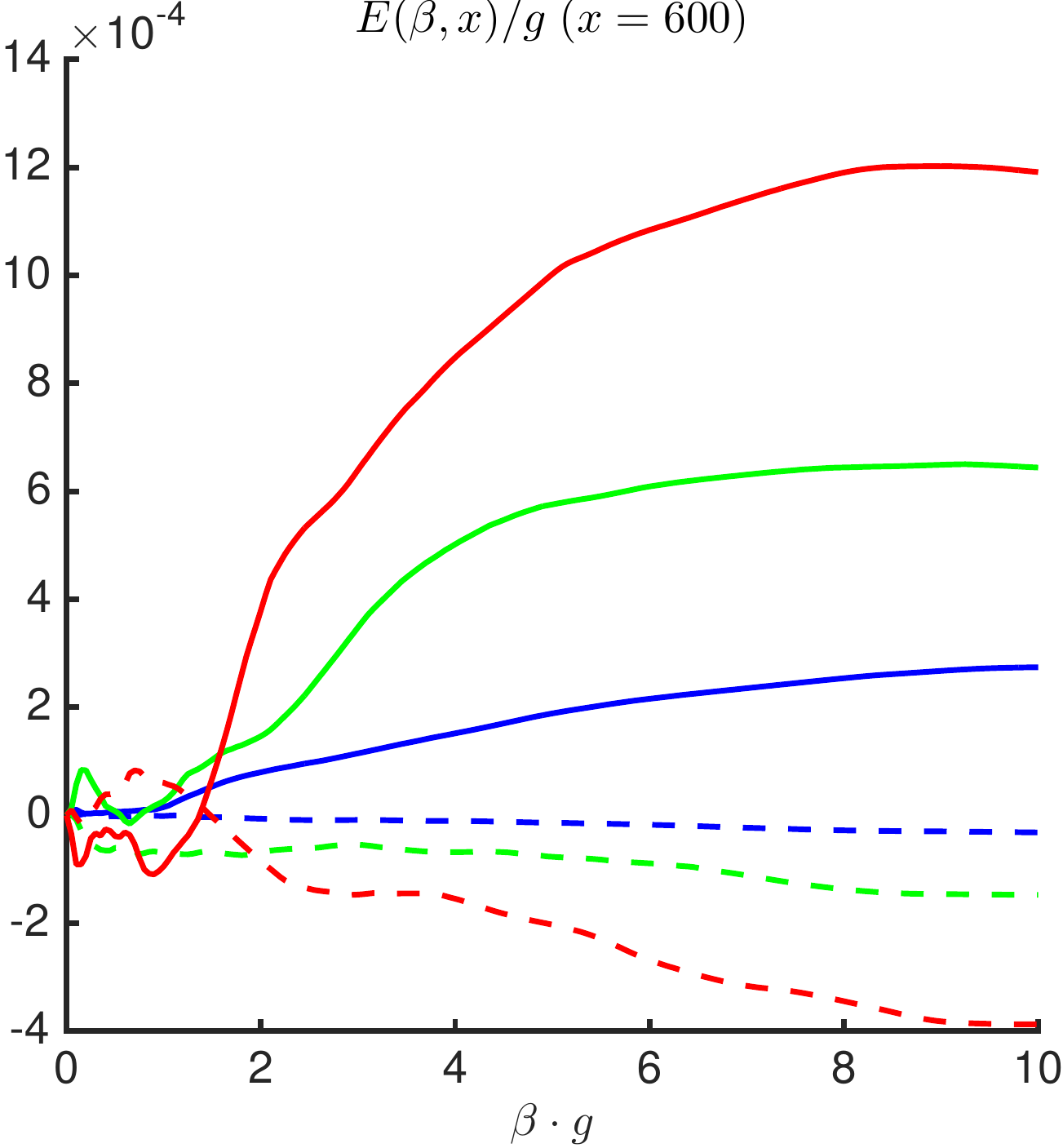}
\caption{\label{fig:diffbetadifftolEFb}}
\end{subfigure}\hfill\null
\vskip\baselineskip
\null\hfill
\begin{subfigure}[b]{.40\textwidth}
\includegraphics[width=\textwidth]{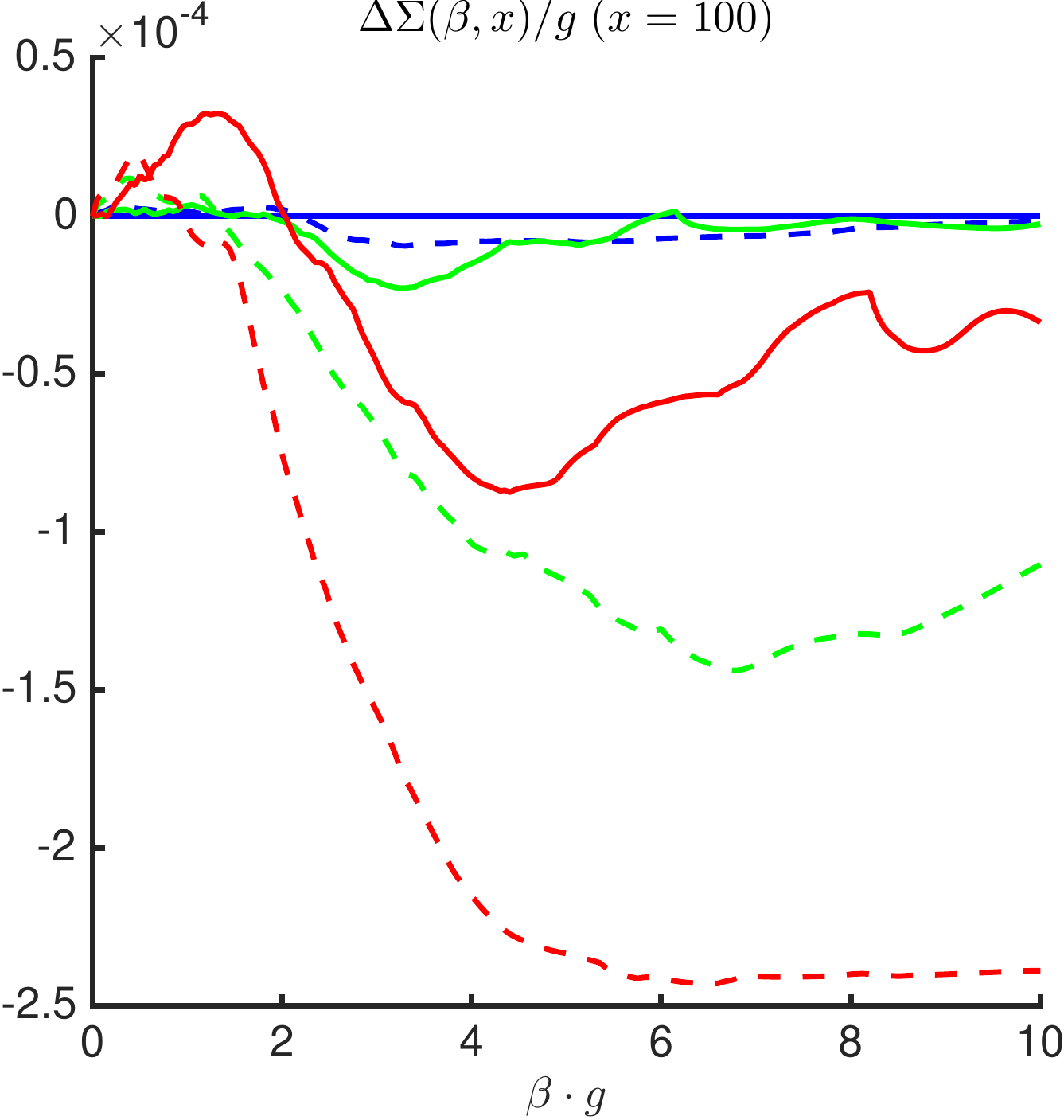}
\caption{\label{fig:diffbetadifftolEFc}}
\end{subfigure}\hfill
\begin{subfigure}[b]{.40\textwidth}
\includegraphics[width=\textwidth]{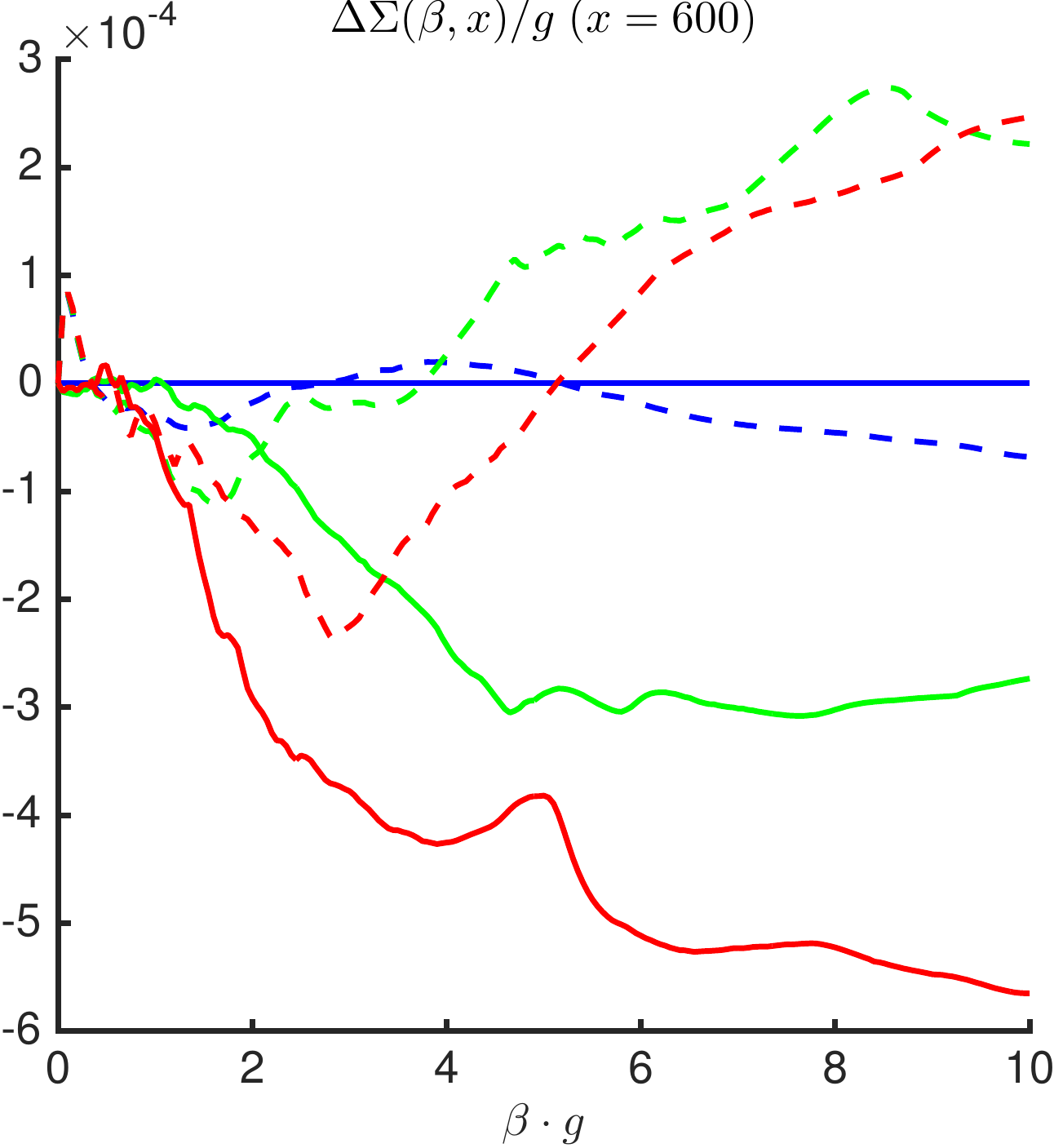}
\caption{\label{fig:diffbetadifftolEFd}}
\end{subfigure}\hfill\null
\vskip\baselineskip
\captionsetup{justification=raggedright}
\caption{\label{figapp:diffbetadifftolEF} $m/g = 0.25$. Simulations for different values of $(\epsilon,d\beta)$. (a-b) Electric field which should be zero for $\alpha = 0$. (a) $x = 100$. (b) $ x= 600$. (c-d) Chiral condensate $\Sigma(\beta,x)$ with respect to its value for $(\epsilon,d\beta) = (10^{-6},0.05)$. (c) $x = 100$. (d) $x = 600$.}
\end{figure}

\noindent In this subsection we address the errors originating from taking nonzero values for $\epsilon$ and $d\beta$. Recall that after the application of a Trotter gate we discard all the Schmidt values of the purification $\ket{\Psi[A(\beta)]}$ smaller than $\epsilon^2$. The fourth-order Trotter decomposition also produces an error $(d\beta)^5$ for each step $d\beta$. In the limits $d\beta, \epsilon \rightarrow 0$ the thermal evolution should lead to the exact representation of the thermal state (note that the error of truncating the spectrum of the electric field is under control, see subsection \ref{subsec:iTEBD}). In figs. \ref{fig:diffbetadifftolEFa} and \ref{fig:diffbetadifftolEFb} we observe indeed that the electric field $E(\beta,x)$ goes to zero when $d\beta \rightarrow 0$ and $\epsilon \rightarrow 0$, as it should for $\alpha = 0$. Not surprisingly, we find that the condition $E(\beta,x) \approx 0$ is best fulfilled for the smallest values of $d\beta$ and $\epsilon$, in our case $\epsilon = 10^{-6}$ and $d\beta = 0.01$. Note also that for fixed values of $d\beta$ and $\epsilon$ we have better approximations for smaller values of $x$, indeed for $x = 600$ the errors are larger than for $x = 100$.

In our simulations, we will use the Gibbs state obtained from the simulations with $\epsilon = 10^{-6}$ and $d\beta = 0.05$. The value $\Sigma(\beta,x)$ is computed with respect to this state. To estimate an error originating from taking a nonzero value for $\epsilon$ and $d\beta$, we also compute $\Sigma(\beta,x)$ for $(\epsilon,d\beta) = (5\times 10^{-6}, 0.05)$ and $(\epsilon,d\beta) = (10^{-6}, 0.01)$. The error $\Delta^{(\epsilon,d\beta)}\Sigma(\beta,x)$ is then estimated as twice the sum of the differences in magnitude of $\Sigma(\beta,x)$ for $(\epsilon,d\beta) = (10^{-6}, 0.05)$ with the values of $\Sigma(\beta,x)$ for $(\epsilon,d\beta) = (10^{-6}, 0.01)$ and $(5\times 10^{-6}, 0.05)$:
\begin{multline}\label{eq:defDeltaepsdbSigma} \Delta^{(\epsilon,d\beta)}\Sigma(\beta,x) = 2\left\vert\left(\Sigma(\beta,x)\Biggl\vert_{\epsilon = 10^{-6}, d\beta = 0.05} - \Sigma(\beta,x)\Biggl\vert_{\epsilon = 10^{-6}, d\beta = 0.01}  \right)\right\vert 
\\ + 2\left\vert\left(\Sigma(\beta,x)\Biggl\vert_{\epsilon = 10^{-6}, d\beta = 0.05} - \Sigma(\beta,x)\Biggl\vert_{\epsilon = 5 \times 10^{-6}, d\beta = 0.05}  \right)\right\vert
\end{multline}

In figs. \ref{fig:diffbetadifftolEFc} and \ref{fig:diffbetadifftolEFd} we show the chiral condensate with respect to the chiral condensate obtained for $\epsilon = 10^{-6}$ and $d\beta = 0.05$, i.e. 
$$ \Delta\Sigma(\beta,x) = \Sigma(\beta,x) - \Sigma(\beta,x)\Biggl\vert_{\epsilon = 10^{-6}, d\beta = 0.05}.$$
From this figures the error $\Delta^{(\epsilon,d\beta)}\Sigma(\beta,x)$ is then estimated as twice the sum of the difference of the dashed blue line with the full blue line and the difference of the full green line with the full blue line. 

\begin{figure}[t]
\null\hfill
\begin{subfigure}[b]{.40\textwidth}
\includegraphics[width=\textwidth]{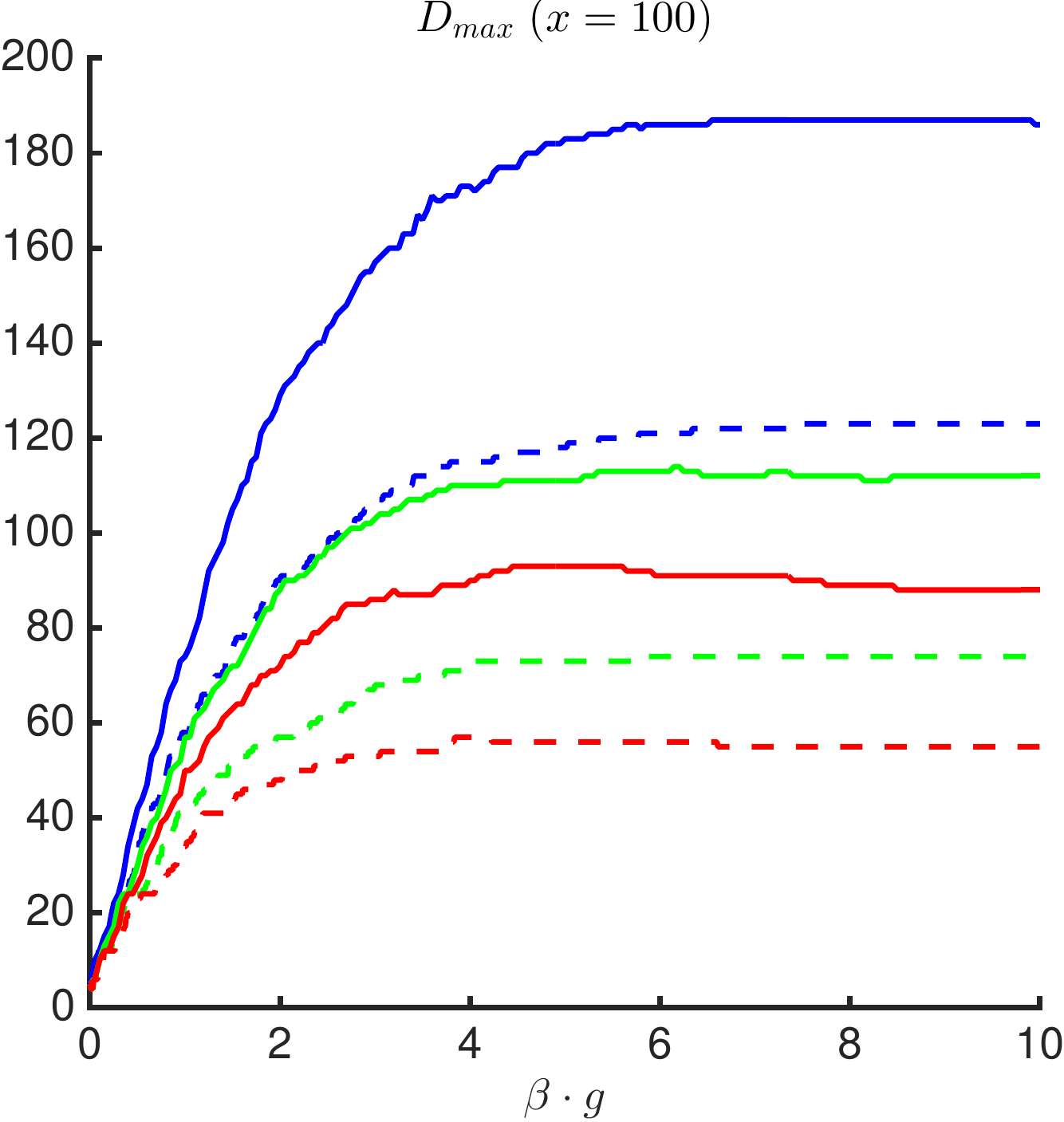}
\caption{\label{fig:diffbetadifftolDa}}
\end{subfigure}\hfill
\begin{subfigure}[b]{.40\textwidth}
\includegraphics[width=\textwidth]{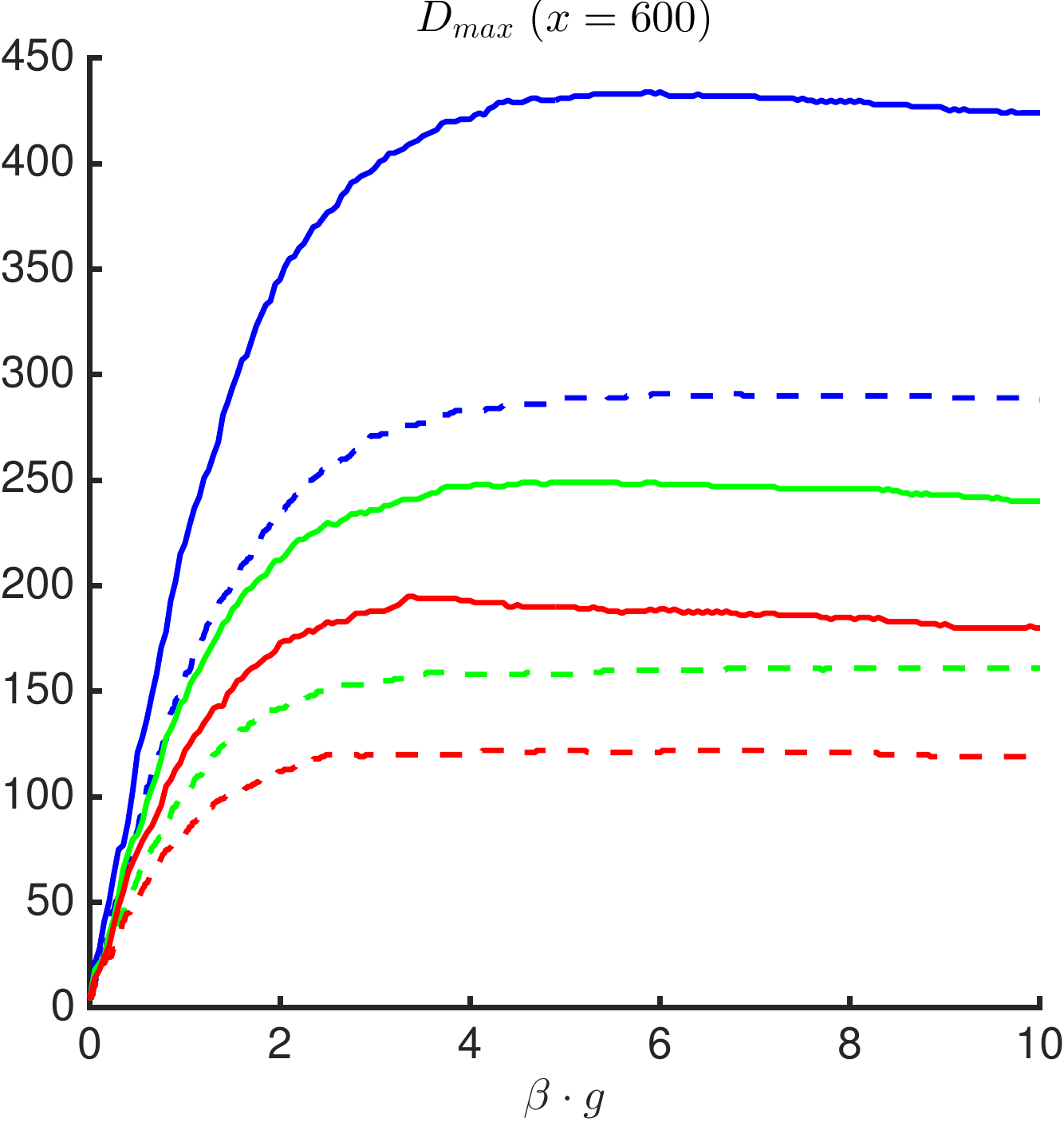}
\caption{\label{fig:diffbetadifftolDb}}
\end{subfigure}\hfill\null
\vskip\baselineskip
\captionsetup{justification=raggedright}
\caption{\label{figapp:diffbetadifftolD} $m/g = 0.25, \alpha = 0$. Evolution of the maximum bond dimension $D_{max} = \max_q D_q$ over the charge sectors for different values of $(\epsilon,d\beta)$. Colors correspond to legend of fig. \ref{figapp:diffbetadifftolEF}: dashed line: $d\beta = 0.01$, full line: $d\beta = 0.05$; red: $\epsilon = 10^{-5}$, green: $\epsilon = 5\times 10^{-6}$, blue: $\epsilon = 10^{-6}$. (a) $x = 100$. (b) $x = 600$.}
\end{figure}

As is obvious from figs. \ref{fig:diffbetadifftolDa} and \ref{fig:diffbetadifftolDb}, taking a nonzero value for $\epsilon$ corresponds to taking finite values for the virtual dimensions $D_q$ of the MPS representation. As one should expect, taking smaller values of $\epsilon$ implies that we need to take larger values of the virtual dimensions (and thus longer computation time). In contrast, taking a smaller value of the step $d\beta$ leads to the need of less variational freedom for a fixed value of $\epsilon$. This is explained by the fact that for a fixed interval $\Delta \beta$ more Trotter gates are applied when $d\beta$ is smaller and hence more Schmidt values are discarded when $d\beta$ is smaller. Hence, the price we need to pay to take larger steps is that we need to take into account more variational freedom and thus longer computation time for each step $d\beta$. Although, the fact that we need to take less steps for $d\beta = 0.05$ implied that in practice computations with $d\beta = 0.05$ are still faster than for simulations with $d\beta = 0.01$. 

\subsection{Continuum extrapolation for the chiral condensate for $\alpha = 0$}\label{subsec:appendixCCcontinuuma}
\noindent Assume we computed the chiral condensate $\Sigma(\beta,x)$ for the $x$-values $x = x_1,\ldots, x_M$ and we want to obtain a continuum value $\Sigma(\beta) = \lim_{x \rightarrow + \infty}\Sigma(\beta,x)$. When $\beta \rightarrow + \infty$ the chiral condensate diverges logarithmically in $x$ for $m/g \neq 0$. Perturbative computations and numerical simulations pointed out \cite{Adam1998,Banuls2013b,Buyens2014,Banuls2016} that this can be traced back to the free theory ($g = 0$). By subtracting the free chiral condensate \cite{Banuls2013b}
\be\label{eq:freeCC} \Sigma_{free}(x)  = -\frac{m}{\pi g}\frac{1}{\sqrt{1 +  \frac{m^2}{g^2 x}}}K\Biggl(\frac{1}{1 +  \frac{m^2}{g^2 x}}\Biggl), \ee
where $K$ is the complete elliptic integral of the first kind, the logarithmic divergence is removed. At finite temperature, for a fixed value of $x$ ,we subtract this contribution, i.e. we consider
\be \label{eq:subtrT02} \Sigma_{sub}(\beta,x) = \Sigma(\beta,x) - \Sigma_{free}(x). \ee
Note that in \cite{Buyens2014} we already obtained $\Sigma_{sub}$ to sufficient precision. 

As in \cite{Banuls2015,Banuls2016}, for a fixed value of $\beta$, we fit $\Sigma_{sub}(\beta,x)$ against the following functions: 
\begin{subequations}\label{eq:fitfunctionapp}
\be\label{eq:fitfunctionaappa} f_1(x) = A_1 + B_1 \frac{\log(x)}{\sqrt{x}} + C_1\frac{1}{\sqrt{x}} \ee
\be\label{eq:fitfunctionbappb}f_2(x) = A_2 + B_2 \frac{\log(x)}{\sqrt{x}} + C_2\frac{1}{\sqrt{x}} + D_2\frac{1}{x} \ee
and
\be\label{eq:fitfunctioncappv}f_3(x) = A_3 + B_3 \frac{\log(x)}{\sqrt{x}} + C_3\frac{1}{\sqrt{x}} + D_3\frac{1}{x} + E_3 \frac{1}{x^{3/2}}.\ee
\end{subequations}

Let us discuss in more detail how we obtain (i) a continuum estimate for each of the fitting ans\"atze $f_n$ and (ii) a final continuum estimate. 

\subsubsection*{(i) Obtaining a continuum estimate for the fitting ansatz $f_n$}
\noindent For every type of fitting ansatz, i.e. a particular $f_n$ $(n = 1,2,3)$ eq. (\ref{eq:fitfunctionapp}), we will determine an estimate $\Sigma_{sub}^{(n)}(\beta)$ for the continuum value and an error $\Delta^{(n)}\Sigma_{sub}(\beta)$ which originates from the choice of fitting interval. 

Given our dataset $\{(x_j,\Sigma_{sub}(\beta,x_j)): j = 1,\ldots, M\}$ of $M$ points with $x_j \in [9,800]$ we perform all possible fits of $f_n$ against at least $n+5$ consecutive data points where
\begin{itemize}
\item[(a)] we discard the data points where the error $\Delta^{(\epsilon,d\beta)}\Sigma(\beta,x)$ originating from taking finite values for $\epsilon$ and $d\beta$, see subsection \ref{subsec:errorsFinD}, is larger than $4 \times 10^{-4}$, 
\item[(b)] the coefficients ${A_n,B_n,C_n,D_n,E_n}$ ($D_n = 0$ if $n < 2$, $E_n = 0$ if $n < 3$) are estimated using an iterative generalized least-squares algorithm.
\end{itemize}

By taking at least $n+5$ consecutive data points we reduce the problem of overfitting: the fitted function $f_n$ fits the considered points extremely well, but fails to fit the overall data. Furthermore we also discard the fits that give statistically insignificant coefficients (p-value $\geq$ 0.05). In practice, this means that we discard the fits $f_n$ where the error on one of its coefficients $(A_n,B_n,C_n,\ldots)$ is larger than approximately half of its value. This is in contrast with \cite{Banuls2016} where a more conservative criterium is used and they only discard a fit where for at least one of the coefficients its error is larger than the coefficient itself. Note however that in \cite{Banuls2016} they fix the maximum $x$ that enters a fit whereas we allow all possible intervals of $n+5$ consecutive $x$-values. 

For every fit $\theta$ of $f_n$ against a subset of at least $n+5$ consecutive $x-$values, say $\{x_j\}_{j \in \mbox{fit}\theta}$, which produces statistically significant coefficients we obtain values
$$A_n^{(\theta)},B_n^{(\theta)},C_n^{(\theta)},D_n^{(\theta)}, E_n^{(\theta)} (D_n^{(\theta)} = 0\mbox{ for } n < 2, E_n^{(\theta)} = 0\mbox{ for } n < 3), $$
and a corresponding fitting function $g_\theta(x)$.
 $$g_\theta(x)  = A_n^{(\theta)} + B_n^{(\theta)} \frac{\log(x)}{\sqrt{x}} + C_n^{(\theta)}\frac{1}{\sqrt{x}} + D_n^{(\theta)}\frac{1}{x} + E_n^{(\theta)} \frac{1}{x^{3/2}}$$

All the values $A_n^{(\theta)}$ we obtain are an estimate for the continuum value of the chiral condensate for the fitting ansatz $f_n$. Let us denote with $\{A_n^{(\theta)}\}_{\theta = 1\ldots R_n}$ all the $A_n$'s obtained from a fit $\theta$ against $f_n$ which produces significant coefficients with
$$A_n^{(1)} \leq A_n^{(2)} \leq \ldots \leq A_n^{(R_n)} .$$
For each fit $\theta$ we also compute its $\chi^2$ value:
\be \label{eq:chisq} \chi_\theta^2 = \sum_{j \in \mbox{fit} \theta} \left(\frac{g_\theta(x_j) - \Sigma_{sub}(\beta,x_j)}{\Delta^{(\epsilon,d\beta)}\Sigma_{sub}(\beta,x_j)}\right)^2. \ee
When our dataset is large enough the quantity $\chi_\theta^2/N_{dof}^{\theta}$, with $N_{dof}^{\theta}$ the number of degrees of freedom of the fit (here the number of data points used in the fit minus $n+2$), gives an indication whether $g_\theta$ fits the dataset well ($\chi_\theta^2/N_{dof}^{\theta} \ll 1$) or not ($\chi_\theta^2/N_{dof}^{\theta} \gg 1$). 

If we have at least 10 fits $\theta$ with $\chi_\theta^2/N_{dof}^{\theta} \leq 1$ we can obtain a reliable continuum estimate by taking the median of $\{A_n^{(1)},  \ldots, A_n^{(R_n)}\}$ weighed by $\exp(-\chi_\theta^2/N_{dof}^{\theta})$, see also \cite{Banuls2013a}. More specifically we build the cumulative distribution $X_\theta$,  
$$ X_\theta = \frac{\sum_{\kappa = 1}^\theta  \exp(-\chi_\kappa^2/N_{dof}^{\kappa})}{\sum_{\kappa = 1}^{R_n} \exp(-\chi_\kappa^2/N_{dof}^{\kappa})},$$
and take as our continuum estimate $\Sigma^{(n)}(\beta)$ for the fitting ansatz $f_n$:  $\Sigma^{(n)}(\beta) = A_n^{(\theta_0)}$ where $\theta_0$ corresponds to the value for which $X_{\theta_0}$ is the closest to $1/2$, i.e.
$$\theta_0 = \mbox{arg}\min_\theta \vert X_{\theta} - 1/2 \vert. $$
The systematic error $\Delta^{(n)}\Sigma(\beta,x)$ from the choice of $x$-interval comes from the $\% 68,3$-confidence interval, it is computed as
$$ \Delta^{(n)}\Sigma(\beta,x) = \frac{1}{2}\left(A_n^{(\theta_2)} - A_n^{(\theta_1)}\right)$$
with 
$$\theta_1 = \mbox{arg}\min_\theta \vert X_{\theta} - 0.85 \vert,  \theta_2 = \mbox{arg}\min_\theta \vert X_{\theta} - 0.15 \vert. $$

If we have less than 10 fits $\theta$ with $\chi_\theta^2/N_{dof}^{\theta} \leq 1$, only a few fits will dominate the histogram of the $\chi^2$-distribution. Therefore we will adopt the more conservative approach from \cite{Banuls2016}. We only consider the fits with statistically significant coefficients and with $\chi_\theta^2/N_{dof}^{\theta} \leq 1$; the corresponding continuum estimates are
$$A_n^{(1)} \leq A_n^{(2)} \leq \ldots \leq A_n^{(R'_n)}, \mbox{ with }R'_n \leq R_n.$$
Of these estimates we take the $A_n^{\theta_0}$ which corresponds to the $\theta$ for which the mean squared of the error in $(d\beta,\epsilon)$ is minimal, i.e.
$$\theta_0 = \mbox{arg}\min_\theta \frac{1}{\vert \mbox{fit} \theta \vert }\left(\sqrt{\sum_{j \in \mbox{fit} \theta}\left(\Delta^{(\epsilon,d\beta)}\Sigma_{sub}(\beta,x_j)\right)^2}\right). $$
As the systematic error originating from the choice of fitting range we take the difference in magnitude of this estimate with the most outlying $A_n^{(\theta)}$ (for the same type of fitting ansatz):
$$ \Delta^{(n)}\Sigma_{sub}(\beta) = \max_{1 \leq \theta \leq R'_n} \vert A_n^{(\theta_0)} - A_n^{(\theta)} \vert.$$

\subsubsection*{(ii) Final continuum estimate and uncertainty}
\noindent Using the method discussed in (i) we now have three estimates for $\Sigma_{sub}(\beta)$:
$$\Sigma_{sub}^{(1)}(\beta),\Sigma_{sub}^{(2)}(\beta),\Sigma_{sub}^{(3)}(\beta) $$
which correspond to the fitting functions $f_1$, $f_2$ and $f_3$. As our final estimate we take the estimate from the fitting function $f_{n_0}$ which had the most statistically significant fits with $\chi_\theta^2/N_{dof}^{\theta} \leq 1$. The error originating from the choice of fitting function is then computed as the maximum of the difference with the continuum estimates from the other fitting functions. As our final result we thus report $\Sigma_{sub}(\beta) = \Sigma_{sub}^{(n_0)}(\beta)$ and the error $\Delta \Sigma_{sub}(\beta)$ is the maximum of
\begin{itemize}
\item[a.] the error originating from taking nonzero $d\beta$ and $\epsilon$: 
$$\max\left(\max_{j} \left(\Delta^{(\epsilon,d\beta)}\Sigma_{sub}(\beta,x_j)\right), 4 \times 10^{-4}\right),$$
\item[b.] the error originating from the choice of $x$-range: $\Delta^{(n_0)}\Sigma(\beta),$
\item[c.] the error originating from the choice of fitting ansatz: $\max_{n = 1,2,3}\vert \Sigma_{sub}(\beta) - \Sigma_{sub}^{(n)}(\beta)\vert.$
\end{itemize}

\begin{figure}[t]
\null\hfill
\begin{subfigure}[b]{.40\textwidth}
\includegraphics[width=\textwidth]{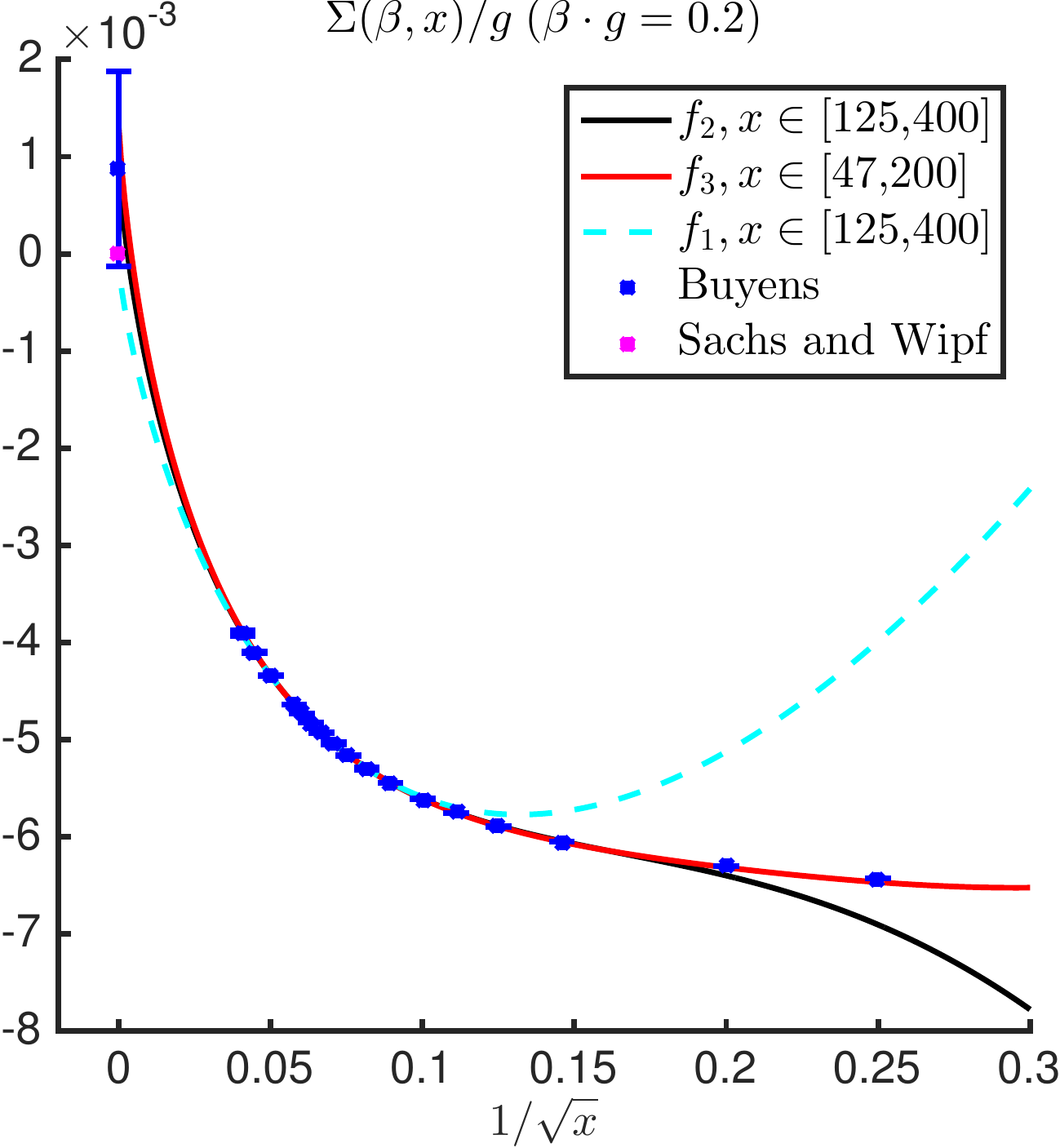}
\caption{\label{fig:extrapolationCCm0a}}
\end{subfigure}\hfill
\begin{subfigure}[b]{.40\textwidth}
\includegraphics[width=\textwidth]{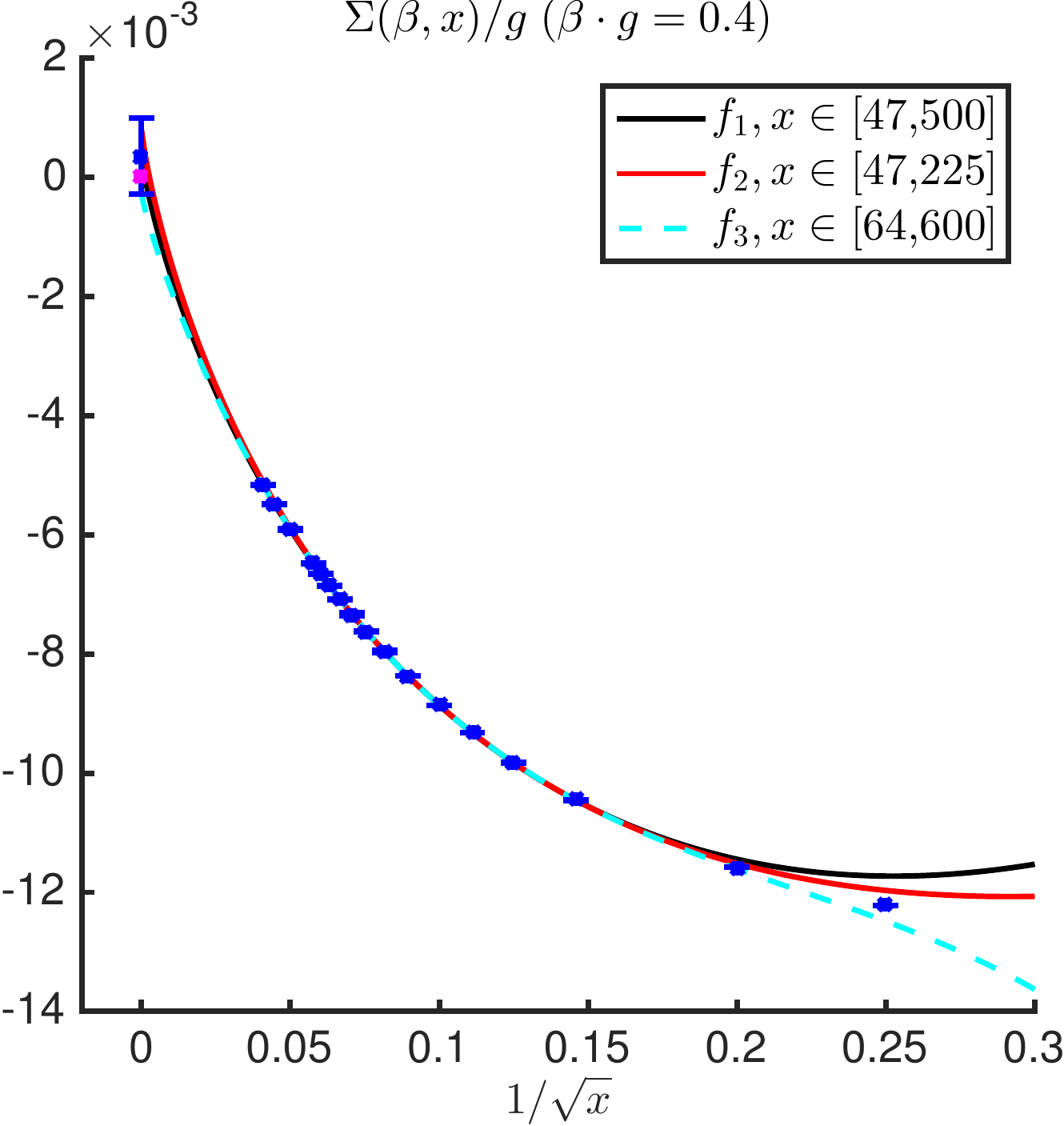}
\caption{\label{fig:extrapolationCCm0b}}
\end{subfigure}\hfill\null
\vskip\baselineskip
\null\hfill
\begin{subfigure}[b]{.40\textwidth}
\includegraphics[width=\textwidth]{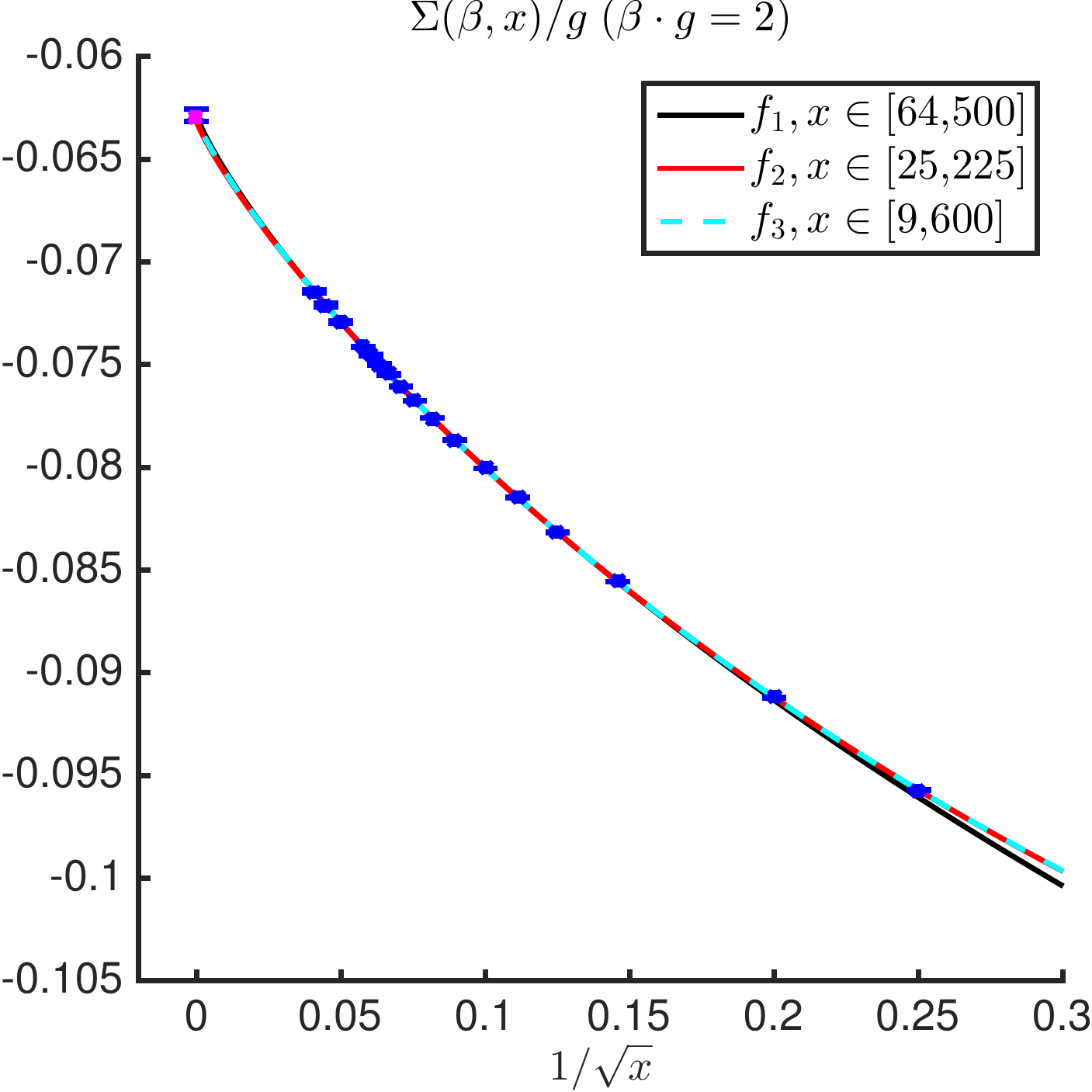}
\caption{\label{fig:extrapolationCCm0c}}
\end{subfigure}\hfill
\begin{subfigure}[b]{.40\textwidth}
\includegraphics[width=\textwidth]{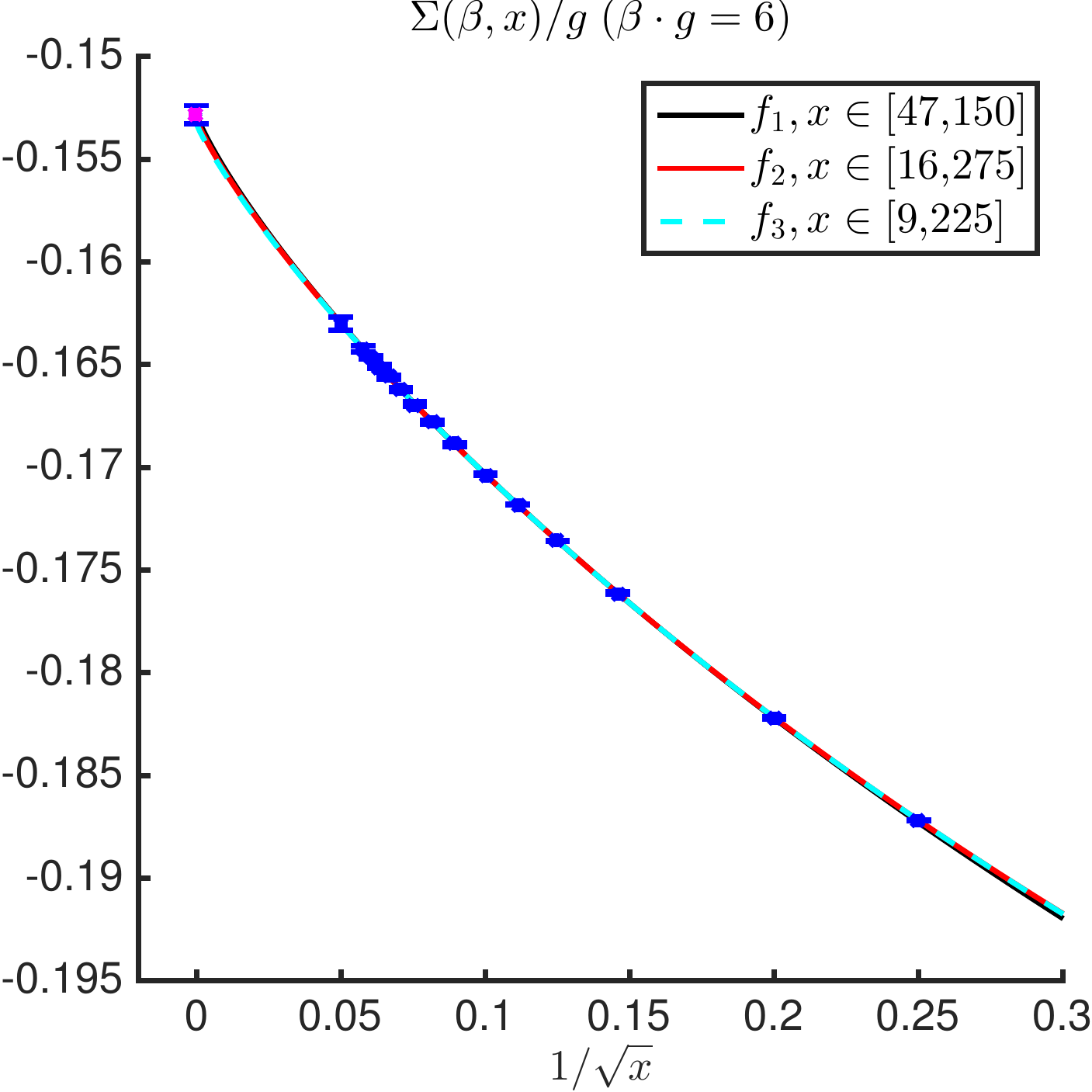}
\caption{\label{fig:extrapolationCCm0d}}
\end{subfigure}\hfill\null
\vskip\baselineskip
\captionsetup{justification=raggedright}
\caption{\label{fig:extrapolationCCm0a0} $m/g = 0$. Continuum extrapolation of $\Sigma(\beta,x)$ for different values of $\beta g$. The black line is the fit which determines the continuum estimate. The red line and dashed blue lines are the best fits for the other fitting functions $f_n$. The error bars on our data ($1/\sqrt{x} \neq 0$) represent the errors $\Delta^{(\epsilon,d\beta)}\Sigma(\beta,x)$ originating from taking nonzero $\epsilon$ and $d\beta$, as explained in subsection \ref{subsec:errorsFinD}. The blue error bar at $1/\sqrt{x} = 0$ represents the error on our continuum estimate. The magenta star is the analytical result of Sachs and Wipf \cite{Sachs1991}.}
\end{figure}

\noindent\textbf{Example 1.} Let us now illustrate this for $m/g = 0$. When $m/g = 0$ we have that $\Sigma_{sub}(\beta,x) = \Sigma(\beta,x)$. We did simulations for $\beta g \in [0,10]$ with steps $d\beta = 0.05$ for
\be \label{eq:xvalueCCmg0} x = 9, 16, 25, 47, 64, 81, 100, 125, 150, 175, 200, 225, 250, 275, 300, 400, 500, 600.\ee
Let us discuss the case $\beta g = 0.2$, fig. \ref{fig:extrapolationCCm0a}. For all the $x-$values the errors originating from taking a nonzero value $\epsilon$ and $d\beta$ were smaller than $5 \times 10^{-5}$ so we could include all the data points. For all the three fitting function $f_n$ we found enough fits with statistically significant coefficients that satisfy $\chi_\theta^2/N_{dof}^{\theta} \leq 1$. By weighing the fits with $\exp(-\chi_\theta^2/N_{dof})$ as discussed in (i) we found
\begin{itemize}
\item[-] $\Sigma^{(1)}(0.2/g)/g = -1(4) \times 10^{-4}$ (22 fits with statistically significant coefficients and $\chi_\theta^2/N_{dof}^{\theta} \leq 1$)
\item[-] $\Sigma^{(2)}(0.2/g)/g = 9(1) \times 10^{-4}$ (51 fits with statistically significant coefficients and $\chi_\theta^2/N_{dof}^{\theta} \leq 1$)
\item[-] $\Sigma^{(3)}(0.2/g)/g = 1.3 (3) \times 10^{-3}$ (33 fits with statistically significant coefficients and $\chi_\theta^2/N_{dof}^{\theta} \leq 1$)
\end{itemize}
where the number between brackets is the error originating from the choice of fitting interval (as discussed in (i)). Because we found the most fits with statistically significant coefficients and $\chi_\theta^2/N_{dof}^{\theta} \leq 1$ for $n = 2$ we take $\Sigma(0.2/g)/g = \Sigma^{(2)}(0.2/g)/g = 9(1) \times 10^{-4}$ as our estimate. To estimate an error originating from the choice of fitting function we take the maximum of the difference with $\Sigma^{(1)}$ and $\Sigma^{(3)}$:
$$\Delta \Sigma(0.2/g)/g = \max\left( \left\vert \Sigma^{(2)}(0.2/g)/g - \Sigma^{(1)}(0.2/g)/g \right\vert, \left\vert \Sigma^{(2)}(0.2/g)/g - \Sigma^{(3)}(0.2/g)/g\right\vert \right) \approx 1 \times 10^{-3}.$$
As this error is larger than the error originating from the choice of fitting interval we report
$$\Sigma(0.2/g)/g  = 9 (10) \times 10^{-4} $$
which is consistent with the analytical result $\Sigma_{SW}(\beta g)/g = -8 \times 10^{-12}$.
In fig. \ref{fig:extrapolationCCm0a} we show $\Sigma(0.2/g,x)/g$ for the values of $x$ displayed in eq. (\ref{eq:xvalueCCmg0}), the error bars represent the error $\Delta^{(\epsilon,d\beta)}\Sigma(0.2/g,x)/g$ (see subsection \ref{subsec:errorsFinD}). The black line shows the quadratic fit through the points which determined our continuum estimate $\Sigma(0.2/g) = \Sigma^{(2)}(0.2/g)$. The coefficients were found by fitting $f_2$ against the data for $x = 125,150,\ldots,300,400$. The red line shows the `best' cubic fit (as discussed above) which is obtained by fitting $f_3$ against the data for $x = 47,64,81,100,125,\ldots,200$. This fit gives us the estimate $\Sigma^{(3)}(0.2/g)/g$. Finally, the blue dashed line shows the `best' linear fit which is obtained by fitting $f_1$ against the data points for $x = 125,150,\ldots,300,400$. Note that when comparing this fit with the best $f_2$ fit (which turns out to fit the same data points) that this fit performs bad for the lower $x-$values. However when comparing $\Sigma^{(1)}(0.2/g)/g$, $\Sigma^{(2)}(0.2/g)/g$ and $\Sigma^{(3)}(0.2/g)/g$ with the exact result of Sachs and Wipf \cite{Sachs1991}(magenta star), $\Sigma^{(1)}(0.2/g)/g$ actually seems to give the best result. This indicates that there are significant cutoff effects in $x$: for a reliable continuum estimate we need to perform the continuum extrapolation with relatively large $x-$values ($x \gtrsim 100$) or we need to fit a higher-order function in $1/\sqrt{x}$ (in this case a quadratic or cubic one). This explains the relatively large error bar at high temperatures (small values of $\beta g$). 

Also for $\beta g = 0.4$, see fig. \ref{fig:extrapolationCCm0b}, there are still relatively large cutoff effects in $x$. Now we found for the continuum estimates corresponding to the fitting functions $f_1$, $f_2$ and $f_3$
\begin{itemize}
\item[-] $\Sigma^{(1)}(0.4/g)/g = 4 (1) \times 10^{-4}$ (55 fits with statistically significant coefficients and $\chi_\theta^2/N_{dof}^{\theta} \leq 1$)
\item[-] $\Sigma^{(2)}(0.4/g)/g = 8(4) \times 10^{-4}$ (18 fits with statistically significant coefficients and $\chi_\theta^2/N_{dof}^{\theta} \leq 1$)
\item[-] $\Sigma^{(3)}(0.4/g)/g = -3 (1) \times 10^{-4}$ (16 fits with statistically significant coefficients and $\chi_\theta^2/N_{dof}^{\theta} \leq 1$)
\end{itemize}
Because we found the most fits with statistically significant coefficients and $\chi_\theta^2/N_{dof}^{\theta} \leq 1$ for $n = 1$ we take $\Sigma(0.2/g) = \Sigma^{(1)}(0.2/g)/g = 4(1) \times 10^{-4}$ as our estimate. To estimate an error originating from the choice of fitting function we take the maximum of the difference with $\Sigma^{(2)}$ and $\Sigma^{(3)}$:
$$\Delta \Sigma(0.4/g)/g = \max\left( \left\vert \Sigma^{(2)}(0.4/g)/g - \Sigma^{(1)}(0.4/g)/g \right\vert, \left\vert \Sigma^{(2)}(0.4/g)/g - \Sigma^{(3)}(0.4/g)/g\right\vert \right) \approx 7 \times 10^{-4}.$$
As this error is larger than the error originating from the choice of fitting interval we report
$$\Sigma(0.2/g)/g  = 4 (7) \times 10^{-4} $$
which can be compared with the exact value $\Sigma_{SW}(\beta g)/g = -4.5 \times 10^{-6}$.

\begin{table}
\begin{tabular}{| c| |  c |   c | c | c || c | c |}
        \hline
     $\beta g$ &  $x$-range &  $\Sigma_{sub}^{(1)}(\beta)/g  \{\# \mbox{ fits}\}$ &    $\Sigma_{sub}^{(2)}(\beta)/g \{\# \mbox{ fits}\}$   & $\Sigma_{sub}^{(3)}(\beta)/g  \{\# \mbox{ fits}\}$ & $\Sigma_{sub}(\beta)/g$ & $\Sigma_{sub}(\beta)/g $ \cite{Banuls2016}\\
     \hline 
     0.2& $[9,800]$ & $0.26 (1) \{17\}$  & $0.230 (6) \{34\}  $& $0.299 (4) \{42\}$& $0.30 (3) $ & $0.298 (7)$  \\ 
0.4 & $[9,800]$ & $0.2243 (3) \{38\} $ &$ 0.228 (4) \{42\}$ & $0.221 (8) \{32\} $ &$0.228 (7) $ & $0.23 (1)$ \\
2 &  $[9,700]$ & $0.00598 (8) \{67\}$ &$ 0.0056 (1) \{63\}$  &$ 0.0059 (4) \{14\} $& $0.0060 (4) $ & $0.0078 (38) $ \\
6 & $[9,500]$ & $-0.0661 (1) \{67\} $ & $-0.06631(9) \{39\}$  & $-0.0664(1) \{16\} $&$ -0.0661 (4) $ & $-0.0657 (43) $ \\
\hline 
\end{tabular}
\captionsetup{justification=raggedright}
\caption{\label{table:extrapolationmdivg25} $m/g = 0.25$. Details on the continuum extrapolation of the subtracted chiral condensate $\Sigma_{sub}(\beta)$. }
\end{table}

When going to lower temperatures, larger values of $\beta g$, we find that cutoff effects in $x$ are less apparent. Indeed in figs. \ref{fig:extrapolationCCm0c} and \ref{fig:extrapolationCCm0d} we find that the `best' fits $f_1$, $f_2$ and $f_3$ lie almost on top of each other, although they are obtained by fitting a different $x-$range. Also the error of the choice of fitting interval is way smaller, for instance for $\beta g = 2$ we have,
\begin{itemize}
\item[-] $\Sigma^{(1)}(2/g)/g = -0.0629 (1)$(66 fits with statistically significant coefficients and $\chi_\theta^2/N_{dof}^{\theta}\leq 1$)
\item[-] $\Sigma^{(2)}(2/g)/g = -0.0632 (1)$ (50 fits with statistically significant coefficients and $\chi_\theta^2/N_{dof}^{\theta} \leq 1$)
\item[-] $\Sigma^{(3)}(2/g)/g = -0.0630 (0) $ (1 fit with statistically significant coefficients and $\chi_\theta^2/N_{dof}^{\theta} \leq 1$).
\end{itemize}
Note that for the fitting ansatz $f_3$ we only found one statistically significant fit with $\chi_\theta^2/N_{dof}^{\theta} \leq 1$. This suggests that the coefficient $E_3$ of $1/x^{3/2}$ is indeed irrelevant and that a quadratic fit ($f_2$) is sufficient. Because there is only one `good' fit of $f_3$ there is of course no estimate for the error on $\Sigma^{(3)}(2/g)$ originating from the choice of fitting interval. As the most fits with statistically significant coefficients and $\chi_\theta^2/N_{dof}^{\theta} \leq 1$ are found for $f_1$ we report 
$$\Sigma(2/g)/g  = -0.0629 (3) $$
where the error is obtained as 
$$\Delta \Sigma(2/g)/g = \max\left( \left\vert \Sigma^{(2)}(2/g)/g - \Sigma^{(1)}(2/g)/g \right\vert, \left\vert \Sigma^{(2)}(2/g)/g - \Sigma^{(3)}(2/g)/g\right\vert \right) \approx 3 \times 10^{-4}.$$
For the analytical result we have $\Sigma_{SW}(2/g) = -0.0630$, which is again in agreement with our result. 

\begin{figure}
\null\hfill
\begin{subfigure}[b]{.40\textwidth}
\includegraphics[width=\textwidth]{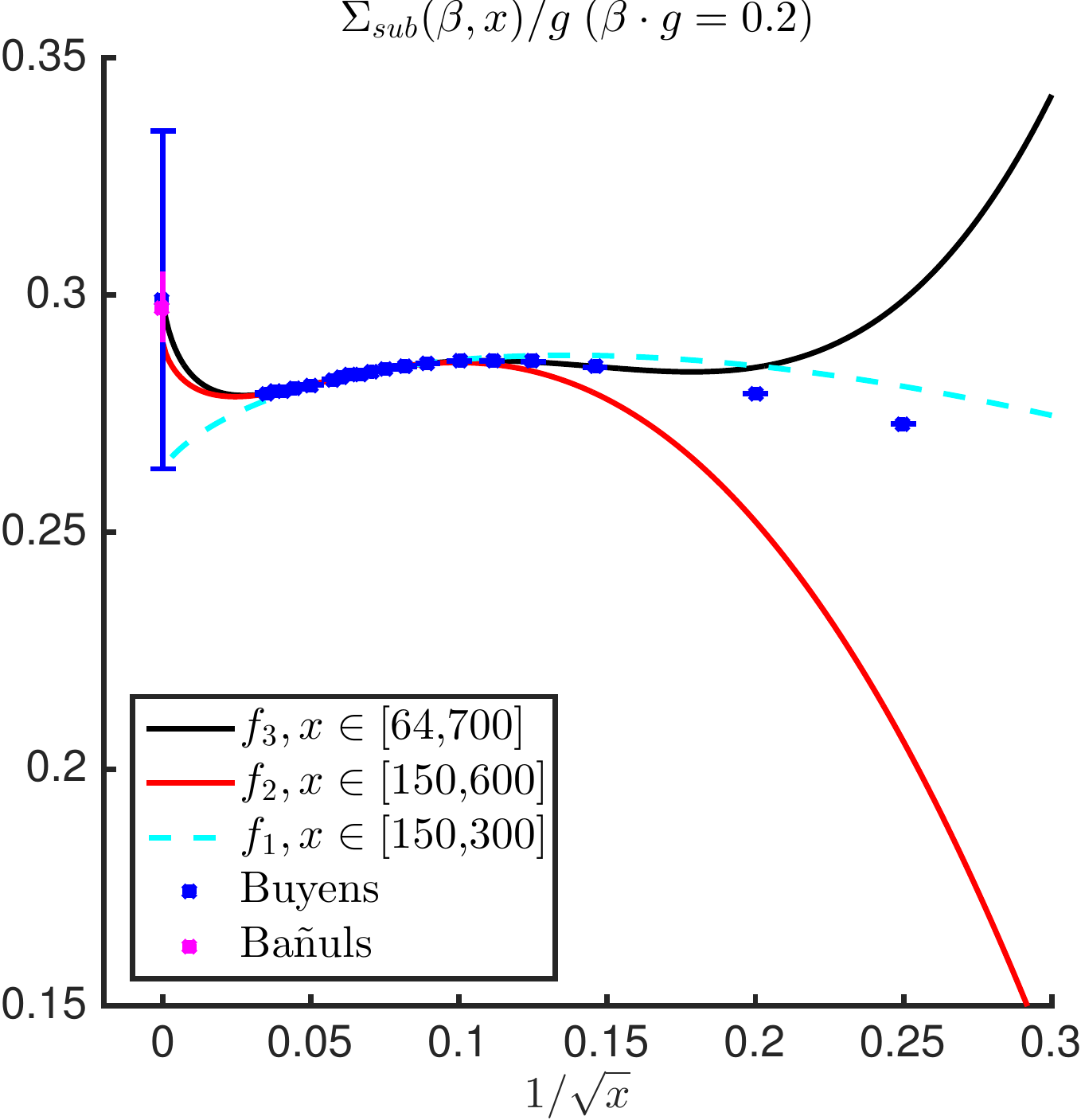}
\caption{\label{fig:extrapolationCCe}}
\end{subfigure}\hfill
\begin{subfigure}[b]{.40\textwidth}
\includegraphics[width=\textwidth]{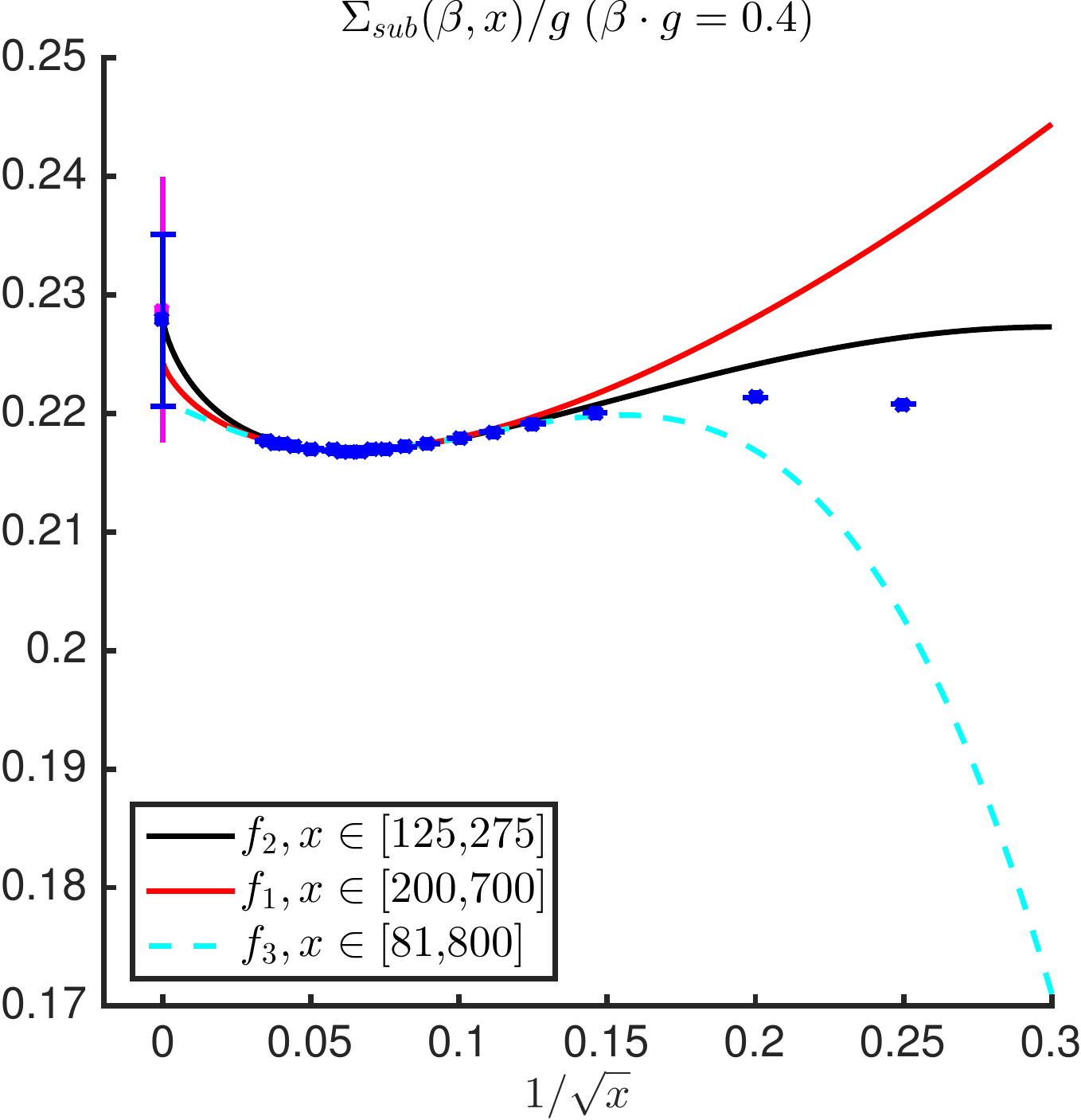}
\caption{\label{fig:extrapolationCCf}}
\end{subfigure}\hfill\null
\vskip\baselineskip
\null\hfill
\begin{subfigure}[b]{.40\textwidth}
\includegraphics[width=\textwidth]{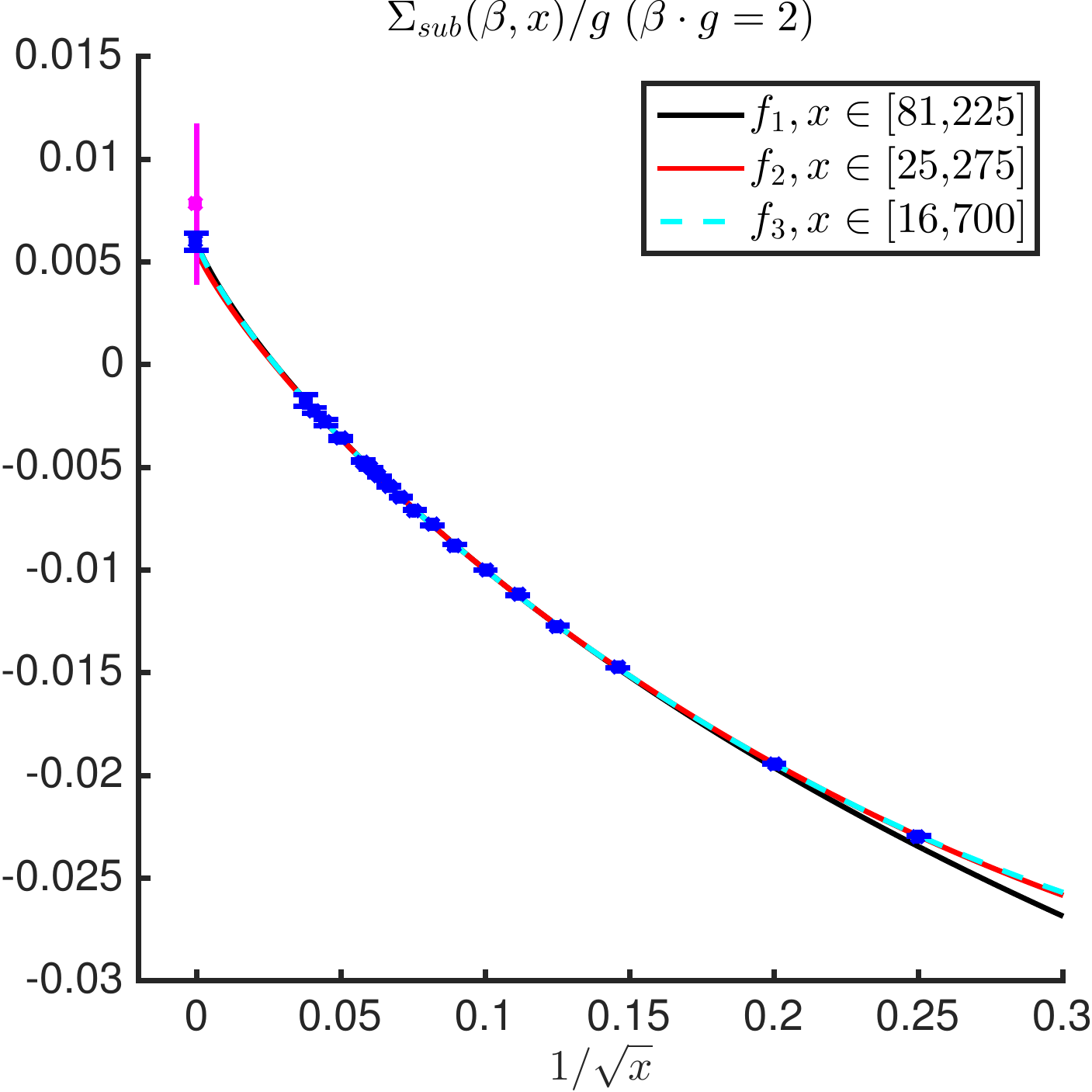}
\caption{\label{fig:extrapolationCCg}}
\end{subfigure}\hfill
\begin{subfigure}[b]{.40\textwidth}
\includegraphics[width=\textwidth]{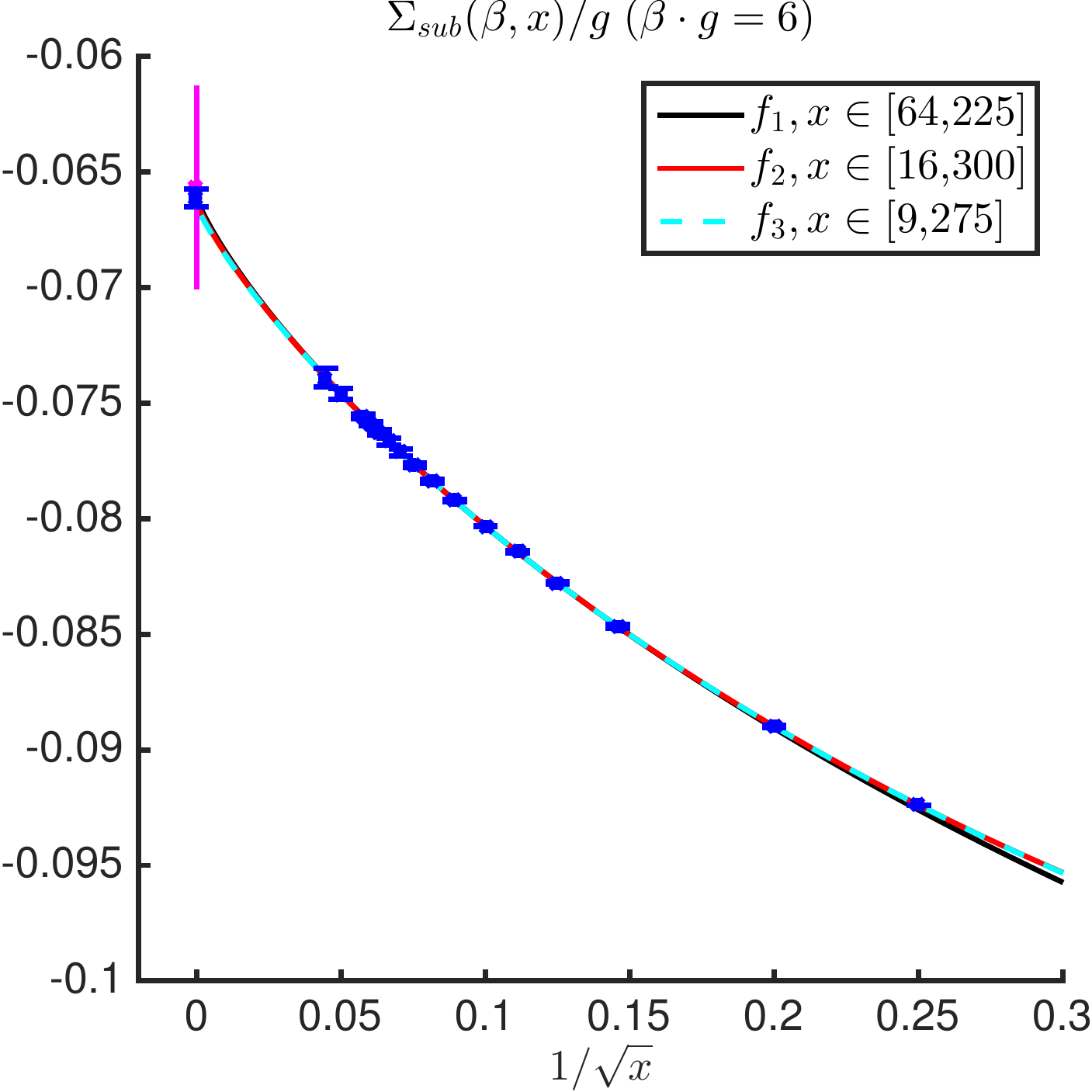}
\caption{\label{fig:extrapolationCCh}}
\end{subfigure}\hfill\null
\vskip\baselineskip
\captionsetup{justification=raggedright}
\caption{\label{fig:extrapolationCCm25e2a0} $m/g = 0.25$.  Continuum extrapolation of $\Sigma_{sub}(\beta,x)$ for different values of $\beta g$. The black line is the fit which determines the continuum estimate. The red line and dashed blue lines are the best fits for the other fitting functions $f_n$. The error bars on our data ($1/\sqrt{x} \neq 0$) represent the errors $\Delta^{(\epsilon,d\beta)}\Sigma(\beta,x)$ originating from taking nonzero $\epsilon$ and $d\beta$, as explained in subsection \ref{subsec:errorsFinD}. The blue error bar at $1/\sqrt{x} = 0$ represents the error on our continuum estimate. The magenta star and error bar are the result resp. error of Ba\-{n}us \emph{et al}.\cite{Banuls2016}.}
\end{figure}

For $\beta g = 6$ we find more or less the same behavior. Now the error $\Delta^{(\epsilon,d\beta)}\Sigma(\beta,x)$ originating of taking nonzero values of ($\epsilon,d\beta$) is larger than $4\times 10^{-4}$ for $x = 500$ and $x = 600$. Therefore these $x-$values are excluded from our data for $\beta g = 6$. Now we find for all the fitting functions:
\begin{itemize}
\item[-] $\Sigma^{(1)}(6/g)/g = -0.1528 (2)$(66 fits with statistically significant coefficients and $\chi_\theta^2/N_{dof}^{\theta}\leq 1$),
\item[-] $\Sigma^{(2)}(6/g)/g = -0.1531 (1)$ (32 fits with statistically significant coefficients and $\chi_\theta^2/N_{dof}^{\theta}\leq 1$),
\item[-] $\Sigma^{(3)}(6/g)/g = -0.1533 (0) $ (11 fits with statistically significant coefficients and $\chi_\theta^2/N_{dof}^{\theta} \leq 1$),
\end{itemize}
which lead to our final estimate $\Sigma(6/g) = -0.1528 (4)$. This is again in agreement with the analytical result $\Sigma_{SW}(6/g) = -0.15277$.
\\  
\\ \noindent\textbf{Example 2.} As another example we consider the case $m/g = 0.25$. We performed simulations for 
$$x = 9, 16, 25, 47, 64, 81, 100, 125, 150, 175, 200, 225, 250, 275, 300, 400, 500, 600, 700, 800.$$ 
The procedure is exactly the same but now applied to $\Sigma_{sub}(\beta,x) = \Sigma(\beta,x) - \Sigma_{free}(x)$. As can be observed from figs. \ref{fig:extrapolationCCe} and \ref{fig:extrapolationCCf} cutoff effects in $x$ are even more obvious for small values of $\beta g$. Again we find that reliable fits need to be fitted against large $x-$values or need to include higher-order terms in $1/\sqrt{x}$. In contrast, for larger values of $\beta g$, see figs. \ref{fig:extrapolationCCg} and \ref{fig:extrapolationCCh}, we find relatively small cutoff effects: the fits are lying almost on top of each other, independent of the chosen $x-$range and fitting function $f_n$.  The details can be found in table \ref{table:extrapolationmdivg25} where we show for $\beta g = 0.2, 0.4, 2, 6$:
\begin{itemize}
\item[-] the range of $x-$values we considered. Recall that $x-$values for which $\Delta^{(\epsilon,d\beta)}\Sigma(\beta,x) \geq 4 \times 10^{-4}$ were discarded,
\item[-] the estimates $\Sigma_{sub}^{(n)}(\beta)$ obtained for each of the fitting functions $f_n$ with between brackets $(\ldots)$ the error $\Delta^{(n)}\Sigma_{sub}(\beta)$ originating from the choice of fitting interval. In curly brackets $\{\ldots\}$ we denoted the number of significant fits we had with $\chi_\theta^2/N_{dof}^{\theta} \leq 1$,
\item[-] the final estimate for $\Sigma_{sub}(\beta)$ (with its error in brackets),
\item[-] the result of Ba\~nuls \emph{et al.} \cite{Banuls2016} for comparison. 
\end{itemize}

\newpage
\section{Asymptotic confinement: $\alpha \neq 0$}\label{sub:appendixCCcontinuumb}
\noindent In this section we explain how to obtain reliable continuum estimates for the quantities discussed in section \ref{sec:asCon} from our simulations at finite $x$. First, in subsection \ref{subsec:errFinDalphaneq0}, we discuss the errors originating from truncating the entanglement spectrum ($\epsilon > 0$) and taking finite steps for the imaginary time evolution ($d\beta > 0$). In subsection \ref{subsec:alphaclosetohalf} we consider the case when $\alpha  \approx 1/2$ because simulations are more hard in this regime. Afterward in subsection \ref{app:alphaneq0continuum} we discuss the continuum extrapolation. 

\begin{figure}[t]
\null\hfill
\begin{subfigure}[b]{.40\textwidth}
\includegraphics[width=\textwidth]{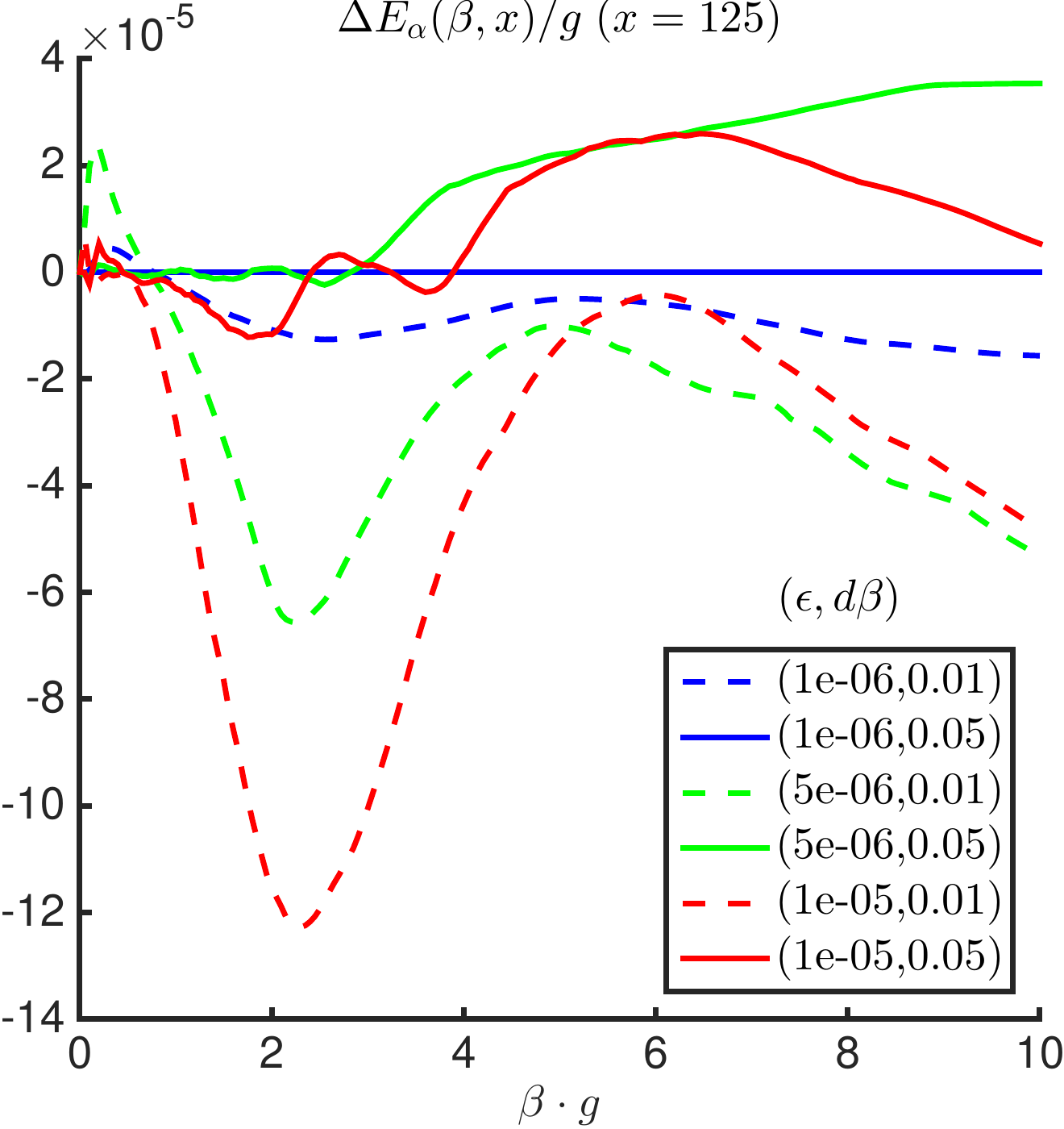}
\caption{\label{fig:diffbetadifftolEFa25e2a}}
\end{subfigure}\hfill
\begin{subfigure}[b]{.40\textwidth}
\includegraphics[width=\textwidth]{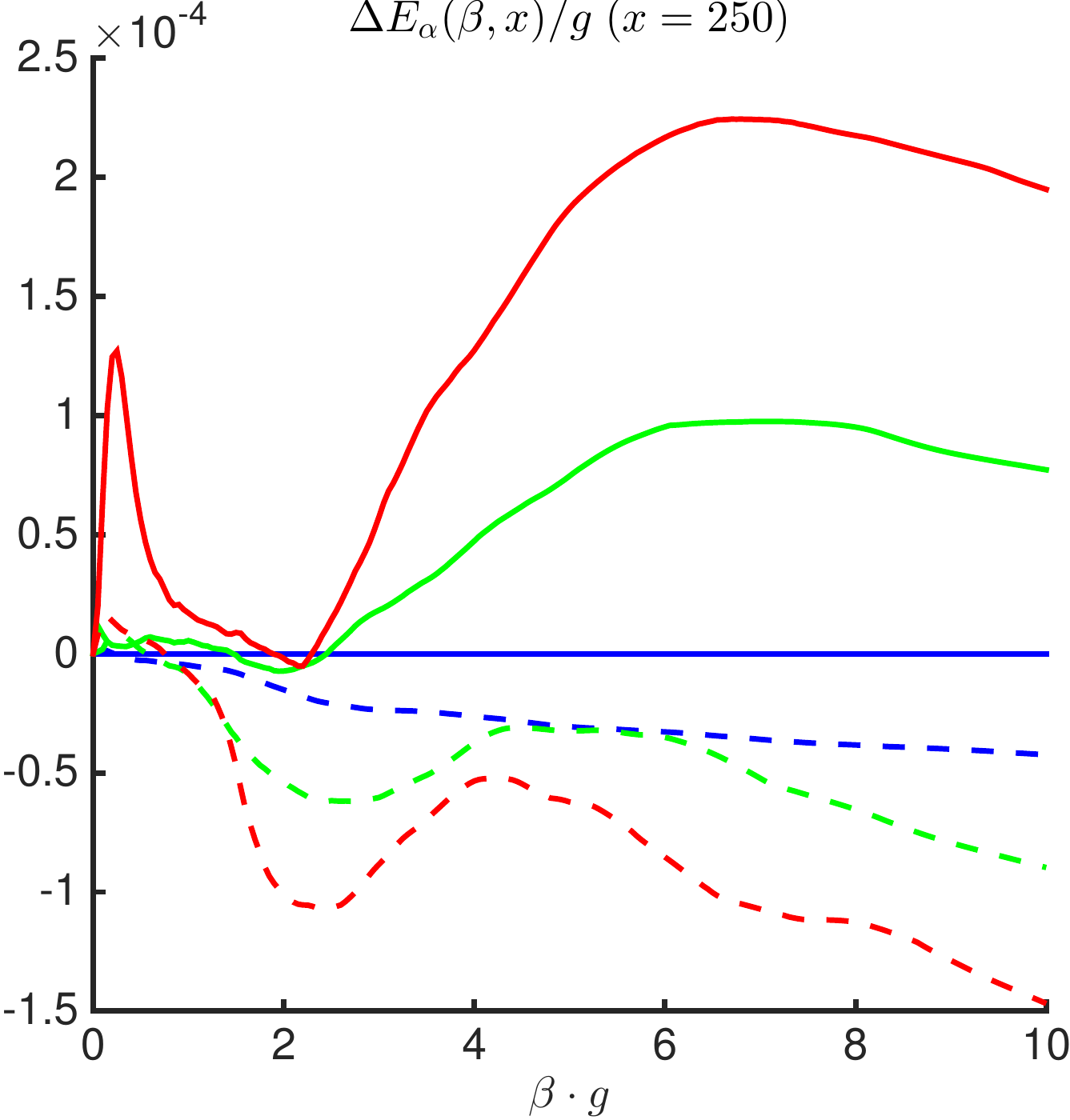}
\caption{\label{fig:diffbetadifftolEFa25e2b}}
\end{subfigure}\hfill\null
\vskip\baselineskip
\null\hfill
\begin{subfigure}[b]{.40\textwidth}
\includegraphics[width=\textwidth]{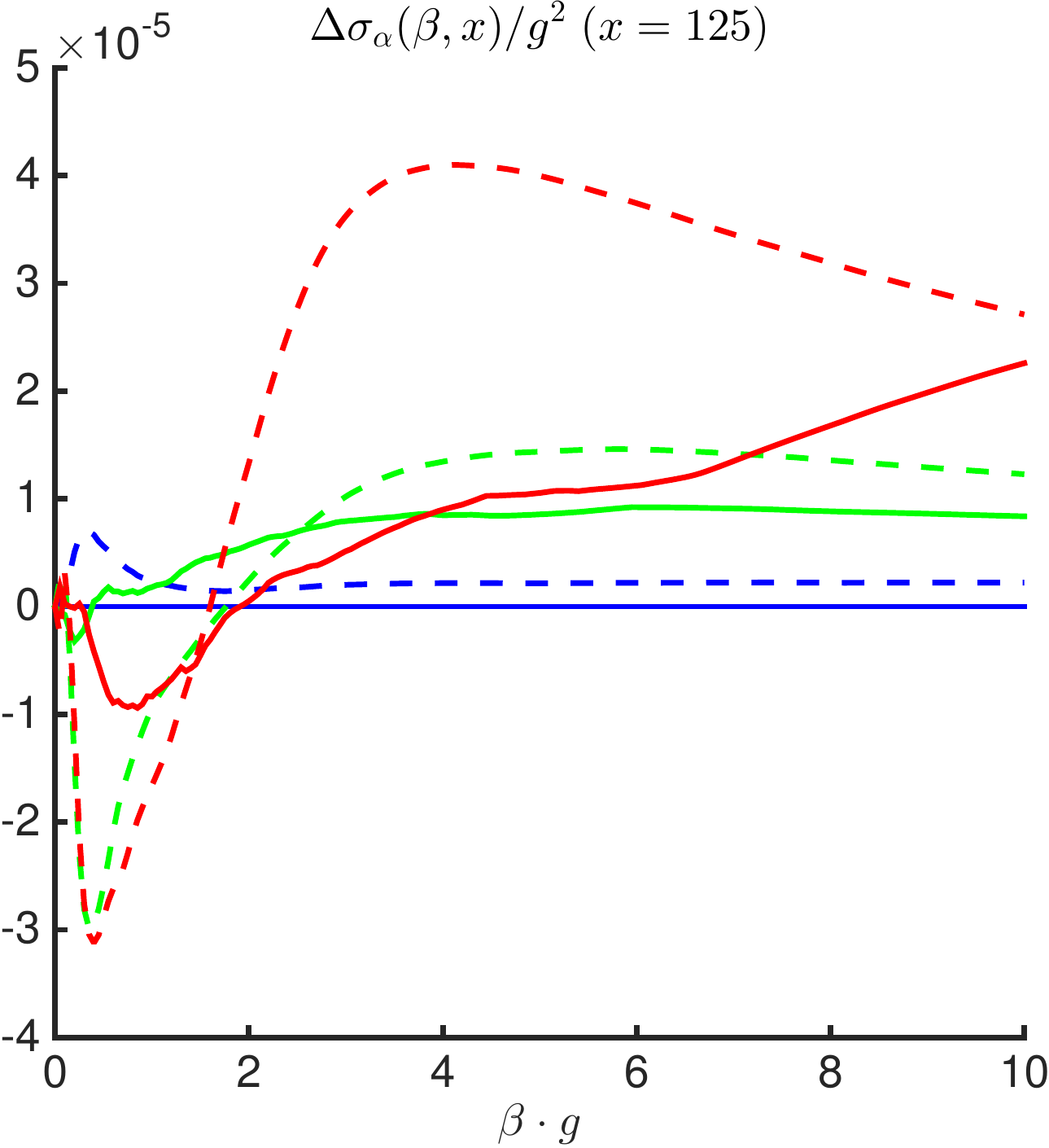}
\caption{\label{fig:diffbetadifftolEFa25e2c}}
\end{subfigure}\hfill
\begin{subfigure}[b]{.40\textwidth}
\includegraphics[width=\textwidth]{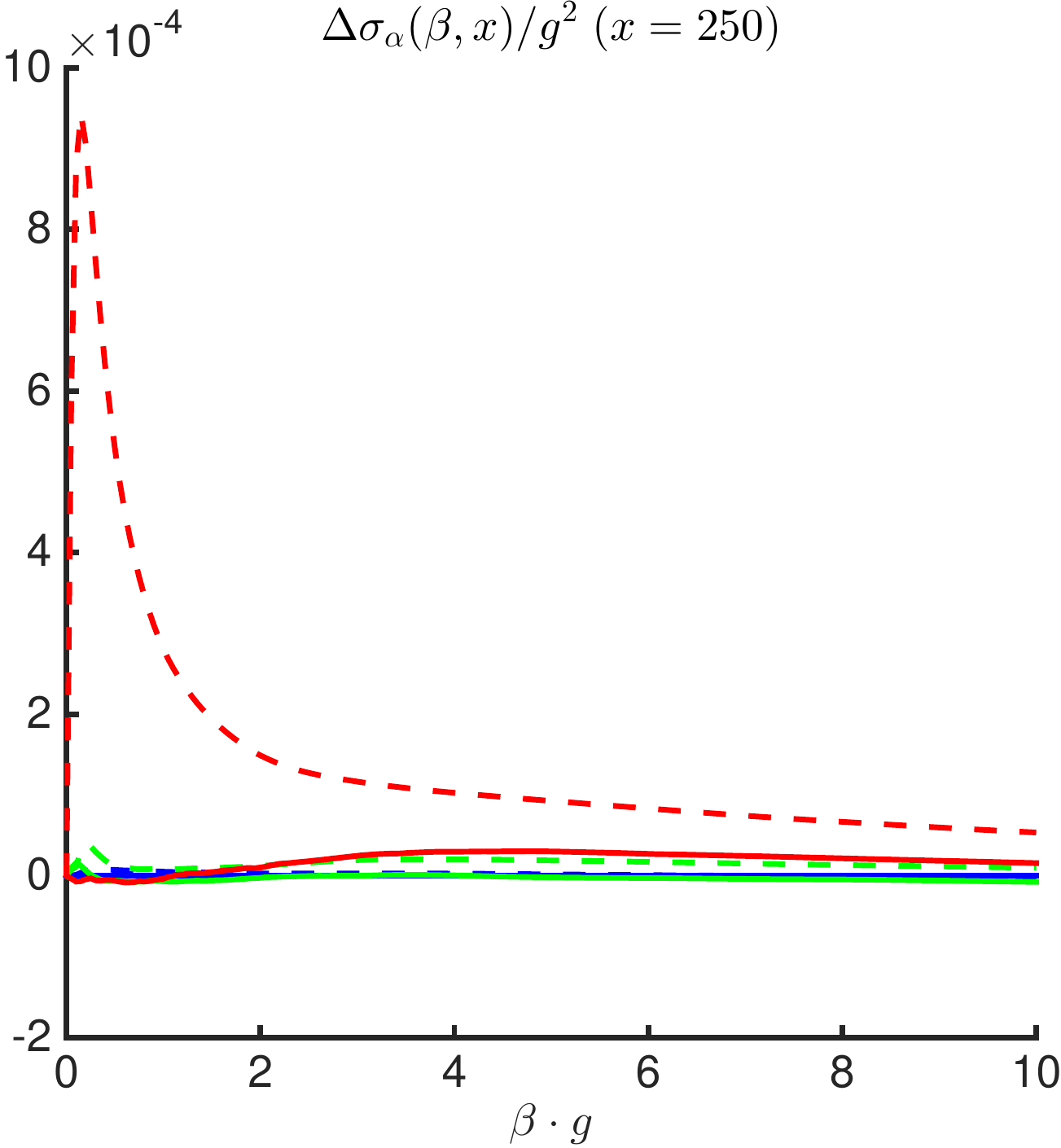}
\caption{\label{fig:diffbetadifftolEFa25e2d}}
\end{subfigure}\hfill\null
\vskip\baselineskip
\captionsetup{justification=raggedright}
\caption{\label{figapp:diffbetadifftolEFa25e2} $m/g = 0.25, \alpha = 0.25$. Simulations for different values of $(\epsilon,d\beta)$. (a-b) Electric field with respect to its value for $(\epsilon,d\beta) = (10^{-6},0.05)$. (a) $x = 125$. (b) $ x= 250$. (c-d) Free energy with respect to its value for $(\epsilon,d\beta) = (10^{-6},0.05)$. (c) $x = 125$. (d) $x = 250$.}
\end{figure}

\subsection{Errors originating from taking finite values for $\epsilon$ and $d\beta$}\label{subsec:errFinDalphaneq0}
\noindent Here we address the errors originating from taking nonzero values for $\epsilon$ and $d\beta$. Similar as in subsection \ref{subsec:errorsFinD}, we will use the Gibbs state obtained from the simulations with $\epsilon = 10^{-6}$ and $d\beta = 0.05$. The expectation value $\mathcal{Q}(\beta,x)$ of a given observable is computed with respect to this state. Again the error originating from taking a nonzero value for $\epsilon$ and $d\beta$ is estimated as follows: we compute $\mathcal{Q}(\beta,x)$ for $(\epsilon,d\beta) = (5\times 10^{-6}, 0.05)$ and $(\epsilon,d\beta) = (10^{-6}, 0.01)$. The error $\Delta^{(\epsilon,d\beta)}\mathcal{Q}(\beta,x)$ is then estimated as twice the sum of the differences in magnitude of $\mathcal{Q}(\beta,x)$ for $(\epsilon,d\beta) = (10^{-6}, 0.05)$ with the values of $\mathcal{Q}(\beta,x)$ for $(\epsilon,d\beta) = (10^{-6}, 0.01)$ and $(5\times 10^{-6}, 0.05)$:

\begin{multline}\label{eq:defDeltaepsdbQ} \Delta^{(\epsilon,d\beta)}\mathcal{Q}(\beta,x) = 2\left\vert\left(\mathcal{Q}(\beta,x)\Biggl\vert_{\epsilon = 10^{-6}, d\beta = 0.05} - \mathcal{Q}(\beta,x)\Biggl\vert_{\epsilon = 10^{-6}, d\beta = 0.01}  \right)\right\vert 
\\ + 2\left\vert\left(\mathcal{Q}(\beta,x)\Biggl\vert_{\epsilon = 10^{-6}, d\beta = 0.05} - \mathcal{Q}(\beta,x)\Biggl\vert_{\epsilon = 5 \times 10^{-6}, d\beta = 0.05}  \right)\right\vert
\end{multline}

An example for $m/g = 0.25$ and $\alpha = 0.25$ is shown in figs. \ref{fig:diffbetadifftolEFa25e2a} and \ref{fig:diffbetadifftolEFa25e2b} where we plot the electric field obtained for different values of $(\epsilon, d\beta)$ with respect to the electric field for $\epsilon = 10^{-6}$ and $d\beta = 0.05$, i.e. 
$$ \Delta E_\alpha (\beta,x) = E_\alpha(\beta,x) - E_\alpha(\beta,x)\Biggl\vert_{\epsilon = 10^{-6}, d\beta = 0.05}.$$
The error $\Delta^{(\epsilon,d\beta)}E_\alpha(\beta,x)$ is then estimated as twice the sum of the difference of the dashed blue line with the full blue line and the difference of the full green line with the full blue line. We also show the free energy for $\epsilon = 10^{-6}$ and $d\beta = 0.05$ with respect to the free energy obtained for $\epsilon = 10^{-6}$ and $d\beta = 0.05$:
$$ \Delta\sigma_\alpha (\beta,x) = \sigma_\alpha(\beta,x) - \sigma_\alpha(\beta,x)\Biggl\vert_{\epsilon = 10^{-6}, d\beta = 0.05},$$ 
see figs. \ref{fig:diffbetadifftolEFa25e2c} and \ref{fig:diffbetadifftolEFa25e2d}. The error $ \Delta^{(\epsilon,d\beta)}\sigma_\alpha(\beta,x)$ is then computed in a similar manner. We find that the errors in $(\epsilon,d\beta)$ are of order $10^{-4}$ and $10^{-5}$ and are under control for the values of $\beta g$ computed here. This holds for all simulated values of $m/g$ and $\alpha \lesssim 0.45$. When $\alpha$ becomes close to $1/2$ we need to be more careful, this will be discussed in \ref{subsec:sima1half}.
Note that, not surprisingly, errors increase when approaching the continuum limit ($x \rightarrow + \infty$).

\subsection{Simulations for $\alpha \approx 1/2$}\label{subsec:sima1half}
\begin{figure}[t]
\null\hfill
\begin{subfigure}[b]{.40\textwidth}
\includegraphics[width=\textwidth]{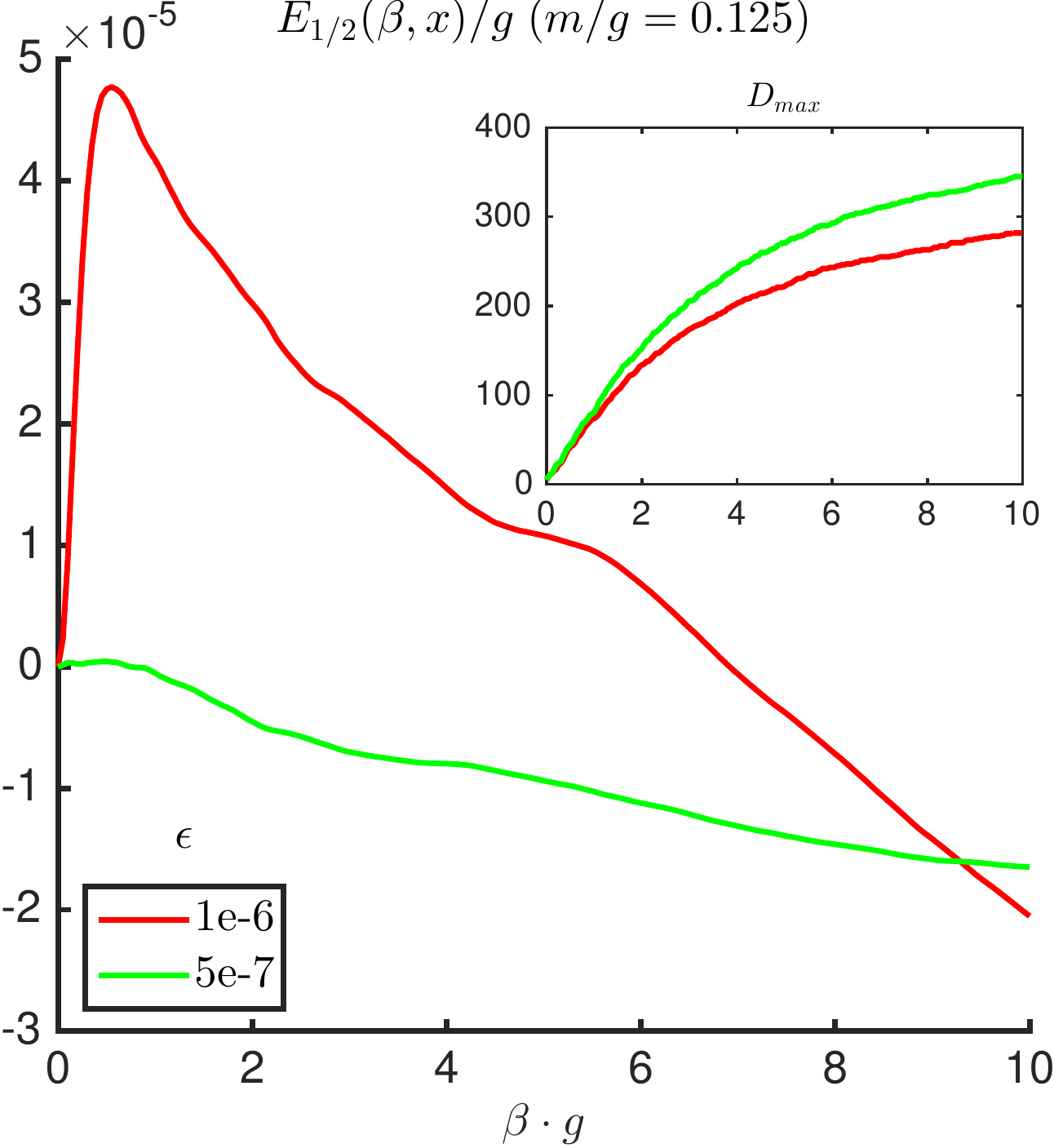}
\caption{\label{fig:EFa5e1a}}
\end{subfigure}\hfill
\begin{subfigure}[b]{.40\textwidth}
\includegraphics[width=\textwidth]{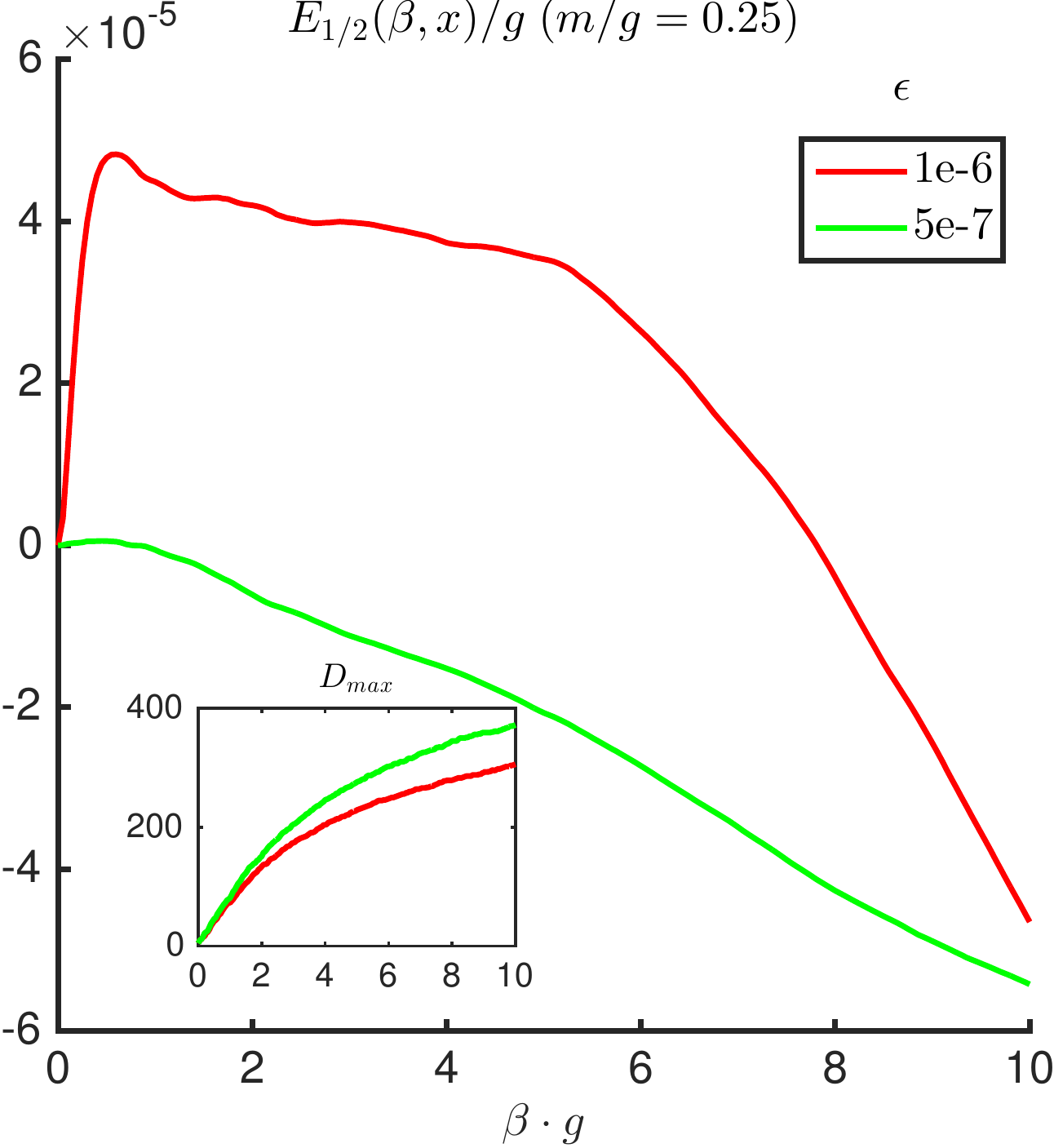}
\caption{\label{fig:EFa5e1b}}
\end{subfigure}\hfill\null
\vskip\baselineskip
\null\hfill
\begin{subfigure}[b]{.40\textwidth}
\includegraphics[width=\textwidth]{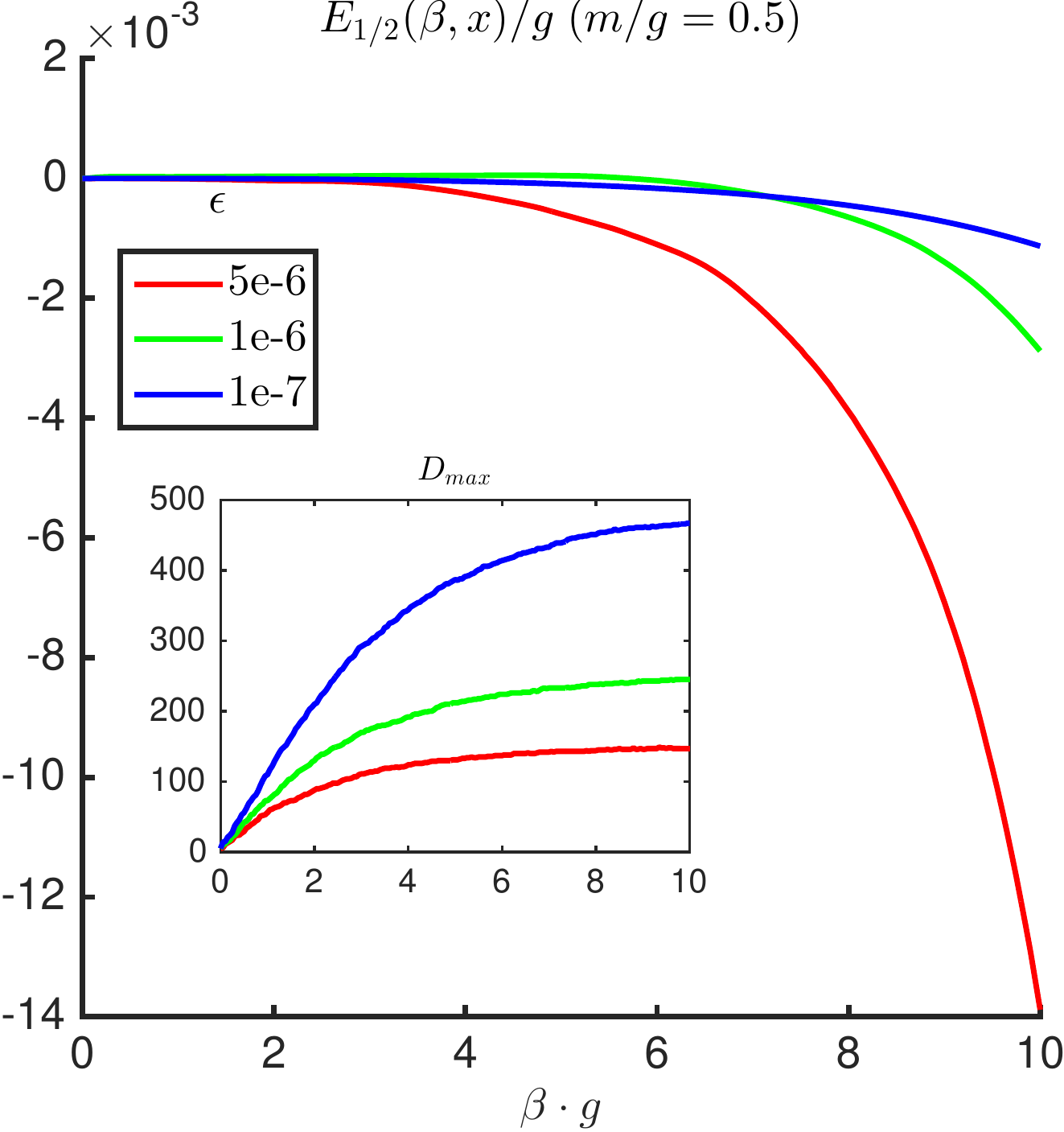}
\caption{\label{fig:EFa5e1c}}
\end{subfigure}\hfill
\begin{subfigure}[b]{.40\textwidth}
\includegraphics[width=\textwidth]{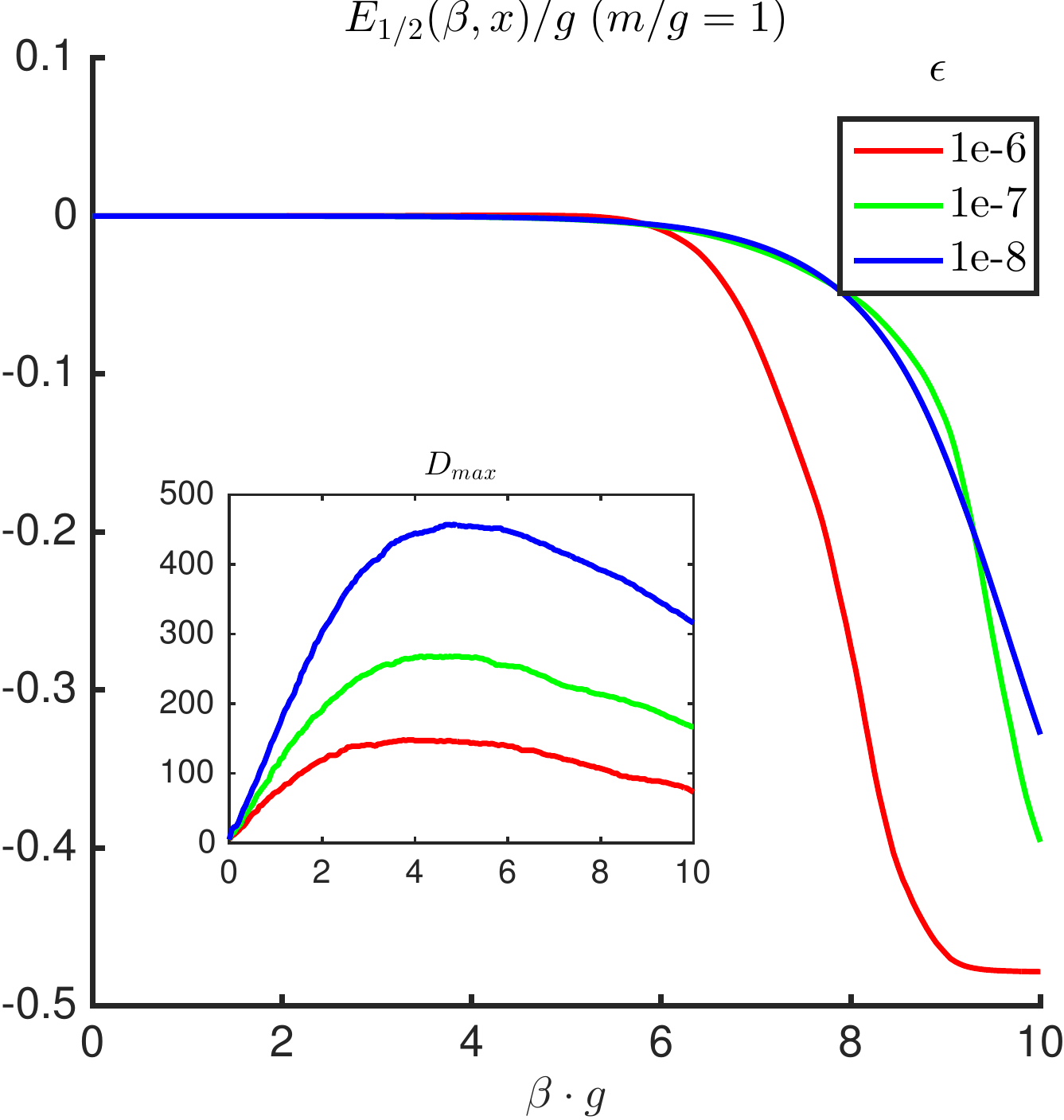}
\caption{\label{fig:EFa5e1d}}
\end{subfigure}\hfill\null
\vskip\baselineskip
\captionsetup{justification=raggedright}
\caption{\label{fig:resa5e1EF} $x = 100$, $\alpha = 0.5$.  Electric field $E_{1/2}(\beta,x)$ for different values of the tolerance $\epsilon$. Inset: the maximum bond dimension over all charge sectors as a function of $\beta g$. (a): $m/g = 0.125$. (b): $m/g = 0.25$. (c) $m/g = 0.5$. (d) $m/g = 1$.}
\end{figure}

\noindent As discussed in subsection \ref{subsec:alphaclosetohalf}, for $m/g \lesssim (m/g)_c \approx 0.33$, we already saw convergence of $E_{1/2- \delta}(\beta,x)$ to zero for $\delta \stackrel{>}{\rightarrow} 0$ for all values of $\beta g \in [0,10]$. When performing simulations for $\alpha = 1/2$ we indeed observe that $E_{1/2}(\beta,x) = 0$, see figs. \ref{fig:EFa5e1a} and \ref{fig:EFa5e1b}. When imposing higher accuracy, which is obtained by lowering the tolerance $\epsilon$ (see subsection \ref{subsec:iTEBD}) we find that $E_{1/2}(\beta,x)$ becomes smaller in magnitude. However, even for $\epsilon = 10^{-6}$, we already have that $ \vert E_{1/2}(\beta,x)\vert \lesssim 5 \times 10^{-5}$. This was expected because the Hamiltonian has for these values of $m/g$ a unique CT invariant ground state which has a zero expectation value for the electric field. 

\begin{figure}[t]
\null\hfill
\begin{subfigure}[b]{.40\textwidth}
\includegraphics[width=\textwidth]{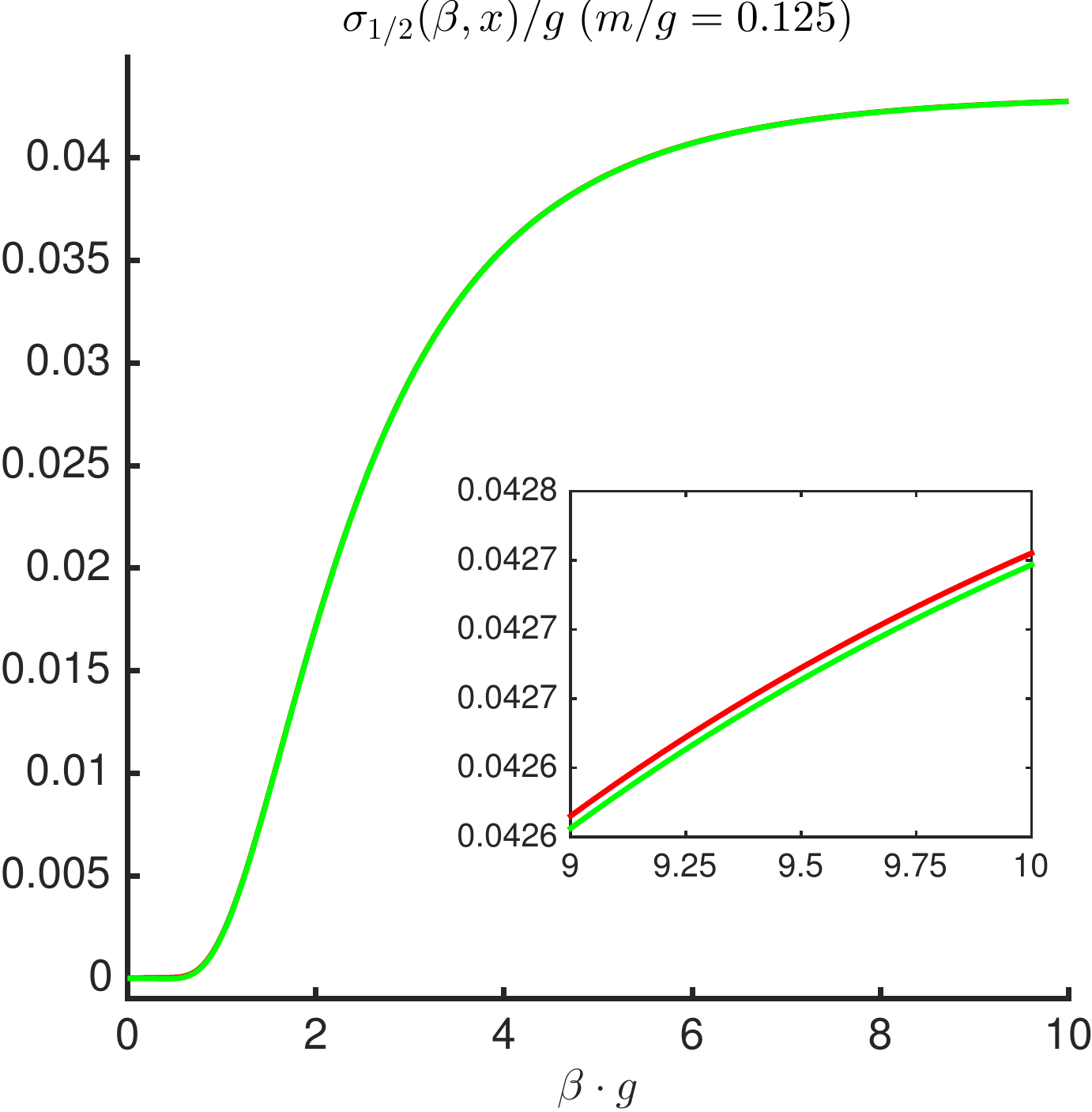}
\caption{\label{fig:EFa5e1e}}
\end{subfigure}\hfill
\begin{subfigure}[b]{.40\textwidth}
\includegraphics[width=\textwidth]{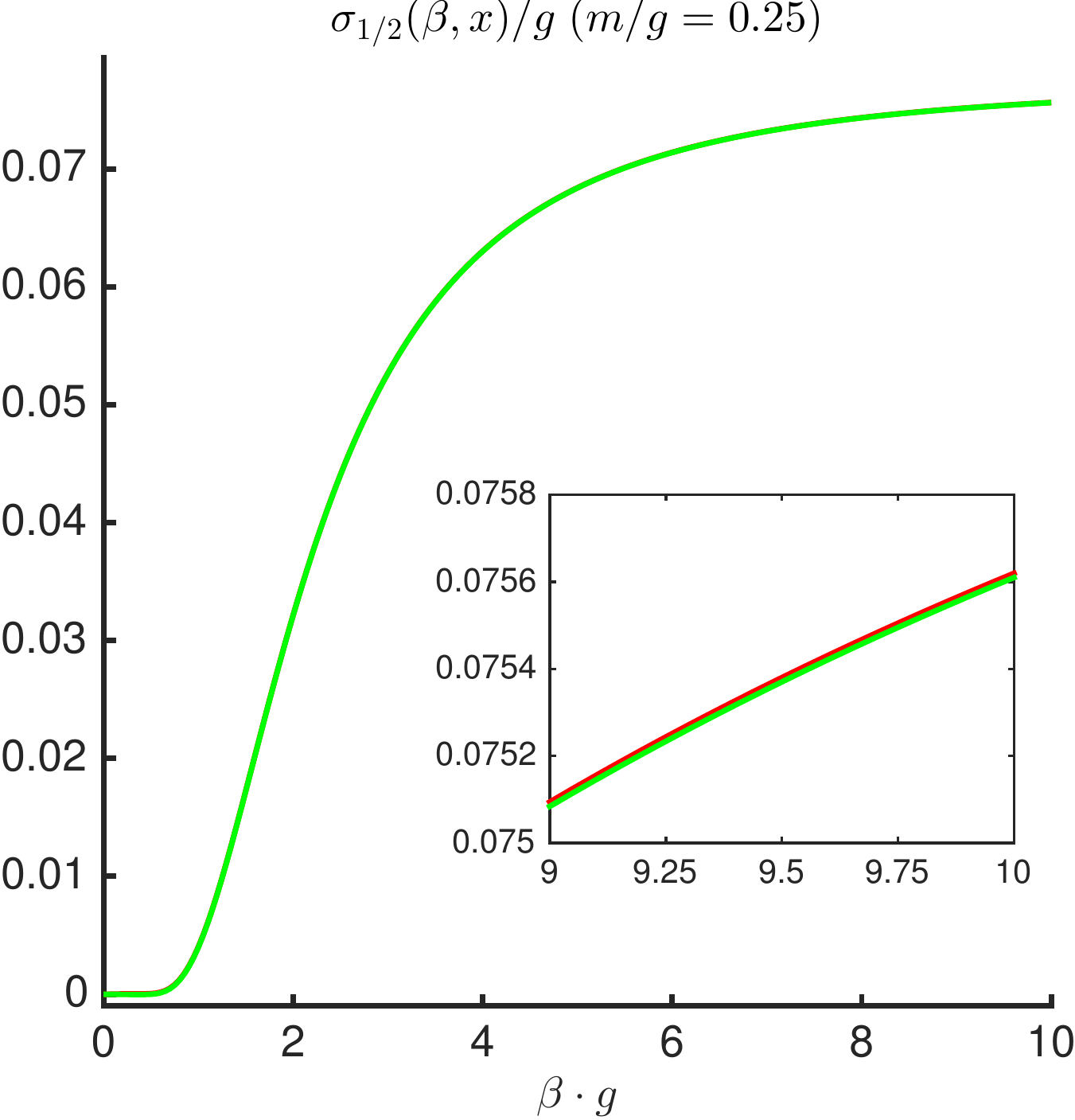}
\caption{\label{fig:EFa5e1f}}
\end{subfigure}\hfill\null
\vskip\baselineskip
\null\hfill
\begin{subfigure}[b]{.40\textwidth}
\includegraphics[width=\textwidth]{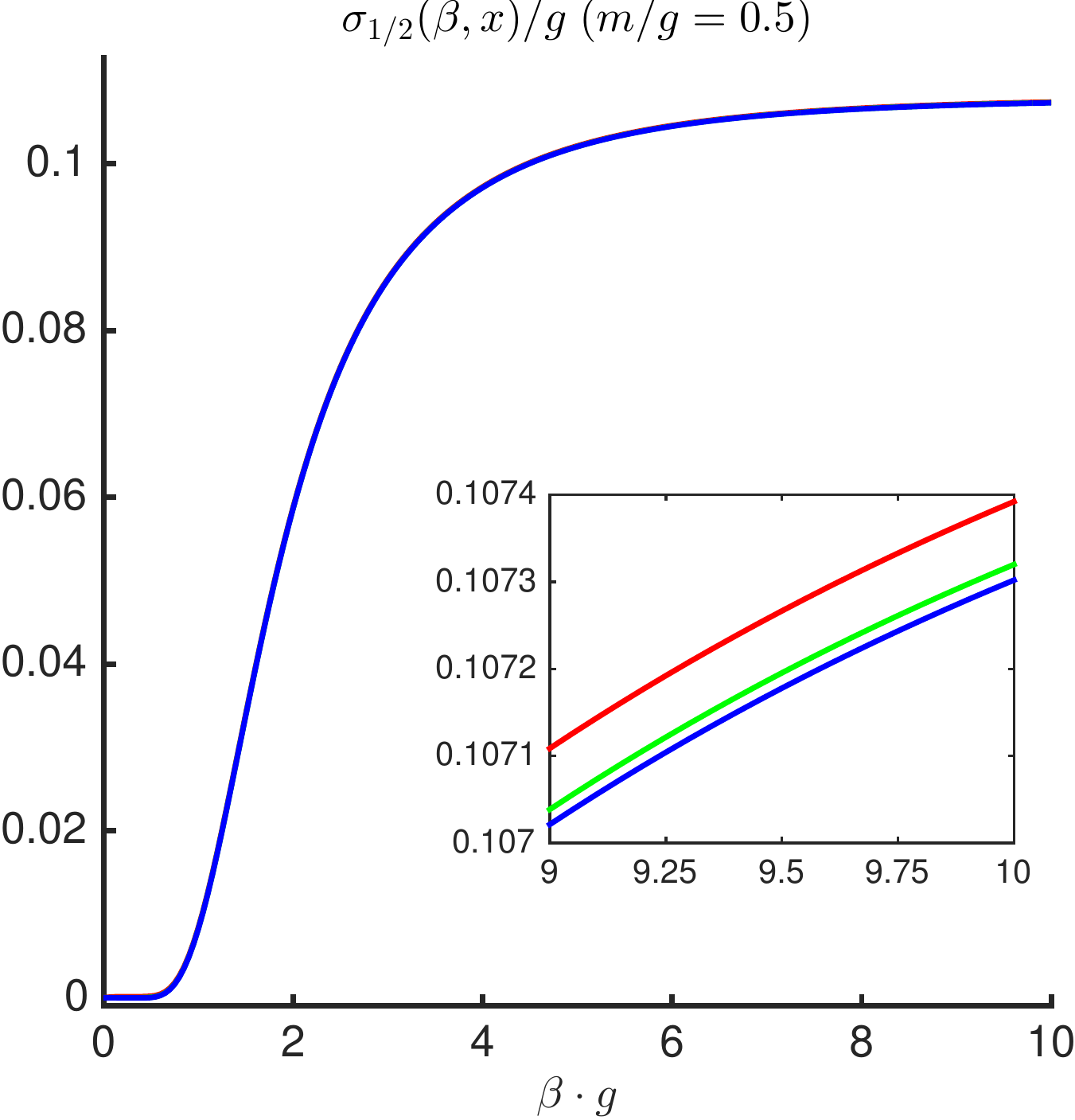}
\caption{\label{fig:EFa5e1g}}
\end{subfigure}\hfill
\begin{subfigure}[b]{.40\textwidth}
\includegraphics[width=\textwidth]{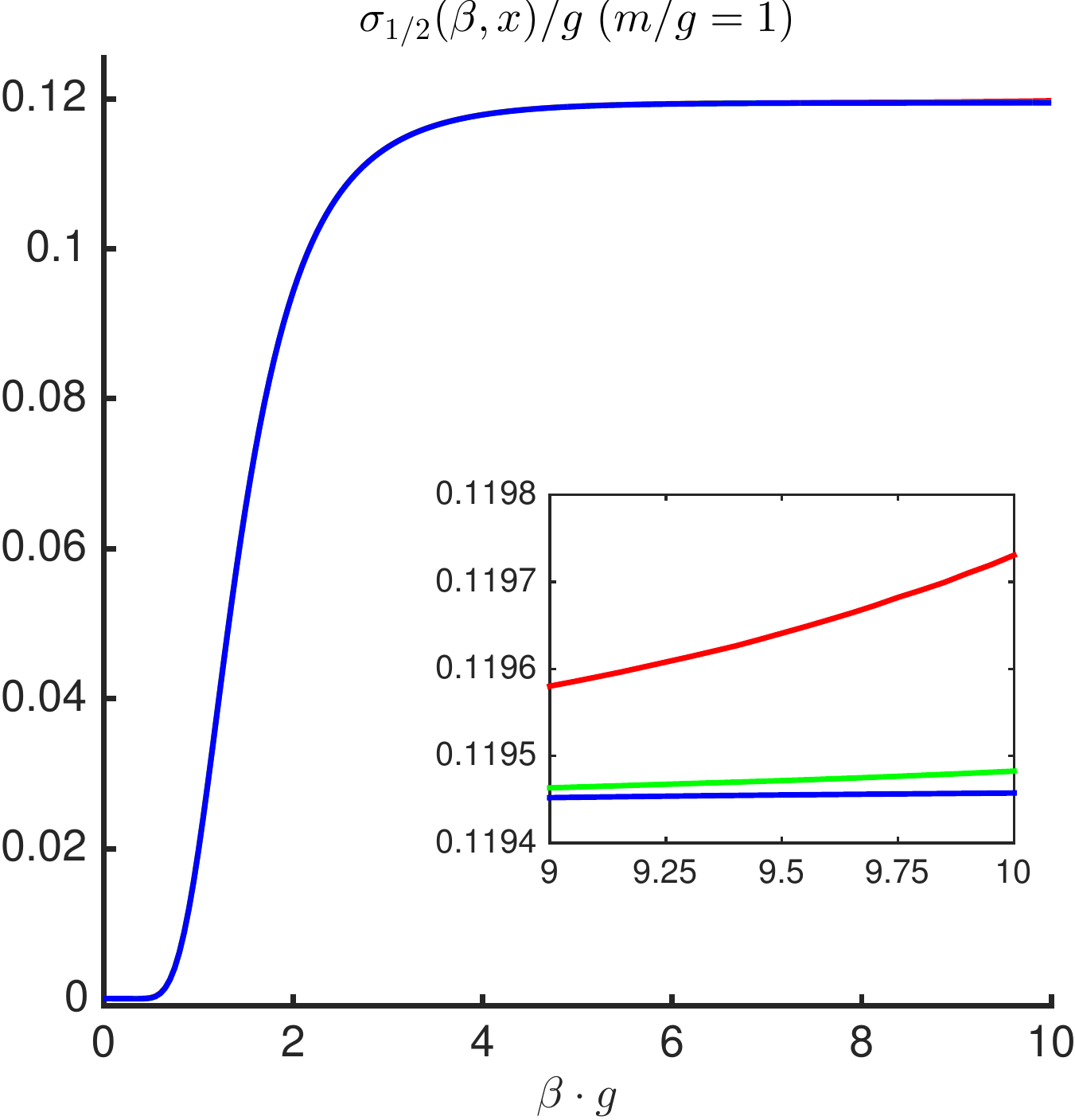}
\caption{\label{fig:EFa5e1h}}
\end{subfigure}\hfill\null
\vskip\baselineskip
\captionsetup{justification=raggedright}
\caption{\label{fig:resa5e1EFree} $x = 100$, $\alpha = 0.5$. String tension $\sigma_{1/2}(\beta,x)$ for different values of the tolerance $\epsilon$. Inset: zooming in on the interval $\beta g \in [9,10]$. (a) $m/g = 0.125$. (b) $m/g = 0.25$. (c) $m/g = 0.5$. (d) $m/g = 1$}
\end{figure}

In contrast, for $m/g \gtrsim (m/g)_c$ the electric field is not stable under variation of $\epsilon$, see figs. \ref{fig:EFa5e1c} and \ref{fig:EFa5e1d}. Because the ground state is two-fold degenerate for $\alpha = 1/2$ and $m/g \gtrsim 0.33$ for a certain value of $\beta g$ the evolution `picks' out the ground state $\ket{\Psi_{1/2 -}}$ corresponding to $\alpha = 1/2 - \delta$ in the limit $\delta \rightarrow 0$. The Gibbs states has evolved then to 
$$ \rho_{1/2-}(\beta) \varpropto \ket{\Psi_{1/2 -}}\bra{\Psi_{1/2 - }} + \mathcal{O}(e^{-\beta \Delta})$$ 
for $\beta g$ large where $\Delta$ is the mass gap of the Hamiltonian $H_{1/2 - \delta}$ in the limit $\delta \rightarrow 0$. This artifact originates mainly from the fact that the iTEBD follows a path with minimal entanglement. Clearly, the state $\rho_{1/2-}(\beta)$ has less entanglement than the exact state $\rho$. One can also observe this by investigating the maximum bond dimension $D_{max}$ over the charge sectors, see insets figs. \ref{fig:resa5e1EF} (a) - (d). We expect that $D$ increases with $\beta g$ and saturates when the system is effectively at zero temperature. For $m/g = 1$ we observe for $\beta g \gtrsim 6$ that $D_{max}$ decreases with $\beta g$. This indicates that the iTEBD algorithm converges to a state with less entanglement. It is clear that this leads to huge errors in the expectation values. Only for CT invariant observables, e.g. the free energy, the average energy and the chiral condensate, we can still find accurate results. In figs. \ref{fig:EFa5e1g} and \ref{fig:EFa5e1h} we observe that for instance the free energy is stable under variation of $\epsilon$.  

\begin{figure}[t]
\null\hfill
\begin{subfigure}[b]{.40\textwidth}
\includegraphics[width=\textwidth]{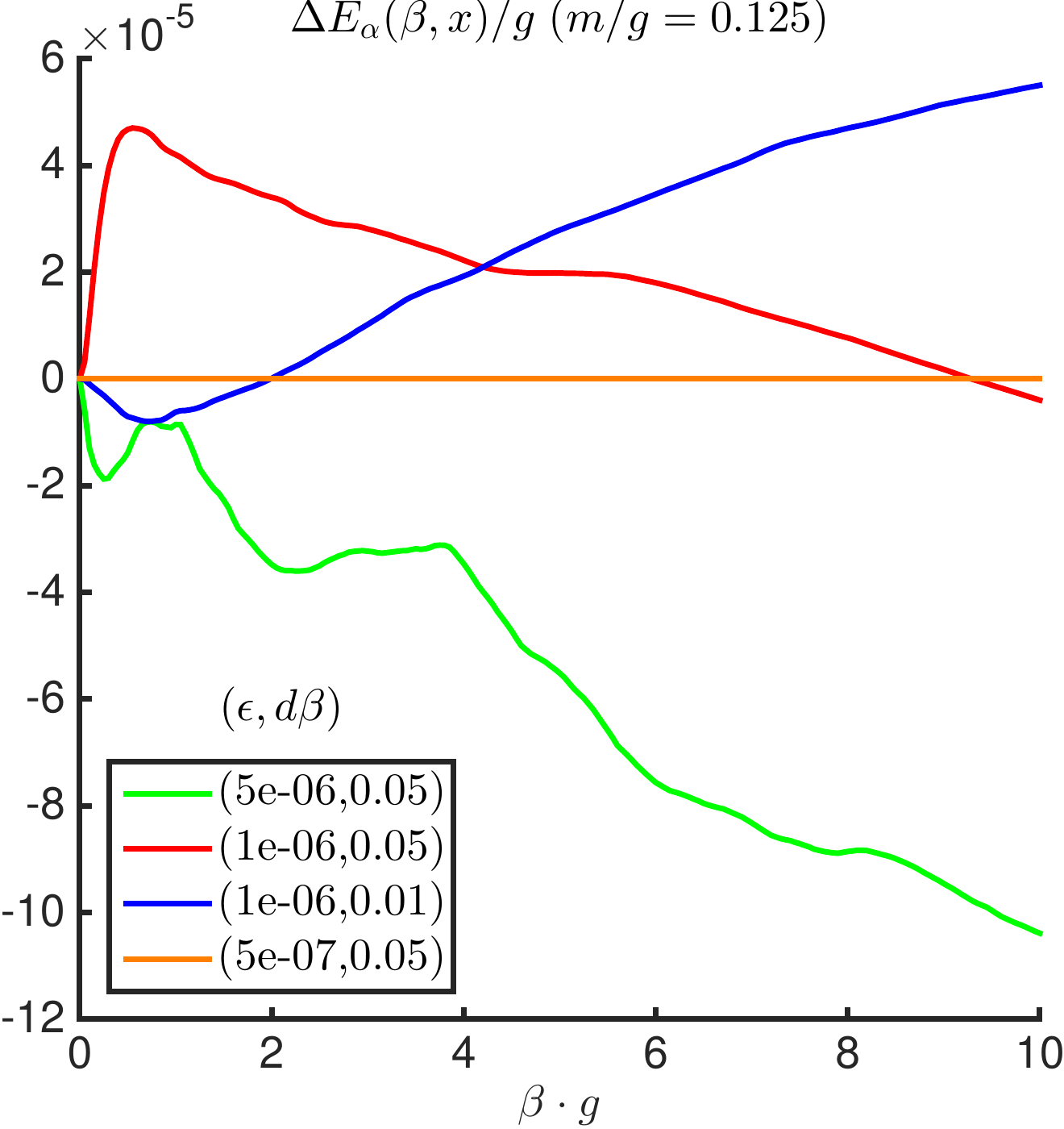}
\caption{\label{fig:EFa495e3a}}
\end{subfigure}\hfill
\begin{subfigure}[b]{.40\textwidth}
\includegraphics[width=\textwidth]{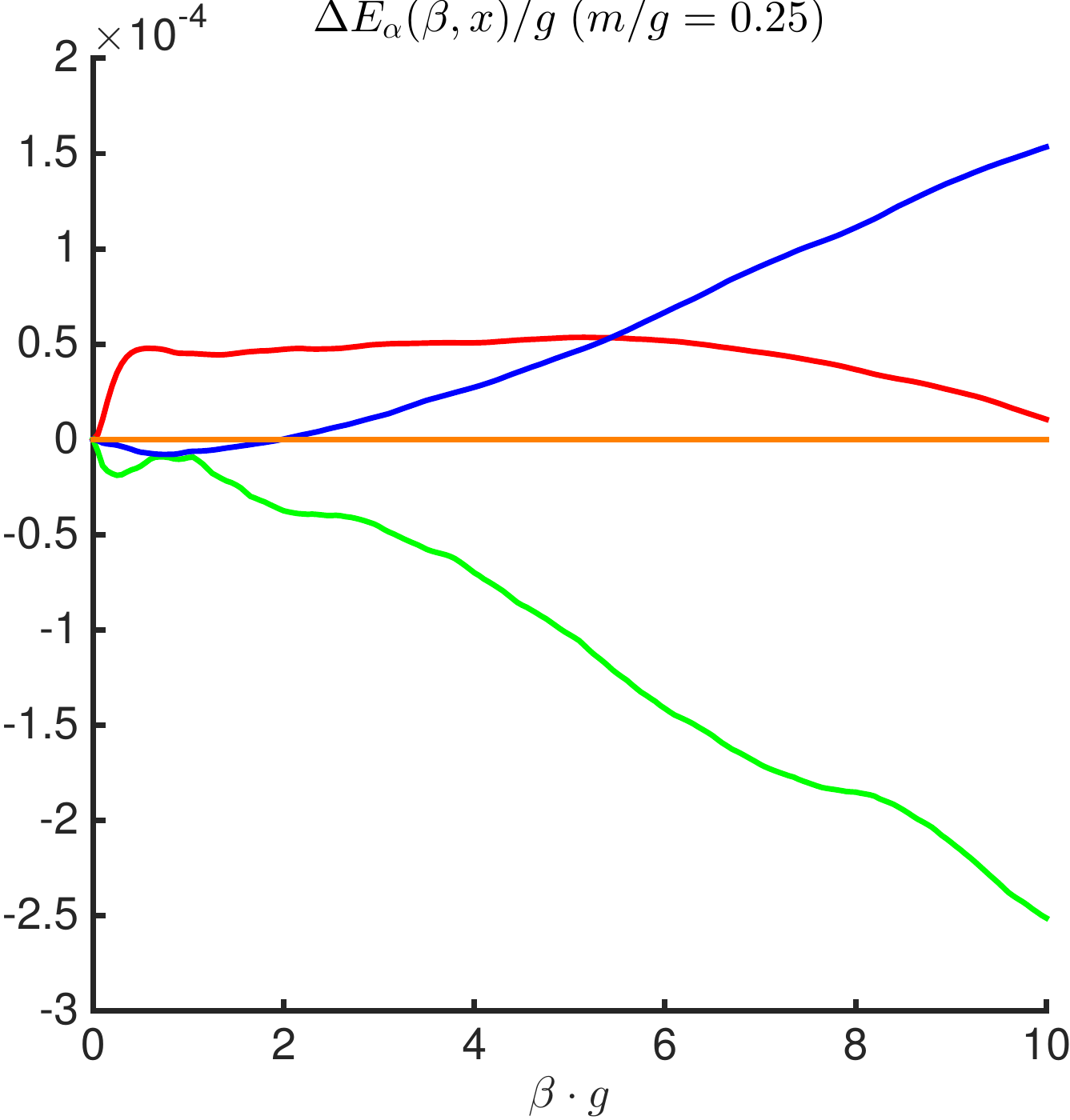}
\caption{\label{fig:EFa495e3b}}
\end{subfigure}\hfill\null
\vskip\baselineskip
\null\hfill
\begin{subfigure}[b]{.40\textwidth}
\includegraphics[width=\textwidth]{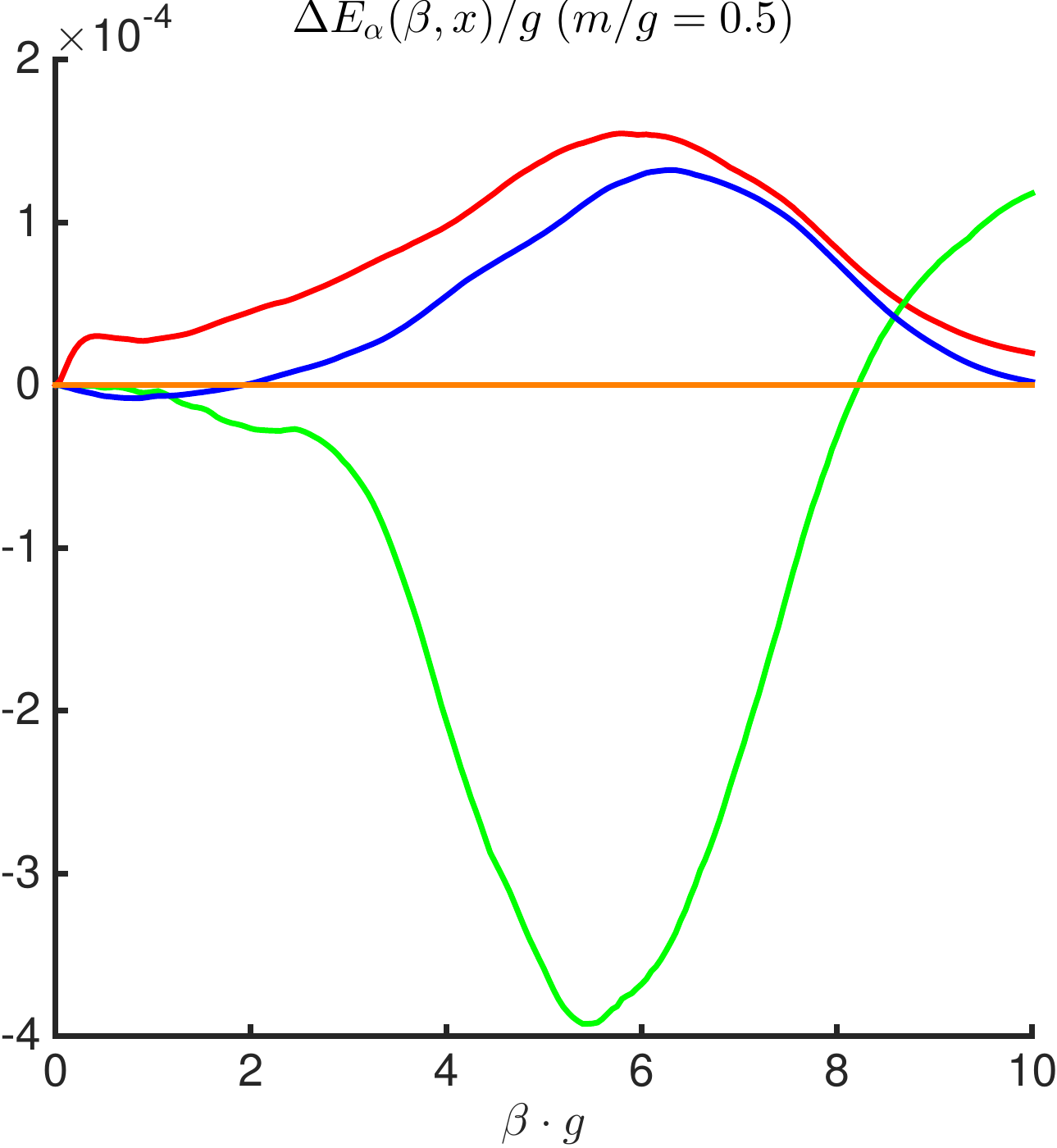}
\caption{\label{fig:EFa495e3c}}
\end{subfigure}\hfill
\begin{subfigure}[b]{.40\textwidth}
\includegraphics[width=\textwidth]{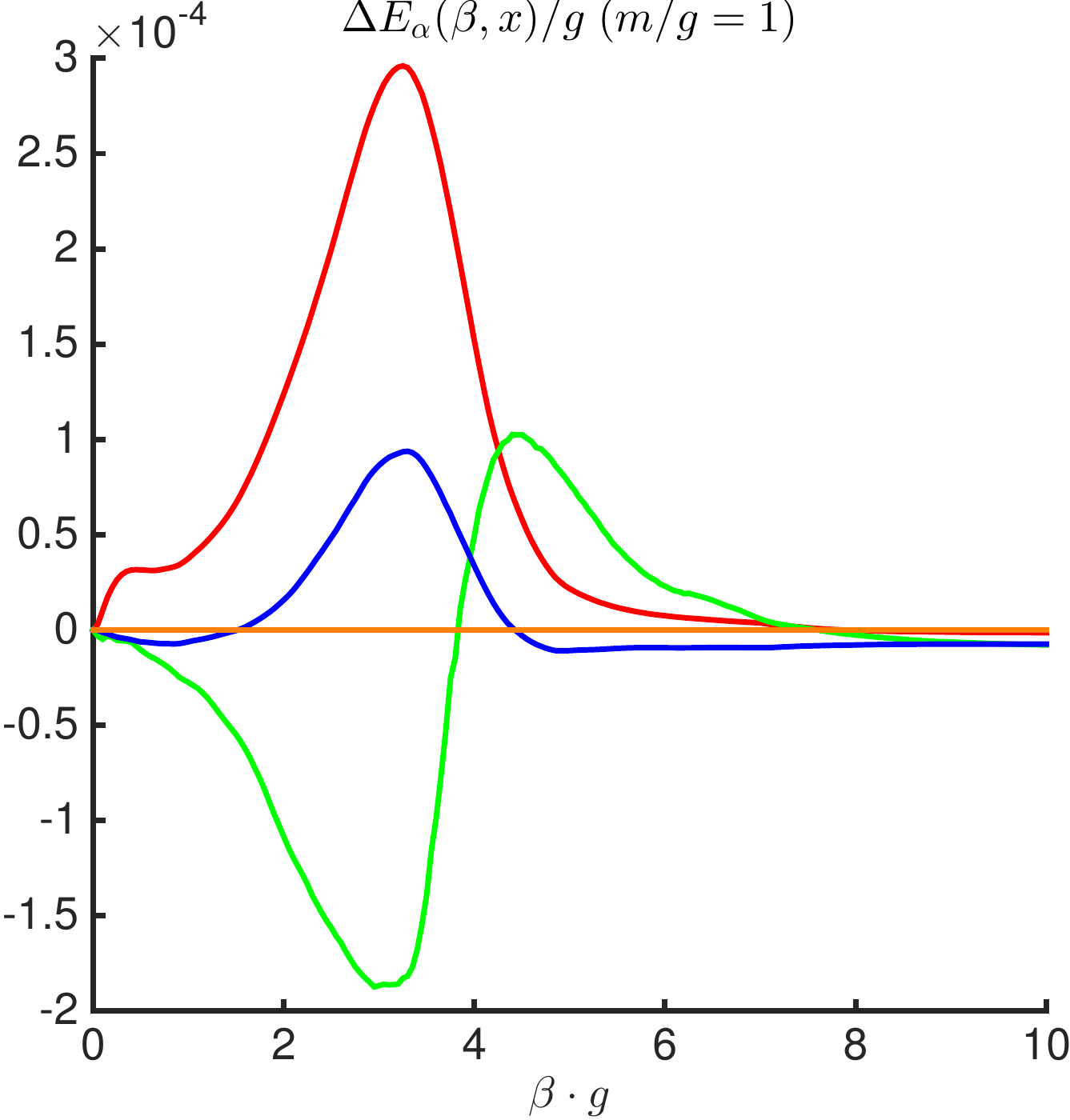}
\caption{\label{fig:EFa495e3d}}
\end{subfigure}\hfill\null
\vskip\baselineskip
\captionsetup{justification=raggedright}
\caption{\label{fig:resa495e3EF} $x = 100$, $\alpha = 0.495$. Electric field $E_{0.495}(\beta,x)$ for different values of the tolerance $\epsilon$ and step $d\beta$ with respect to the simulation for $(\epsilon,d\beta) = (5 \times 10^{-7},0.05)$. (a) $m/g = 0.125$. (b) $m/g = 0.25$. (c) $m/g = 0.5$. (d) $m/g = 1$}
\end{figure}

When $\alpha \neq 1/2$ but is close to $\alpha = 1/2$ the electric field is also less stable under variation of $\epsilon$. Therefore, when $\alpha > 0.45$ we also perform simulations for $(\epsilon,d\beta) = (5\times 10^{-7},0.05)$ and take this state as our reference state. In fig. \ref{fig:resa495e3EF} we show the electric field for $\alpha = 0.495$ for different values of $(\epsilon,d\beta)$ where we subtract the value obtained for $(\epsilon,d\beta) = (5\times 10^{-7},0.05)$:
$$\Delta E_\alpha(\beta,x) = E_{\alpha}(\beta,x) - E_{\alpha}(\beta,x)\Biggl\vert_{\epsilon = 5 \times 10^{-7}, d\beta = 0.05}.$$ 
By comparing the difference between the simulations for
\begin{itemize}
\item[-] $(\epsilon,d\beta) = (10^{-6}, 0.05)$ (red line) and $(\epsilon,d\beta) = (10^{-6},0.01)$ (blue line) (call this difference $\Delta' E_{\alpha}(\beta,x)$)
\item[-] $(\epsilon,d\beta) = (5\times 10^{-7}, 0.05)$ (orange line) and $(\epsilon,d\beta) = (10^{-6},0.05)$ (red line) (call this difference $\Delta'' E_{\alpha}(\beta,x)$)
\end{itemize}
we estimate the error introduced by taking a nonzero $\epsilon$ and $d\beta$ as 
$$2\left(\biggl\vert \Delta' E_{\alpha}(\beta,x) \biggl\vert + \biggl\vert \Delta'' E_{\alpha}(\beta,x) \biggl \vert \right). $$
As one can estimate from fig. \ref{fig:resa495e3EF} we then find this error to be no larger than of order $10^{-3}$ which is still reasonable, although this is at least one order of magnitude worser than the error we found for $\alpha = 0.25$. 

\subsection{Continuum extrapolation}\label{app:alphaneq0continuum}
\begin{figure}[t]
\null\hfill
\begin{subfigure}[b]{.40\textwidth}
\includegraphics[width=\textwidth]{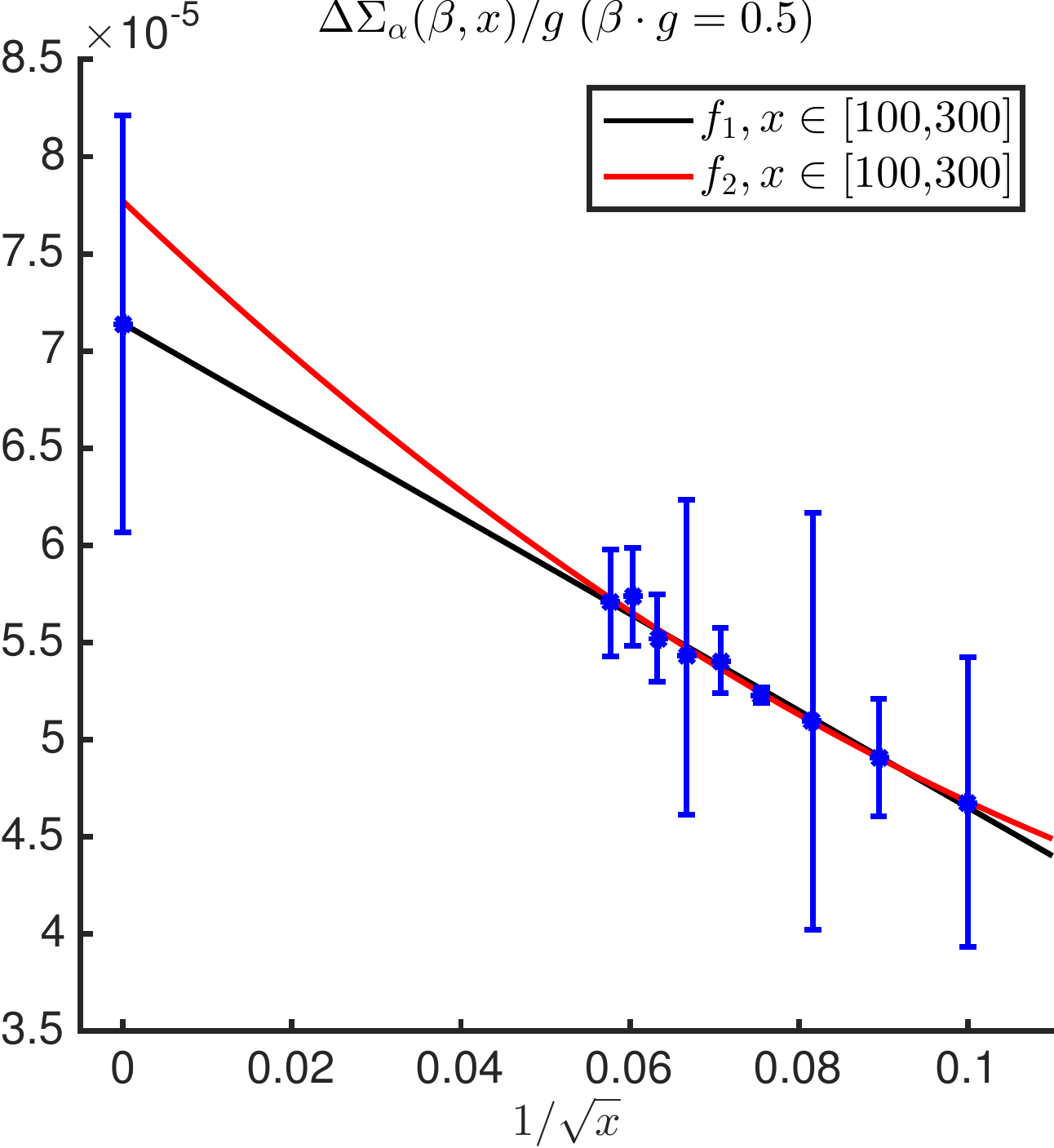}
\caption{\label{fig:CCm25a25e2ExtrCCrena}}
\end{subfigure}\hfill
\begin{subfigure}[b]{.40\textwidth}
\includegraphics[width=\textwidth]{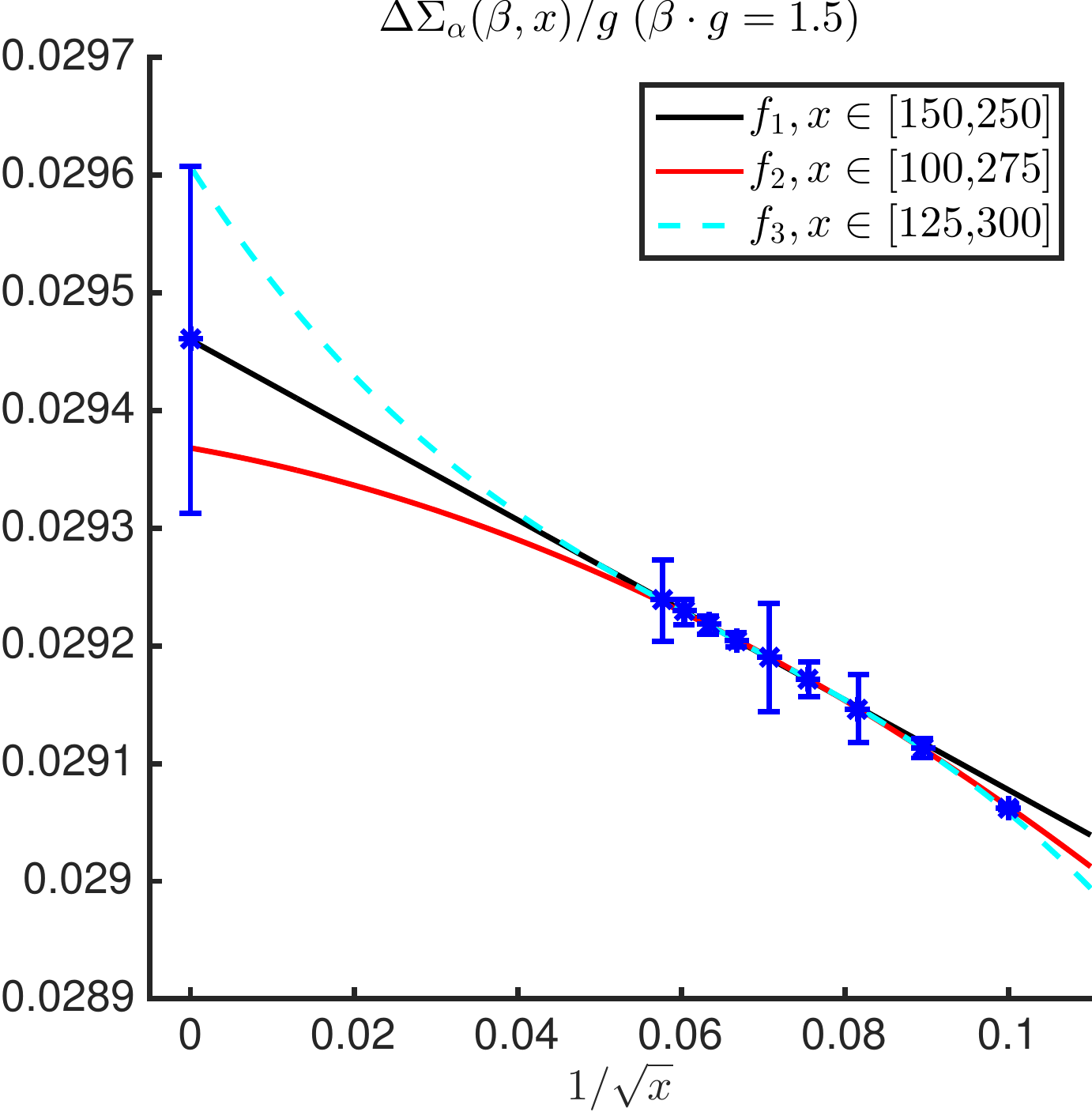}
\caption{\label{fig:CCm25a25e2ExtrCCrenb}}
\end{subfigure}\hfill\null
\vskip\baselineskip
\null\hfill
\begin{subfigure}[b]{.40\textwidth}
\includegraphics[width=\textwidth]{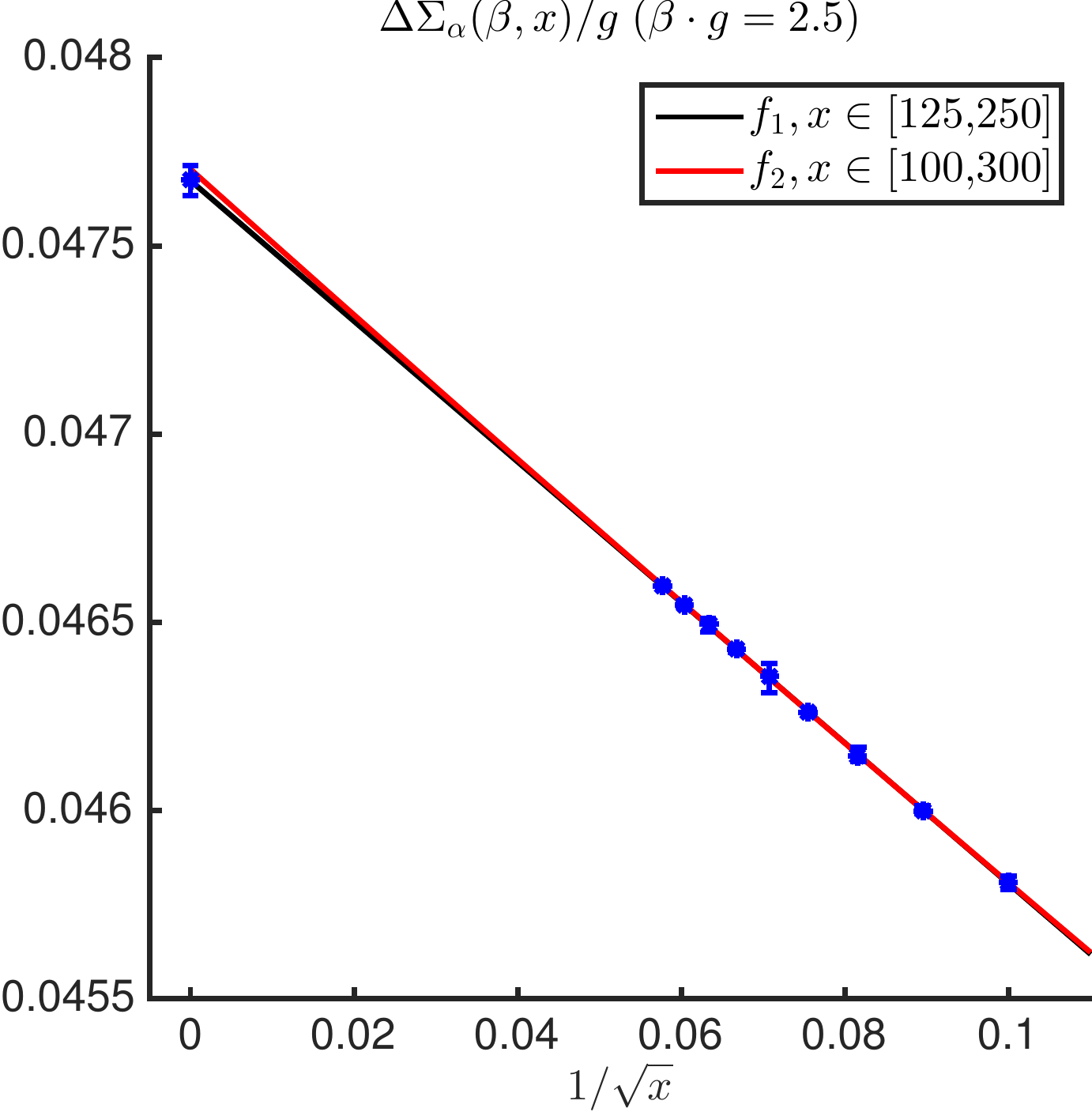}
\caption{\label{fig:CCm25a25e2ExtrCCrenc}}
\end{subfigure}\hfill
\begin{subfigure}[b]{.40\textwidth}
\includegraphics[width=\textwidth]{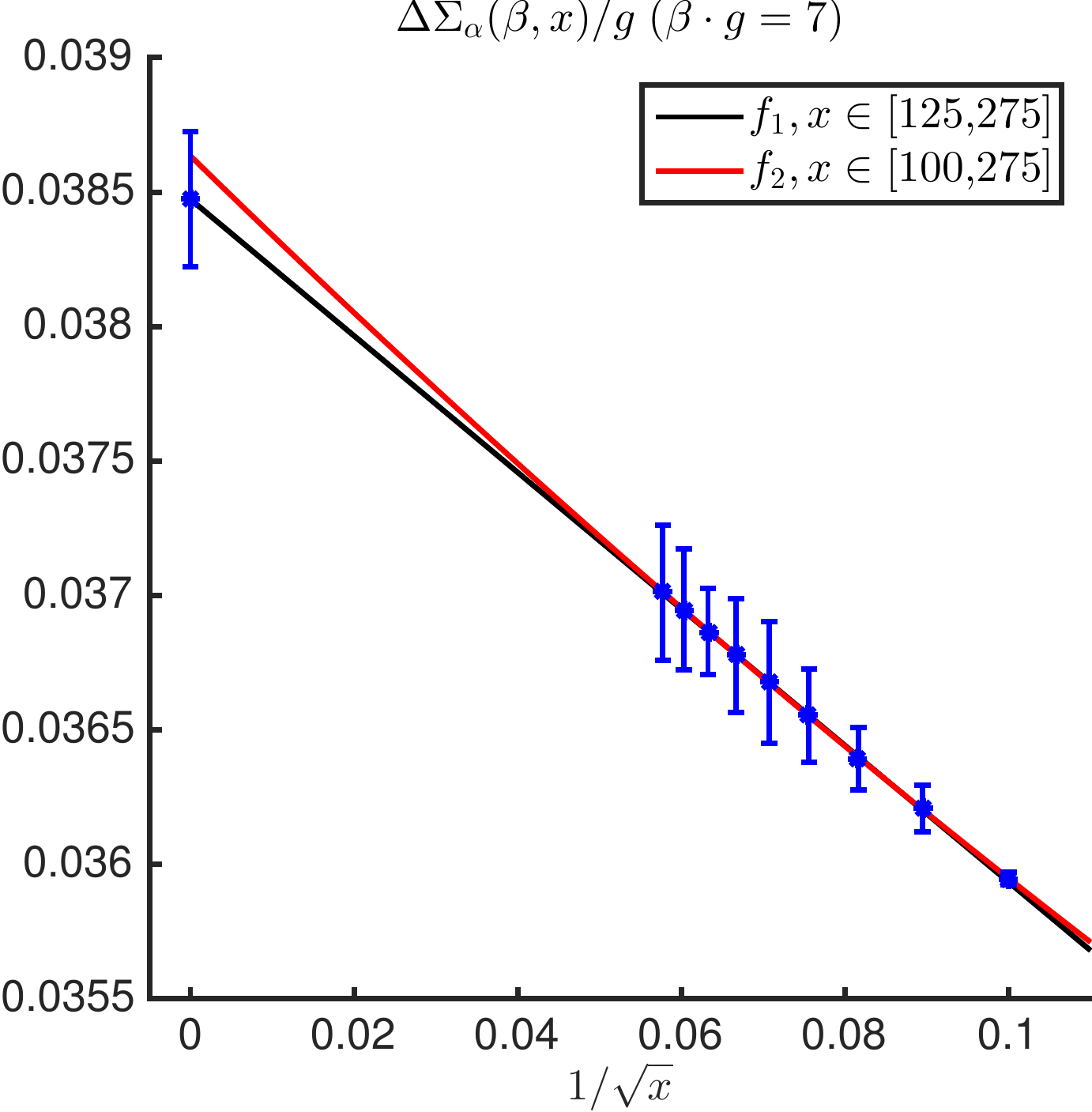}
\caption{\label{fig:CCm25a25e2ExtrCCrend}}
\end{subfigure}\hfill\null
\vskip\baselineskip
\captionsetup{justification=raggedright}
\caption{\label{figapp:CCm25a25e2ExtrCC} $m/g = 0.25,\alpha = 0.25$. Continuum extrapolation of $\Delta\Sigma_{\alpha}(\beta,x)$ for different values of $\beta g$. The black line is the fit which determines the continuum estimate. The red line and blue line are the most appropriate fits through the data for the other fitting functions. The error bars on our data for $1/\sqrt{x} \neq 0$ represent the errors $\Delta^{(\epsilon,d\beta)}\Sigma_\alpha(\beta,x)$ originating from taking nonzero $\epsilon$ and $d\beta$, as explained in subsection \ref{subsec:errFinDalphaneq0}. The error bar at $1/\sqrt{x} = 0$ is the estimated error on the continuum result.}
\end{figure}

\noindent In this subsection we discuss the continuum extrapolation of several quantities for the case $\alpha \neq 0$. For a quantity $\mathcal{Q}_\alpha$ we will subtract its $(\alpha = 0)$-value at finite temperature and thus consider $\Delta \mathcal{Q}_{\alpha}(\beta,x) \equiv \mathcal{Q}_{\alpha}(\beta,x) -  \mathcal{Q}_{\alpha = 0}(\beta,x)$. The quantities we will consider here are the chiral condensate, the average energy and the free energy (see section \ref{sec:asCon}). For these quantities we find that they scale almost linearly in $1/\sqrt{x}$ as we approach the continuum limit $x \rightarrow + \infty$. For the average energy and the free energy this is supported by their scaling at $\beta g = \infty$ which was also found to be polynomial in $1/\sqrt{x}$ \cite{Buyens2015}. Therefore, instead of the fitting functions in eq. (\ref{eq:fitfunctionapp}), we will now fit $\Delta \mathcal{Q}_{\alpha}(\beta,x)$ against 
\begin{subequations}\label{eq:polfitapp}
\be f_1(x) = A_1 + B_1\frac{1}{\sqrt{x}}  \ee
\be f_2(x) = A_2 + B_2\frac{1}{\sqrt{x}} + C_2\frac{1}{x}  \ee
\be f_3(x) = A_3 + B_3\frac{1}{\sqrt{x}} + C_3\frac{1}{x} + D_3 \frac{1}{x^{3/2}}  \ee
\end{subequations}
to obtain a continuum estimate ($x \rightarrow + \infty$). Now we performed simulations for 
\be\label{eq:xvaluesquench} x = 100,125,150,\ldots,300.\ee
The method to obtain the continuum limit and an error which includes the uncertainty in $(\epsilon,d\beta)$, the choice of fitting interval and the choice of fitting function is found in the same as for the subtracted chiral condensate for $\alpha = 0$, but now by considering the fitting functions in eq. (\ref{eq:polfitapp}) instead of the functions in eq. (\ref{eq:fitfunctionapp}), see subsection \ref{subsec:appendixCCcontinuuma} for the details.
\\
\\
\noindent \textbf{Example 1.} Let us explain this by discussing a specific example from our simulations: the continuum extrapolation of $\Delta\Sigma_\alpha(\beta)/g$ for $m/g = 0.25,\alpha = 0.25$ and $\beta g = 0.5,1.5,2.5,7$, see table \ref{table:extrapolationmdivg25CondensateRen} and fig. \ref{figapp:CCm25a25e2ExtrCC}. 

\begin{table}
\begin{tabular}{| c| |  c |   c | c | c || c | }
        \hline
     $\beta g$ &  $x$-range &  $\Delta\Sigma_{\alpha}^{(1)}(\beta)/g  \{\# \mbox{ fits}\}$ &    $\Delta\Sigma_{\alpha}^{(2)}(\beta)/g \{\# \mbox{ fits}\}$   & $\Delta\Sigma_{\alpha}^{(3)}(\beta)/g  \{\# \mbox{ fits}\}$ & $\Delta\Sigma_{\alpha}(\beta)/g$\\
     \hline 
     0.5& $[100,300]$ & $7.1 (2) \times 10^{-5} \{15\}$  & $7.8  \times 10^{-5}  \{0\} $ & $ \{0\}$& $7 (1) \times 10^{-5}$    \\ 
1.5 & $[100,300]$ & $0.02946 (1) \{15\} $ &$ 0.02937 (2) \{7\}$ & $0.02961 (17) \{2\} $ &$0.0295 (1) $  \\
2.5 &  $[100,300]$& $0.047673 (5) \{15\}$ &$ 0.04770 (2) \{3\}$  &$ \{0\} $& $0.04767 (4) $  \\
7 & $[100,300]$ & $0.03847 (3) \{15\} $ & $0.03863(3) \{3\}$  & $\{0\} $&$ 0.0385 (2) $ \\
\hline 
\end{tabular}
\captionsetup{justification=raggedright}
\caption{\label{table:extrapolationmdivg25CondensateRen} $m/g = 0.25,\alpha = 0.25$. Details on the continuum extrapolation of $\Delta\Sigma_{\alpha}(\beta)$. }
\end{table}

Consider first the case when $\beta g = 0.5$, fig. \ref{fig:CCm25a25e2ExtrCCrena}. For all values of $x$ in eq. (\ref{eq:xvaluesquench}) the estimated error $\Delta^{(\epsilon,d\beta)} \Sigma_\alpha(\beta,x)$ (see subsection \ref{subsec:errFinDalphaneq0}) was smaller than $4 \times 10^{-4}$ as can be observed from the error bars in fig. \ref{fig:CCm25a25e2ExtrCCrena}. Therefore we included all these values in our analysis. Similar as for the subtracted chiral condensate in subsection \ref{subsec:appendixCCcontinuuma}, given the fitting function $f_n$ eq. (\ref{eq:polfitapp}), we perform all possible fits of against at least $n+4$ consecutive $x-$values (we take at least $n+4$ data point such that the number of degrees of freedom $N_{dof}$ is larger or equal to 3 which lowers the risk of overfitting the data). For each of fit $\theta$ we get an estimate for the coefficients $A_n,B_n,C_n,D_n$ (with $C_n = 0$ for $n = 1$ and $D_n = 0$ for $n = 1,2$): $A_n^{(\theta)}$, $B_n^{(\theta)}$, $C_n^{(\theta)}$, $D_n^{(\theta)}$. 

Here also we discard the fits that give statistically insignificant coefficients (p-value $\geq$ 0.05). In practice, this means that we discard the fits $f_n$ where the error on one of its coefficients $(A_n^{(\theta)},B_n^{(\theta)},C_n^{(\theta)},\ldots)$ is larger than approximately half of its value. 
 $$g_\theta(x)  = A_n^{(\theta)} + B_n^{(\theta)} \frac{1}{\sqrt{x}} + C_n^{(\theta)}\frac{1}{x} + D_n^{(\theta)} \frac{1}{x^{3/2}}$$
(with $C_n^{(\theta)} = 0$ for $n =1$ and $D_n^{(\theta)} = 0$ for $n = 1,2$).

All the values $A_n^{(\theta)}$ we obtain are an estimate for the continuum value of $\Delta\Sigma_\alpha^{(n)}(\beta)$ for the fitting ansatz $f_n$. Let us denote with $\{A_n^{(\theta)}\}_{\theta = 1\ldots R_n}$ all the $A_n$'s obtained from a fit $\theta$ against $f_n$ which produces significant coefficients with
$$A_n^{(1)} \leq A_n^{(2)} \leq \ldots \leq A_n^{(R_n)} .$$
For each fit $\theta$ we also compute its $\chi^2$ value (see eq. (\ref{eq:chisq})):
$$ \chi_\theta^2 = \sum_{j \in \mbox{fit} \theta} \left(\frac{g_\theta(x_j) - \Delta\Sigma_{\alpha}(\beta,x_j)}{\Delta^{(\epsilon,d\beta)}(\Delta\Sigma_{\alpha}(\beta,x_j))}\right)^2. $$
When our dataset is large enough the quantity $\chi_\theta^2/N_{dof}^{\theta}$, with $N_{dof}^{\theta}$ the number of degrees of freedom of the fit (here the number of data points used in the fit minus $n+1$), gives an indication whether $g_\theta$ fits the dataset well ($\chi_\theta^2/N_{dof}^{\theta} \ll 1$) or not ($\chi_\theta^2/N_{dof}^{\theta} \gg 1$). 

As can be found in table \ref{table:extrapolationmdivg25CondensateRen} (curly brackets $\{\ldots\}$) for $\beta g = 0.5$ we found 15 significant fits with $\chi_\theta^2/N_{dof}^{\theta} \leq 1$ for $f_1$ while for $f_2$ and $f_3$ there were no statistically significant fits with $\chi_\theta^2/N_{dof}^{\theta} \leq 1$. As explained in subsection \ref{subsec:appendixCCcontinuuma}, because there are more than $10$ statistically significant fits with $\chi_\theta^2/N_{dof}^{\theta} \leq 1$, we deduce the continuum estimate for the fitting function by taking the median of the distribution of these $A_1^{(\theta)}$ weighed by $\exp(-\chi_\theta^2/N_{dof})$. The systematic error of the choice of $x-$interval comes from the $68.3 \%$ confidence interval. As can be read from table \ref{table:extrapolationmdivg25CondensateRen} we have $\Delta\Sigma_\alpha^{(1)}(\beta)/g = 7.1(2) \times 10^{-5}$. The fit which corresponds to this estimate for $\Delta\Sigma_\alpha^{(1)}(\beta)/g$ is show in fig. \ref{fig:CCm25a25e2ExtrCCrena} by the black line and corresponds to a fit through all the data points. 

Although there are no statistically significant fits with $\chi_\theta^2/N_{dof}^{\theta} \leq 1$ for $f_2$ and $f_3$ we still want to have a rough idea of the error originating from the choice of fitting function. Therefore we perform a fit of $f_2$ against all data points, red line in fig. \ref{fig:CCm25a25e2ExtrCCrena}, and take the $A_2^{(\theta)}$ that comes out of that as the estimate for $\Delta\Sigma_\alpha^{(2)}(\beta)/g \approx 7.8 \times 10^{-5}$. The error for the choice of fitting function is now the difference $\Bigl\vert\Delta\Sigma_\alpha^{(1)}(\beta)/g - \Delta\Sigma_\alpha^{(2)}(\beta)/g \Bigl\vert \approx 7 \times 10^{-6}$. Note however that both this error and the error originating from the choice of fitting interval is smaller than the error originating from taking nonzero $(\epsilon,d\beta)$ which equals 
$$\max_{x_j} \left(\Delta^{(\epsilon,d\beta)}\Sigma_{sub}(\beta,x_j)\right) \approx 1\times 10^{-5} $$
(it comes from the relatively large error bar for $x = 150$, i.e. $1/\sqrt{x} \approx 0.08$, in fig. \ref{fig:CCm25a25e2ExtrCCrena}). The relatively large error is due to the fact that $\Delta\Sigma_\alpha(\beta,x)$ is very small here (only of order $1\times 10^{-4}$) and hence errors of order $10^{-5}$ lead to relatively large errors. This problem is resolved by decreasing $\epsilon$ (and $d\beta$) as has been done in subsection  \ref{subsec:deconfinement} for $x = 100$. There it is found that for $\beta g \lesssim 0.5$ the expectation values are exponentially suppressed with temperature $T/g = 1/\beta g$. 

Consider now the case $\beta g = 1.5$, fig. \ref{fig:CCm25a25e2ExtrCCrenb}. Again we find for $f_1$ 15 significant fits with $\chi_\theta^2/N_{dof}^{\theta} \leq 1$. By weighing the distribution of the estimates with $\exp(-\chi_\theta^2/N_{dof}^{\theta})$ and taking the mean we obtain the estimate $\Delta\Sigma_\alpha^{(1)}(\beta)/g = 0.02946(1)$. The corresponding fit which gives the estimate $\Delta\Sigma_\alpha^{(1)}(\beta)/g$ is shown in fig. \ref{fig:CCm25a25e2ExtrCCrenb} (black line). For $f_2$ and $f_3$ we have fewer than 10 significant fits with $\chi_\theta^2/N_{dof}^{\theta} \leq 1$ and the histogram of $\exp(-\chi_\theta^2/N_{dof}^{\theta})$ might be dominated by only a few fits. Therefore we use a more conservative approach: we consider all the significant fits $\theta$ with $\chi_\theta^2/N_{dof}^{\theta} \leq 1$ and take the $A_n^{(\theta)}$ coming from the fit which has the least error in $(\epsilon,d\beta)$, i.e. the fit $\theta$ for which 
$$\frac{1}{\vert \mbox{fit} \theta \vert}\sqrt{\sum_{j \in \mbox{fit} \theta}\left(\Delta^{(\epsilon,d\beta)}[\Delta\Sigma_{\alpha}(\beta,x_j)]\right)^2}$$
is minimal. These fits are also displayed in fig. \ref{fig:CCm25a25e2ExtrCCrenb} (red line for $f_2$ and dashed blue line for $f_3$). The error on the estimates is obtained by comparing $\Delta\Sigma_\alpha^{(n)}(\beta,x)$ with the most outlying $A_n^{(\theta)}$ obtained from a statistically significant fit against $f_n$ with $\chi_\theta^2/N_{dof}^{\theta} \leq 1$. This lead to the estimates $\Delta\Sigma_\alpha^{(2)}(\beta)/g = 0.02937 (2)$ and $\Delta\Sigma_\alpha^{(3)}(\beta)/g = 0.02961 (17)$. 
Finally, the error for the choice of fitting ansatz is 
$$ \max\left( \left\vert \Delta\Sigma_\alpha^{(2)}(1.5/g)/g - \Delta\Sigma_\alpha^{(1)}(1.5/g)/g \right\vert, \left\vert \Delta\Sigma_\alpha^{(3)}(1.5/g)/g - \Delta\Sigma_\alpha^{(1)}(1.5/g)/g\right\vert \right) \approx 1.5 \times 10^{-4}.$$
Therefore our final result is $\Delta\Sigma_\alpha(1.5/g)/g = 0.0295 (1)$. 

Finally, let us also briefly discuss the cases $\beta g = 2.5$ and $\beta g = 7$, see figs. \ref{fig:CCm25a25e2ExtrCCrenc} and \ref{fig:CCm25a25e2ExtrCCrend}. In both cases the continuum estimate for the fit $f_1$ is obtained as before. Because there are only three statistically significant fits with $\chi_\theta^2/N_{dof}^{\theta} \leq 1$ we have to use the more conservative approach to obtain the estimate $\Delta\Sigma_{\alpha}^{(2)}(\beta)$ (i.e. taking the fit which has the least error in $(\epsilon,d\beta)$ and comparing this with the most outlying other estimate). For the cubic fitting ansatz $f_3$ there are statistically significant fits with $\chi_\theta^2/N_{dof}^{\theta} \leq 1$. Hence to obtain an error for the choice of fitting ansatz we can only compare $\Delta\Sigma_\alpha^{(1)}(\beta)$ and $\Delta\Sigma_\alpha^{(2)}(\beta)$, the error is then estimated as
$$ \vert\Delta\Sigma_\alpha^{(1)}(\beta) - \Delta\Sigma_\alpha^{(2)}(\beta) \vert. $$
The values can be found in table \ref{table:extrapolationmdivg25CondensateRen}. 

\begin{table}
\begin{tabular}{| c| |  c |   c | c | c || c | }
        \hline
     $\beta g$ &  $x$-range &  $\sigma_{\alpha}^{(1)}(\beta)/g^2  \{\# \mbox{ fits}\}$ &    $\sigma_{\alpha}^{(2)}(\beta)/g^2\{\# \mbox{ fits}\}$   & $\sigma_{\alpha}^{(3)}(\beta)/g^2  \{\# \mbox{ fits}\}$ & $\sigma_{\alpha}(\beta)/g^2$\\
     \hline 
     0.5& $[100,300]$ & $2.3 (2) \times 10^{-5} \{2\}$  & $7 (6)  \times 10^{-5}  \{4\} $ & $ -2.5 (5) \times 10^{-4}\{2\}$& $2 (31) \times 10^{-5}$    \\ 
1.5 & $[100,300]$ & $0.007816 (3) \{15\} $ &$ 0.007801 \{ 0\}$ & $ \{0\} $ &$0.0078 (1) $  \\
2.5 &  $[100,300]$& $0.017105 (1) \{15\}$ &$ 0.017105  \{0\}$  &$ \{0\} $& $0.01710 (8) $  \\
7 & $[100,300]$ & $0.021619 (8) \{15\} $ & 0.02166 (2)$\{8\} $ &$ \{0\} $ &$ 0.02162 (5) $ \\
\hline 
\end{tabular}
\captionsetup{justification=raggedright}
\caption{\label{table:extrapolationmdivg25EFreeRen} $m/g = 0.25,\alpha = 0.25$. Details on the continuum extrapolation of $\sigma_{\alpha}(\beta)$. }
\end{table}

\begin{figure}[t]
\null\hfill
\begin{subfigure}[b]{.40\textwidth}
\includegraphics[width=\textwidth]{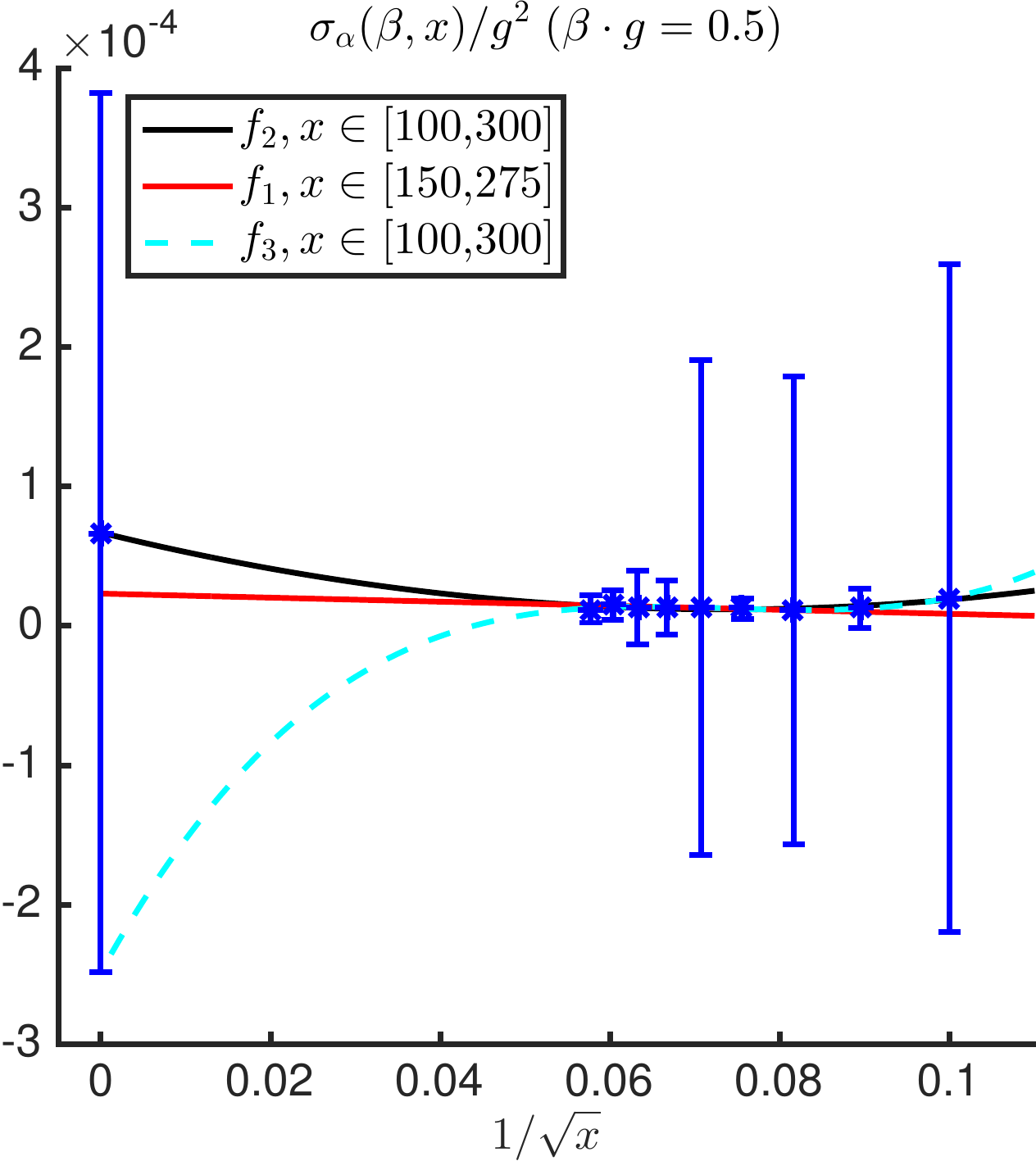}
\caption{\label{fig:continuumExtrapolationEFreea25e2a}}
\end{subfigure}\hfill
\begin{subfigure}[b]{.40\textwidth}
\includegraphics[width=\textwidth]{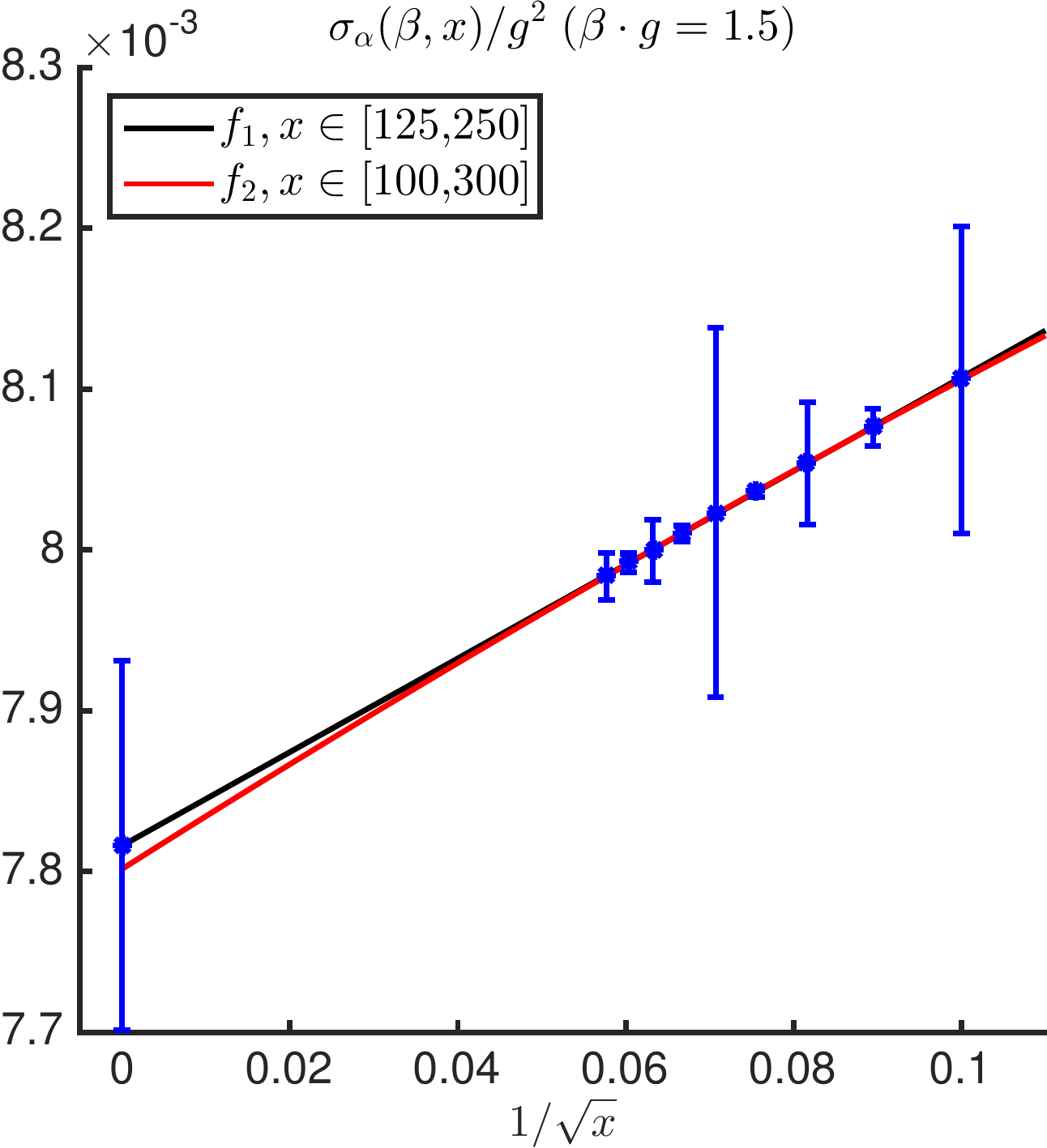}
\caption{\label{fig:continuumExtrapolationEFreea25e2b}}
\end{subfigure}\hfill\null
\vskip\baselineskip
\null\hfill
\begin{subfigure}[b]{.40\textwidth}
\includegraphics[width=\textwidth]{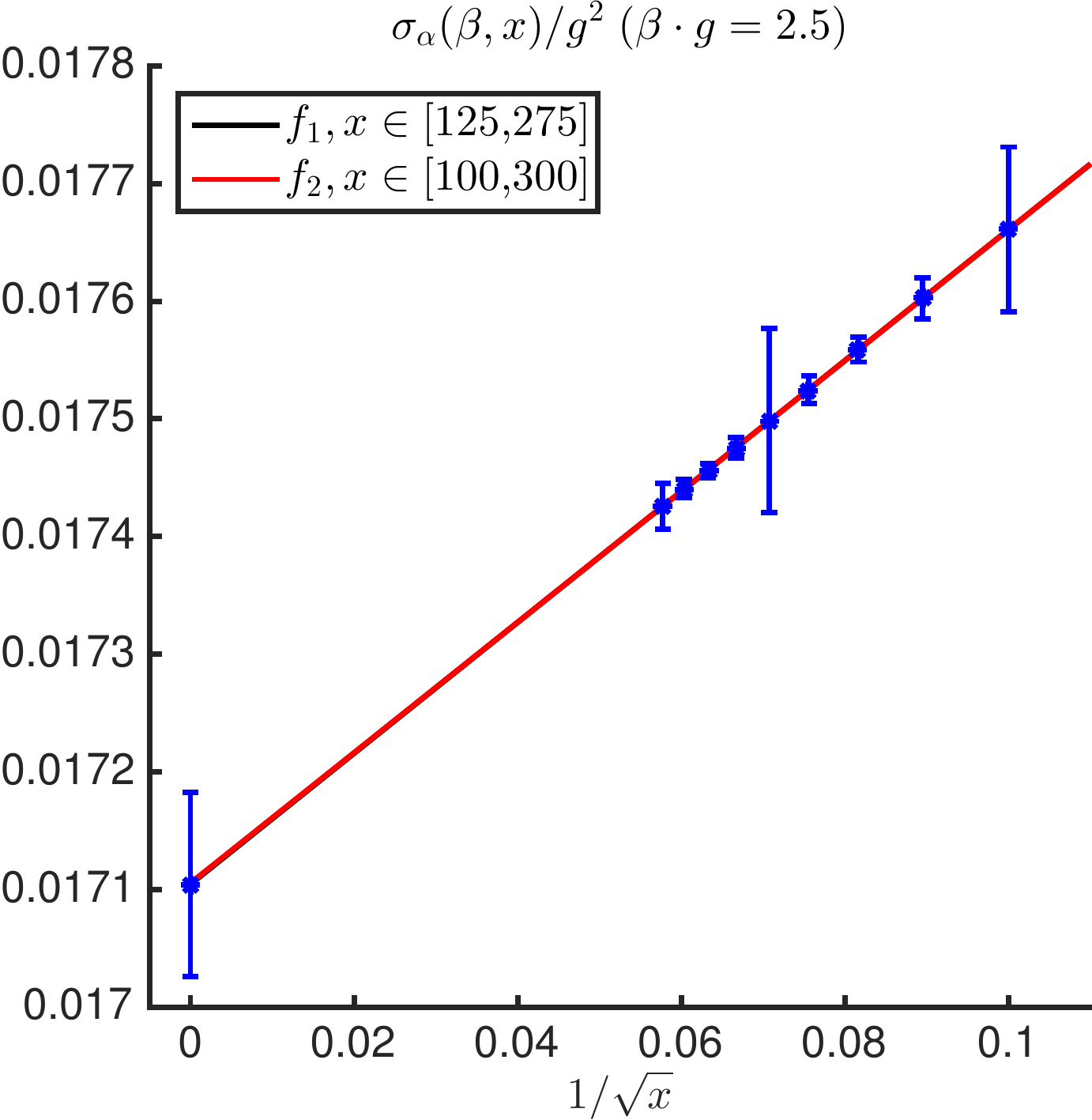}
\caption{\label{fig:continuumExtrapolationEFreea25e2c}}
\end{subfigure}\hfill
\begin{subfigure}[b]{.40\textwidth}
\includegraphics[width=\textwidth]{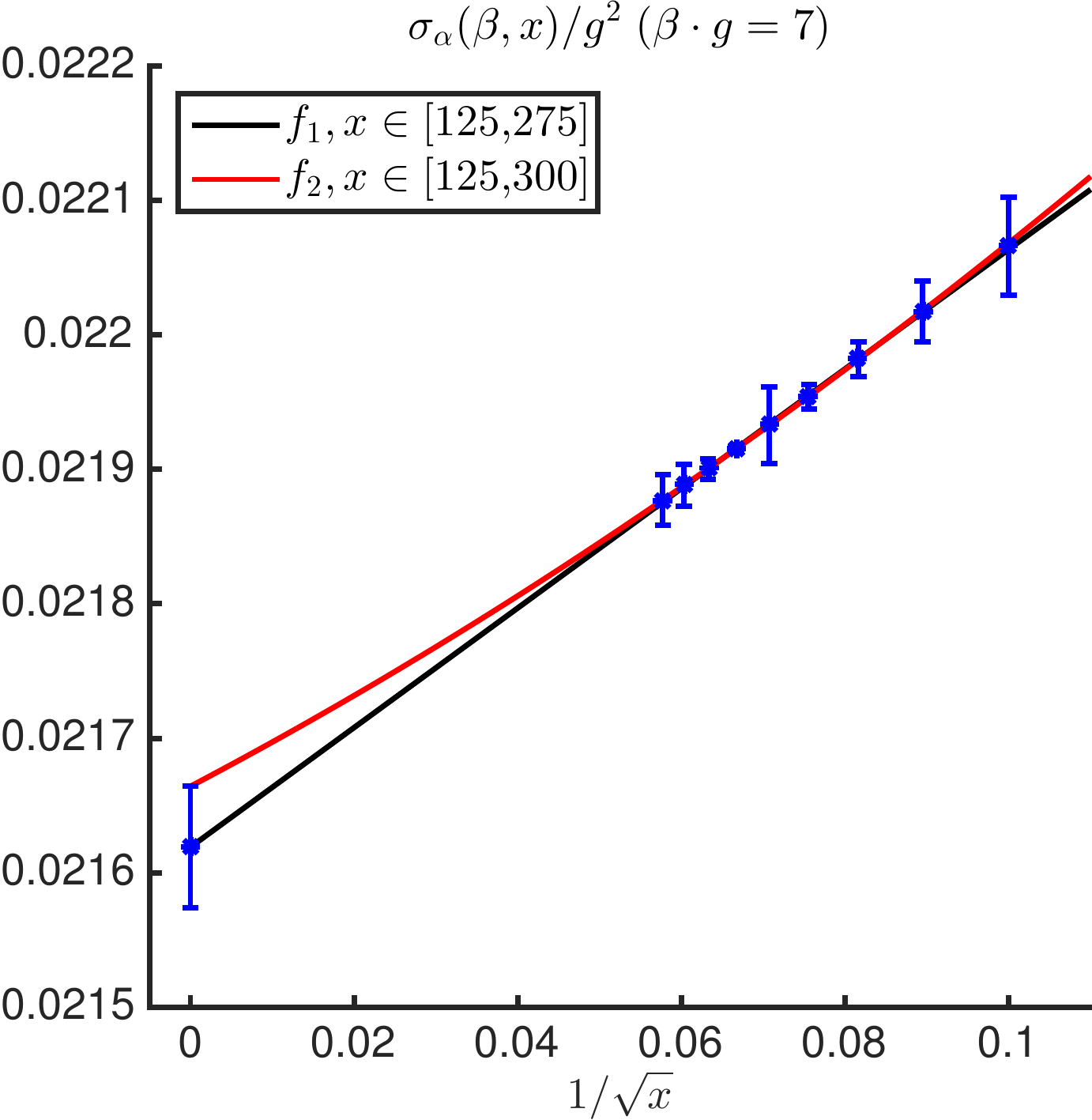}
\caption{\label{fig:continuumExtrapolationEFreea25e2d}}
\end{subfigure}\hfill\null
\vskip\baselineskip
\captionsetup{justification=raggedright}
\caption{\label{fig:continuumExtrapolationEFreea25e2} $m/g = 0.25,\alpha = 0.25$. Continuum extrapolation of $\sigma_{\alpha}(\beta,x)$ for different values of $\beta g$. The black line is the fit which determines the continuum estimate. The red line and blue line are the most appropriate fits through the data for the other fitting functions. The error bars on our data for $1/\sqrt{x} \neq 0$ represent the errors $\Delta^{(\epsilon,d\beta)}\sigma_\alpha(\beta,x)$ originating from taking nonzero $\epsilon$ and $d\beta$, as explained in subsection \ref{subsec:errFinDalphaneq0}. The error bar at $1/\sqrt{x} = 0$ is the estimated error on the continuum result.}
\end{figure}

To conclude this example we note that in figure \ref{figapp:CCm25a25e2ExtrCC} one observes that for $\beta g = 2$ and $\beta g = 7$ the results are quite robust against the choice of fitting function and the choice of fitting interval. In contrast, for smaller values of $\beta g$ cutoff effects in $x$ are larger. We also observe that for $\beta g = 7$ the error in $(\epsilon,d\beta)$ is larger than for $\beta g = 2.5$ which is a consequence of the accumulation of errors in $d\beta$ and $\epsilon$ during the evolution. 
\\
\\ \noindent \textbf{Example 2.} Let us now examine the continuum extrapolation of the string tension $\sigma_\alpha$ for the same values of $\beta g$ as in example 1: $\beta g = 0.5, 1.5, 2.5, 7$. The details can be found in table \ref{table:extrapolationmdivg25EFreeRen} and fig. \ref{fig:continuumExtrapolationEFreea25e2}. 

In subsection \ref{subsec:deconfinement} we argue that for high temperatures, small values of $\beta g$, the string tension decays exponentially with the temperature. Hence, for small values of $\beta g$ the string tension is very small. This in turn leads to relatively large systematic errors in $\epsilon$ and $d\beta$ for $\beta g = 0.5$, see fig. \ref{fig:continuumExtrapolationEFreea25e2a}. Given this fact it should not come as a surprise that only a few of the fits are statistically significant and have $\chi_\theta^2/N_{dof}^{\theta} \leq 1$. More specifically, we could only use 2 fits of our data against $f_1$ and $f_3$ and only 4 fits of our data against $f_2$. As for $\Delta\Sigma_\alpha (\beta)$, because there were less than 10 `good' fits, we chose the fit through the data points with the least error in $(\epsilon,d\beta)$. They are shown in fig. \ref{fig:continuumExtrapolationEFreea25e2a}. The most `good' fits were found for $f_2$ and therefore we took as our continuum estimate $\sigma_{\alpha}(\beta)/g = \sigma_{\alpha}^2(\beta)/g$. Here the error is obtained by taking the difference with the estimate $\sigma^{(3)}_{\alpha}(\beta)$ corresponding to $f_3$. Not quite unexpected this error is relatively large.

\begin{figure}[t]
\null\hfill
\begin{subfigure}[b]{.40\textwidth}
\includegraphics[width=\textwidth]{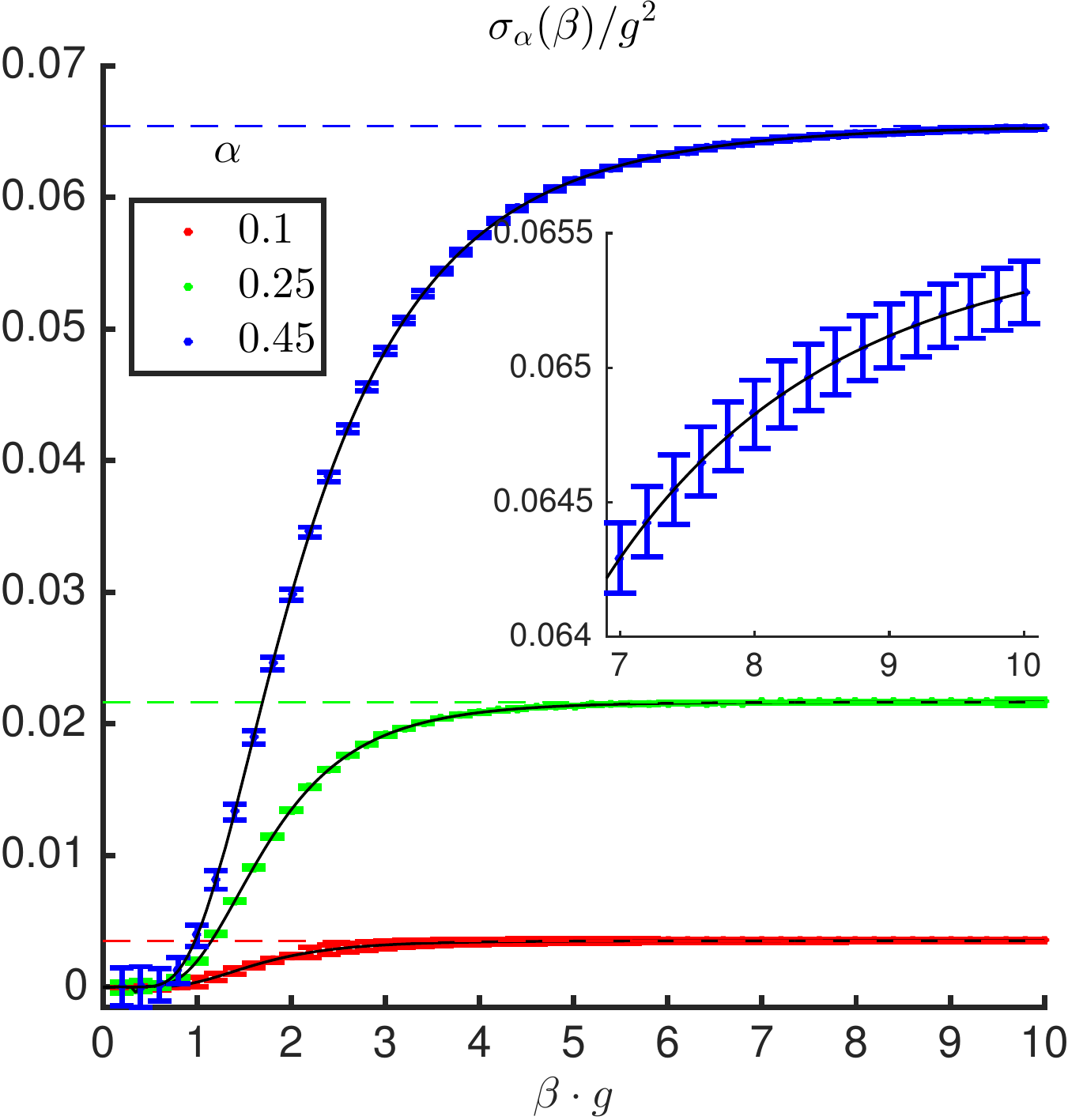}
\caption{\label{fig:continuumResmdivg25e2a}}
\end{subfigure}\hfill
\begin{subfigure}[b]{.40\textwidth}
\includegraphics[width=\textwidth]{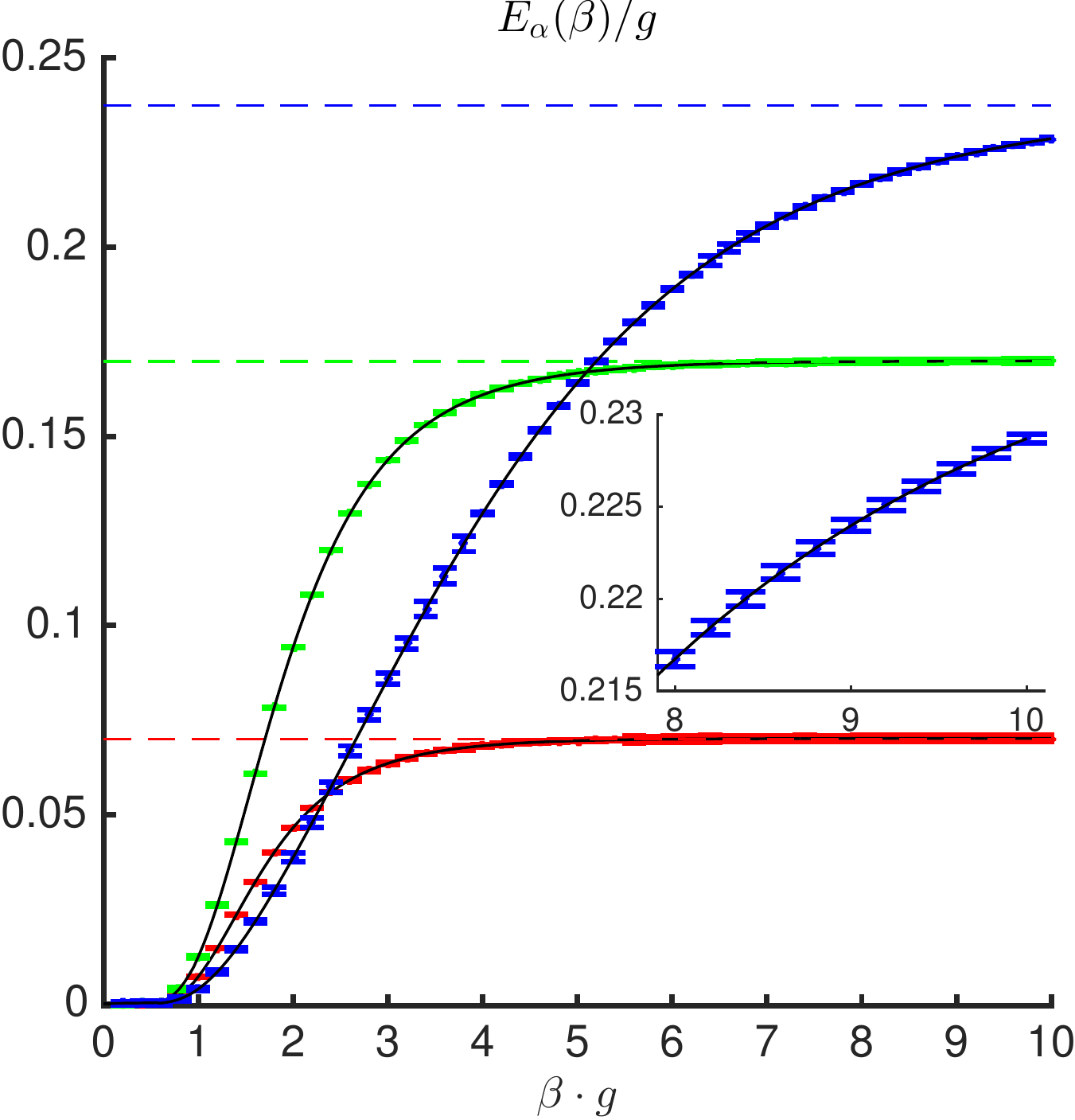}
\caption{\label{fig:continuumResmdivg25e2b}}
\end{subfigure}\hfill\null
\vskip\baselineskip
\null\hfill
\begin{subfigure}[b]{.40\textwidth}
\includegraphics[width=\textwidth]{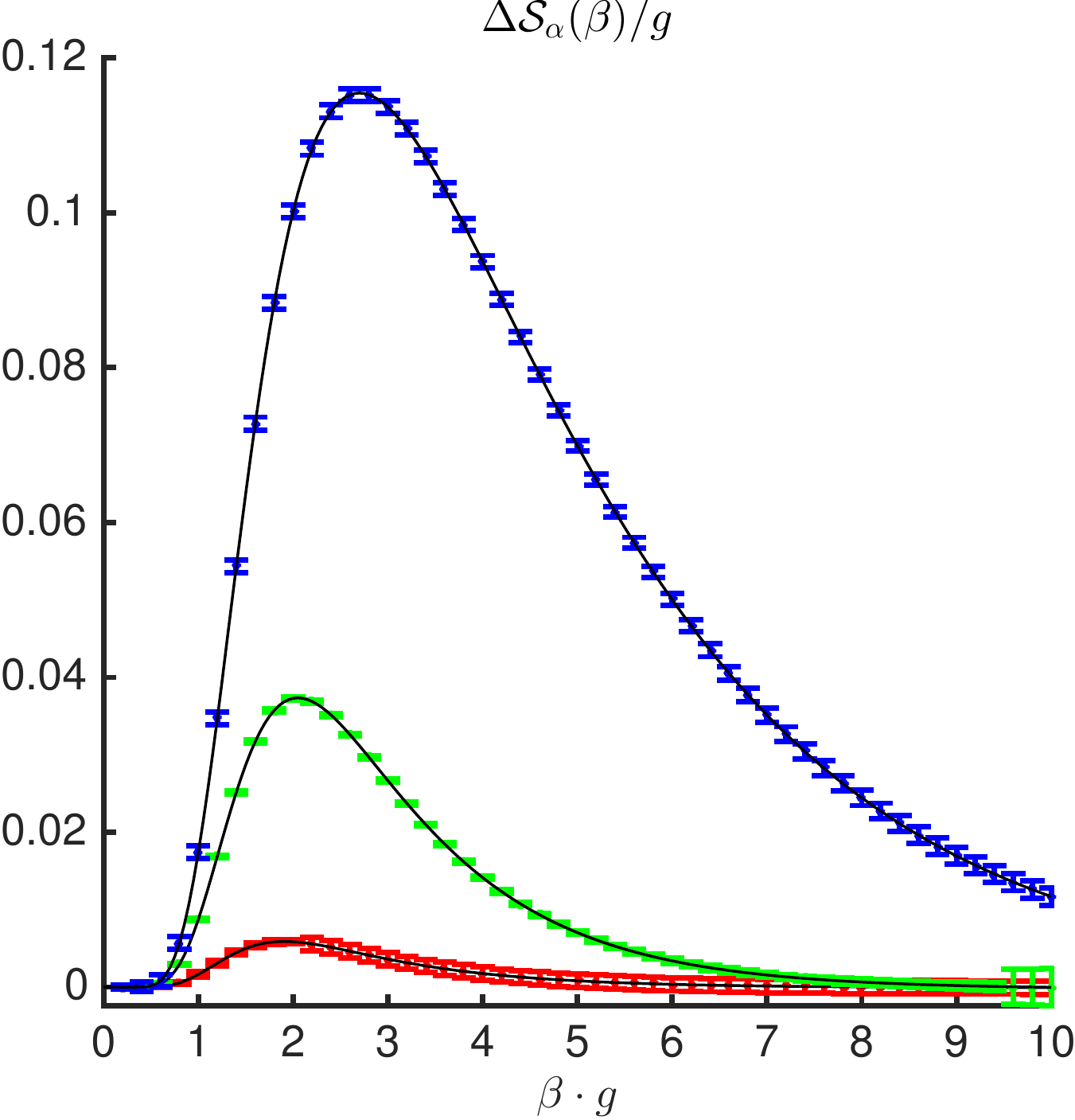}
\caption{\label{fig:continuumResmdivg25e2c}}
\end{subfigure}\hfill
\begin{subfigure}[b]{.40\textwidth}
\includegraphics[width=\textwidth]{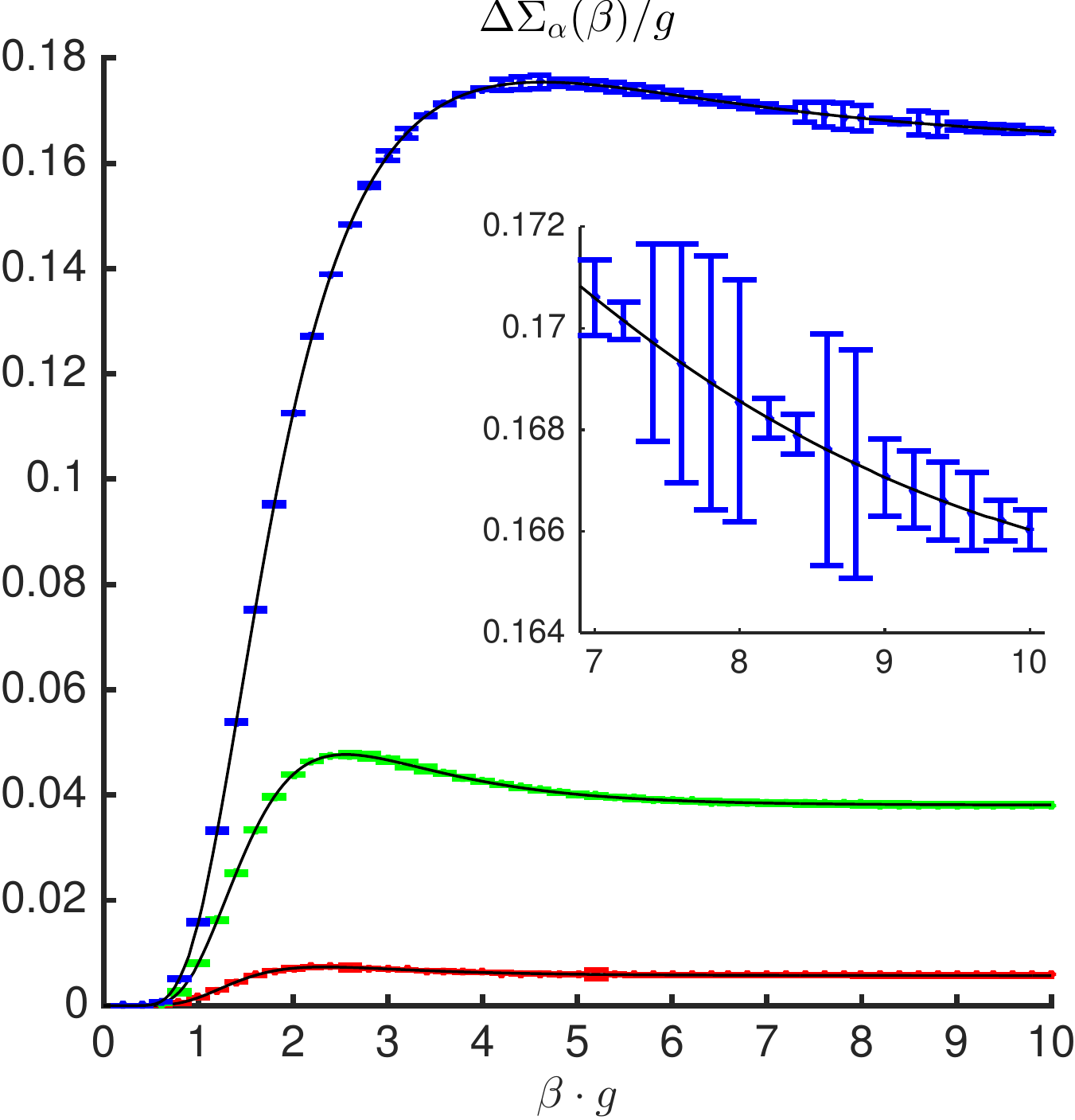}
\caption{\label{fig:continuumResmdivg25e2d}}
\end{subfigure}\hfill\null
\vskip\baselineskip
\captionsetup{justification=raggedright}
\caption{\label{fig:continuumResmdivg25e2} $m/g = 0.25$. Continuum results with error bars for $\alpha = 0.1$ (red), $\alpha = 0.25$ (green) and $\alpha = 0.45$ (blue). (a) String tension. The dashed lines are the results for $\beta g = +\infty$ obtained in \cite{Buyens2015}. (b) Electric field. (c) Renormalized entropy. (d) $\Delta\Sigma_\alpha(\beta)$.  }
\end{figure}

For the other values of $\beta g$ the errors in $(\epsilon,d\beta)$ are better under control and the data shows almost exactly linear scaling in $1/\sqrt{x}$, see figs. \ref{fig:continuumExtrapolationEFreea25e2}(b)-(d). This is also confirmed by the fact that only the fits $f_1$ have statistically significant coefficients and $\chi_\theta^2/N_{dof}^{\theta} \leq 1$ for $\beta g = 1.5$ and $\beta g = 2.5$, see table \ref{table:extrapolationmdivg25EFreeRen}. In this case, to still have an idea of the error of the fitting ansatz we compare the result obtained from the linear fit $f_1$ to the continuum result we would obtain by a quadratic fit through all our data (red line in figs. \ref{fig:continuumExtrapolationEFreea25e2b} and \ref{fig:continuumExtrapolationEFreea25e2c}). Note however that this does not increase the errors on our result: the largest uncertainty seems to originate from taking nonzero $(\epsilon, d\beta)$. For $\beta g = 7$ a similar picture holds, but now there are also quadratic fits with statistically significant fits with $\chi_\theta^2/N_{dof}^{\theta}$. From the two sets of estimates for $\sigma_\alpha(\beta)$ (each of the sets corresponding to $f_n, n = 1,2$) we estimate the $\sigma_\alpha(\beta)$ and the error as before. 
\\
\begin{figure}[t]
\null\hfill
\begin{subfigure}[b]{.40\textwidth}
\includegraphics[width=\textwidth]{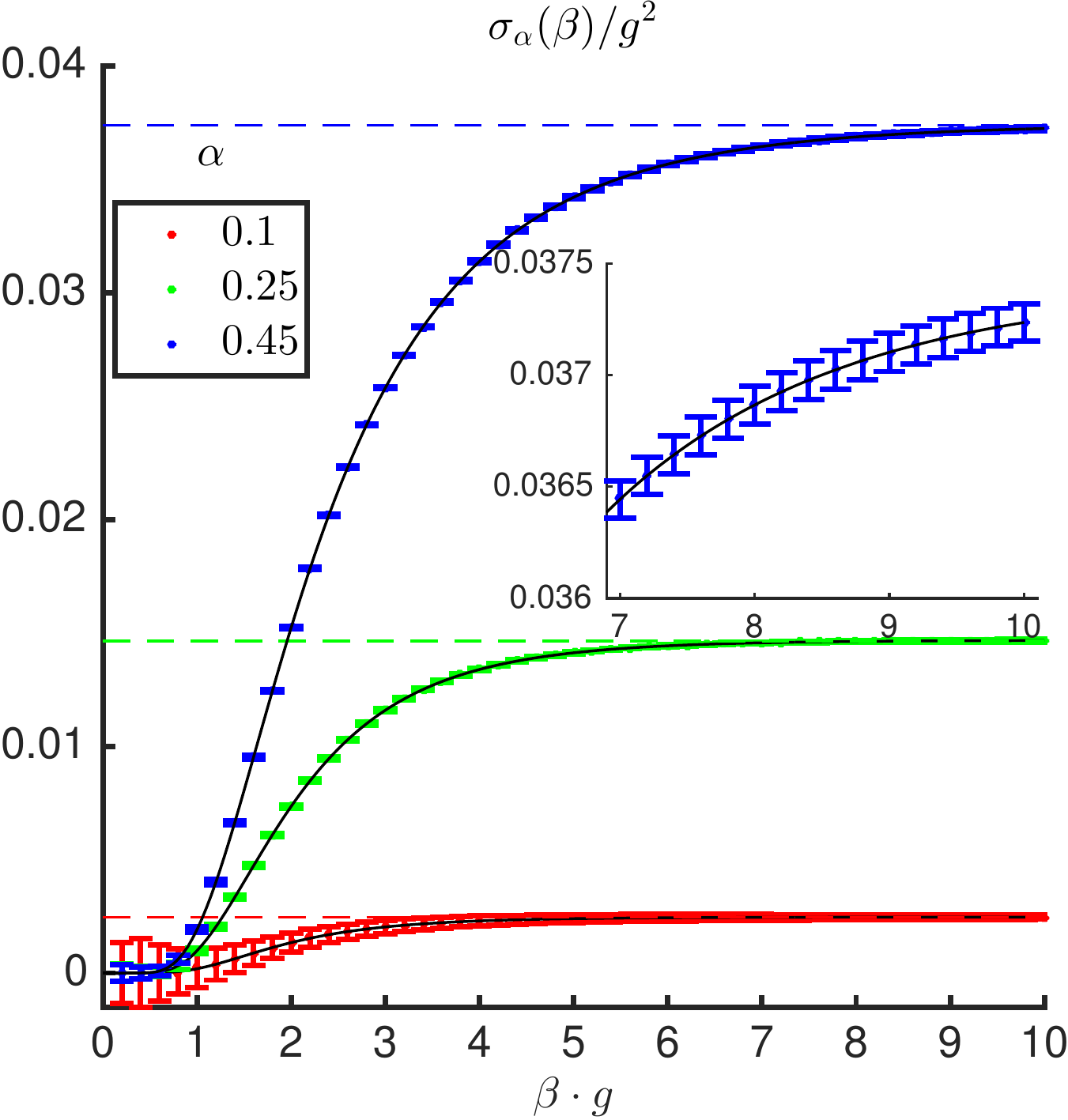}
\caption{\label{fig:continuumResmdivg125e3a}}
\end{subfigure}\hfill
\begin{subfigure}[b]{.40\textwidth}
\includegraphics[width=\textwidth]{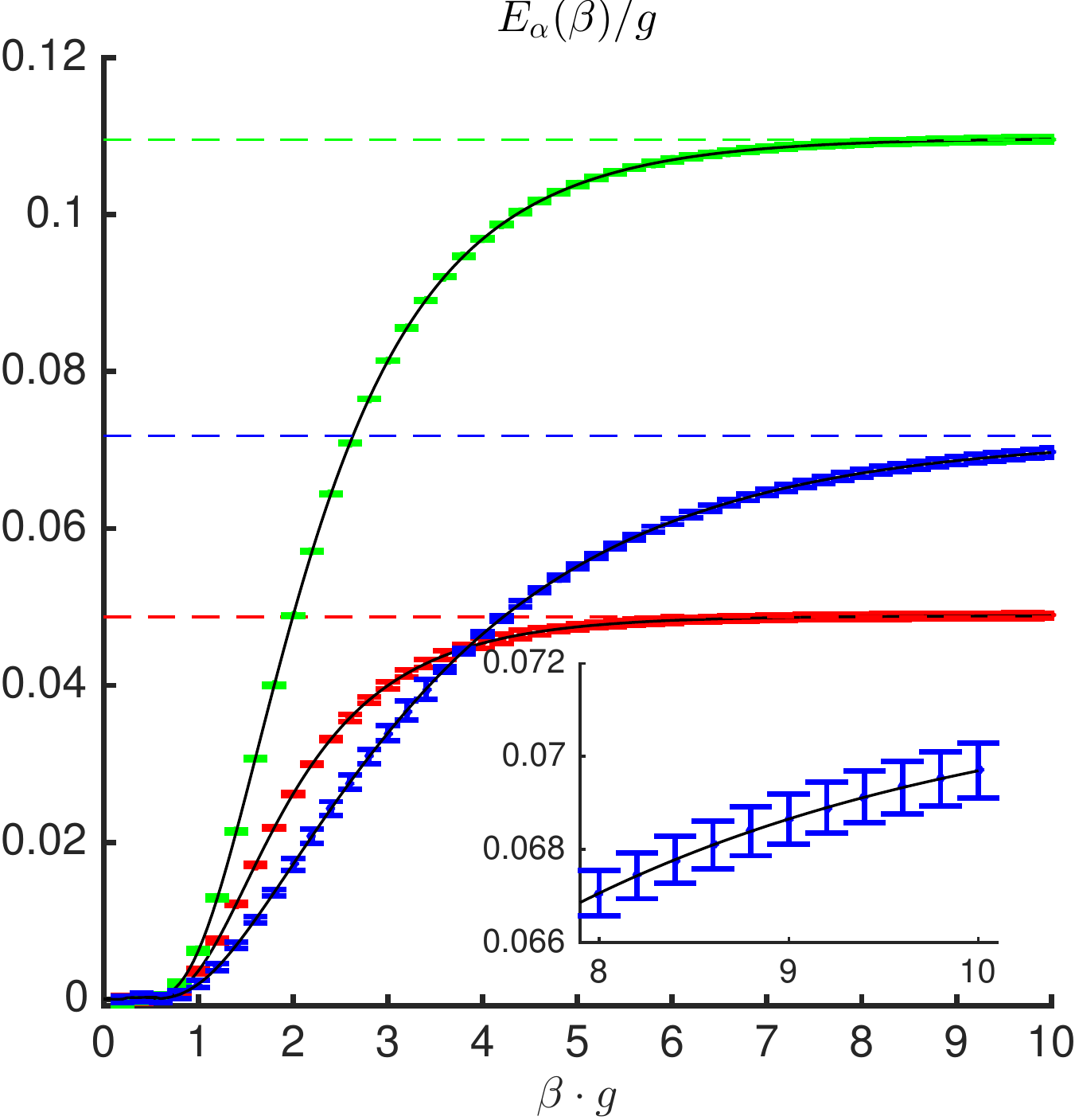}
\caption{\label{fig:continuumResmdivg125e3b}}
\end{subfigure}\hfill\null
\vskip\baselineskip
\null\hfill
\begin{subfigure}[b]{.40\textwidth}
\includegraphics[width=\textwidth]{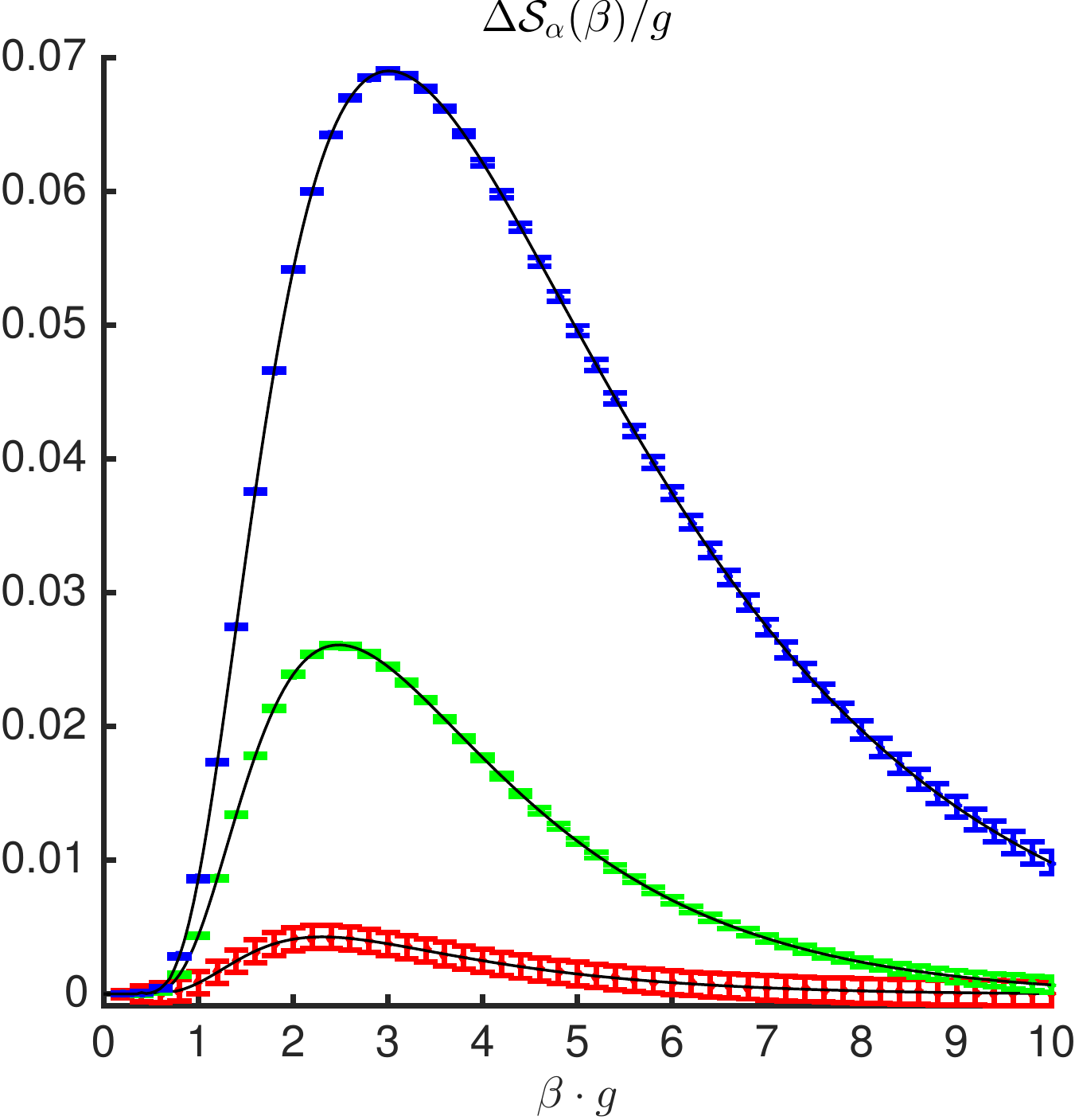}
\caption{\label{fig:continuumResmdivg125e3c}}
\end{subfigure}\hfill
\begin{subfigure}[b]{.40\textwidth}
\includegraphics[width=\textwidth]{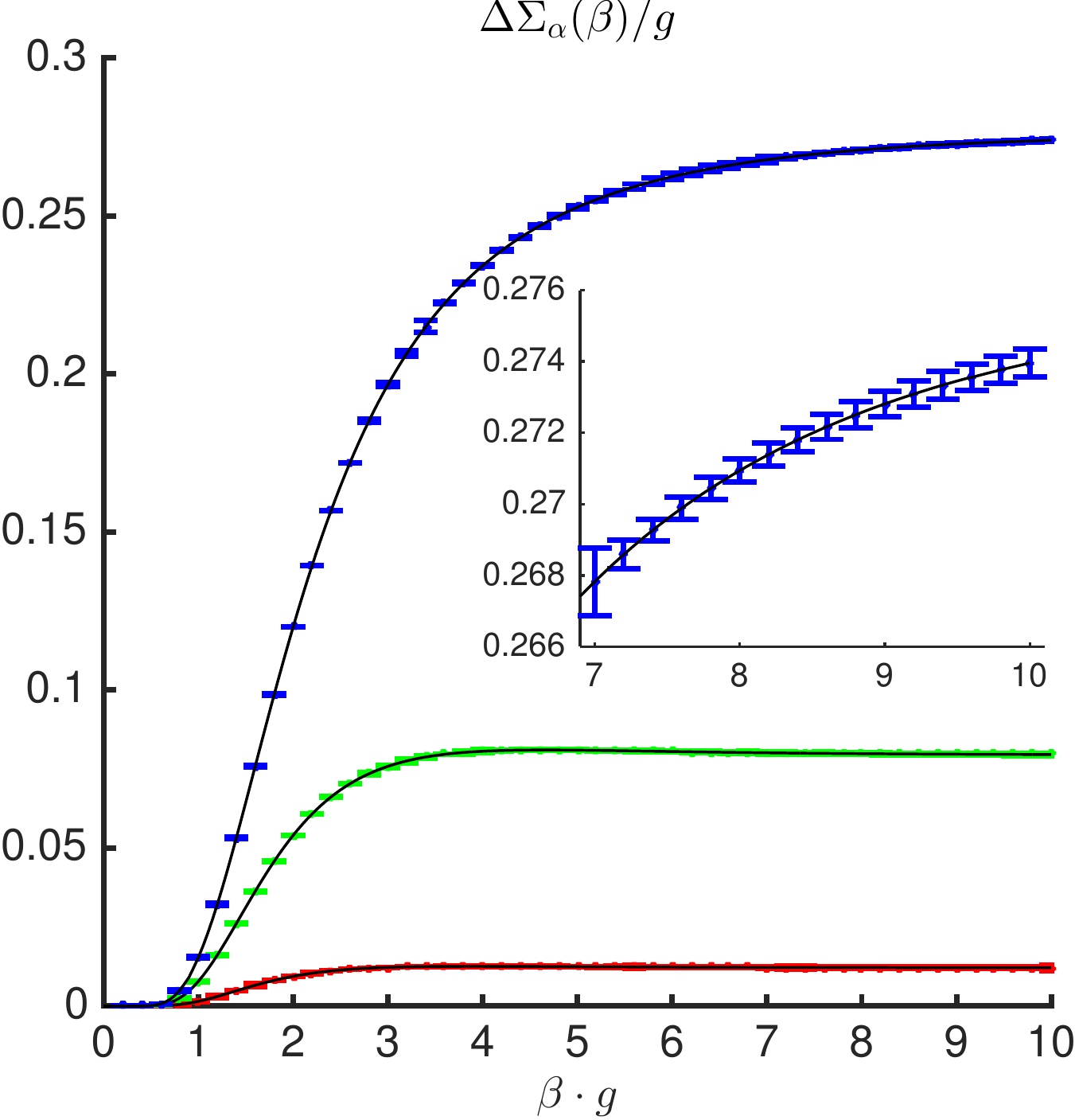}
\caption{\label{fig:continuumResmdivg125e3d}}
\end{subfigure}\hfill\null
\vskip\baselineskip
\captionsetup{justification=raggedright}
\caption{\label{fig:continuumResmdivg125e3} $m/g = 0.125$. Continuum results with error bars for $\alpha = 0.1$ (red), $\alpha = 0.25$ (green) and $\alpha = 0.45$ (blue). (a) String tension. The dashed lines are the results for $\beta g = +\infty$ obtained in \cite{Buyens2015}. (b) Electric field. (c) Renormalized entropy. (d) $\Delta\Sigma_\alpha(\beta)$. }
\end{figure}
\\
\noindent \textbf{Results.} In figs. \ref{fig:continuumResmdivg125e3} and \ref{fig:continuumResmdivg25e2} we show the results for $m/g = 0.125$ (fig. \ref{fig:continuumResmdivg125e3}), $0.25$ (fig. \ref{fig:continuumResmdivg25e2}) and $\alpha = 0.1, 0.25, 0.45$ in the continuum limit. We plot the string tension (a), the electric field (b), the renormalized entropy per unit of length (c) and the renormalized chiral condensate (d). The entropy $\mathcal{S}_\alpha$ is obtained from the free energy per unit of length $\mathcal{F}_{\alpha}(\beta)$ and the average energy per unit of length $\mathcal{E}_{\alpha}(\beta)$ via the relation
$$\mathcal{S}_{\alpha}(\beta) = -\beta\Bigl(\sigma_{\alpha}(\beta) - \mathcal{E}_{\alpha}(\beta)\Bigl). $$
We computed the quantities for $\beta g \in [0,10]$ with steps $d\beta = 0.05$. For convenience we only show the error bars with steps $d\beta  = 0.2$. One observes that the errors on the continuum extrapolation are small enough. Only for very small values of $\beta g$, the deconfinement at $T \rightarrow + \infty$ (subsection \ref{subsec:deconfinement}) cannot precisely be described for the simulated values: it suggests that the quantities are zero for $\beta g \lesssim 0.5$, rather than it displays exponential decay with temperature as $T\rightarrow + \infty$. 

The dashed lines in figs. \ref{fig:continuumResmdivg125e3c}, \ref{fig:continuumResmdivg125e3d}, \ref{fig:continuumResmdivg25e2c} and \ref{fig:continuumResmdivg125e3d} are the values at $\beta g = + \infty$ computed in \cite{Buyens2015}. For $\alpha = 0.1$ and $\alpha = 0.25$ we are at $\beta g = 10$ effectively at zero temperature while for $\alpha = 0.45$ there are still thermal fluctuations left. This correlates with the renormalized entropy $\Delta\mathcal{S}_\alpha$ which is obviously non-zero for $\beta g \approx 10$ and $\alpha = 0.45$ and which is very close to zero for $\beta g \gtrsim 10$ and $\alpha = 0.1, 0.25$, see figs. \ref{fig:continuumResmdivg125e3c} and \ref{fig:continuumResmdivg25e2c}. This shows that the renormalized entropy is a good quantity to measure thermal fluctuations in a Gibbs state.  

\begin{figure}[t]
\null\hfill
\begin{subfigure}[b]{.40\textwidth}
\includegraphics[width=\textwidth]{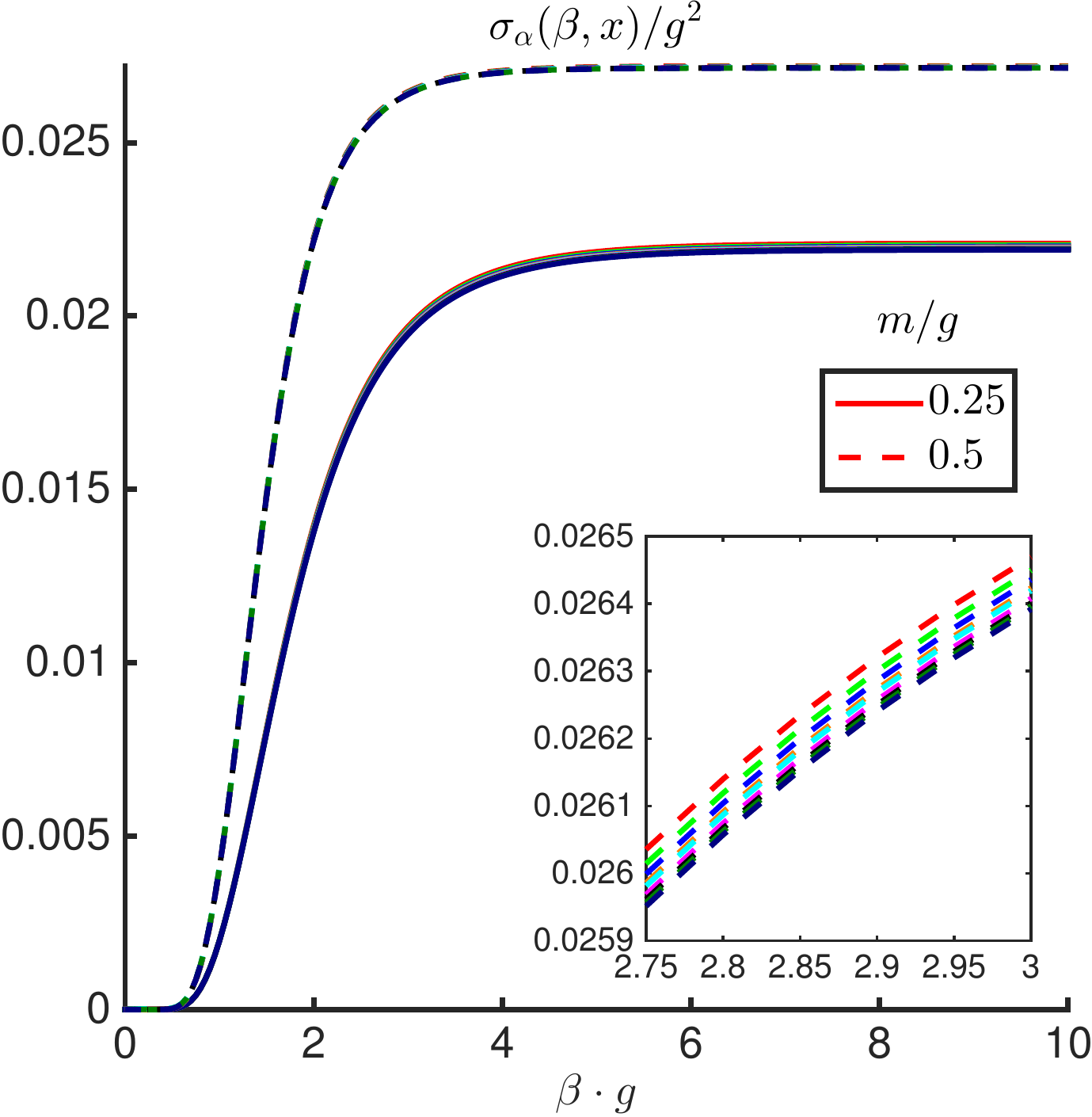}
\caption{\label{fig:varDiffxa}}
\end{subfigure}\hfill
\begin{subfigure}[b]{.40\textwidth}
\includegraphics[width=\textwidth]{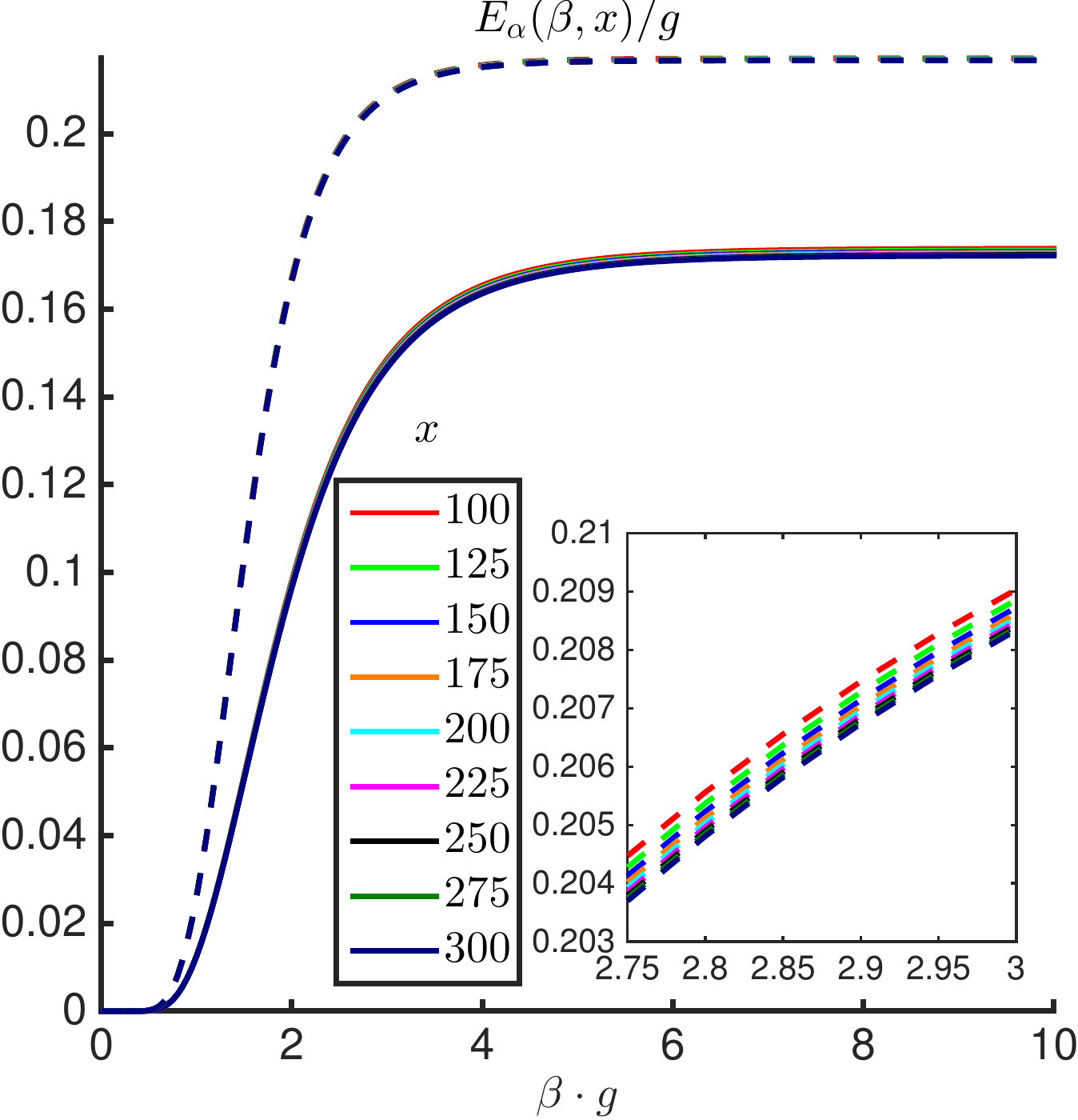}
\caption{\label{fig:varDiffxb}}
\end{subfigure}\hfill\null
\vskip\baselineskip
\null\hfill
\begin{subfigure}[b]{.40\textwidth}
\includegraphics[width=\textwidth]{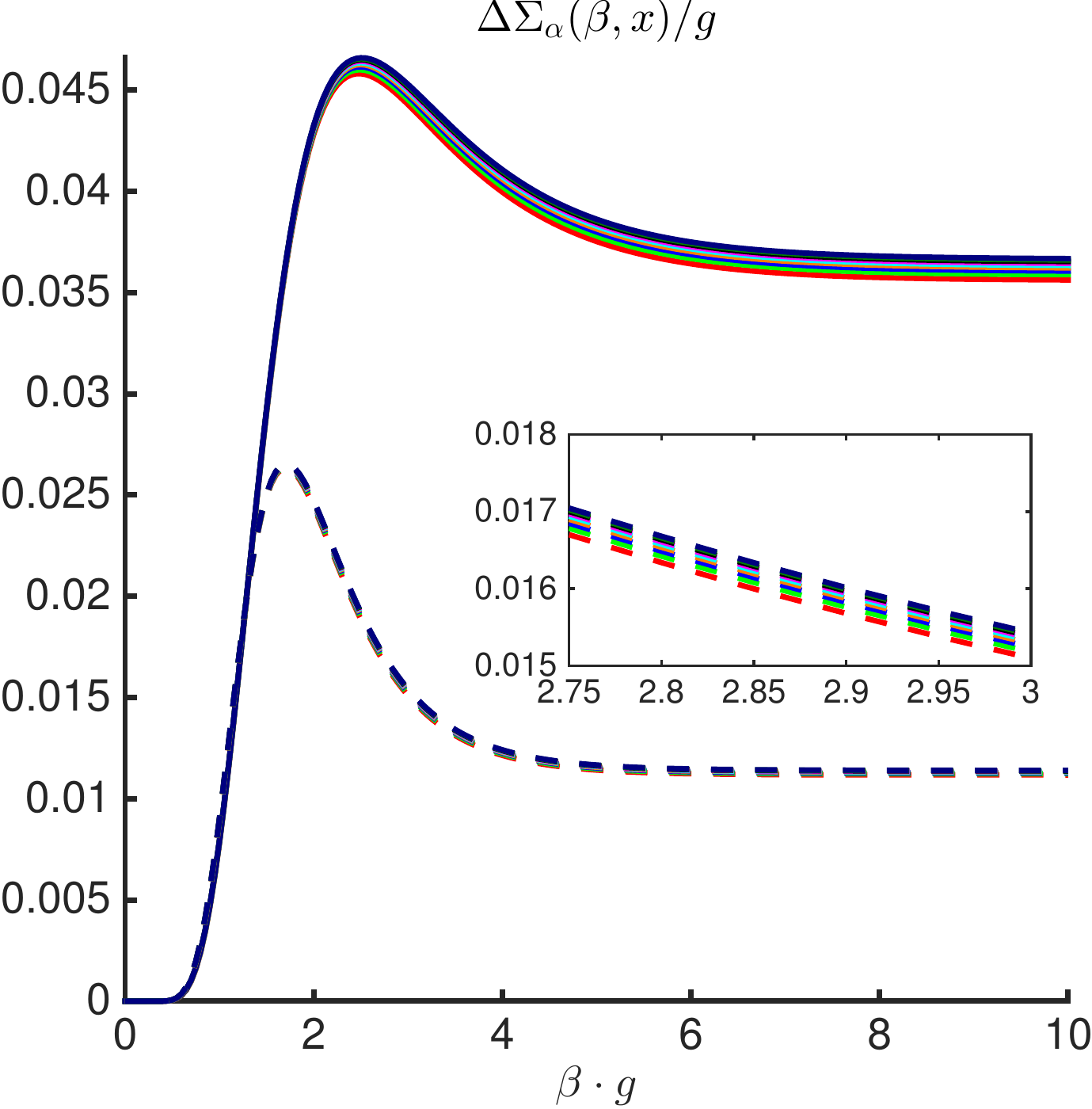}
\caption{\label{fig:varDiffxc}}
\end{subfigure}\hfill
\begin{subfigure}[b]{.40\textwidth}
\includegraphics[width=\textwidth]{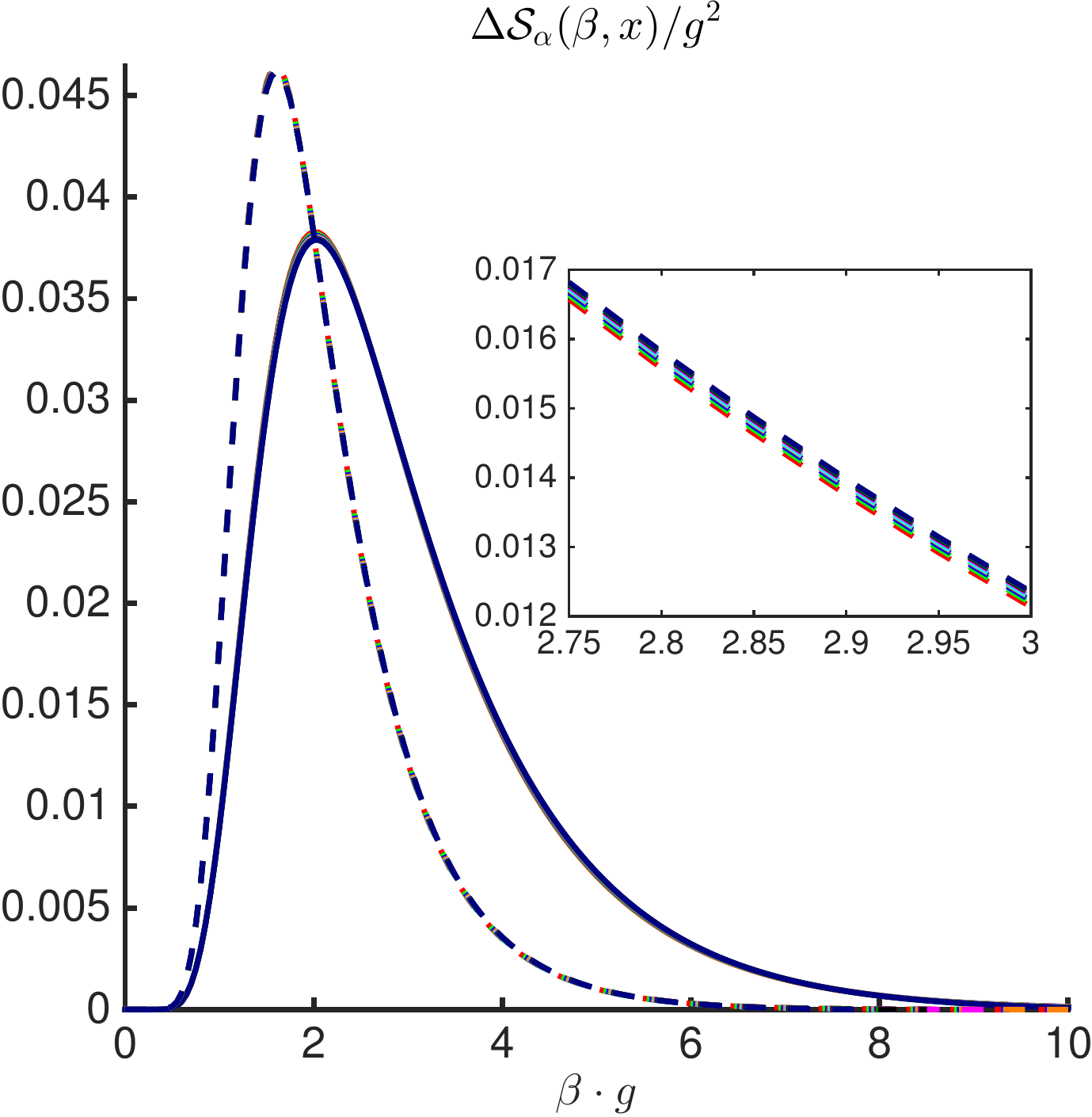}
\caption{\label{fig:varDiffxd}}
\end{subfigure}\hfill\null
\vskip\baselineskip
\captionsetup{justification=raggedright}
\caption{\label{fig:varDiffx}$\alpha = 0.25$. $m/g = 0.25$ (full line) and $m/g = 0.5$ (dashed line). Quantities for  $x = 100,125,150,\ldots, 300$. (a) String tension. (b) Electric field. (c) Renormalized chiral condensate. (d) Entropy.}
\end{figure}

Note also that already at $x = 100$ our results are differ by $10 \%$ or less from their continuum value. For $m/g \gtrsim 0.5$ the results for different $x-$values are even closer to each other, see fig. \ref{fig:varDiffx}. This suggests that although we limit our analysis section \ref{sec:asCon} to $x = 100$, we can be confident to be close to the continuum limit.

\section{Thermal corrections in the weak coupling limit}\label{sec:thermalwc}
\noindent The Lagrangian for the Schwinger model is:
$$ \mathcal{L} = \bar{\psi}\left(\gamma^\mu(i\partial_\mu+g A_\mu) - m\right) \psi - \frac{1}{4}F_{\mu\nu}F^{\mu\nu}\,. $$
In the weak coupling limit ($m/g \gg 1$) with an electric background field $g\alpha$ Coleman \cite{Coleman1976} considered the Hamiltonian for this Lagrangian in the semi-classical approximation and where he restricted to the two-particle subspace:
$$ H_{\alpha} \approx \int_{-\infty}^{+\infty} dp \; 2\sqrt{p^2 + m^2} + \int_{-\infty}^{+\infty} dx\; \frac{g^2}{2}\left(\vert x \vert  -2\alpha x\right) + \mathcal{O}\left(\hbar,\frac{g}{m}\right), [x,p] = i.$$

 The first term is the total energy of a fermion-antifermion pair. The second term gives the energy due to the separation of the fermion and the antifermion and yields an infinite number of bound states. Within this semi-classical approximation Coleman then argued that the number of two-particle states with energy smaller than energy $\mathcal{E}$ is \cite{Coleman1976}
$$ N(\mathcal{E}) \approx \frac{\mathcal{E}^2}{g^2\pi (1 - 4\alpha^2)}\Theta(\mathcal{E} - 2m) + \mathcal{O}\left(\frac{g}{m}\right),$$
where $\Theta$ is the Heaviside-function: $\Theta(x) = 1$ if $x > 0$ and $\Theta(x) = 0$ if $x < 0$. Therefore, thermal fluctuations to the ground state are only relevant if 
$$\int_{2m}^{+\infty} d\mathcal{E} \frac{dN}{d\mathcal{E}}(\mathcal{E}) \; e^{-\beta \mathcal{E}}  \sim C_m$$
for some constant $C_m$ of order $1$ depending on $m$ but not on $\beta$. Hence, for a large fixed value of $\beta g$ we will only observe significant thermal fluctuations to ground-state expectation values if $\delta \lesssim K_m e^{-2\beta m }/\beta$ on, with $K_m$ some positive constant which depends on $m$ but is independent of $\beta$. 

\end{document}